\documentclass[rmp,twocolumn,showpacs,bm,amssymb,amsmath,floatfix]{revtex4}
\usepackage{epsfig}
\begin{document}
\title{Exact solution of the Falicov-Kimball model with dynamical mean-field 
theory}
\date{\today}
\author{J.~K.~Freericks}
\email{freericks@physics.georgetown.edu}
\homepage{http://www.physics.georgetown.edu/~jkf}
\affiliation{Department of Physics, Georgetown University, Washington, DC
20057}
\author{V.~Zlati\' c}
\email{zlatic@ifs.hr}
\affiliation{Institute of Physics, Zagreb, Croatia}

\begin{abstract}
The Falicov-Kimball model was introduced in 1969 as a statistical model for
metal-insulator transitions; it includes itinerant and localized electrons that 
mutually interact with a local Coulomb interaction and is the simplest model
of electron correlations.  It can be
solved exactly with dynamical mean-field theory in the limit of large spatial
dimensions which provides an interesting benchmark for the
physics of locally correlated systems.  In this review, we develop the
formalism for solving the Falicov-Kimball model from a path-integral perspective,
and provide a number of expressions for single and two-particle properties.  We
examine many important theoretical results that show
the absence of fermi-liquid features and provide 
a detailed description of the static and dynamic correlation 
functions and of transport properties. The parameter space is rich and one 
finds a variety of many-body features like metal-insulator transitions, 
classical valence fluctuating transitions,  metamagnetic transitions, 
charge density wave order-disorder transitions, and phase separation.  
At the same time, a number of experimental systems have been
discovered that show anomalies related to Falicov-Kimball
physics [including YbInCu$_4$,
EuNi$_2$(Si$_{1-x}$Ge$_x$)$_2$, NiI$_2$ and Ta$_x$N]. 

\end{abstract}
\maketitle

\tableofcontents

\section{Introduction}

\subsection{Brief History}

The Falicov-Kimball model~\cite{falicov_kimball_1969} was introduced in
1969 to describe metal-insulator transitions in a number of rare-earth
and transition-metal compounds [but see Hubbard's earlier 
work~\cite{hubbard_I_1963}
where the spinless version of the Falicov-Kimball
model was introduced as an approximate solution to the Hubbard model;
one assumes that one species of spin does not hop and is frozen on the
lattice].
The initial work by Falicov and collaborators focused primarily on analyzing
the thermodynamics of the metal-insulator transition with a static mean-field theory
approach~\cite{falicov_kimball_1969,ramirez_falicov_kimball_1970}.  
The resulting solutions displayed both continuous and discontinuous
metal-insulator phase transitions, and they could fit the conductivity of
a wide variety of transition-metal and rare-earth compounds with their
results. Next,~\textcite{ramirez_falicov_1971}
applied the model to describe the $\alpha-\gamma$ phase transition in Cerium.
Again, a number of thermodynamic quantities were approximated well by the 
model, but it did not examine any effects associated with Kondo screening
of the $f$-electrons [and subsequently was discarded in favor of the Kondo
volume collapse picture~\cite{allen_martin_1982}].  

Interest in the model waned once~\textcite{plischke_1972} showed that 
when the coherent-potential-approximation 
(CPA)~\cite{soven_1967,velicky_kirkpatrick_ehrenreich_1968} 
was applied to it, all first-order phase transitions disappeared and the 
solutions only displayed smooth crossovers from a metal to an insulator [this claim
was strongly refuted by Falicov's group~\cite{dasilva_falicov_1972} but interest
in the model was limited for almost 15 years].

The field was revitalized by mathematical physicists in the mid 1980's, who
realized that the spinless version of this model is the simplest correlated 
system that displays long-range order at low temperatures and for dimensions
greater than one.  Indeed, two groups produced independent proofs of the
long-range 
order~\cite{brandt_schmidt_1986,brandt_schmidt_1987,kennedy_lieb_1986,lieb_1986}.
The work by Kennedy and Lieb rediscovered Hubbard's original approximation
that yields the spinless Falicov-Kimball model, and also provided a new
interpretation of the model for the physics of crystallization. A number of 
other exact results followed including (i) a proof of no quantum-mechanical
mixed valence (or spontaneous hybridization) at finite 
$T$~\cite{subrahmanyam_barma_1988} [based on the presence of a local gauge symmetry
and Elitzur's theorem~\cite{elitzur_1975}]; (ii) proofs of phase separation
and of periodic ordering in one dimension (with large interaction 
strength)~\cite{lemberger_1992}; (iii) proofs about ground-state properties
in two dimensions (also at large interaction 
strength)~\cite{kennedy_1994,kennedy_1998,haller_2000,haller_kennedy_2001};
(iv) a proof of phase separation in one dimension and small interaction
strength~\cite{freericks_gruber_macris_1996}; and (v) a proof of 
phase separation for large interaction strength and all 
dimensions~\cite{freericks_lieb_ueltschi_2002a,freericks_lieb_ueltschi_2002b}.
Most of these rigorous results have already been summarized in 
reviews~\cite{gruber_macris_1996,gruber_1999}. In addition, a series of 
numerical calculations were performed in one and two dimensions~\cite{%
freericks_falicov_1990,gruber_ueltschi_jedrezejewski_1994,watson_lemanski_1995,%
devries_michielsen_deraedt_1993,michielsen_1993,devries_michielsen_deraedt_1994}.
While not providing 
complete results for the model, the numerics do illustrate a number of
important trends in the physics of the FK model.

At about the same time, there was a parallel development of the dynamical
mean-field theory (DMFT), which is what we concentrate on in this review.  The
DMFT was invented by~\textcite{metzner_vollhardt_1989}.
Almost immediately after the idea that in large spatial dimensions the 
self energy becomes local, Brandt and collaborators showed how to solve the
static problem exactly (requiring no quantum Monte Carlo), thereby providing the
exact solution of the Falicov-Kimball 
model~\cite{brandt_mielsch_1989,brandt_mielsch_1990,brandt_fledderjohann_1990,%
brandt_mielsch_1991,brandt_fledderjohann_1992,brandt_urbanek_1992}.  This work
is the extension of Onsager's famous solution for the transition temperature
of the two-dimensional Ising model to the fermionic case (and large dimensions).
These series of papers revolutionized
Falicov-Kimball model physics and provided the only exact quantitative results for 
electronic phase transitions in the thermodynamic limit for all values of
the interaction strength.   They showed how to solve the infinite-dimensional
DMFT model, illustrated how to determine the order-disorder transition
temperature for a checkerboard (and incommensurate) charge-density-wave phase, 
showed how to find the free energy (including a first study of phase
separation), examined properties of the spin-one-half model, and calculated
the $f$-particle spectral function.

Further work concentrated on static properties such as charge-density-wave
order~\cite{vandongen_vollhardt_1990,vandongen_1991,vandongen_1992,freericks_1993a,%
freericks_1993b,gruber_macris_royer_2001} and 
phase separation~\cite{letfulov_1999,freericks_gruber_macris_1999,%
freericks_lemanski_2000}.  
The original Falicov-Kimball problem of the
metal-insulator transition~\cite{chung_freericks_1998}
was solved, as was the problem of classical intermediate
valence~\cite{chung_freericks_2000} [both using the spin-one-half 
generalization~\cite{brandt_fledderjohann_1990,freericks_zlatic_1998}]. 
The ``Mott-like'' metal-insulator transition~\cite{vandongen_leinung_1997}
and the non-fermi-liquid behavior~\cite{si_kotliar_1992}
were also investigated.
Dynamical properties and transport have been determined ranging from
the charge susceptibility~\cite{shvaika_2000,freericks_miller_2000,%
shvaika_2002}, to the optical 
conductivity~\cite{moeller_ruckenstein_schmittrink_1992},
to the Raman response~\cite{freericks_devereaux_2001a,%
freericks_devereaux_2001b,freericks_devereaux_2001c,%
devereaux_mccormack_freericks_2003}, to an evaluation of the
$f$-spectral function~\cite{si_kotliar_1992,brandt_urbanek_1992,%
zlatic_review_2001}.  
Finally, the static susceptibility for spontaneous
polarization was also determined~\cite{subrahmanyam_barma_1988,%
si_kotliar_1992,portengen_oestreich_1996a,portengen_oestreich_1996b,%
zlatic_review_2001}.

These solutions have allowed the FK model to be applied to a number
of different experimental systems ranging from valence-change-transition
materials~\cite{zlatic_freericks_2001a,zlatic_freericks_2001b}
like YbInCu$_4$ and EuNi$_2$(Si$_{1-x}$Ge$_x$)$_2$ to materials that can be doped
through a metal-insulator transition like Ta$_x$N [used as a barrier in Josephson
junctions~\cite{miller_freericks_2001,freericks_nikolic_miller_2001a,%
freericks_nikolic_miller_2002}], to Raman scattering in materials on
the insulating side of the metal-insulator 
transition~\cite{freericks_devereaux_2001b} like FeSi or SmB$_6$. The model,
and some straightforward modifications appropriate for double-exchange,
have been used to describe the CMR materials~\cite{allub_alascio_1996,%
allub_alascio_1997,letfulov_freericks_2001,ramakrishnan_2002}.

Generalizations of the FK model to the static Holstein model were first
carried out by~\textcite{millis_littlewood_shraiman_1995,%
millis_mueller_shraiman_1996} and also applied to the CMR materials.  Later 
more fundamental properties relating to the transition temperature
for the harmonic \cite{ciuchi_depasquale_1999,blawid_millis_2000} and anharmonic 
cases \cite{freericks_zlatic_jarrell_2000} and relating to the gap ratio
for the harmonic \cite{blawid_millis_2001} and anharmonic
cases \cite{freericks_zlatic_2001a} were 
completed.  Modifications to examine
diluted magnetic semiconductors have also 
appeared~\cite{chattopadhyay_dassarma_millis_2001,hwang_millis_dassarma_2002}.
A new approach to DMFT, which allows the correlated hopping Falicov-Kimball
model to be solved has also been completed recently~\cite{schiller_1999,%
shvaika_2002b}.

\subsection{Hamiltonian and Symmetries}

The Falicov-Kimball model is the simplest model of correlated electrons.
The original version~\cite{falicov_kimball_1969} involved
spin-one-half-electrons.  Here, we will generalize to the case of an
arbitrary degeneracy of the itinerant and localized electrons.  The
general Hamiltonian is then
\begin{eqnarray}
\mathcal{H}&=&-t\sum_{\langle ij\rangle}\sum_{\sigma=1}^{2s+1}c^{\dagger}_{i\sigma}
c_{j\sigma}+\sum_i\sum_{\eta=1}^{2S+1}E_{f\eta}f^{\dagger}_{i\eta}f_{i\eta}\cr
&+&U\sum_i\sum_{\sigma=1}^{2s+1}\sum_{\eta=1}^{2S+1}c^{\dagger}_{i\sigma}
c_{i\sigma}f^{\dagger}_{i\eta}f_{i\eta}\cr
&+&\sum_i\sum_{\eta\eta^\prime=1}^{2S+1}
U^{ff}_{\eta\eta^\prime}f^{\dagger}_{i\eta}f_{i\eta}f^\dagger_{i\eta^\prime}
f_{i\eta^\prime}\cr
&-&g\mu_B H \sum_i\sum_{\sigma=1}^{2s+1}m_\sigma c^\dagger_{i\sigma}c_{i\sigma}
\cr
&-&g_f\mu_B H \sum_i\sum_{\eta=1}^{2S+1}m_\eta f^\dagger_{i\eta}f_{i\eta}.
\label{eq: hamiltonian}
\end{eqnarray}
The symbols $c^\dagger_{i\sigma}$ ($c_{i\sigma}$) denote the itinerant
electron creation (annihilation) operators at site $i$ in state $\sigma$
(the index $\sigma$ takes $2s+1$ values).  Similarly, the symbols
$f^\dagger_{i\eta}$ ($f_{i\eta}$) denote the localized
electron creation (annihilation) operators at site $i$ in state $\eta$
(the index $\eta$ takes $2S+1$ values). Customarily, we identify the index
$\sigma$ and $\eta$ with the $z$-component of spin, but the index could
denote other quantum numbers in more general cases.
The first term is the kinetic energy (hopping) of the conduction electrons
(with $t$ denoting the nearest-neighbor hopping integral); the summation is over
nearest neighbor sites $i$ and $j$ (we count each pair twice to guarantee
hermiticity).  The second term is the localized electron site energy
(which we allow to depend on the index $\eta$ to include crystal-field
effects [without spin-orbit coupling for simplicity]); in most applications
the site-energy is taken to be $\eta$-independent.  The third term is the
Falicov-Kimball interaction term (of strength $U$) which represents the 
local Coulomb interaction when itinerant and localized electrons occupy
the same lattice site (we could make $U$ depend on $\sigma$ or $\eta$, but this
complicates the formulas and is not normally
needed).  The fourth term is the $ff$ Coulomb interaction
energy  of strength $U^{ff}_{\eta\eta^\prime}$
(which can be chosen to depend on $\eta$ if desired; the term with
$\eta=\eta^\prime$ is unnecessary and can be absorbed into $E_{f\eta}$).  
Finally, the fifth and sixth terms represent the magnetic energy due to the
interaction with an external
magnetic field $H$, with $\mu_B$ the Bohr magneton, $g$ ($g_f$) the
respective Land\'e g-factors, and $m_\sigma$ ($m_\eta$) the $z$-component of
spin for the respective states. Chemical potentials $\mu$ and $\mu_f$
are employed to adjust the itinerant and localized electron concentrations,
respectively (in cases where the localized particle is
fixed independently of the itinerant electron concentration, the
localized particle chemical potential $\mu_f$ can be absorbed into 
the site-energy $E_f$; in cases where the localized particles are 
electrons, they share a common
chemical potential with the conduction electrons $\mu=\mu_f$).

The spinless case corresponds to the case where $s=S=0$ and there is no
$ff$ interaction term because of the Pauli exclusion
principle.  The original Falicov-Kimball model corresponds to
the case where $s=S=1/2$, with spin-one-half electrons for both itinerant
and localized cases (and the limit $U^{ff}\rightarrow\infty$).

The Hamiltonian in Eq.~(\ref{eq: hamiltonian})
possesses a number of different symmetries.  The
partial particle-hole symmetry holds on a bipartite lattice
in no magnetic field ($H=0$), where
the lattice sites can be organized into two sublattices $A$ and $B$,
and the hopping integral only connects different sublattices.  In this
case, one performs a partial particle-hole symmetry transformation on either
the itinerant or localized electrons~\cite{kennedy_lieb_1986}.  The
transformation includes a phase factor of $(-1)$ for electrons on the $B$
sublattice. When the partial particle-hole transformation is applied to the
itinerant electrons, 
\begin{equation}
c_{i\sigma}\rightarrow c^{h\dagger}_{i\sigma}(-1)^{p(i)},
\label{eq: c_ph}
\end{equation}
with $p(i)=0$ for $i\in A$ and $p(i)=1$ for $i\in B$
and $h$ denoting the hole operators,
then the Hamiltonian maps onto itself (when expressed in terms of the 
hole operators for the itinerant electrons), up to a numerical shift,
with $U\rightarrow -U$, $E_{f\eta}\rightarrow E_{f\eta}+U$, and $\mu\rightarrow
-\mu$.  When applied to the localized electrons, 
\begin{equation}
f_{i\eta}\rightarrow f^{h\dagger}_{i\eta}(-1)^{p(i)},
\label{eq: f_ph}
\end{equation}
the Hamiltonian maps onto itself (when expressed in terms of the hole operators
for the localized electrons), up to a numerical shift
with $U\rightarrow -U$, $\mu\rightarrow \mu+U$,  
$E_{f\eta}\rightarrow -E_{f\eta}-\sum_{\eta^\prime} U^{ff}_{\eta\eta^\prime}
-\sum_{\eta^\prime} U^{ff}_{\eta^\prime\eta}$, and $\mu_f\rightarrow -\mu_f$.

These particle-hole symmetries are particularly useful when $E_{f\eta}=0$,
$\sum_{\eta^\prime}U^{ff}_{\eta\eta^\prime}$ does not depend on $\eta$,
and we work in the canonical formalism with fixed values of $\rho_e$ and
$\rho_f$, the total itinerant and localized electron densities.  Then, one can
show that the ground-state energies of $\mathcal{H}$ are simply related
\begin{eqnarray}
E_{g.s.}(\rho_e,\rho_f,U)&=&E_{g.s.}(2s+1-\rho_e,\rho_f,-U)\cr
&=&E_{g.s.}(\rho_e,2S+1-\rho_f,-U)\cr
&=&E_{g.s.}(2s+1-\rho_e,2S+1-\rho_f,U),\cr
&&
\label{eq: gs_ph}
\end{eqnarray}
(up to constant shifts or shifts proportional to $\rho_e$ or $\rho_f$)
and one can restrict the phase space to $\rho_e\le s+\frac{1}{2}$ and 
$\rho_f\le S+\frac{1}{2}$.  

When one or more of the Coulomb interactions are infinite, then there
are additional symmetries to the 
Hamiltonian~\cite{freericks_gruber_macris_1999,freericks_lieb_ueltschi_2002b}.  
In the case where all $U^{ff}=\infty$, then we are restricted to the subspace
$\rho_f\le 1$.  This system is formally identical to the case of
spinless localized electrons, and we will develop a full solution
of this limit using DMFT below.  The extra symmetry
is precisely that of Eq.~(\ref{eq: gs_ph}), but now with $S=0$ (\textit{
regardless of the number of} $\eta$ \textit{states}).  
Similarly, when both $U^{ff}=\infty$ and $U=\infty$, then Eq.~(\ref{eq: gs_ph})
holds for $s=S=0$ as well (\textit{regardless of the number of} $\sigma$ 
\textit{and} $\eta$ \textit{states}).  These infinite-$U$ symmetries are also
related to particle-hole symmetry, but now restricted to the lowest Hubbard
band in the system, since all upper Hubbard bands are pushed out to infinite
energy.

The Falicov-Kimball model also possesses a local symmetry, related to the
localized particles.  One can easily show that $[\mathcal{H},f^\dagger_{i\eta}
f_{i\eta}]=0$, implying that the \textit{local occupancy} of the
$f$-electrons is conserved.  Indeed, this leads to a local $U(1)$ symmetry, as 
the phase of the localized electrons can be rotated at will, without any effect
on the Hamiltonian.  Because of this local gauge symmetry, Elitzur's theorem
requires that there can be no quantum-mechanical mixing of the $f$-particle
number at finite temperature~\cite{elitzur_1975,subrahmanyam_barma_1988},
hence the system can never develop a spontaneous hybridization (except
possibly at $T=0$).

There are a number of different ways to provide a physical interpretation
of the Falicov-Kimball model.
In the original idea~\cite{falicov_kimball_1969}, we think of having 
itinerant and localized electrons,
that can change their statistical occupancy as a function of temperature
(maintaining a constant total number of electrons).  This is the
interpretation that leads to a metal-insulator transition due to the
change in the occupancy of the different electronic levels, rather than
via a change in the character of the electronic states themselves (the
Mott-Hubbard approach).
Another interpretation~\cite{kennedy_lieb_1986}, is to consider the localized
particles as ions, which have an attractive interaction with the electrons.
Then one can examine how the Pauli principle forces the system to minimize
its energy by crystallizing into a periodic arrangement of ions and electrons
(as seen in nearly all condensed-matter systems at low temperature).  Finally,
we can map onto a binary alloy problem~\cite{freericks_falicov_1990}, where
the presence of an ``ion'' denotes the $A$ species, and the absence of an ``ion''
denotes a $B$ species, with $U$ becoming the difference in site energies for
an electron on an $A$ or a $B$ site.  In these latter two interpretations,
the localized particle number is always a constant, and a canonical
formalism is most appropriate.  In the first interpretation, a grand canonical
ensemble is the best approach, with a common chemical potential ($\mu=\mu_f$)
for the itinerant and localized electrons.

In addition to the traditional Falicov-Kimball model, in which conduction
electrons interact with a discrete set of classical variables (the localized
electron-number operators), there is another class of static models that
can be solved using the same kind of techniques---the static anharmonic
Holstein model~\cite{holstein_1959,millis_littlewood_shraiman_1995}.  This
is a model of classical phonons interacting with conduction electrons and can
be viewed as replacing the discrete spin variable of the FK model by a 
continuous classical field.
The phonon is an Einstein mode, with infinite mass (and hence zero
frequency), but nonzero spring constant.  One can add any form of local
anharmonic potential for the phonons into the system as well.  The
Hamiltonian becomes (in the spin one-half case for the conduction
electrons, with one phonon mode per site)
\begin{eqnarray}
\mathcal{H}_{Hol}&=&-t\sum_{\langle ij\rangle\sigma}c^{\dagger}_{i\sigma}
c_{j\sigma}+g_{ep}\sum_{i\sigma}
x_i(c^{\dagger}_{i\sigma}c_{i\sigma}-\rho_{e\sigma})\cr
&+&\frac{1}{2}\kappa\sum_i x_i^2+\beta_{an}\sum_i x_i^3+\alpha_{an}\sum_i x_i^4,
\label{eq: holstein_hamiltonian}
\end{eqnarray}
where, for concreteness, we assumed a quartic phonon potential.  The phonon
coordinate at site $i$ is $x_i$, $g_{ep}$ is the electron-phonon interaction
strength (the so-called deformation potential), and the coefficients $\beta_{an}$
and $\alpha_{an}$ measure the strength of the (anharmonic)
cubic and quartic contributions to the
local phonon potential.  Note that the phonon couples to the fluctuations
in the local electronic charge (rather than the total charge).  This makes no 
difference for a harmonic system,
where the shift in the phonon coordinate can always be absorbed, but it does
make a difference for the anharmonic case, where such shifts cannot
be transformed away.  The particle-hole symmetry of this model is similar
to that of the discrete Falicov-Kimball model, described above, except the
particle-hole transformation on the phonon coordinate requires us to
send $x_i\rightarrow -x_i$.  Hence, the presence of a cubic contribution
to the phonon potential $\beta_{an}\ne 0$, breaks the particle-hole symmetry of the
system~\cite{hirsch_1993}; in this case the phase diagram is not symmetric
about half filling for the electrons.  We will discuss some results of the
static Holstein model, but we will not discuss any further extensions
(such as including double exchange for colossal magnetoresistance materials or
including interactions with classical spins to describe diluted 
magnetic semiconductors).

\subsection{Outline of the Review}

The Falicov-Kimball model (and the static Holstein model) become exactly
solvable in the limit of infinite spatial dimensions (or equivalently
when the coordination number of the lattice becomes large).  This occurs 
because both the self energy and the (relevant) irreducible two-particle vertices
are local.  The procedure involves a mapping of the infinite-dimensional
lattice problem onto a single-site impurity problem in the presence of
a time-dependent (dynamical) mean field.  The path integral for the
partition function can be evaluated exactly via the so-called static
approximation (in an arbitrary time-dependent field).  Hence the problem
is reduced to one of ``quadratures'' to determine the correct self-consistent 
dynamical mean field for the quantum system.  One can next
employ the Baym-Kadanoff conserving approximation approach to determine
the self energies and the irreducible charge vertices (both static and 
dynamic).  Armed with these quantities, one can calculate essentially
all many-body correlation functions imaginable, ranging from static
charge-density order to a dynamical Raman response.  Finally, one can also
calculate the properties of the $f$-electron spectral function, and with
that, one can calculate the susceptibility for spontaneous hybridization
formation.  The value of the Falicov-Kimball model lies in the fact
that all of these many-body properties can be determined exactly, and
thereby form a useful benchmark for the properties of correlated electronic
systems.

In Section II, we review the formalism that develops the exact solution
for all of these different properties employing DMFT.
Our attempt is to provide all details of the most important derivations,
and summarizing formulae for some of the more complicated results, which
are treated fully in the literature.  We believe that this review provides
a useful starting point for interested researchers to then understand that
literature.
Section III presents a summary of the results for a number of different
properties of the model, concentrating mainly on the spinless and spin-one-half
cases.  In Section IV, we provide a number of examples where the Falicov-Kimball
model can be applied to model real materials, concentrating mainly on 
valence-change systems like YbInCu$_4$.  We discuss a number of interesting
new directions in Section V, followed by our conclusions in Section VI.

\section{Formalism}

\subsection{Limit of Infinite Spatial Dimensions}

In 1989, Metzner and Vollhardt demonstrated that the many-body problem
simplified in the limit of large dimensions~\cite{metzner_vollhardt_1989};
equivalently, this observation could be noted to be a simplification when
the coordination number $Z$ on a lattice becomes large.  Such ideas find
their origin in the justification of the inverse coordination number $1/Z$
as the small parameter governing the convergence of the coherent
potential approximation~\cite{schwartz_siggia_1972}.
\textcite{metzner_vollhardt_1989} introduced an
important scaling of the hopping matrix element
\begin{equation}
t=t^*/2\sqrt{d}=t^*/\sqrt{2Z}
\label{eq: hopping_scale}
\end{equation}
where $d$ is the spatial dimension.  In the limit where $d\rightarrow \infty$,
the hopping to nearest neighbors vanishes, but the coordination number becomes
infinite---this is the only scaling that produces a nontrivial electronic
density of states in the large dimensional limit.  Since there is a 
noninteracting ``band'', one can observe the effects of the competition of
kinetic-energy delocalization, with potential energy localization, which forms
the crux of the many-body problem.  Hence, this limit provides 
an example where the many-body problem can be solved exactly, and these
solutions can be analyzed for correlated-electron behavior.

Indeed, the central limit theorem shows that the noninteracting
density of states on a hybercubic lattice $\rho_{hyp}(\epsilon)$
satisfies~\cite{metzner_vollhardt_1989}
\begin{equation}
\rho_{hyp}(\epsilon)=\frac{1}{t^*\sqrt{\pi}\Omega_{uc}}
\exp (-\epsilon^2/t^{*2}),
\label{eq: hypercubic_dos}
\end{equation}
which follows from the fact that the band structure is a sum of cosines,
which are distributed between $-1$ and $1$ for a ``general'' wave-vector
in the Brillouin zone ($\Omega_{uc}$ is the volume of the unit cell, which we normally
take to be equal to one).  Adding together $d$ cosines will produce a sum
that typically grows like $\sqrt{d}$, which is why the hopping is chosen to scale
like $1/\sqrt{d}$.  The central limit theorem then states that the distribution
of these energies is in a Gaussian [an alternate derivation relying
on tight-binding Green's functions and the
properties of Bessel functions can be found 
in~\textcite{mueller-hartmann_1989a}].
Another common lattice that is examined is the infinite-coordination
Bethe lattice (which can be thought of as the interior of a large Cayley
tree).  The noninteracting density of states is~\cite{economou_1983}
\begin{equation}
\rho_{Bethe}(\epsilon)=\frac{1}{2\pi t^{*2}\Omega_{uc}}\sqrt{4t^{*2}-\epsilon^2},
\label{eq: bethe_dos}
\end{equation}
where we used the number of neighbors $Z=4d$ and the scaling in 
Eq.~(\ref{eq: hopping_scale}).

The foundation for DMFT comes from two facts: 
first the self energy is a local quantity (possesses temporal but not spatial
fluctuations) and second it is a functional of the local interacting Green's
function.  These observations hold for any ``impurity'' model as well, where
the self energy can be extracted by a functional derivative of the 
Luttinger-Ward skeleton expansion for the self-energy generating
functional~\cite{luttinger_ward_1960}. 
Hence, a solution of
the impurity problem provides the functional relationship between the
Green's function and the self energy.  A second relationship is found from
Dyson's equation, which expresses the local Green's function as a summation
of the momentum-dependent Green's functions over all momenta in the
Brillouin zone.  Since the self energy has no momentum dependence, this
relation is a simple integral relation (called the Hilbert 
transformation) of the noninteracting density of states.  Combining these two 
ideas together, in a self-consistent fashion provides the basic strategy of 
DMFT.

\begin{figure}[htb]
\epsfxsize=3.0in
\epsffile{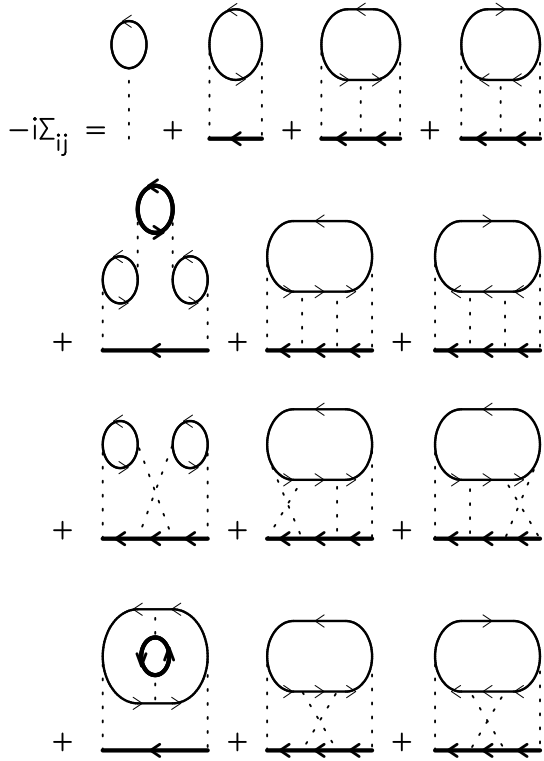}
\caption{\label{fig: pert_feynman} Skeleton expansion for the itinerant electron
self energy $\Sigma_{ij}$ through fourth 
order.  The wide solid lines denote itinerant electron
Green's functions and the thin solid lines denote localized electron Green's
functions; the dotted lines denote the Falicov-Kimball interaction $U$.  The 
series is identical to that of the Hubbard model, except we
must restrict the localized electron propagator to be diagonal in real space
(which reduces the number of off-diagonal diagrams significantly).  The first 
diagrams in the second and third rows are the only diagrams that contribute when
$i\ne j$ for the Falicov-Kimball model.}
\end{figure}

For the Falicov-Kimball model, we need to establish these two facts.
The locality of the self energy is established most directly from an
examination of the perturbation series, where one can show nonlocal
self energies are smaller by powers of $1/\sqrt{d}$. The skeleton expansion
for the self energy (determined by the functional derivative of the 
Luttinger-Ward self-energy generating
functional with respect to $G$), appears in Fig.~\ref{fig: pert_feynman}
through fourth order in $U$. This
expansion is identical to the expansion for the Hubbard model, except we 
explicitly note the localized and itinerant Green's functions graphically.
Since the localized electron propagator is local (i.e. no off-diagonal spatial
components), many of the diagrams in Fig.~\ref{fig: pert_feynman} are also
purely local.  The only nonlocal diagrams through fourth order are the 
first diagrams in the second and third rows.  If we suppose $i$ and $j$ correspond
to nearest neighbors, then we immediately conclude that the diagram has three
factors of $G_{ij}\propto 1/\sqrt{d^3}$ (each $1/\sqrt{d}$ factor comes from
$t_{ij}$).  Summing over all of the $2d$ nearest
neighbors, still produces a result that scales like $1/\sqrt{d}$ in the large
dimensional limit, which vanishes.  A similar argument can be extended to all
nonlocal diagrams~\cite{brandt_mielsch_1989,metzner_1991}.  Hence, the self 
energy is local in the infinite-dimensional
limit.  The functional dependence on the local Green's function then follows from
the skeleton expansion, restricted to the local self energy $(i=j)$.

\subsection{Single-Particle Properties (Itinerant Electrons)}

We start our analysis by deriving the set of equations satisfied by the
single-particle lattice Green's function for the itinerant electrons
defined by the time-ordered product
\begin{equation}
G_{ij\sigma}(\tau)=-\frac{1}{\mathcal{Z}_L}{\textrm{Tr}}_{cf}\langle e^{-\beta 
(\mathcal{H} -\mu N-\mu_fN_f)}\mathcal{T}_\tau c_{i\sigma}(\tau) 
c^{\dagger}_{j\sigma} (0)\rangle,
\label{eq: green_def}
\end{equation}
where $\beta=1/T$ is the inverse temperature, $0\le\tau\le\beta$ is the imaginary
time, $\mathcal{Z}_L$ is the lattice partition function, $N$ is the total itinerant electron 
number, $N_f$ is the total localized electron number, and 
$\mathcal{T}_\tau$ denotes imaginary time ordering (earlier times to the right).  
The time dependence of the electrons is
\begin{equation}
c_{i\sigma}(\tau)=e^{\tau (\mathcal{H} -\mu N)}c_{i\sigma}(0) e^{-\tau (\mathcal{H}
-\mu N)},
\label{eq: time-dependence}
\end{equation}
and the trace is over all of the itinerant and localized electronic states.
It is convenient to introduce a path-integral formulation using Grassman
variables $\bar\psi_{i\sigma}(\tau)$ and $\psi_{i\sigma}(\tau)$ for the
itinerant electrons at site-$i$ with spin $\sigma$. Using the 
Grassman form for the coherent states, then produces the following
path integral
\begin{eqnarray}
G_{ij\sigma}(\tau)&=&-\frac{1}{\mathcal{Z}_L}{\textrm{Tr}}_{f} e^{-\beta(\mathcal{H}_f-\mu_f
N_f)}\cr
&\int& \mathcal{D}\bar\psi\mathcal{D}\psi
\psi_{i\sigma}(\tau)\bar\psi_{j\sigma}(0)e^{-S^L},
\label{eq: g-path-integral}
\end{eqnarray}
for the Green's function, where $\mathcal{H}_f$ 
denotes the $f$-electron-only piece of the
Hamiltonian [corresponding to the second, fourth and sixth terms in
Eq.~(\ref{eq: hamiltonian})].  The lattice action associated with the Hamiltonian
in Eq.~(\ref{eq: hamiltonian}) is
\begin{eqnarray}
S^L&=&\sum_{ij}\sum_{\sigma=1}^{2s+1}\int_0^\beta d\tau^\prime\bar\psi_{i\sigma}
(\tau^\prime)\cr
&\times&
\left [ \delta_{ij}\frac{\partial}{\partial\tau^\prime}-\frac{t^*_{ij}}
{2\sqrt{d}}+\delta_{ij}(UN_{fi}-\mu-g\mu_BHm_\sigma)\right ]\cr
&\times& \psi_{j\sigma}(\tau^\prime),
\label{eq: lattice-action}
\end{eqnarray}
with $N_{fi}$ the total number of $f$-electrons at site $i$.
Since the Grassman variables are antiperiodic on the interval
$0\le \tau\le\beta$, we can expand them in Fourier modes, indexed by the
fermionic Matsubara frequencies $i\omega_n=i\pi T(2n+1)$:
\begin{equation}
\psi_{i\sigma}(\tau)=T
\sum_{n=-\infty}^{\infty}e^{-i\omega_n\tau}\psi_{i\sigma} (i\omega_n)
\label{eq: psi_matsubara}
\end{equation}
and
\begin{equation}
\bar\psi_{i\sigma}(\tau)=T
\sum_{n=-\infty}^{\infty}e^{i\omega_n\tau}\bar\psi_{i\sigma} (i\omega_n).
\label{eq: psibar_matsubara}
\end{equation}
In terms of this new set of Grassman variables, the lattice action becomes
$S^L=T\sum_{n=-\infty}^{\infty}S^L_n$, with
\begin{eqnarray}
S_n^L&=&\sum_{ij}\sum_{\sigma=1}^{2s+1}
\Biggr [ (-i\omega_n-\mu-g\mu_BHm_\sigma+UN_{fi})
\delta_{ij} \cr
&-&\frac{t^*_{ij}}{2\sqrt{d}}\Biggr ]
\bar\psi_{i\sigma}(i\omega_n)\psi_{j\sigma}(i\omega_n)
\label{eq: lattice-action-matsubara}
\end{eqnarray}
and the Fourier component of the local Green's function becomes
\begin{eqnarray}
G_{ii\sigma}(i\omega_n)&=&\int_0^\beta d\tau 
e^{i\omega_n\tau}G_{ii\sigma}(\tau)\cr
&=&-\frac{T}{\mathcal{Z}_L}{\textrm{Tr}}_fe^{-\beta(\mathcal{H}_f-\mu_fN_f)}\cr
&\times&\int\mathcal{D}\bar\psi\mathcal{D}\psi \psi_{i\sigma}(i\omega_n)
\bar\psi_{i\sigma}(i\omega_n)e^{-T\sum_{n^\prime}S^L_{n^\prime}}.\cr
&&
\label{eq: green-path-matsubara}
\end{eqnarray}

Now we are ready to begin the derivation of the DMFT equations for the Green's
functions.  We start with the many-body-variant of the cavity 
method~\cite{georges_kotliar_krauth_rozenberg_1996,gruber_macris_royer_2001}, 
where we separate the
path integral into pieces that involve site-$i$ only and all other terms.
The action, is then broken into three pieces: (i) the local piece at site $i$,
$S_n(i,i)$;
(ii) the piece that couples to site $i$, $S_n(i,j)$; and the piece that does not 
involve site $i$ at all, $S_n({\textrm{cavity}})$ (called the cavity piece).  Hence,
\begin{equation}
S_n=S_n(i,i)+S_n(i,j)+S_n({\textrm{cavity}})
\label{eq: action-decomposition}
\end{equation}
with
\begin{eqnarray}
S_n(i,i)&=&\sum_{\sigma=1}^{2s+1}
(-i\omega_n-\mu-g\mu_BHm_\sigma+UN_{fi})\cr
&\times&\bar\psi_{i\sigma}(i\omega_n) \psi_{i\sigma}(i\omega_n),
\label{eq: s_local}
\end{eqnarray}
\begin{eqnarray}
S_n(i,j)&=&-\sum_{j,t^*_{ij}\ne 0}\sum_{\sigma=1}^{2s+1}
\frac{t^*_{ij}}{2\sqrt{d}}\cr
&\times&[\bar
\psi_{i\sigma}(i\omega_n)\psi_{j\sigma}(i\omega_n)
+\bar\psi_{j\sigma}(i\omega_n)\psi_{i\sigma}(i\omega_n)],\cr
&&
\label{eq: s_couple}
\end{eqnarray}
and
\begin{equation}
S_n({\textrm{cavity}})=S_n-S_n(i,i)-S_n(i,j).
\label{eq: s_not_i}
\end{equation}

Let $\mathcal{Z}_{\textrm{cavity}}$ and $\langle -\rangle_{\textrm{cavity}}$ denote the partition
function and path-integral average associated with the cavity action
$S_{\textrm{cavity}}=T\sum_nS_n({\textrm{cavity}})$ (the path-integral average
for the cavity is divided by the cavity partition 
function).  Then the Green's function can be written as
\begin{widetext}
\begin{eqnarray}
G_{ii\sigma}(i\omega_n)&=&-T
\frac{\mathcal{Z}_{\textrm{cavity}}}{\mathcal{Z}_L}
{\textrm{Tr}}_fe^{-\beta(\mathcal{H}_f-\mu_f N_f)}
\int \mathcal{D}\bar\psi_i\mathcal{D}\psi_i\psi_{i\sigma}(i\omega_n)
\bar\psi_{i\sigma}(i\omega_n)e^{-T\sum_{n^{\prime\prime}=-\infty}%
^{\infty}S_{n^{\prime\prime}}(i,i)}\cr
&\times&
\Biggr\langle \exp \Bigr \{
T\sum_{n^\prime=-\infty}^{\infty}\sum_{j,t^*_{ij}\ne 0}\sum_{\sigma^\prime=1}^{2s+1}
\frac{t^*_{ij}}{2\sqrt{d}}
[\bar\psi_{i\sigma^\prime}(i\omega_{n^\prime})\psi_{j\sigma^\prime}
(i\omega_{n^\prime})
+\bar\psi_{j\sigma^\prime}(i\omega_{n^\prime})\psi_{i\sigma^\prime}
(i\omega_{n^\prime})]\Bigr\}\Biggr\rangle_{\textrm{cavity}}.
\label{eq: green-cavity}
\end{eqnarray}
A simple power-counting argument, shows that only the second moment of the
exponential factor in the cavity average contributes as 
$d\rightarrow\infty$~\cite{georges_kotliar_krauth_rozenberg_1996}, which then
becomes
\begin{eqnarray}
G_{ii\sigma}(i\omega_n)&=&-T
\frac{\mathcal{Z}_{\textrm{cavity}}}{\mathcal{Z}_L}
{\textrm{Tr}}_fe^{-\beta(\mathcal{H}_f-\mu_f N_f)}
\int \mathcal{D}\bar\psi_i\mathcal{D}\psi_i\psi_{i\sigma}(i\omega_n)
\bar\psi_{i\sigma}(i\omega_n)\cr
&\times&\exp[T\sum_{n^\prime=-\infty}^{\infty}
\sum_{\sigma^\prime=1}^{2s+1}\{ i\omega_{n^\prime}+
\mu+g\mu_BHm_{\sigma^\prime} 
-UN_{fi}-\lambda_{i\sigma^\prime}(i\omega_{n^\prime})\}
\bar\psi_{i\sigma^\prime}(i\omega_{n^\prime})\psi_{i\sigma^\prime}
(i\omega_{n^\prime})].
\label{eq: green-cavity-2}
\end{eqnarray}
with $\lambda_{i\sigma}(i\omega_n)$ the function that results from the
second-moment average that is called the dynamical mean
field.  On a Bethe lattice, one finds $\lambda_{i\sigma}(i\omega_n)
=t^{*2}G_{ii\sigma}(i\omega_n)$, while on a general lattice, one finds
$\lambda_{i\sigma}
(i\omega_n)=\sum_{jk}t_{ij}t_{ik}[G_{jk\sigma}(i\omega_n)-G_{ji\sigma}
(i\omega_n)G_{ik\sigma}(i\omega_n)/G_{ii\sigma}(i\omega_n)]$~%
\cite{georges_kotliar_krauth_rozenberg_1996}.  But the exact form is
not important for deriving the DMFT equations.  Instead we simply need to 
note that after integrating over the cavity, we find an impurity path integral
for the Green's function (after defining the impurity partition function
via $\mathcal{Z}_{imp}=\mathcal{Z}_L/\mathcal{Z}_{\textrm{cavity}}$)
\begin{eqnarray}
G_{ii\sigma}(i\omega_n)&=&-\frac{\partial\ln \mathcal{Z}_{imp}}{\partial\lambda_{i\sigma}
(i\omega_n)},\cr
\mathcal{Z}_{imp}&=&{\textrm{Tr}}_fe^{-\beta(\mathcal{H}_f-\mu_fN_f)}\int 
\mathcal{D}\bar\psi_i\mathcal{D} \psi_i
\exp[T\sum_{n=-\infty}^{\infty}\sum_{\sigma=1}^{2s+1}\{ i\omega_n+\mu+
g\mu_BHm_\sigma
-UN_{fi}-\lambda_{i\sigma}(i\omega_n)\}
\bar\psi_{i\sigma}(i\omega_n)\psi_{i\sigma}(i\omega_n)].\cr
&&
\label{eq: z_imp}
\end{eqnarray}
\end{widetext}
The impurity partition function is easy to calculate, because the effective action
is quadratic in the Grassman variables.  Defining $\mathcal{Z}_{0\sigma}(\mu)$ by
\begin{equation}
\mathcal{Z}_{0\sigma}(\mu)=2e^{\beta\mu/2}\prod_{n=-\infty}^{\infty}
\frac{i\omega_n+\mu-\lambda_{n\sigma}}{i\omega_n},
\label{eq: z0}
\end{equation}
[where we used the notation $\lambda_{n\sigma}=\lambda_{i\sigma}(i\omega_n)$ and 
adjusted the prefactor to give the noninteracting result] allows
us to write the partition function as
\begin{eqnarray}
\mathcal{Z}_{imp}&=&{\textrm{Tr}_f}\exp[-\beta(\mathcal{H}_{fi}-\mu_fN_{fi})]\cr
&\times&\prod_{\sigma=1}^{2s+1}\mathcal{Z}_{0\sigma}(\mu+g\mu_BHm_\sigma-UN_{fi}).
\label{eq: z_imp-2}
\end{eqnarray}
[$\mathcal{Z}_{0\sigma}$ is the generating functional for the $U=0$ impurity 
problem and satisfies the relation 
$\mathcal{Z}_{0\sigma}=\textrm{Det}G^{-1}_{0\sigma}$, with $G^{-1}_{0\sigma}$
defined below in frequency space in Eq.~(\ref{eq: g0def}).]
It is cumbersome to write out the trace over the fermionic states in 
Eq.~(\ref{eq: z_imp-2}) for the general case.  The spin-one-half case
appears in \textcite{brandt_fledderjohann_1990}.  Here we consider only the
strong-interaction limit, where $U^{ff}\rightarrow\infty$, so that there
is no double occupancy of the $f$-electrons.  In this case, we find
\begin{eqnarray}
\mathcal{Z}_{imp}&=&\prod_{\sigma=1}^{2s+1}\mathcal{Z}_{0\sigma}(\mu+g\mu_BHm_\sigma)\cr
&+&\sum_{\eta=1}^{2S+1}e^{-\beta (E_{f\eta}-\mu_f-g_f\mu_BHm_\eta)}\cr
&\times&\prod_{\sigma=1}^{2s+1}\mathcal{Z}_{0\sigma}(\mu+g\mu_BHm_\sigma-U).
\label{eq: z_imp-final}
\end{eqnarray}
Evaluating the derivative yields
\begin{eqnarray}
G_{n\sigma}&=&\frac{w_0}{i\omega_n+\mu+g\mu_BHm_\sigma-\lambda_{n\sigma}}\cr
&+&\frac{w_1}{i\omega_n+\mu+g\mu_BHm_\sigma-\lambda_{n\sigma}-U},
\label{eq: green_imp}
\end{eqnarray}
with 
\begin{equation}
w_0=\prod_{\sigma=1}^{2s+1}\mathcal{Z}_{0\sigma}(\mu+g\mu_BHm_\sigma)/
\mathcal{Z}_{imp},
\label{eq: w0}
\end{equation}
and $w_1=1-w_0$. The weight $w_1$ equals the average $f$-electron concentration
$\rho_f$.
If we relaxed the restriction $U^{ff}\rightarrow\infty$, then we would have
additional terms corresponding to $w_i$, with $1<i\le 2S+1$.  Defining the 
effective medium (or bare Green's function) via 
\begin{equation}
[G_{0\sigma}(i\omega_n)]^{-1}=
i\omega_n+\mu+g\mu_BHm_\sigma-\lambda_{n\sigma},
\label{eq: g0def}
\end{equation}
allows us to re-express Eq.~(\ref{eq: green_imp}) as
\begin{equation}
G_{n\sigma}=\frac{w_0}{G_{0\sigma}^{-1}(i\omega_n)}+
\frac{w_1}{G_{0\sigma}^{-1}(i\omega_n)-U},
\label{eq: green_imp-2}
\end{equation}
which is a form that often appears in the literature.  Dyson's equation for the
impurity self energy is
\begin{equation}
\Sigma_{n\sigma}=\Sigma_\sigma(i\omega_n)=[G_{0\sigma}(i\omega_n)]^{-1}-
[G_\sigma(i\omega_n)]^{-1}.
\label{eq: self_dyson}
\end{equation}
This impurity self energy is equated with the local self energy of the
lattice (since they satisfy the same skeleton expansion with respect to
the local Green's function). We can calculate the local Green's function
directly from this self energy by performing a spatial Fourier transform
of the momentum-dependent Green's function on the lattice:
\begin{eqnarray}
G_{n\sigma}&=&\sum_{\textbf{k}}G_{n\sigma}(\textbf{k})\cr
&=&\sum_{\textbf{k}}\frac{1}{i\omega_n+\mu+
g\mu_BHm_\sigma -\Sigma_{n\sigma}-\epsilon_{\textbf{k}}}\cr
&=&\int d\epsilon \rho(\epsilon)\frac{1}{i\omega_n+\mu+g\mu_BHm_\sigma
-\Sigma_{n\sigma}-\epsilon}
\label{eq: green_hilbert}
\end{eqnarray}
where $\epsilon_{\textbf{k}}$ is the noninteracting bandstructure.  This 
relation is called the Hilbert transformation of the noninteracting
density of states.  Equating the Green's
function in Eq.~(\ref{eq: green_hilbert}) to that in Eq.~(\ref{eq: green_imp-2})
is the self consistency relation of DMFT.  Substituting Eq.~(\ref{eq: self_dyson})
into Eq.~(\ref{eq: green_imp-2}) to eliminate the bare Green's function,
produces a quadratic equation for the self energy, solved by~\textcite{%
brandt_mielsch_1989,brandt_mielsch_1990}
\begin{equation}
\Sigma_{n\sigma}=\frac{1}{2}\left [ U-\frac{1}{G_{n\sigma}}\pm
\sqrt{\left ( U-\frac{1}{G_{n\sigma}}\right )^2+4w_1\frac{U}{G_{n\sigma}}}\right ],
\label{eq: self_sqrt}
\end{equation}
which is the \textit{
exact summation of the skeleton expansion for the self energy in
terms of the interacting Green's function} (the sign in front of the square
root is chosen to preserve analyticity of the self energy). Note that because
$w_1$ is a complicated functional of $G_n$ and $\Sigma_n$, 
Eq.~(\ref{eq: self_sqrt}) actually corresponds to a highly nonlinear functional
relation between the Green's function and the self energy.

There are two independent strategies that one can employ to calculate the FK
model Green's functions.  The original method of Brandt and Mielsch is to 
substitute Eq.~(\ref{eq: self_sqrt}) into Eq.~(\ref{eq: green_hilbert}), which
produces a transcendental equation for $G_{n\sigma}$ in the complex plane~\cite{%
brandt_mielsch_1989,brandt_mielsch_1990} (for fixed $w_1$).
This equation can be solved by a one-dimensional \textit{complex} root-finding
technique (typically Newton's method or Mueller's method is employed), as long
as one pays attention to maintaining the analyticity of the self energy by choosing
the proper sign for the square root [for large $U$, the sign changes at a critical
value of the Matsubara frequency~\cite{freericks_1993a}].  The other technique is 
the
iterative technique first introduced by~\textcite{jarrell_1992}, 
which is the
most commonly used method.  One starts the algorithm either with 
$\Sigma_{n\sigma}=0$ or with it chosen appropriately from an earlier calculation.  
Evaluating Eq.~(\ref{eq: green_hilbert}) for $G_{n\sigma}$ then allows one to
calculate $G_0$ from Eq.~(\ref{eq: self_dyson}).  If we work at fixed chemical
potential and fixed $E_f$, then we must determine $w_0$ and $w_1$ from 
Eq.~(\ref{eq: w0}); this step is not necessary if $w_0$ and $w_1$ are fixed in
a canonical ensemble.  Eq.~(\ref{eq: green_imp-2}) is employed to find the
new Green's function, and Eq.~(\ref{eq: self_dyson}) is used to extract the new
self energy.  The algorithm then iterates to convergence by starting with this
new self energy.  Typically, the algorithm converges to eight decimal points
in less than 100 iterations, but for some regions of parameter space the
equations can either converge very slowly, or not converge at all.  Convergence
can be accelerated by averaging the last iteration with the new result for 
determining the new self energy, but there are regions of parameter space where the
iterative technique does not appear to converge.

Once $w_0$, $w_1$, and $\mu$  are known from the imaginary-axis calculation, one
can employ the analytic continuation of Eqs.~(\ref{eq: green_imp-2}$-$%
\ref{eq: green_hilbert}) (with $i\omega_n\rightarrow \omega+i0^+$) to calculate
$G_{\sigma}(\omega)$ on the real axis.  In general, the convergence is slower
on the real axis, than on the imaginary axis, with the spectral weight slowest
to converge near correlation-induced ``band edges''.
A stringent consistency test of this
technique is a comparison of the Green's function calculated directly on the
imaginary axis to that found from the spectral formula
\begin{equation}
G_\sigma(z)=\int d\omega \frac{A_\sigma(\omega)}{z-\omega+i0^+},
\label{eq: g_spectral}
\end{equation}
with
\begin{equation}
A_\sigma(\omega)=\int d\epsilon \rho(\epsilon)A_\sigma(\epsilon,\omega),
\label{eq: interacting_dos}
\end{equation}
and
\begin{equation}
A_\sigma(\epsilon,\omega)=-\frac{1}{\pi}{\textrm{Im}}\frac{1}{\omega+\mu+g\mu_BHm_\sigma
-\Sigma_\sigma(\omega)-\epsilon+i0^+}
\label{eq: spectral}
\end{equation}
[the infinitesimal $0^+$ is needed only when ${\textrm{Im}}\Sigma_\sigma
(\omega)=0$]. Note that this definition of the interacting DOS has the
chemical potential located at $\omega=0$.

In zero magnetic field, one need perform these calculations for one $\sigma$
state only, since all $G_{n\sigma}$'s are equal, but in a magnetic field,
one must perform $2s+1$ parallel calculations to determine the $G_{n\sigma}$'s.

Our derivation of the single-particle Green's functions has followed 
a path-integral approach throughout.  One could have used an equation of
motion approach instead.  This technique has been reviewed in~\textcite{%
zlatic_review_2001}.

The formalism for the static Holstein model is 
similar~\cite{millis_mueller_shraiman_1996}.  Taking $\eta$ to be 
a continuous variable $(x)$ and $H=0$, we find
\begin{eqnarray}
\mathcal{Z}_{imp}&=&\int_{-\infty}^{\infty} dx\prod_{\sigma=1}^{2s+1}
\mathcal{Z}_{0\sigma}
(\mu-g_{ep}x)\cr
&\times&\exp(-\beta [\frac{1}{2}\kappa x^2+\beta_{an}x^3+\alpha_{an}x^4]),
\label{eq: z_holst}
\end{eqnarray}
\begin{equation}
G_{n\sigma}=\int_{-\infty}^{\infty}dx \frac{w(x)}
{i\omega_n+\mu-g_{ep}x-\lambda_{n\sigma}},
\label{eq: g_holst}
\end{equation}
and
\begin{eqnarray}
w(x)&=&\prod_{\sigma=1}^{2s+1}\mathcal{Z}_{0\sigma}(\mu-g_{ep}x)\cr
&\times&\exp(-\beta [\frac{1}{2}\kappa x^2+\beta_{an}x^3+\alpha_{an}x^4])/
\mathcal{Z}_{imp}.
\label{eq: w_x}
\end{eqnarray}
These equations (\ref{eq: z_holst}--{\ref{eq: w_x}) and Eqs.~(\ref{eq: self_dyson})
and (\ref{eq: green_hilbert}) are all that are needed to determine the Green's
functions by using the iterative algorithm.

The final single-particle quantity of interest is the Helmholtz free energy per
lattice site
$F_{Helm.}$.  We restrict our discussion to the case of zero magnetic field
$H=0$ and to $E_{f\eta}=E_f$ (no $\eta$ dependence to $E_f$).  There are two
equivalent ways to calculate the free energy.  The original method 
uses the noninteracting functional form for the free energy, with the interacting
density of states $A(\epsilon)$
replacing the noninteracting density of states $\rho(\epsilon)$~\cite{%
ramirez_falicov_kimball_1970,plischke_1972}
\begin{eqnarray}
&~&F_{Helm.}(\textrm{lattice})
=(2s+1)\int d\epsilon f(\epsilon )A(\epsilon)(\epsilon+\mu) +E_fn_f\cr
&+&(2s+1)T\int d\epsilon \Big \{ f(\epsilon)\ln f(\epsilon)\cr
&+&[1-f(\epsilon)]\ln [1-f(\epsilon)]\Big \} A(\epsilon)\cr
&+&T[w_1\ln w_1+w_0\ln w_0 -w_1\ln (2S+1)],
\label{eq: free_fk}
\end{eqnarray}
where $f(\epsilon)=1/[1+\exp(\beta\epsilon)]$ is the Fermi-Dirac distribution
function.  This form has the itinerant electron energy  (plus interactions)
and the localized electron energy on the first line [the shift by $\mu$ is 
needed because $A(\epsilon)$ is defined to have $\epsilon=0$ lie at the
chemical potential $\mu$], the itinerant electron
entropy on the second and third lines, and the localized electron entropy on the 
fourth
line.  Note that the total energy does not need the standard many-body correction
to remove double counting of the interaction because the localized particles
commute with the Hamiltonian~\cite{fetter_walecka_1971}.

The Brandt-Mielsch approach is different~\cite{brandt_mielsch_1991} and is based
on the equality of the impurity and the lattice Luttinger-Ward self-energy
generating functionals $\Phi$.  A general conserving analysis~\cite{baym_1962}
shows that the lattice free energy satisfies
\begin{eqnarray}
&~&F_{Helm.}(\textrm{lattice})=T\Phi_{latt}
-T\sum_{n\sigma} \Sigma_{n\sigma}G_{n\sigma}\cr
&+&T\sum_{n\sigma}\int d\epsilon
\rho(\epsilon)
\ln \left [ \frac{1}{i\omega_n+\mu+g\mu_BHm_\sigma-
\Sigma_{n\sigma}-\epsilon}\right ]\cr
&+&\mu \rho_e+\mu_f w_1
\label{eq: lattice_free}
\end{eqnarray}
while the impurity (or atomic) free energy satisfies~\cite{brandt_mielsch_1991}
\begin{eqnarray}
&~&F_{Helm.}(\textrm{impurity})=T\Phi_{imp}
-T\sum_{n\sigma} \Sigma_{n\sigma}G_{n\sigma}\cr
&+&T\sum_{n\sigma}\ln G_{n\sigma} +\mu\rho_e+\mu_f w_1.
\label{eq: impurity_free}
\end{eqnarray}
The first two terms on the right hand side of
Eqs.~(\ref{eq: lattice_free}) and (\ref{eq: impurity_free}) are equal, so we 
immediately learn that
\begin{eqnarray}
&~&F_{Helm.}(\textrm{lattice})=-T\ln \mathcal{Z}_{imp}-T\sum_{n\sigma}\int d\epsilon
\rho(\epsilon)\cr
&\times&\ln [(i\omega_n+\mu+g\mu_BHm_\sigma-
\Sigma_{n\sigma}-\epsilon)G_{n\sigma}]
+\mu\rho_e+\mu_f w_1\cr
&&
\label{eq: free_bm}
\end{eqnarray}
since $F_{Helm.}(\textrm{impurity})=
-T\ln \mathcal{Z}_{imp}+\mu\rho_e+\mu_f w_1$.
The equivalence of
Eqs.~(\ref{eq: free_fk}) and (\ref{eq: free_bm}) has been explicitly 
shown~\cite{shvaika_freericks_2002}. 
When calculating these terms numerically, one needs to use caution to ensure
that sufficient Matsubara frequencies are employed to guarantee convergence
of the summation in Eq.~(\ref{eq: free_bm}).

\subsection{Static Charge, Spin, or Superconducting Order}

The FK model undergoes a number of different phase transitions as
a function of the parameters of the system.  Many of these transitions
are continuous (second-order) transitions, that can be described by the
divergence of a static susceptibility at the transition temperature $T_c$.
Thus, it is useful to examine how one can calculate different susceptibilities
within the FK model.  In this section, we will examine the charge susceptibility
for arbitrary spin, and then will consider the spin and superconducting
susceptibilities for the spin-one-half model.  In addition, we will examine
how one can perform an ordered phase calculation when such an ordered phase
exists.  Our discussion follows closely that of~\textcite{brandt_mielsch_1989,%
brandt_mielsch_1990,freericks_zlatic_1998} and we consider only the case
of vanishing external magnetic field $H=0$ and the case when
$E_{f\eta}$ has no $\eta$ dependence.

We begin with the static itinerant-electron
charge susceptibility in real space defined by (our extra factor of $2s+1$
makes the normalization simpler)
\begin{eqnarray}
\chi^{cc}(\textbf{R}_i-\textbf{R}_j)&=&\frac{1}{2s+1}\int_0^\beta d\tau
[\textrm{Tr}_{cf}\frac{\langle e^{-\beta \mathcal{H}}
n_i^c(\tau)n_j^c(0)\rangle}{\mathcal{Z}_L}\cr
&-&\textrm{Tr}_{cf}\frac{\langle e^{-\beta \mathcal{H}}
n_i^c\rangle}{\mathcal{Z}_L}\textrm{Tr}_{cf}\frac{
\langle e^{-\beta \mathcal{H}} n_j^c\rangle}{\mathcal{Z}_L}],
\label{eq: chi_cc_ij}
\end{eqnarray}
where $\mathcal{Z}_L$ is the lattice partition function,
$n_i^c=\sum_{\sigma=1}^{2s+1} c^\dagger_{i\sigma}c_{i\sigma}$ and
$n_i^c(\tau)=\exp[\tau (\mathcal{H}-\mu N)]n_i^c(0)\exp[-\tau(\mathcal{H}-\mu N)]$. 
If we imagine introducing a symmetry breaking field $-\sum_i\bar h_in_i^c$ to the
Hamiltonian, then we can evaluate the susceptibility as a derivative
with respect to this field (in the limit where $\bar h =0$)
\begin{widetext}
\begin{equation}
\chi^{cc}(\textbf{R}_i-\textbf{R}_j)=\frac{T}{2s+1}\sum_n\sum_\sigma
\frac{dG_{jj\sigma}(i\omega_n)}{d\bar h_i}
=-\frac{T}{2s+1}\sum_n\sum_\sigma\sum_{kl}G_{jk\sigma}(i\omega_n)
\frac{dG_{kl\sigma}^{-1}(i\omega_n)}{d\bar h_i}G_{lj\sigma}(i\omega_n).
\label{eq: ggg}
\end{equation}
Since $G_{kl\sigma}^{-1}=[i\omega_n+\mu+\bar h_k-\Sigma_{kk\sigma}(i\omega_n)]
\delta_{kl}-t^*_{kl}/2\sqrt{d}$, we find
\begin{equation}
\chi^{cc}(\textbf{R}_i-\textbf{R}_j)=\frac{T}{2s+1}\sum_n\sum_\sigma 
\Big [ -G_{ij\sigma}(i\omega_n)G_{ji\sigma}(i\omega_n)
+\sum_k \sum_{\sigma^\prime} \sum_m
G_{jk\sigma}(i\omega_n)G_{kj\sigma}(i\omega_n)
\frac{d\Sigma_{kk\sigma}(i\omega_n)}{dG_{kk\sigma^\prime}(i\omega_m)}
\frac{d G_{kk\sigma^\prime}(i\omega_m)}{d\bar h_i} \Big ],
\label{eq: dyson_real}
\end{equation}
\end{widetext}
where we used the chain rule to relate the derivative of the local
self energy with
respect to the field to a derivative with respect to $G$ times a derivative
of $G$ with respect to the field (recall, the self energy is a functional
of the local Green's function). It is easy to verify that both $G_{ij\sigma}
G_{ji\sigma}$ and $dG_{jj\sigma}/d\bar h_i$ are independent of $\sigma$ (indeed,
this is why we set the external magnetic field to zero). If we now
perform a spatial Fourier transform of Eq.~(\ref{eq: dyson_real}), we find
Dyson's equation
\begin{equation}
\chi^{cc}_n(\textbf{q})=\chi^{cc0}_n(\textbf{q})-T\sum_m\chi_n^{cc0}(\textbf{q})
\Gamma_{nm}^{cc} \chi_m^{cc}(\textbf{q}),
\label{eq: dyson_q}
\end{equation}
where we have defined 
\begin{eqnarray}
\chi^{cc}_n(\textbf{q})&=&\frac{1}{V}\sum_{\textbf{R}_i-\textbf{R}_j}\frac{1}
{2s+1}\sum_\sigma \frac{dG_{ii\sigma}(i\omega_n)}{d\bar h_j}
e^{i\textbf{q}\cdot(\textbf{R}_i-\textbf{R}_j)},\cr
\chi^{cc0}_n(\textbf{q})&=&-\frac{1}{V}\sum_{\textbf{R}_i-\textbf{R}_j}\frac{1}
{2s+1}\sum_\sigma G_{ij\sigma}(i\omega_n)G_{ji\sigma}(i\omega_n)\cr
&\times& e^{i\textbf{q}\cdot(\textbf{R}_i-\textbf{R}_j)},\cr
\Gamma_{nm}^{cc}&=&\frac{1}{T}\frac{1}{2s+1}\sum_{\sigma\sigma^\prime}
\frac{d\Sigma_{\sigma}(i\omega_n)}{dG_{\sigma^\prime}(i\omega_m)},
\label{eq: chidefs}
\end{eqnarray}
$V$ is the number of lattice sites and $\chi^{cc}(\textbf{q})=T\sum_n\chi^{cc}_n
(\textbf{q})$. 
The fact that we have taken a Fourier transform, implies that we are considering
a periodic lattice here (like the hypercubic lattice); we will discuss below
where these results are applicable to the Bethe lattice.
Note that it seems like we have made an assumption that the irreducible 
vertex is local.  Indeed, the vertex for the lattice is not local, because
the second functional derivative of the lattice Luttinger-Ward
functional with respect 
to $G$ does have nonlocal contributions as $d\rightarrow\infty$.  These
nonlocal corrections are only for a set of measure zero of \textbf{q}-values
\cite{zlatic_horvatic_1990,georges_kotliar_krauth_rozenberg_1996,%
hettler_mukherjee_jarrell_2000}, and
one can safely replace the vertex by its local piece within any momentum
summations, which is why Eqs.~(\ref{eq: dyson_q}) and (\ref{eq: chidefs}) 
are correct.

The local irreducible vertex function $\Gamma$ can be determined by taking the
relevant derivatives of the skeleton expansion for the impurity self energy in
Eq.~(\ref{eq: self_sqrt}).  The self energy depends explicitly on $G_n$
and implicitly through $w_1$. It is because $w_1$ has $G$ dependence that the
vertex function differs from that of the coherent potential approximation
(where the derivative of $w_1$ with respect to $G$ would be zero).
The irreducible vertex becomes
\begin{eqnarray}
\Gamma^{cc}_{nm}&=&\frac{1}{(2s+1)T}\sum_{\sigma\sigma^\prime} \Biggr \{
\left ( \frac{\partial\Sigma_{n\sigma}}{\partial G_{n\sigma}}
\right )_{w_1}\delta_{\sigma\sigma^\prime}\delta_{mn}\cr
&+& \left( \frac{\partial\Sigma_{n\sigma}}{\partial w_1}\right )_{G_{n\sigma}}
\left ( \frac{\partial w_1}{\partial G_{m\sigma^\prime}}\right ) \Biggr \}.
\label{eq: vertex}
\end{eqnarray}
Substituting the irreducible vertex into the Dyson equation 
[Eq.~(\ref{eq: dyson_q})], then yields
\begin{equation}
\chi_n^{cc}(\textbf{q})=\chi^{cc0}_n(\textbf{q})\frac{1-(\partial
\Sigma_{n\sigma_1}/\partial w_1)_{G_{n\sigma_1}}\gamma(\textbf{q})}
{1+\chi_n^{cc0}(\textbf{q})(\partial\Sigma_{n\sigma_1}/\partial G_{n\sigma_1}
)_{w_1}},
\label{eq: chi2}
\end{equation}
with the function $\gamma(\textbf{q})$ defined by
\begin{equation}
\gamma(\textbf{q})=\sum_n\chi_n^{cc}(\textbf{q})\sum_\sigma
\left (\frac{\partial w_1}{\partial G_{n\sigma}}\right ),
\label{eq: gamma_def}
\end{equation}
and we have chosen a particular spin state $\sigma_1$ to evaluate the
derivatives in Eq.~(\ref{eq: chi2}) since they do not depend on
$\sigma_1$.  Multiplying Eq.~(\ref{eq: chi2}) by $\sum_\sigma(\partial w_1/
\partial G_{n\sigma})$ and summing over $n$ yields an equation for
$\gamma(\textbf{q})$.  Defining $Z_{n\sigma}=i\omega_n+\mu-\lambda_{n\sigma}=
G_{n\sigma}^{-1}+\Sigma_{n\sigma}$ and noting that $\partial w_1/\partial
G_n=\sum_m[\partial w_1/\partial Z_m][\partial Z_m/\partial G_n]$, allows us
to replace $\partial w_1/\partial G_n$ by
\begin{equation}
\frac{\partial w_1}{\partial G_n}=-\frac{\frac{\partial w_1}{\partial Z_n}
\left [ 1-G_n^2\left (\frac{\partial \Sigma_n}{\partial G_n}\right )_{w_1}\right ]}
{G_n^2\left [
1-\sum_m\frac{\partial w_1}{\partial Z_m}\left ( \frac{\partial \Sigma_m}
{\partial w_1} \right )_{G_n} \right ]},
\label{eq: dw_dG}
\end{equation}
and solve the equation for $\gamma(\textbf{q})$ to yield
\begin{widetext}
\begin{equation}
\gamma(\textbf{q})=\frac{\sum_{n\sigma}\partial w_1/\partial Z_{n\sigma}
[1-G_n^2(\partial \Sigma_n/\partial G_n)_{w_1}]/[1+G_n\eta_n(\textbf{q})
-G_n^2(\partial\Sigma_n/\partial G_n)_{w_1}]}
{1-\sum_{n\sigma}\partial w_1/\partial Z_{n\sigma}G_n\eta_n(\textbf{q})
(\partial\Sigma_n/\partial w_1)_{G_n}/[1+G_n\eta_n(\textbf{q})-
G_n^2(\partial\Sigma_n/\partial G_n)_{w_1}]},
\label{eq: gamma_solved}
\end{equation}
\end{widetext}
with $\eta_n(\textbf{q})$ defined by
\begin{equation}
\eta_n(\textbf{q})=G_n\left [ -\frac{1}{G_n^2}-\frac{1}
{\chi_n^{cc0}(\textbf{q})} \right ],
\label{eq: eta_def}
\end{equation}
and we have dropped the explicit $\sigma$ dependence for $G$ and $\Sigma$.
The full charge-density-wave susceptibility then follows 
\begin{equation}
\chi^{cc}(\textbf{q})=-T\sum_n\frac{[1-\gamma(\textbf{q})(\partial\Sigma_n/
\partial w_1)_{G_n}]G_n^2}{1+G_n\eta_n(\textbf{q})-G_n^2(\partial\Sigma_n/
\partial G_n)_{w_1}}.
\label{eq: chicc_final}
\end{equation}

The derivatives needed in Eqs.~(\ref{eq: gamma_solved}) and 
(\ref{eq: chicc_final}) can be determined straightforwardly:
\begin{equation}
\sum_{\sigma=1}^{2s+1}\frac{\partial w_1}{\partial Z_{n\sigma}}=
\frac{(2s+1)w_1(1-w_1)UG_n^2}{(1+G_n\Sigma_n)[1+G_n(\Sigma_n-U)]},
\label{eq: dw_dz}
\end{equation}
\begin{equation}
1-G_n^2\left ( \frac{\partial \Sigma_n}{\partial G_n}\right )_{w_1}
=\frac{(1+G_n\Sigma_n)[1+G_n(\Sigma_n-U)]}{1+G_n(2\Sigma_n-U)},
\label{eq: dsig_dg}
\end{equation}
and
\begin{equation}
G_n^2\left (\frac{\partial \Sigma_n}{\partial w_1}\right )_{G_n}=
\frac{UG_n^2}{1+G_n(2\Sigma_n-U)}.
\label{eq: dsig_dw}
\end{equation}
The final expression for the susceptibility and for $\gamma(\textbf{q})$
appears in Table~\ref{table: susceptibilities}.

The mixed $cf$ susceptibilities can be calculated by taking derivatives
of $w_1$ with respect to $\bar h_i$
as shown in~\textcite{brandt_mielsch_1989} and
\textcite{freericks_zlatic_1998}. We won't
repeat the details here, just the end result in 
Table~\ref{table: susceptibilities}. The calculation of the $ff$ 
susceptibilities is similar.  By recognizing that we could have calculated
the $cf$ susceptibility by adding a local $f$-electron chemical potential
and taking the derivative of the itinerant electron concentration with
respect to the local field, we can derive a relation between the $cf$
and $ff$ susceptibilities. These results are also summarized in the table.

In addition to charge susceptibilities, we also can calculate spin and 
pair-field susceptibilities for $s>0$.  We consider the spin-one-half case
in detail.  The spin susceptibility vertex simplifies, since the
off-diagonal terms now cancel, and one finds a relatively simple result.
The mixed $cf$ spin susceptibility vanishes (because the Green's function
depends only on the total $f$-electron concentration).  The $ff$ spin 
susceptibility is difficult to determine in general, but it assumes
a Curie form for $\textbf{q}=0$.  The pair-field susceptibility is determined
by employing a Nambu-Gor'kov formalism and taking the limit where the
pair-field vanishes.  There is an off-diagonal dynamical mean field
analogous to $\lambda_n$ and $w_1$ depends quadratically on this
off-diagonal field.  Hence, in the normal state, the irreducible pair-field
vertex is diagonal in  the Matsubara frequencies, just like the spin
vertex.  This means the susceptibility is easy to calculate as a function
of the bare pair-field susceptibility
\begin{equation}
\bar\chi^{cc0}_n(\textbf{q})=-\frac{1}{V}\sum_{\textbf{R}_i-\textbf{R}_j}
G_{ij\uparrow}(i\omega_n)G_{ji\downarrow}(-i\omega_n)e^{i\textbf{q}\cdot
(\textbf{R}_i-\textbf{R}_j)}.
\label{eq: chi0_pf}
\end{equation}
The result appears in Table~\ref{table: susceptibilities}.

\begin{table*}
\caption{Static charge, spin, and pair-field susceptibilities for the
Falicov-Kimball model.  The charge susceptibility is given for the general
case, the spin susceptibility for spin-one-half (and only \textbf{q}=0 for the
$ff$ spin susceptibility), and the pair-field susceptibility only for 
spin-one-half
in the $cc$ channel. Note that the $cf$ charge susceptibility is equal to 
$\gamma(\textbf{q})$.
\label{table: susceptibilities}}
\begin{ruledtabular}
\begin{tabular}{ll}
Charge&$\chi^{cc}(\textbf{q})=-T\sum_{n=-\infty}^{\infty}
\frac{[1+G_n(2\Sigma_n-U)
-\gamma(\textbf{q})U]G_n^2}{[1+G_n(2\Sigma_n-U)]G_n\eta_n(\textbf{q})+
(1+G_n\Sigma_n)[1+G_n(\Sigma_n-U)]}$\\
 &$\chi^{cf}(\textbf{q})=\gamma(\textbf{q})$\\
 &$=\frac{\sum_{n=-\infty}^\infty
(2s+1)w_1(1-w_1)UG_n^2/\{ [1+G_n(2\Sigma_n-U)]G_n\eta_n(\textbf{q})+
(1+G_n\Sigma_n)[1+G_n(\Sigma_n-U)]\} }
{1-\sum_{n=-\infty}^{\infty}(2s+1)w_1(1-w_1)U^2G_n^3\eta_n(\textbf{q})/
(1+G_n\Sigma_n)[1+G_n(\Sigma_n-U)]\{ [1+G_n(2\Sigma_n-U)]G_n\eta_n(\textbf{q})+
(1+G_n\Sigma_n)[1+G_n(\Sigma_n-U)]\} }$\\
 &$\chi^{ff}(\textbf{q})=\frac{w_1(1-w_1)/T}
{1-\sum_{n=-\infty}^{\infty}(2s+1)w_1(1-w_1)U^2G_n^3\eta_n(\textbf{q})/
(1+G_n\Sigma_n)[1+G_n(\Sigma_n-U)]\{ [1+G_n(2\Sigma_n-U)]G_n\eta_n(\textbf{q})+
(1+G_n\Sigma_n)[1+G_n(\Sigma_n-U)]\} }$\\
 & \\
\colrule
& \\
Spin&$\chi^{\prime cc}(\textbf{q})=-T\sum_{n=-\infty}^{\infty}
\frac{G_n^2[1+G_n(2\Sigma_n-U)]}{[1+G_n(2\Sigma_n-U)]G_n\eta_n(\textbf{q})+
(1+G_n\Sigma_n)[1+G_n(\Sigma_n-U)]}$\\
 & $\chi^{\prime cf}(\textbf{q})=0$\\
 & $\chi^{\prime ff}(\textbf{q}=0)=w_1/2T$\\
 & \\
\colrule
 & \\
Pair-field&$\bar\chi^{cc}(\textbf{q})=T\sum_{n=-\infty}^\infty\bar\chi^{cc0}_n
(\textbf{q})\left \{ 1-\frac{w_1(1-w_1)U^2|1+G_n\Sigma_n|^2
|1+G_n(\Sigma_n-U)|^2}
{|1+G_n(\Sigma_n-[1-w_1]U)|^2[(1-w_1)|1+G_n(\Sigma_n-U)|^2+
w_1|1+G_n\Sigma_n|^2]}\right \}$\\
\end{tabular}
\end{ruledtabular}
\end{table*}

The $cc$ charge
susceptibility diverges whenever $\gamma(\textbf{q})$ diverges, which
happens when the denominator in Table~\ref{table: susceptibilities} vanishes.
Since this denominator is identical for the $cc$, $cf$, and $ff$ 
susceptibilities, all three diverge at the same transition temperature
as we would expect. 
This yields the same result as found in~\textcite{brandt_mielsch_1989,%
brandt_mielsch_1990} for the spinless case, except there is an additional
factor of $2s+1$ multiplying the sum in the denominator
arising from the $2s+1$ derivatives of $w_1$, which are all equal. This factor
``essentially'' increases $T_c$ by $2s+1$ over that found in the spinless case.
The existence of a $T_c$ is easy to establish, since the summation in the
denominator goes to zero like $1/T^4$ for large $T$ and diverges like $C/T$
for small $T$.  If $C>0$, then there is a transition.

The \textbf{q}-dependence of the charge 
susceptibility comes entirely from $\eta_n
(\textbf{q})$ and hence from the bare susceptibility
\begin{eqnarray}
\chi^{cc0}_n(\textbf{q})&=&-\sum_{\textbf{k}}G_n(\textbf{k}+\textbf{q})
G_n(\textbf{k})\cr
&=&-\frac{1}{\sqrt{1-X^2}}\int_{-\infty}^\infty d\epsilon
\frac{\rho(\epsilon)}{i\omega_n+\mu-\Sigma_n-\epsilon}\cr
&\times&F_\infty\left [ \frac{i\omega_n+\mu-\Sigma_n-X\epsilon}{\sqrt{1-X^2}}
\right ],
\label{eq: chi0_int}
\end{eqnarray}
where all of the \textbf{q}-dependence can be summarized in a single parameter
$X(\textbf{q})=\lim_{d\rightarrow\infty}\sum_{i=1}^d\cos q_i/d$
\cite{mueller-hartmann_1989a} and we use $F_\infty(z)=\int d\epsilon 
\rho(\epsilon)/(z-\epsilon)$ to denote the Hilbert transform.  The results
for $\chi_n^{cc0}(\textbf{q})$ and $\eta_n(\textbf{q})$ simplify for three
general cases~\cite{brandt_mielsch_1989}:
$X=-1$, which corresponds to the ``checkerboard''
zone-boundary point $\textbf{Q}=(\pi,\pi,\pi,...)$; $X=1$, which corresponds
to the uniform zone-center point $\textbf{q}=0$; and $X=0$, which
corresponds to a general momentum vector in the Brillouin zone [since the
value of the cosine will look like a random number for a general wave vector
and the summation will grow like $\sqrt{d}$, implying $X\rightarrow 0$;
only a set of measure zero of momenta have $X(\textbf{q})\ne 0$].  The
results for $\chi_n^{cc0}(\textbf{q})$ and $\eta_n(\textbf{q})$ appear
in Table~\ref{table: Xspecial} for the hypercubic lattice.  Both the
uniform ($X=1$) and the ``checkerboard'' ($X=-1$) susceptibilities
can be defined for the Bethe lattice, but there does not seem to be any
simple way to extend the definition to all $X$.  Indeed, higher-period
ordered phases on the Bethe lattice seem to have first-order (discontinuous)
phase transitions~\cite{gruber_macris_royer_2001} so such a generalization
is not needed.

\begin{table}[htb]
\caption{Values of $\chi_n^{cc0}(\textbf{q})$ and $\eta_n(\textbf{q})$
for the special $X$ points $1$, $0$, and $-1$ on the hypercubic lattice.
\label{table: Xspecial}}
\begin{ruledtabular}
\begin{tabular}{rll}
$X(\textbf{q})$&$\chi_n^{cc0}(\textbf{q})$ & $\eta_n(\textbf{q})$\\
\colrule
$-1$&$-G_n/(i\omega_n+\mu-\Sigma_n)$&$\lambda_n$\\
$0$&$-G_n^2$&$0$\\
$1$&$2[1-(i\omega_n+\mu-\Sigma_n)G_n]$&$-\frac{1}{G_n}+\frac{1}{2\lambda_n}$\\
\end{tabular}
\end{ruledtabular}
\end{table}

We will find that near half filling, the $X=-1$ charge
susceptibility diverges and
far away from half filling the $X=1$ charge susceptibility diverges. The $X=0$
susceptibility never diverges at finite $T$.  This is also true for the
spin and pair-field susceptibilities.  They are always finite at finite $T$.
One can easily understand why the spin susceptibility does not diverge---it
arises simply from the fact that the spins are independent of each other and
do not interact.  Similarly, one can understand why the pair-field 
susceptibility does not diverge---Anderson's theorem~\cite{anderson_1959,%
bergmann_rainer_1974}
states that one cannot have superconductivity with a static electron-electron
interaction; the interaction must be dynamic.

Since the $(X=-1)$ ``checkerboard'' charge susceptibility diverges, we can also
examine the ordered state~\cite{brandt_mielsch_1990}.  In this case, we
have a charge-density wave, with different electronic densities on each of the
two sublattices of the bipartite lattice.  Hence both $G_{n\sigma}^A$ 
($G_{n\sigma}^B$) and $\Sigma_{n\sigma}^A$ ($\Sigma_{n\sigma}^B$) are different
on the two sublattices.  Evaluating the momentum-dependent Green's functions
yields
\begin{widetext}
\begin{equation}
G_{n\sigma}^{A,B}(\textbf{q})=\frac{i\omega_n+\mu+g\mu_BHm_\sigma-
\Sigma_{n\sigma}^{B,A}+\epsilon_{\textbf{q}}}{(i\omega_n+\mu+g\mu_BHm_\sigma-
\Sigma_{n\sigma}^{A})(i\omega_n+\mu+g\mu_BHm_\sigma-
\Sigma_{n\sigma}^{B})-\epsilon_{\textbf{q}}^2},
\label{eq: bipartite_q}
\end{equation}
which can be summed over \textbf{q} to yield
\begin{eqnarray}
G_{n\sigma}^A&=&\int d\epsilon\frac{\rho(\epsilon)}{\bar Z_{n\sigma}-\epsilon}
\frac{i\omega_n+\mu+g\mu_BHm_\sigma-\Sigma_{n\sigma}^B}{\bar Z_{n\sigma}},\quad
G_{n\sigma}^B=\int d\epsilon\frac{\rho(\epsilon)}{\bar Z_{n\sigma}-\epsilon}
\frac{i\omega_n+\mu+g\mu_BHm_\sigma-\Sigma_{n\sigma}^A}{\bar Z_{n\sigma}},\cr
\bar Z_{n\sigma}&=&\sqrt{(i\omega_n+\mu+g\mu_BHm_\sigma-\Sigma_{n\sigma}^A)
(i\omega_n+\mu+g\mu_BHm_\sigma-\Sigma_{n\sigma}^B)}.
\label{eq: bipartite_final}
\end{eqnarray}
\end{widetext}
The algorithm to solve for the Green's functions is modified to the following:
(i) start with a guess for $\Sigma_{n\sigma}^A$ and $\Sigma_{n\sigma}^B$
(or set both to zero); (ii) evaluate Eq.~(\ref{eq: bipartite_final}) to
find $G_{n\sigma}^A$ and then determine $G_{0\sigma}^A(i\omega_n)$ from
Eq.~(\ref{eq: self_dyson}) evaluated on the A sublattice [$\Sigma_{n\sigma}^A=
\{G_{0\sigma}^A(i\omega_n)\}^{-1}-(G_{n\sigma}^A)^{-1}$]; (iii) determine
$w_0^A$  and $w_1^A$
from the appropriate A-sublattice generalization of Eq.~(\ref{eq: w0});
(iv) evaluate Eq.~(\ref{eq: green_imp-2}) on the A-sublattice to find
$G_{n\sigma}^A$ and Eq.~(\ref{eq: self_dyson}) to find $\Sigma_{n\sigma}^A$;
(v) now find $G_{n\sigma}^B$ from Eq.~(\ref{eq: bipartite_final}) and
$G_{0\sigma}^B(i\omega_n)$ from Eq.~(\ref{eq: self_dyson}); (vi) determine
$w_0^B$ and $w_1^B$ from Eq.~(\ref{eq: w0}) on the B-sublattice;
and (vii) evaluate Eq.~(\ref{eq: green_imp-2}) on the B-sublattice to find
$G_{n\sigma}^B$ and Eq.~(\ref{eq: self_dyson}) to find $\Sigma_{n\sigma}^B$.
Now repeat (ii--vii) until convergence is reached.  We will need to adjust
$E_f-\mu_f$ until $(w_1^A+w_1^B)/2=w_1$.  The calculations are then finished
and the order parameter is $(w_1^A-w_1^B)/2$.  A similar generalization can
be employed for the static Holstein model~\cite{ciuchi_depasquale_1999}. Since 
the skeleton expansion
for the self energy in terms of the local Green's function is unknown for the
static Holstein model (and hence it is not obvious how to determine 
$\Gamma$), this is the most direct way to search for the ordered
checkerboard phase.

\subsection{Dynamical Charge Susceptibility \label{sec: dyn_charge}}

We now examine the dynamical charge susceptibility~\cite{shvaika_2000,%
freericks_miller_2000,shvaika_2002}, defined by
\begin{eqnarray}
\chi^{cc}(\textbf{q},i\nu_l)&=&\frac{1}{2s+1}\int_0^\beta d\tau e^{i\nu_l\tau}
\frac{1}{V}\sum_{\textbf{R}_i-\textbf{R}_j}e^{\textbf{q}\cdot
(\textbf{R}_i-\textbf{R}_j)}
\cr
&\times&
\Bigr [\textrm{Tr}_{cf}\frac{\langle e^{-\beta \mathcal{H}}
n_i^c(\tau)n_j^c(0)\rangle}{\mathcal{Z}_L}\cr
&-&\textrm{Tr}_{cf}\frac{\langle e^{-\beta \mathcal{H}}n_i^c\rangle}{\mathcal{Z}_L}
\textrm{Tr}_{cf}\frac{ \langle e^{-\beta \mathcal{H}}n_j^c\rangle}{\mathcal{Z}_L}\Bigr ],
\label{eq: chi_dyn}
\end{eqnarray}
with $i\nu_l=2i\pi lT$ the bosonic Matsubara frequency.  Once again we assume we 
are in zero magnetic field $H=0$ and the $f$-electron site-energy is 
$\eta$-independent $E_{f\eta}=E_f$.  Our analysis follows the path-integral
approach--one can also employ a strong-coupling perturbation theory to derive
these formulae~\cite{shvaika_2000,shvaika_2002}. 
The dynamical
susceptibilities of the FK model have some subtle properties to them
that arise from the fact that the local $f$-electron number is conserved.
In particular, the isothermal susceptibilities (calculated by adding
an external field to the system, and determining how it modifies the system
in the limit where the field vanishes) and the so-called isolated (Kubo)
susceptibilities [which assume the system starts in equilibrium at zero field,
then is removed from the thermal bath (isolated) and the field is turned
on slowly] differ from each
other (the response of the isolated system to the field is the
isolated susceptibility) due to the conserved nature of the
$f$-electrons~\cite{wilcox_1968,shvaika_2002}.  In particular, the isolated
susceptibility vanishes for the $ff$ and mixed $cf$ susceptibilities
[due to the fact that the local $f$-electron
number is conserved (\textit{i.e.}, $[\mathcal{H},n^f_i]=0$)],
but is nonzero for the $cc$ susceptibility.  The divergence of the
static charge susceptibility arises entirely from the coupling between the
itinerant and localized electronic systems; 
the pure itinerant electron response (isolated susceptibility) never diverges.
Hence the isothermal
susceptibility is discontinuous at zero frequency (i.e., it is not
analytic).  The continuous (analytic) susceptibility
is the isolated susceptibility, and we will spend our time discussing it
for the conduction electrons.  

If we express the susceptibility as a matrix in Matsubara frequency space,
we find the following Dyson equation:
\begin{eqnarray}
&~&\chi^{cc}(\textbf{q},i\omega_m,i\omega_n;i\nu_l)\cr
&=&\chi^{cc0}(\textbf{q},i\omega_m;i\nu_l)\delta_{mn}-T\sum_{n^\prime}
\chi^{cc0}(\textbf{q},i\omega_m;i\nu_l)\cr
&\times&\Gamma(i\omega_m,i\omega_{n^\prime};i\nu_l)
\chi^{cc}(\textbf{q},i\omega_{n^\prime},i\omega_n;i\nu_l),
\label{eq: chi_dyn_dyson}
\end{eqnarray}
and the susceptibility is found by summing over the Matsubara frequencies
$\chi^{cc}(\textbf{q},i\nu_l)=T\sum_{mn}
\chi^{cc}(\textbf{q},i\omega_m,i\omega_n;i\nu_l)$.  Once again, the bare
susceptibility depends only on the momentum parameter $X$, and takes the 
form
\begin{eqnarray}
\chi^{cc0}(X,i\omega_m;i\nu_l)&=&-\frac{1}{2s+1}\sum_{\textbf{k}\sigma}
G_{m\sigma}(\textbf{k})G_{m+l\sigma}(\textbf{k+q})\cr
&=&-\frac{1}{\sqrt{1-X^2}}\int_{-\infty}^{\infty}d\epsilon\frac{\rho(\epsilon)}
{i\omega_m+\mu-\Sigma_{m}-\epsilon}\cr
&\times&F_{\infty}\left [\frac{i\omega_{m+l}+\mu-\Sigma_{m+l}-X\epsilon}
{\sqrt{1-X^2}}\right ],
\label{eq: chi0_dyn}
\end{eqnarray}
which reduces to our static results for $l=0$.  The bare dynamical
susceptibility simplifies in three cases that are summarized in 
Table~\ref{table: chi0_dyn_special}. Note that one needs to evaluate
$\chi^{cc0}$ with l'H\^opital's rule whenever the denominator vanishes
and we dropped the spin subscripts in the Table.  The irreducible charge vertex
is calculated for the impurity, and it satisfies
\begin{equation}
\Gamma(i\omega_m,i\omega_n;i\nu_l)=\frac{1}{2s+1}\sum_\sigma
\frac{1}{T}\frac{\delta\Sigma_\sigma(i\omega_m,i\omega_{m+l})}
{\delta G_\sigma(i\omega_n,i\omega_{n+l})},
\label{eq: dyn_vertex}
\end{equation}
where we now have both a self energy and a Green's function that depend
on \textit{two} Matsubara frequencies, because these functions are not
time-translation-invariant in imaginary time.  This occurs because we need to
add a time-dependent charge field $-\int_0^\beta d\tau \chi(\tau)\sum_\sigma 
c_\sigma^\dagger(\tau)c_\sigma(\tau)$ to the action in order to evaluate
the dynamic charge susceptibility and this time-dependent field removes
time-translation invariance from the system (it does not depend on the 
time difference of the arguments of the fermionic variables).

\begin{table}[h]
\caption{Values of $\chi^{cc0}(X,i\omega_m;i\nu_l)$
for the special $X$ points $1$, $0$, and $-1$ on the hypercubic lattice.
\label{table: chi0_dyn_special}}
\begin{ruledtabular}
\begin{tabular}{rl}
$X(\textbf{q})$&$\chi^{cc0}(X,i\omega_m;i\nu_l)$\\
\colrule
$-1$&$-(G_m+G_{m+l})/(i\omega_m+i\omega_{m+l}+2\mu-\Sigma_m-\Sigma_{m+l})$\\
$0$&$-G_mG_{m+l}$\\
$1$&$-(G_m-G_{m+l})/(i\nu_l+\Sigma_m-\Sigma_{m+l})$\\
\end{tabular}
\end{ruledtabular}
\end{table}    

It is not easy to perform calculations for Green's functions that depend on
two time variables.  The original work by~\textcite{brandt_urbanek_1992} 
illustrates
how to proceed.  We start with the definition of an auxiliary Green's function
\begin{widetext}
\begin{equation}
g^{aux}_\sigma(\tau,\tau^\prime)=-\frac{\textrm{Tr}_{c}\mathcal{T}_\tau
\left \langle e^{-\beta\mathcal{H}_0}\exp\left [ \sum_{\bar\sigma}
\int_0^\beta d\bar\tau
\chi(\bar\tau)c^\dagger_{\bar\sigma}(\bar\tau)c_{\bar\sigma}(\bar\tau)\right ]
c_\sigma(\tau)c^\dagger_\sigma(\tau^\prime)\right \rangle}
{\left \{1+e^{\beta\mu}\exp\left [ \int_0^\beta d\bar\tau\chi(\bar\tau) 
\right ] \right \}^{2s+1}},
\label{eq: gaux}
\end{equation}
\end{widetext}
where $\mathcal{H}_0=-\mu\sum_\sigma c^\dagger_\sigma c_\sigma$ and the 
time-dependence is with respect to $\mathcal{H}_0$ [the auxiliary time-dependent
field $\chi(\tau)=\sum_l\chi(i\nu_l)\exp(-i\nu_l\tau)$ should not be confused 
with any susceptibility].  It is easy to show that
this auxiliary Green's function is antiperiodic with respect to either $\tau$
variable being increased by $\beta$.  Hence, we can perform a double
Fourier transform to yield
\begin{eqnarray}
&~&g^{aux}_\sigma(i\omega_m,i\omega_n)\cr
&~&=T\int_0^\beta d\tau 
\int_{-\beta+\tau}^{\tau} d\tau^\prime
e^{i\omega_m\tau}g^{aux}_\sigma(\tau,\tau^\prime)e^{-i\omega_n\tau^\prime},
\label{eq: gaux_mats}
\end{eqnarray}
and this is the same Matsubara-frequency dependence as in 
Eq.~(\ref{eq: dyn_vertex}).  Substituting in the fermionic Grassman variables
from Eq.~(\ref{eq: psi_matsubara})  and restricting ourselves to $\chi$
fields that satisfy $\chi(i\nu_0)=0$ produces the following
path-integral form for the auxiliary Green's function:
\begin{eqnarray}
g^{aux}_\sigma(i\omega_m,i\omega_n)&=&-\frac{T}{\mathcal{Z}^{aux}}\int\mathcal{D}\bar\psi
\mathcal{D}\psi \psi_{m\sigma}\bar\psi_{n\sigma}\cr
&\times&\exp \Biggr [ T\sum_{m^\prime n^\prime}
\sum_{\sigma^\prime}\{(i\omega_{m^\prime}
+\mu)\delta_{m^\prime n^\prime}\cr
&+&\chi(i\omega_{n^\prime}-i\omega_{m^\prime})\}
\bar\psi_{n^\prime\sigma^\prime}\psi_{m^\prime\sigma^\prime}\Biggr ],
\label{eq: gaux_mats_path}
\end{eqnarray}
where $\mathcal{Z}^{aux}=[1+e^{\beta\mu} ]^{2s+1}$ is the auxiliary partition function
for $\chi(i\nu_0)=0$.  We calculate the Green's function by adding 
an infinitesimal field $T\bar\chi_{mn\sigma}\bar\psi_{n\sigma}\psi_{m\sigma}$
and noting that $g^{aux}_\sigma(i\omega_m,i\omega_n)=\partial\ln \mathcal{Z}^{aux}/
\partial \bar\chi_{mn\sigma}$.  We will restrict our discussion to the case where
\textit{only one} Fourier component $\chi(i\nu_l)$ is nonzero.  The
matrix then has a nonzero diagonal, and one nonzero off-diagonal, whose
elements are all equal to $\chi(i\nu_l)$. Now the partition function is the
determinant of the matrix that appears in the action of 
Eq.~(\ref{eq: gaux_mats_path}), which assumes the simple form of the product
of all diagonal elements (multiplied by a constant to assure the correct
limiting behavior)
\begin{equation}
\mathcal{Z}^{aux}=\left [
2e^{\beta\mu/2}\prod_{n=-\infty}^{\infty}
\frac{i\omega_n+\mu}{i\omega_n}\right ]^{2s+1}.
\label{eq: zaux}
\end{equation}
When we add the extra field $\bar\chi_\sigma$
to the action, the eigenvalues of the matrix change
only for $m=n$, $m+l=n$, and the $\sigma$ value of the $\bar\chi_\sigma$
field.  The former case is the diagonal matrix element,
which is simply shifted by $\bar\chi_{mm\sigma}$.  The shift in the latter case
can be worked out via perturbation theory to lowest order in $\bar\chi$.
We find the $m$th and $m+l$th eigenvalues change to $i\omega_m+\mu-
\bar\chi_{m+lm}\chi(i\nu_l)/i\nu_l$ and $i\omega_{m+l}+\mu+
\bar\chi_{m+lm}\chi(i\nu_l)/i\nu_l$, respectively.  Taking the relevant
derivatives to calculate the auxiliary Green's function, then yields
\begin{eqnarray}
g^{aux}_\sigma(i\omega_m,i\omega_n)&=&\frac{\delta_{mn}}{i\omega_m+\mu}+\frac%
{\delta_{m+ln}\chi(i\nu_l)}{i\nu_l}\cr
&\times&\left ( \frac{1}{i\omega_{m+l}+\mu}-\frac{1}{i\omega_m+\mu}\right ).
\label{eq: gaux_final}
\end{eqnarray}

As before, we define $G_0$ by adding a $\lambda_\sigma$ field to the action
in Eq.~(\ref{eq: gaux_mats_path}) $T\lambda_{n\sigma}\bar\psi_{n\sigma}
\psi_{n\sigma}$ (and change the partition function normalization 
accordingly).  It
is easy to show~\cite{brandt_urbanek_1992,zlatic_review_2001} that due to the
fact that one can write the partition function of a path integral with 
a quadratic action as $\mathcal{Z}=\det g^{-1}$, with $g$ the corresponding Green's
function, one discovers
\begin{equation}
G_0^{-1}=[g^{aux}]^{-1}-\lambda\openone.
\label{eq: g0_matrix_dyson}
\end{equation}
The full Green's function involves an additional trace over $f$-states.  The
``itinerant-electron'' piece of the impurity Hamiltonian equals $-\mu \sum_\sigma
c^\dagger_{\sigma}c_\sigma$ when $n^f=0$ and equals $(U-\mu)\sum_\sigma          
c^\dagger_{\sigma}c_\sigma$ when $n^f=1$.  It is a straightforward exercise
to then show that 
\begin{equation}
G_\sigma(i\omega_m,i\omega_n)=w_0G_{0\sigma}(i\omega_m,i\omega_n)+\
w_1[G_{0\sigma}^{-1}-U\openone]^{-1}_{mn},
\label{eq: green_dyn_final}
\end{equation}
with $w_0$ given by Eq.~(\ref{eq: w0}) for $H=0$ and $w_1=1-w_0$. Note that
all of the Green's functions in Eq.~(\ref{eq: green_dyn_final}) are
matrices, and the $-1$ superscript denotes the inverse of the corresponding
matrix.

Eq.~(\ref{eq: green_dyn_final}) can be rearranged by multiplying on the
left or the right by matrices like $G^{-1}$, $G_0^{-1}$, and $G_0^{-1}-U$
to produce the following two matrix equations (with matrix indices and
spin indices suppressed)
\begin{equation}
G_0^{-2}-(U+G^{-1})G_0^{-1}+(1-w_1)UG^{-1}=0,
\label{eq: green_quad1}
\end{equation}
\begin{equation}
G_0^{-2}-G_0^{-1}(U+G^{-1})+(1-w_1)UG^{-1}=0.
\label{eq: green_quad2}
\end{equation}
Adding these two equations together and collecting terms results in
\begin{equation}
\left [G_0^{-1}-\frac{1}{2}(U+G^{-1})\right ]^2-\frac{1}{4}(U+G^{-1})^2+(1-w_1)
UG^{-1}=0.
\label{eq: green_quad3}
\end{equation}
Now we substitute the matrix self energy $\Sigma$ for $G_0$, with the
self energy defined by Dyson's equation
\begin{equation}
\Sigma(i\omega_m,i\omega_n)=[G_0^{-1}]_{mn}-[G^{-1}]_{mn},
\label{eq: self_dyn_dyson}
\end{equation}
to yield a matrix quadratic equation
\begin{equation}
\left [\Sigma+\frac{1}{2}(G^{-1}-U)\right ]^2=\frac{1}{4}[U^2+2(2w_1-1)UG^{-1}
+G^{-2}],
\label{eq: dyn_quadratic}
\end{equation}
that relates the self energy to the Green's function.  In the case where
only one Fourier component $l\ne 0$ of $\chi$ is nonzero, both the self energy
and the Green's function are nonzero on the diagonal $m=n$ and on the
diagonal shifted by $l$ units $m+l=n$.  Substituting into the quadratic
equation, and solving for the shifted diagonal component of the self-energy
[to first order in $\chi(i\nu_l)$] yields the amazingly simple result
\begin{equation}
\Sigma(i\omega_{m},i\omega_{m+l})=G(i\omega_{m},i\omega_{m+l})\frac{\Sigma_m-
\Sigma_{m+l}}{G_m-G_{m+l}},
\label{eq: sigma_dyn_off}
\end{equation}
after some tedious algebra.  In Eq.~(\ref{eq: sigma_dyn_off}), the symbols
$\Sigma_m$ and $G_m$ denote the diagonal components of the self energy and
the Green's function, respectively, which are equal to the results we already
calculated for the self energy and Green's function when $\chi(i\nu_l)=0$.

We can now calculate the irreducible dynamical charge vertex from 
Eq.~(\ref{eq: dyn_vertex}), which yields
\begin{equation}
\Gamma(i\omega_m,i\omega_n;i\nu_{l\ne 0})=
\delta_{mn}\frac{1}{T}\frac{\Sigma_m-\Sigma_{m+l}}{G_m-G_{m+l}}.
\label{eq: dyn_vertex_final}
\end{equation}
The dynamical charge vertex is then a relatively simple object for $l\ne 0$
[the static vertex is much more complicated, as can be inferred from
Eqs.~(\ref{eq: chidefs}), (\ref{eq: vertex}), and (\ref{eq: dw_dG})].
The original derivation, based on the atomic 
limit~\cite{shvaika_2000,shvaika_2002}, produced a much more complicated
looking result for the vertex, but some tedious algebra shows that the two
forms are indeed identical~\cite{freericks_miller_2000}.
Using this result for the vertex, we immediately derive the final form for
the dynamical charge susceptibility on the imaginary axis
\begin{equation}
\chi^{cc}(X;i\nu_{l\ne 0})=T\sum_m\frac{\chi^{cc0}(X,i\omega_m;i\nu_l)}
{1+\chi^{cc0}(X,i\omega_m;i\nu_l)\frac{\Sigma_m-\Sigma_{m+l}}{G_m-G_{m+l}}}.
\label{eq: chi_dyn_final}
\end{equation}
Note, that if we evaluate the uniform, dynamical charge susceptibility
$\textbf{q}=\textbf{0}$ ($X=1$), we find
\begin{equation}
\chi^{cc}(X=1,i\nu_{l\ne 0})=-T\sum_m\frac{G_m-G_{m+l}}{i\nu_l}=0,
\label{eq: chy_dyn_x=1}
\end{equation}
which is what we expect, because the total conduction electron charge commutes 
with
the Hamiltonian, and hence has no $\tau$-dependence, implying only the static
response can be nonzero. Note that $\chi^{cc0}(X=1)\ne 0$, the vertex corrections
are needed to produce a vanishing total susceptibility.

In general, one is interested in the dynamical charge response on the real
axis.  To find the real (dynamical) response, we need to perform an
analytical continuation of Eq.~(\ref{eq: chi_dyn_final}) to the real
axis.  Since the isothermal susceptibility is discontinuous at $\nu_l=0$, we
can perform the analytic continuation only for the isolated susceptibility.
The procedure is straightforward: we divide the complex plane into
regions where the Green's functions, self energies, and susceptibilities
are all analytic, and we express the Matsubara summations as contour integrals
over the poles of the Fermi-Dirac distribution function.  Then, under the
assumption that there are no poles in the integrands other than those 
determined by the fermi factors, we deform the contours until they
are parallel to the real axis.  After replacing any fermi factors of
the form $f(\omega+i\nu_l)$ by $f(\omega)$, we can make the analytic
continuation $i\nu_l\rightarrow \nu+i0^+$.  The final expressions are
straightforward, but cumbersome.  We summarize them for $X=-1$ and $X=0$
in Table~\ref{table: chi_dyn}. Only the $X=-1$ result is meaningful for the
Bethe lattice. Note that one could just as easily calculate the 
incommensurate dynamical charge response, but the equations become
significantly more complicated.  We will calculate a similar result when
we examine inelastic X-ray scattering below.

One can test the accuracy of the analytic continuation by employing the 
spectral formula for the dynamical charge susceptibility and calculating the
charge susceptibility at each of the bosonic Matsubara frequencies.  Comparing
the spectral form with the result found directly from 
Eq.~(\ref{eq: chi_dyn_final}) is a stringent self-consistency test.  In most
calculations, these two forms agree to at least one part in a thousand.

\begin{table*}
\caption{\label{table: chi_dyn}
Dynamical charge susceptibility on the real axis for $X=-1$ and $X=0$ (the
dynamical susceptibility vanishes for $X=1$).}
\begin{ruledtabular}
\begin{tabular}{rl}
$X$&Dynamical susceptibility $\chi^{cc}(X,\nu)$\\
\colrule
$-1$&$\frac{i}{2\pi}\int_{-\infty}^{\infty}
d\omega\Biggr \{ f(\omega)\frac{[G(\omega)+G(\omega+\nu)]
/[2\omega+2\mu+\nu+\Sigma(\omega)-\Sigma(\omega+\nu)]}
{1-[G(\omega)+G(\omega+\nu)][\Sigma(\omega)-\Sigma(\omega+\nu)]/
[2\omega+2\mu+\nu+\Sigma(\omega)-\Sigma(\omega+\nu)][G(\omega)-G(\omega+\nu)]}$
\\
&$-f(\omega+\nu)\frac{[G^*(\omega)+G^*(\omega+\nu)]
/[2\omega+2\mu+\nu+\Sigma^*(\omega)-\Sigma^*(\omega+\nu)]}
{1-[G^*(\omega)+G^*(\omega+\nu)][\Sigma^*(\omega)-\Sigma^*(\omega+\nu)]/
[2\omega+2\mu+\nu+\Sigma^*(\omega)-\Sigma^*(\omega+\nu)][G^*(\omega)-
G^*(\omega+\nu)]}$\\
&$-[f(\omega)-f(\omega+\nu)]\frac{[G^*(\omega)+G(\omega+\nu)]
/[2\omega+2\mu+\nu+\Sigma^*(\omega)-\Sigma(\omega+\nu)]}
{1-[G^*(\omega)+G(\omega+\nu)][\Sigma^*(\omega)-\Sigma(\omega+\nu)]/
[2\omega+2\mu+\nu+\Sigma^*(\omega)-\Sigma(\omega+\nu)][G^*(\omega)-
G(\omega+\nu)]}\Biggr \}$\\
\colrule
$0$&$\frac{i}{2\pi}\int_{-\infty}^{\infty}
d\omega\Biggr \{ f(\omega)\frac{G(\omega)G(\omega+\nu) }
{1-G(\omega)G(\omega+\nu)[\Sigma(\omega)-\Sigma(\omega+\nu)]/
[G(\omega)-G(\omega+\nu)]}
-f(\omega+\nu)\frac{G^*(\omega)G^*(\omega+\nu)}
{1-G^*(\omega)G^*(\omega+\nu)[\Sigma^*(\omega)-\Sigma^*(\omega+\nu)]/
[G^*(\omega)-G^*(\omega+\nu)]}$\\
&$-[f(\omega)-f(\omega+\nu)]\frac{G^*(\omega)G(\omega+\nu)}
{1-G^*(\omega)G(\omega+\nu)[\Sigma^*(\omega)-\Sigma(\omega+\nu)]/
[G^*(\omega)-G(\omega+\nu)]}\Biggr \}$\\  
\end{tabular}
\end{ruledtabular}
\end{table*}

\subsection{Static and Dynamical Transport}

Some of the most important many-body correlation functions to determine
are the correlation functions related to transport properties via
their corresponding Kubo formulas~\cite{kubo_1957,greenwood_1958}.  
Transport can also 
be solved exactly
in the FK model, and we illustrate here how to determine the optical
conductivity, the thermopower, the thermal conductivity, and the response
to inelastic light scattering.

The Kubo formula relates the response function to the corresponding 
current-current correlation functions.  Before we discuss how to calculate
such correlation functions, it makes sense to describe the different 
transport currents that we consider.  Each current (except the heat
current) takes the generic form
\begin{equation}
j_a(\textbf{q})=\sum_{\sigma=1}^{2s+1}\sum_{\textbf{k}}\gamma_a(\textbf{k}+
\textbf{q}/2)c^\dagger_{\textbf{k}+\textbf{q}\sigma}c_{\textbf{k}\sigma},
\label{eq: current_general}
\end{equation}
with $\gamma_a(\textbf{k})$ the corresponding current vertex function.
The heat-current operator takes the form
\begin{equation}
j_Q(0)=\sum_{\sigma =1}^{2s+1}\left \{ \sum_{\textbf{k}}\gamma_Q(\textbf{k})
c^\dagger_{\textbf{k}\sigma}c_{\textbf{k}\sigma}+\sum_{\textbf{k}
\textbf{k}^\prime}{\bar\gamma}_Q (\textbf{k},\textbf{k}^\prime)
c^\dagger_{\textbf{k}\sigma}c_{\textbf{k}^\prime\sigma}\right \}
\label{eq: current_heat}
\end{equation}
with $\gamma_Q(\textbf{k})$ the vertex that arises from the kinetic energy
and $\bar\gamma_Q (\textbf{k},\textbf{k}^\prime)$ the vertex from the
potential energy.
For conventional charge, and thermal transport, we are interested only in the
$\textbf{q}\rightarrow 0$ (uniform) limit of the current operators,
but the finite-\textbf{q} dependence is important for inelastic (Raman) scattering
of X-ray light.  The current vertex functions are summarized in 
Table~\ref{table: current_vertex}.

\begin{table}
\caption{Current vertex functions for use in Eqs.~(\ref{eq: current_general})
and (\ref{eq: current_heat}).
The symbols $e^I_\alpha$ and $e^O_\alpha$ are the polarization
vectors for the incident and outgoing light, respectively, and $a$ denotes the
lattice spacing. An overall factor depending on properties of the incident
and outgoing light is neglected for the inelastic light scattering vertex.
\label{table: current_vertex}
}
\begin{ruledtabular}
\begin{tabular}{ll}
Current type&vertex function $\gamma_a(\textbf{k})$\\
\colrule
Charge&$\gamma_n(\textbf{k})=iea\nabla \epsilon(\textbf{k})/\hbar$\\
Heat&$\gamma_Q(\textbf{k})=
ia\nabla \epsilon(\textbf{k})[\epsilon(\textbf{k})-\mu]/\hbar$\\
~&${\bar\gamma}_Q (\textbf{k},\textbf{k}^\prime)=iaU[\nabla \epsilon(\textbf{k})
+\nabla \epsilon(\textbf{k}^\prime)]W(\textbf{k}-\textbf{k}^\prime)/2\hbar$\\
~&$W(\textbf{k})=\sum_j \exp(-i\textbf{k}\cdot\textbf{R}_j)\sum_\eta
f^\dagger_{j\eta}f_{j\eta}/V$\\

Inelastic&$\gamma_L(\textbf{k})=
\sum_{\alpha\beta}e^I_\alpha\frac{\partial^2\epsilon(\textbf{k})}
{\partial \textbf{k}_\alpha\partial\textbf{k}_\beta}e^{O*}_\beta$\\
light scattering&
\end{tabular}
\end{ruledtabular}
\end{table}

The Dyson equation for any current-current correlation function takes the
form shown in Fig.~\ref{fig: curr_curr_dyson}, which is similar to that
given by Eq.~(\ref{eq: chi_dyn_dyson}), but the bare and interacting 
susceptibilities are now different due to the corresponding $\gamma_a$ factors.
Note that there are two coupled equations illustrated in 
Figs.~\ref{fig: curr_curr_dyson}~(a) and (b); these equations differ by the
number of $\gamma_a$ factors in them (of course, there is only one equation
when $\gamma_a=1$ as we saw for the dynamical charge susceptibility).  Note 
further, that one could evaluate
mixed current-current correlation functions where there is a $\gamma_a$ on
the left vertex and a $\gamma_b$ on the right vertex, but we don't describe
that case in detail here. The irreducible vertex function $\Gamma$ is the
dynamical charge vertex of Eq.~(\ref{eq: dyn_vertex_final}) which has the 
full symmetry of the lattice.  If the vertex factor $\gamma_a$ does not have
a projection onto the full symmetry of the lattice, then
there are no vertex corrections from the local dynamical charge 
vertex~\cite{khurana_1990}.  This occurs whenever
\begin{equation}
\sum_{\textbf{k}}\gamma_a(\textbf{k}+\frac{\textbf{q}}{2})G_{n\sigma}(\textbf{k})
G_{n+l\sigma}(\textbf{k}+\textbf{q})=0.
\label{eq: no_verts}
\end{equation}
If Eq.~(\ref{eq: no_verts}) is satisfied, then the corresponding current-current
correlation function is given by the bare bubble depicted by the first
diagram on the right-hand-side of Fig.~\ref{fig: curr_curr_dyson}~(a).

\begin{figure}[htb]
\epsfxsize=3.0in
\epsffile{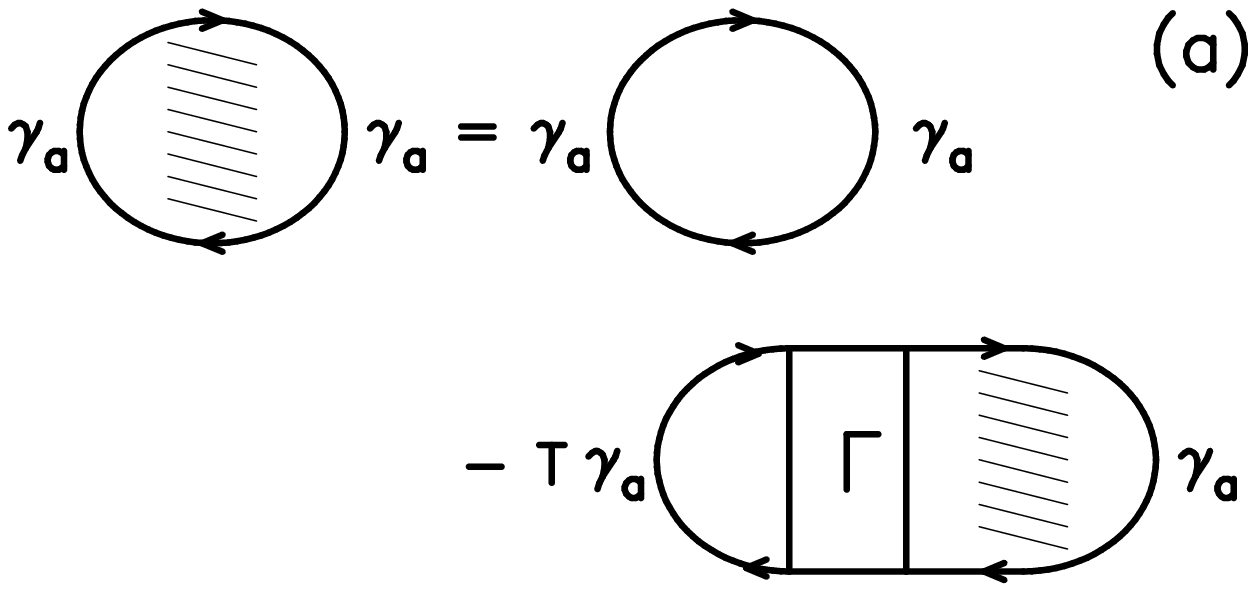}
\epsfxsize=2.7in
\epsffile{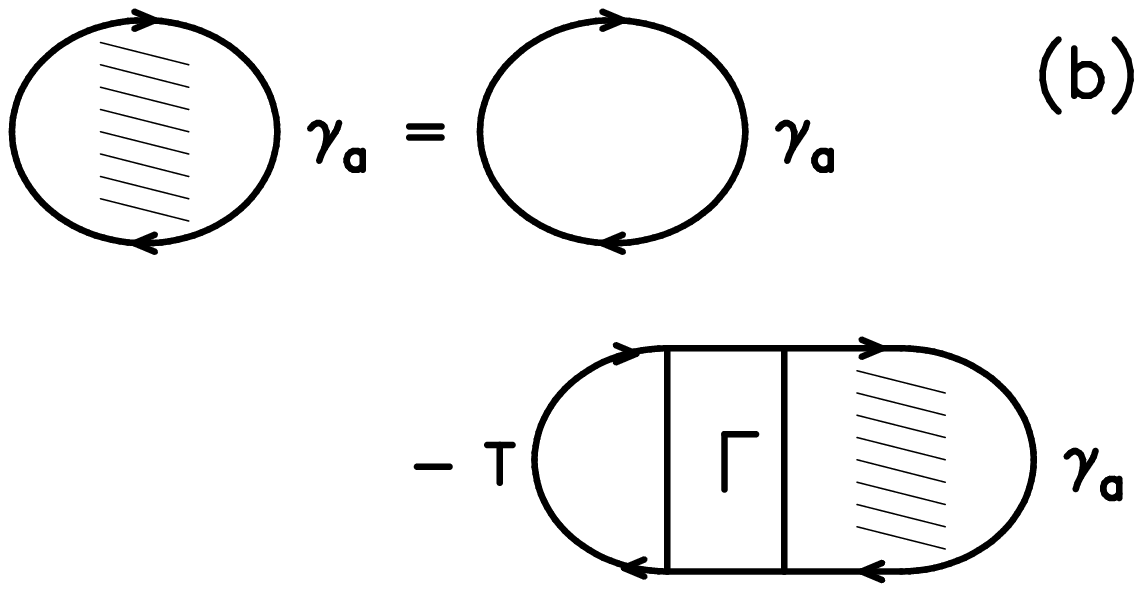}
\caption{\label{fig: curr_curr_dyson} Coupled Dyson equations for current-current
correlation functions described by the vertex function $\gamma_a$.  Panel (a)
depicts the Dyson equation for the interacting correlation function, while
panel (b) is the supplemental equation needed to solve for the correlation
function. The symbol $\Gamma$ stands for the local dynamical irreducible
charge vertex given in Eq.~(\ref{eq: dyn_vertex_final}).
In situations where Eq.~(\ref{eq: no_verts}) is satisfied, there are no 
charge vertex corrections, and the correlation function is simply given by
the first diagram on the right hand side of panel (a).
}
\end{figure}   

One case where there are no vertex corrections is the optical conductivity,
which is constructed from a $\textbf{q}=0$ current-current correlation function.
We consider the case where the only effect of the magnetic field is from the
Zeeman splitting of the states (which is accurate in large dimensions,
because the phase factors induced in the hopping matrix occurs only in one 
dimension).
We illustrate here how to calculate the optical conductivity on the hypercubic 
lattice~\cite{moeller_ruckenstein_schmittrink_1992,pruschke_cox_jarrell_1993a,%
pruschke_cox_jarrell_1993b,pruschke_jarrell_freericks_1995}. The starting point
is the current-current correlation function on the imaginary lattice, which
is equal to the bare bubble diagram in Fig.~\ref{fig: curr_curr_dyson}~(a)
\begin{equation}
\chi^{nn}(i\nu_l)=\frac{e^2}{da^{d-2}\hbar^2}\sum_\sigma \sum_n\sum_{\textbf{k}}
\sin^2 (\textbf{k}_1)G_{n\sigma}(\textbf{k})G_{n+l\sigma}(\textbf{k}),
\label{eq: oc_imag}
\end{equation}
where we have chosen the 1 direction for the velocity operator and the $n$
superscript denotes the particle-number (charge) current operator
(note our energy unit $t^*$ equals 1 so it is suppressed). The average
of $\sin^2(\textbf{k}_1)$ (times a function of $\epsilon_{\textbf{k}}$) is
equal to $1/2$ times the integral over $\epsilon$ of the function times the 
density of states.  Hence, the current-current correlation function becomes
\begin{eqnarray}
\chi^{nn}(i\nu_l)&=&\frac{e^2}{2da^{d-2}\hbar^2}
\sum_{\sigma=1}^{2s+1}
\sum_{n=-\infty}^{\infty}\int_{-\infty}^{\infty}d\epsilon\rho(\epsilon)
\cr
&\times&\frac{1}{i\omega_n+\mu+g\mu_BHm_\sigma-\Sigma_{n\sigma}-\epsilon}\cr
&\times& \frac{1}{i\omega_{n+l}+\mu+g\mu_BHm_\sigma-\Sigma_{n+l\sigma}-\epsilon}.
\label{eq: oc_imag2}
\end{eqnarray}
Note that the optical conductivity is a $1/d$ correction.
The analytic continuation is performed in the same fashion as was done for
the dynamic charge susceptibility.  The optical conductivity is defined to
be $\sigma(\omega)=\textrm{Im}\chi^{nn}(\nu)/\nu$, so the final result is
\begin{eqnarray}
\sigma(\nu)&=&\sigma_0\sum_{\sigma=1}^{2s+1}\int_{-\infty}^{\infty}d\epsilon
\rho(\epsilon)\int_{-\infty}^\infty d\omega A_\sigma(\epsilon,\omega)
A_\sigma(\epsilon,\omega+\nu)\cr
&\times&\frac{f(\omega)-f(\omega+\nu)}{\nu},
\label{eq: oc_final}
\end{eqnarray}
with $\sigma_0=e^2\pi^2/hda^{d-2}$ (which is approximately $4.3\times 10^3
\Omega^{-1}$-cm$^{-1}$ in $d=3$ with $a\approx 3\times 10^{-8}$~cm). Note that
one factor of $\hbar$ is needed to construct the dimensionless energy unit when
multiplied by $\nu$ in the denominator, which is why only one factor of $h$
appears in $\sigma_0$.

Eq.~(\ref{eq: oc_final}) can be simplified further by performing the integral
over $\epsilon$.  This is done by using Eq.~(\ref{eq: spectral}) to rewrite
both spectral functions as an imaginary part of a Green's function, and noting 
that the
imaginary part of $z$ is $(z-z^*)/2i$.  The integral over $\epsilon$ can
then be performed after expanding each integrand by partial fractions to yield
\begin{eqnarray}
\sigma(\nu)&=&\frac{\sigma_0}{2\pi^2}\sum_{\sigma=1}^{2s+1}
\int_{-\infty}^{\infty}d\omega
\textrm{Re}\Biggr \{\frac{G_\sigma(\omega)-G_\sigma^*(\omega+\nu)}
{\nu+\Sigma_\sigma(\omega)-\Sigma_\sigma^*(\omega+\nu)}\cr
&-& \frac{G_\sigma(\omega)-G_\sigma(\omega+\nu)}
{\nu+\Sigma_\sigma(\omega)-\Sigma_\sigma(\omega+\nu)}\Biggr \}
\frac{f(\omega)-f(\omega+\nu)}{\nu}.
\label{eq: oc_integrated}
\end{eqnarray}

On another lattice, the optical conductivity takes the form of 
Eq.~(\ref{eq: oc_final}) but with an extra factor of $v^2(\epsilon)$.
Enforcing the optical sum rule (which relates the integral of the optical
conductivity to the average kinetic energy) \cite{maldague_1977,%
chattopadhyay_millis_dassarma_2000}
produces a differential equation for $v^2(\epsilon)$
\begin{equation}
\frac{d}{d\epsilon}[v^2(\epsilon)\rho(\epsilon)]+\epsilon\rho(\epsilon)=0,
\label{eq: oc_diffeq}
\end{equation}
with the boundary condition that $v^2(\epsilon)\rightarrow 0$ at the band edges
(where appropriate).
On the hypercubic lattice one needs to sum the optical conductivity over all
$d$ axis directions to be able to be put in the above form; solving the
differential equation gives $v^2(\epsilon)=1/2$ as expected.  For the infinite
coordination Bethe lattice, one finds $v^2(\epsilon)=(4-\epsilon^2)/3$. This 
agrees with the conjectured result~\cite{velicky_1969,chung_freericks_1998}, but 
it relies on enforcing the sum rule for the Bethe lattice, which has not been
established independently~\cite{vandongen_blumer_2002}.

If we take the limit $\nu\rightarrow 0$ on the hypercubic lattice, we find
\begin{equation}
\sigma_{dc}=\sigma_0\int_{-\infty}^\infty d\omega\left (-\frac{df(\omega)}
{d\omega}\right ) \sum_{\sigma=1}^{2s+1}\tau_\sigma(\omega),
\label{eq: sigmadc}
\end{equation}
with the spin-dependent relaxation time equal to
\begin{eqnarray}
\tau_\sigma(\omega)&=&\frac{\textrm{Im}G_\sigma(\omega)}
{\textrm{Im}\Sigma_\sigma(\omega)}+2\cr
&-&2\textrm{Re}\{[\omega+\mu+g\mu_BHm_\sigma-
\Sigma_\sigma(\omega)]G_\sigma(\omega)\}.
\label{eq: tau}
\end{eqnarray}
In addition, one is often interested in thermal transport quantities such as
the thermopower $S$ and the electronic contribution to the thermal conductivity
$\kappa_e$.  These three quantities are usually expressed in terms of three
different transport coefficients $L_{11}$, $L_{12}=L_{21}$ and $L_{22}$
as follows:
\begin{equation}
\sigma_{dc}=e^2L_{11},
\label{eq: sigmadc_l}
\end{equation}
\begin{equation}
S=\frac{k_B}{|e|T}\frac{L_{12}}{L_{11}},
\label{eq: s_l}
\end{equation}
and
\begin{equation}
\kappa_e=\frac{k_B^2}{T}\left [ L_{22}-\frac{L_{12}L_{21}}{L_{11}}\right ],
\label{eq: kappa_l}
\end{equation}
with $k_B$ the Boltzmann constant.  The individual transport coefficients
are determined by the zero-frequency limit of the analytic continuation of the
relevant polarization operators $L_{ij}=\lim_{\nu\rightarrow 0}\textrm{Re}[i
\bar L_{ij}(\nu)/\nu]$ with
\begin{equation}
\bar L_{11}(i\nu_l)= \int_0^\beta d\tau e^{i\nu_l\tau}\textrm{Tr}_{cf}
\frac{\langle \mathcal{T}_\tau e^{-\beta\mathcal{H}}j_n(\tau)j_n(0)\rangle}{\mathcal{Z}_L},
\label{eq: l11}
\end{equation}
\begin{equation}
\bar L_{12}(i\nu_l)= \int_0^\beta d\tau e^{i\nu_l\tau}\textrm{Tr}_{cf}
\frac{\langle \mathcal{T}_\tau e^{-\beta\mathcal{H}}j_n(\tau)j_Q(0)\rangle}{\mathcal{Z}_L},
\label{eq: l12}
\end{equation}
and
\begin{equation}
\bar L_{22}(i\nu_l)= \int_0^\beta d\tau e^{i\nu_l\tau}\textrm{Tr}_{cf}
\frac{\langle \mathcal{T}_\tau e^{-\beta\mathcal{H}}j_Q(\tau)j_Q(0)\rangle}{\mathcal{Z}_L},
\label{eq: l22}
\end{equation}
where the subscripts $n$ and $Q$ denote the number (charge) and heat currents
respectively (and we suppressed the Cartesian vector indices).  All of these
correlation functions are determined by their corresponding bare bubbles, 
because there are no vertex corrections for any of them.  Note that our sign
convention for the thermopower is that of \textcite{ashcroft_mermin_1976},
where the coefficient multiplying $L_{12}/L_{11}$
is positive for negatively charged carriers. This leads
to the situation where electron-like transport  has a positive thermopower,
and hole-like transport has a negative thermopower at low temperature.  (The
sign of the thermopower varies depending on what convention is used for
the definition of $S$ in terms of the transport coefficients.)  A 
theorem by~\textcite{jonson_mahan_1980,jonson_mahan_1990} says that even in
a correlated system, there is a simple relation between these different
transport coefficients [that they reproduce the so-called Mott noninteracting
form~\cite{chester_thellung_1961}]
\begin{equation}
L_{ij}=\frac{\sigma_0}{e^2}\int_{-\infty}^\infty d\omega
\left (-\frac{df(\omega)}
{d\omega}\right ) \sum_{\sigma=1}^{2s+1}\tau_\sigma(\omega)\omega^{i+j-2}.
\label{eq: mott}
\end{equation}
What is remarkable, is that in the case of the FK model, one can explicitly 
calculate the relevant correlation functions and verify directly the
Jonson-Mahan theorem~\cite{freericks_zlatic_2001b,%
freericks_zlatic_2001_erratum}.  The derivation is too long
to reproduce here.  Note that we have taken the more modern definitions
of the $L_{ij}$ coefficients here, which has one less power of $T$ than
the normalization used by \textcite{jonson_mahan_1980,jonson_mahan_1990,%
freericks_zlatic_2001b,freericks_zlatic_2001_erratum}.

The final transport property we determine is the scattering of inelastic light. 
When optical photons are used, this corresponds to conventional electronic
Raman scattering: inelastic light scattering off of the charge excitations
of the many-body system.  When higher-energy (X-rays) are employed, one
scatters the photon off of the momentum and frequency-dependent charge
excitations of the system.  The Raman scattering limit results in the
limit $\textbf{q}\rightarrow 0$.  For simplicity, we consider only the
spinless case, with $H=0$.

Inelastic light scattering depends on the polarizations of the incident and 
outgoing light.  As such, it provides some additional symmetry resolution
over and above the elastic scattering of an optical conductivity measurement.
There are traditionally three main symmetries considered in Raman scattering
experiments: (i) $A_{\textrm{1g}}$ which has the full symmetry of the lattice;
(ii) $B_{\textrm{1g}}$ which has a d-wave symmetry and (iii) $B_{\textrm{2g}}$
which is another d-wave symmetry.  Each symmetry is chosen by different 
polarizations for the incident and scattered light.  Here we concentrate on
nonresonant Raman scattering, where the vertex functions are not functions
of the photon energies.  If we sum over the $d$
pairs of polarizations, where $e^I=e^O$ and each 
vector points along each of the
different Cartesian axes, then we have the $A_{\textrm{1g}}$ sector.  If we
choose $e^I=(1,1,1,...)$ and $e^O=(1,-1,1,-1,...)$, then we have the 
$B_{\textrm{1g}}$ sector.  And if we choose $e^I=(1,0,1,0,...)$ and
$e^O=(0,-1,0,-1,...)$ then we have the $B_{\textrm{2g}}$ sector.  If we
have just nearest-neighbor hopping, then the $B_{\textrm{2g}}$ response 
vanishes because $\gamma_{B_{\textrm{2g}}}=0$.  Following the form given
in Table~\ref{table: current_vertex}, we find $\gamma_{A_{\textrm{1g}}}
(\textbf{q})=-\epsilon(\textbf{q})$ and $\gamma_{B_{\textrm{1g}}}(\textbf{q})=
t^*\sum_{j=1}^\infty \cos \textbf{q}_j (-1)^j/\sqrt{d}$.

A straightforward calculation, using Eq.~(\ref{eq: no_verts}), shows that
the $B_{\textrm{1g}}$ response has no vertex corrections on the zone
diagonal $\textbf{q}=(q,q,q,q,...)$.  Hence, the zone-diagonal
$B_{\textrm{1g}}$ response
is the bare bubble.  The $A_{\textrm{1g}}$ response (and the B$_{\textrm{1g}}$
off of the zone diagonal), on  the other hand,
does have vertex corrections, and is more complicated.  The calculation
of each response function is straightforward, but tedious.  One needs to
first solve the coupled equations depicted in Fig.~\ref{fig: curr_curr_dyson}
on the imaginary axis and then perform the analytic continuation as we have
done previously for other 
response functions.  The end result is quite long and is
summarized in Tables~\ref{table: xray_response} and \ref{table: chi0_raman}.

\begin{table*}[htb]
\caption{Response functions for inelastic X-ray scattering by a photon
of momentum $\textbf{q}=(q,q,q,...)$ along the zone diagonal
described by the parameter $X(\textbf{q})=\cos q$ for the $A_{\textrm{1g}}$
and $B_{\textrm{1g}}$ sectors. The symbols with $\chi_0$ appear in 
Table~\ref{table: chi0_raman} and $*$ denotes complex conjugation.
\label{table: xray_response}}
\begin{ruledtabular}
\begin{tabular}{l}
 Scattering response function\\
\colrule
$\chi_{A_{\textrm{1g}}}(\textbf{q},\nu)=\frac{i}{2\pi}\int_{-\infty}^{\infty}
d\omega$\\
$\left \{ f(\omega)\frac{\bar\chi_0(\omega;X,\nu)+\frac{\Sigma(\omega)-
\Sigma(\omega+\nu)}{G(\omega)-G(\omega+\nu)}[\chi_0(\omega;X,\nu)\bar\chi_0(
\omega;X,\nu)-\chi_0^{\prime 2}(\omega;X,\nu)]}
{1+\frac{\Sigma(\omega)-
\Sigma(\omega+\nu)}{G(\omega)-G(\omega+\nu)}\chi_0(\omega;X,\nu)}
-f(\omega+\nu)\frac{\bar\chi_0^*(\omega;X,\nu)+\frac{\Sigma^*(\omega)-
\Sigma^*(\omega+\nu)}{G^*(\omega)-G^*(\omega+\nu)}[\chi_0^*(\omega;X,\nu)
\bar\chi_0^*(\omega;X,\nu)-\chi_0^{\prime *2}(\omega;X,\nu)]}
{1+\frac{\Sigma^*(\omega)-
\Sigma^*(\omega+\nu)}{G^*(\omega)-G^*(\omega+\nu)}\chi_0^*(\omega;X,\nu)}
\right .$\\
$\left . -[f(\omega)-f(\omega+\nu)]
\frac{\tilde{\bar\chi}_0(\omega;X,\nu)+\frac{\Sigma^*(\omega)-
\Sigma(\omega+\nu)}{G^*(\omega)-G(\omega+\nu)}[\tilde\chi_0(\omega;X,\nu)
\tilde{\bar\chi}_0(\omega;X,\nu)-\tilde\chi_0^{\prime 2}(\omega;X,\nu)]}
{1+\frac{\Sigma^*(\omega)-
\Sigma(\omega+\nu)}{G^*(\omega)-G(\omega+\nu)}\tilde\chi_0(\omega;X,\nu)}
\right \}$\\
\colrule
$\chi_{B_{\textrm{1g}}}(\textbf{q},\nu)=\frac{i}{4\pi}\int_{-\infty}^{\infty}
d\omega\left \{ f(\omega)\chi_0(\omega;X,\nu)
-f(\omega+\nu)\chi_0^*(\omega;X,\nu)
-[f(\omega)-f(\omega+\nu)]
\tilde\chi_0(\omega;X,\nu)
\right \}$\\
\end{tabular}
\end{ruledtabular}
\end{table*}

\begin{table*}
\caption{Symbols $\chi_0$, $\chi_0^\prime$, $\bar\chi_0$, $\tilde\chi_0$,
$\tilde\chi_0^\prime$, and $\tilde{\bar\chi}_0$ that appear in 
Table~\ref{table: xray_response}
\label{table: chi0_raman}}
\begin{ruledtabular}
\begin{tabular}{l} 
Symbol\\
\colrule
$\chi_0(\omega;X,\nu)=-\int_{-\infty}^{\infty}d\epsilon\rho(\epsilon)
\frac{1}{\omega+\mu-\Sigma(\omega)-\epsilon}\frac{1}{\sqrt{1-X^2}}F_\infty
\left ( \frac{\omega+\nu+\mu-\Sigma(\omega+\nu)-X\epsilon}{\sqrt{1-X^2}}\right )$
\\
$\tilde\chi_0(\omega;X,\nu)=-\int_{-\infty}^{\infty}d\epsilon\rho(\epsilon)
\frac{1}{\omega+\mu-\Sigma^*(\omega)-\epsilon}\frac{1}{\sqrt{1-X^2}}F_\infty
\left ( \frac{\omega+\nu+\mu-\Sigma(\omega+\nu)-X\epsilon}{\sqrt{1-X^2}}\right )$
\\
$\chi_0^\prime(\omega;X,\nu)=-\sqrt{\frac{1+X}{8}}
\int_{-\infty}^{\infty}d\epsilon \rho(\epsilon)\left \{
\frac{1}{[\omega+\mu-\Sigma(\omega)-\epsilon]^2}\frac{1}{\sqrt{1-X^2}}F_\infty
\left ( \frac{\omega+\nu+\mu-\Sigma(\omega+\nu)-X\epsilon}{\sqrt{1-X^2}}\right )
\right .$\\
$ \left .  +\frac{1}{\omega+\mu-\Sigma(\omega)-\epsilon}\frac{2}{1-X^2}
\left [ -1 +\frac{\omega+\nu+\mu-\Sigma(\omega+\nu)}{\sqrt{1-X^2}}
F_\infty\left ( \frac{\omega+\nu+\mu-\Sigma(\omega+\nu)-X\epsilon}
{\sqrt{1-X^2}}\right ) \right ] \right \}
$
\\
$\tilde\chi_0^\prime(\omega;X,\nu)=-\sqrt{\frac{1+X}{8}}
\int_{-\infty}^{\infty}d\epsilon \rho(\epsilon)\left \{
\frac{1}{[\omega+\mu-\Sigma^*(\omega)-\epsilon]^2}\frac{1}{\sqrt{1-X^2}}F_\infty
\left ( \frac{\omega+\nu+\mu-\Sigma(\omega+\nu)-X\epsilon}{\sqrt{1-X^2}}\right )
\right .$\\
$\left .  +\frac{1}{\omega+\mu-\Sigma^*(\omega)-\epsilon}\frac{2}{1-X^2}
\left [ -1 +\frac{\omega+\nu+\mu-\Sigma(\omega+\nu)}{\sqrt{1-X^2}}
F_\infty\left ( \frac{\omega+\nu+\mu-\Sigma(\omega+\nu)-X\epsilon}
{\sqrt{1-X^2}}\right ) \right ] \right \}
$
\\ 
$\bar\chi_0(\omega;X,\nu)=\frac{1}{2}\chi_0(\omega;X,\nu)-\frac{1+X}{4}
\int_{-\infty}^{\infty}d\epsilon \rho(\epsilon)\left \{
\frac{1}{[\omega+\mu-\Sigma(\omega)-\epsilon]^3}\frac{1}{\sqrt{1-X^2}}F_\infty
\left ( \frac{\omega+\nu+\mu-\Sigma(\omega+\nu)-X\epsilon}{\sqrt{1-X^2}}\right )
\right .$\\
$ +\frac{1}{[\omega+\mu-\Sigma(\omega)-\epsilon]^2}\frac{2}{1-X^2}
\left [ -1 +\frac{\omega+\nu+\mu-\Sigma(\omega+\nu)}{\sqrt{1-X^2}}
F_\infty\left ( \frac{\omega+\nu+\mu-\Sigma(\omega+\nu)-X\epsilon}
{\sqrt{1-X^2}}\right ) \right ] 
$
\\                        
$ \left .+\frac{1}{\omega+\mu-\Sigma(\omega)-\epsilon}\frac{1}{(1-X^2)^{3/2}}
\left [ -F_\infty\left ( \frac{\omega+\nu+\mu-\Sigma(\omega+\nu)-X\epsilon}
{\sqrt{1-X^2}}\right )+\frac{2\{ \omega+\nu-\Sigma(\omega+\nu)
-X\epsilon\}}{\sqrt{1-X^2}}
\{-1 +\frac{\omega+\nu+\mu-\Sigma(\omega+\nu)}{\sqrt{1-X^2}}
F_\infty\left ( \frac{\omega+\nu+\mu-\Sigma(\omega+\nu)-X\epsilon}
{\sqrt{1-X^2}}\right )\}\right ]\right \}$
\\
$\tilde{\bar\chi}_0(\omega;X,\nu)=\frac{1}{2}\tilde\chi_0(\omega;X,\nu)
-\frac{1+X}{4} \int_{-\infty}^{\infty}d\epsilon \rho(\epsilon)\left \{
\frac{1}{[\omega+\mu-\Sigma^*(\omega)-\epsilon]^3}\frac{1}{\sqrt{1-X^2}}F_\infty
\left ( \frac{\omega+\nu+\mu-\Sigma(\omega+\nu)-X\epsilon}{\sqrt{1-X^2}}\right )
\right .$\\
$ +\frac{1}{[\omega+\mu-\Sigma^*(\omega)-\epsilon]^2}\frac{2}{1-X^2}
\left [ -1 +\frac{\omega+\nu+\mu-\Sigma(\omega+\nu)}{\sqrt{1-X^2}}
F_\infty\left ( \frac{\omega+\nu+\mu-\Sigma(\omega+\nu)-X\epsilon}
{\sqrt{1-X^2}}\right ) \right ]
$
\\
$ \left .+\frac{1}{\omega+\mu-\Sigma^*(\omega)-\epsilon}\frac{1}{(1-X^2)^{3/2}}
\left [ -F_\infty\left ( \frac{\omega+\nu+\mu-\Sigma(\omega+\nu)-X\epsilon}
{\sqrt{1-X^2}}\right )+\frac{2\{ \omega+\nu-\Sigma(\omega+\nu) 
-X\epsilon\}}{\sqrt{1-X^2}}  
\{-1 +\frac{\omega+\nu+\mu-\Sigma(\omega+\nu)}{\sqrt{1-X^2}} 
F_\infty\left ( \frac{\omega+\nu+\mu-\Sigma(\omega+\nu)-X\epsilon}
{\sqrt{1-X^2}}\right )\}\right ]\right \}$
\\
\end{tabular}
\end{ruledtabular}
\end{table*}

It is interesting to take the limit of conventional Raman scattering with
optical photons, where $\textbf{q}\rightarrow 0$ ($X\rightarrow 1$).  In this
case, one finds $\chi_0=-\int d\epsilon \rho(\epsilon)/[\omega+\mu-\Sigma(\omega)
-\epsilon][\omega+\nu+\mu-\Sigma(\omega+\nu)-\epsilon]$ and
the $B_{\textrm{1g}}$ response simplifies to
\begin{eqnarray}
\chi_{B_{\textrm{1g}}}(\nu)&=&\frac{i}{4\pi}\int_{-\infty}^{\infty}d\omega
\int_{-\infty}^{\infty}d\epsilon \rho(\epsilon)\cr
&\times&\left \{
f(\omega)\frac{1}{\omega+\nu+\mu-\Sigma(\omega+\nu)-\epsilon}\right .\cr
&\times&
\left [ \frac{1}{\omega+\mu-\Sigma(\omega)-\epsilon}-
\frac{1}{\omega+\mu-\Sigma^*(\omega)-\epsilon}\right ] \cr
&-&f(\omega+\nu)\frac{1}{\omega+\mu-\Sigma^*(\omega)-\epsilon}\cr
&\times&
\left [ \frac{1}{\omega+\nu+\mu-\Sigma^*(\omega+\nu)-\epsilon}\right .\cr
&-&\left . \left .
\frac{1}{\omega+\nu+\mu-\Sigma(\omega+\nu)-\epsilon}\right ]\right \}.
\label{eq: b1g_q=0}
\end{eqnarray}
Now using the definition of the spectral function in Eq.~(\ref{eq: spectral}),
and taking the imaginary part of Eq.~(\ref{eq: b1g_q=0}) produces the
so-called Shastry-Shraiman relation~\cite{shastry_shraiman_1990,%
shastry_shraiman_1991} which relates the imaginary part of the nonresonant
$B_{\textrm{1g}}$ Raman response function to the optical conductivity
\begin{equation}
\textrm{Im}\chi_{B_{\textrm{1g}}}(\nu)\propto \nu\sigma(\nu),
\label{eq: ss_relation}
\end{equation}
which was first proved in~\textcite{freericks_devereaux_2001b}.

Another interesting limit is the $\textbf{q}=(\pi,\pi,\pi,...)$ ($X=-1$) limit,
where we find $\chi_0^\prime=\tilde\chi_0^\prime=0$, $\bar\chi_0=
\chi_0/2$, and $\tilde{\bar\chi}_0=\tilde\chi_0/2$.  In this case, the
renormalization due to the dynamical charge vertex exactly cancels in
the numerator and denominator, and one finds \textit{the same result}
for the inelastic X-ray scattering in the $A_{\textrm{1g}}$ and $B_{\textrm{1g}}$
sectors.  Hence, a polarized measurement of the inelastic X-ray scattering
at the zone boundary determines the relative importance of nonlocal charge 
fluctuations to the strongly correlated system.  This is quantified by comparing
the scattering in the two different symmetry sectors; the difference in the
results is a measure of the nonlocal correlations.

We have produced results for inelastic X-ray scattering only along the 
zone diagonal.  One could choose other directions as well.  In general, we
find that the symmetries from different channels then mix, and both
$A_{\textrm{1g}}$ and $B_{\textrm{1g}}$ channels are renormalized by
the irreducible dynamic charge vertex.  The formulas in that case are 
complicated and will not be presented here.

One could also study resonant (or mixed) Raman response functions.  The
formalism is similar to the nonresonant case, except  it is better to work 
directly on the real axis.
These calculations have not yet been completed by anyone.  But
a simple power-counting analysis of the different symmetry sectors can be 
performed~\cite{freericks_devereaux_2001b}.  What is found is that the
$A_{\textrm{1g}}$ sector has contributions from the nonresonant, the mixed
diagrams, and the resonant diagrams.  The $B_{\textrm{1g}}$ sector is
either nonresonant or resonant, but the mixed contributions disappear as
$1/d$.  The $B_{\textrm{2g}}$ response is purely resonant.

\subsection{Single-Particle Properties (Localized Electrons)}

We now study $f$-electron properties.  Since the local $f$-electron number is 
conserved, one might believe that the localized particle dynamics are trivial.
Indeed, this is true for the particle number, but the \textit{fermionic}
electron degrees of freedom are nontrivial, even if they are restricted to
be local.  As a first step in our analysis, we can imagine writing the
statistical factor for the $f$-occupancy in terms of an effective
$f$-level $E_f^*$~\cite{czycholl_1999}
\begin{equation}
w_1=\frac{1}{1+\exp[-\beta (E_f^*-\mu_f)]}
\label{eq: flevel}
\end{equation}
which we have written explicitly for the spinless case, with suitable 
generalizations for higher-spin cases.  Notice that in many cases, either
$w_1$ changes as a function of $T$, or $\mu_f$ changes as a function of $T$,
or both.  Hence, $E_f^*$ typically has temperature dependence.  If we view
this energy as the centroid of the $f$-spectral function, we can immediately
learn about how the $f$-spectral function may change as a function of $T$.
Of course one can only learn so much from a single number.

What is more interesting is to 
evaluate the local $f$-electron Green's function following the
work of \textcite{brandt_urbanek_1992} and
\textcite{zlatic_review_2001} [see also \textcite{janis_1994} which examined
the corresponding X-ray edge problem].  The local 
$f$-electron Green's function on the imaginary axis is defined by
\begin{equation}
F_\eta(\tau)=-\textrm{Tr}_{cf}\frac{\left \langle e^{-\beta\mathcal{H}_{imp}}
S(\lambda)f_\eta(\tau)f_\eta^\dagger(0)\right\rangle}{\mathcal{Z}_{imp}},
\label{eq: f_green}
\end{equation}
for $\tau>0$ in the interaction representation [$\mathcal{Z}_{imp}$ is given by
Eq.~(\ref{eq: z_imp-final}) with $H=0$]. The impurity Hamiltonian in
zero magnetic field is $\mathcal{H}_{imp}=-\mu n^c+\sum_\eta (E_{f\eta}-\mu_f)
n^f_\eta +Un^cn^f$ and the evolution operator satisfies
\begin{equation}
S(\lambda)=\mathcal{T}_\tau
\exp \left [-\int_0^\beta d\tau \int_0^\beta d\tau^\prime \sum_\sigma
c^\dagger_\sigma(\tau)\lambda_\sigma(\tau-\tau^\prime)c_\sigma(\tau^\prime)
\right ],
\label{eq: evolution}
\end{equation}
with $\lambda_\sigma(\tau)=T\sum_n\exp(-i\omega_n\tau)\lambda_{n\sigma}$. 
The field $\lambda_{n\sigma}$ is the discrete function that
satisfies Eqs.~(\ref{eq: green_imp}--\ref{eq: green_hilbert}). Since
$f_\eta(\tau)=\exp(\tau\mathcal{H}_{imp})f_\eta(0)\exp(-\tau\mathcal{H}_{imp})$,
we find
\begin{equation}
\frac{d}{d\tau}f_\eta(\tau)=\left [ -(E_{f\eta}-\mu_f)-U\sum_\sigma c^\dagger_\sigma
(\tau)c_\sigma(\tau)\right ] f_\eta(\tau),
\label{eq: f_eom}
\end{equation}
which can be integrated to yield
\begin{eqnarray}
f_\eta(\tau)&=&e^{-(E_{f\eta}-\mu_f)\tau}\cr
&\times&\mathcal{T}_\tau \exp \left (
-U\sum_\sigma\int_0^\tau d\tau^\prime c^\dagger_\sigma(\tau^\prime)
c_\sigma(\tau^\prime)\right )f_\eta(0).\cr
&&
\label{eq: f_integrated}
\end{eqnarray}
If we define a time-dependent field 
\begin{equation}
\chi_\tau(\tau^\prime,\tau^{\prime\prime})
=-U\theta(\tau-\tau^\prime)\delta(\tau^\prime-\tau^{\prime\prime}),
\label{eq: chi_tau_def}
\end{equation}
then the $f$-electron Green's function becomes
\begin{equation}
F_\eta(\tau)=-\frac{1}{\mathcal{Z}_{imp}}e^{-(E_{f\eta}-\mu_f)\tau}\textrm{Tr}_c\mathcal{T}_\tau
\langle e^{-\beta\mathcal{H}_{0}}S(\lambda-\chi_\tau)\rangle,
\label{eq: f_green_trace}
\end{equation}
with $\mathcal{H}_0=-\mu n^c$, 
because the trace over $f$ states is restricted to $n^f=0$ for
$\tau>0$ propagation [the operator $f(0)f^\dagger(0)$ projects onto $n^f=0$
and all of the remaining operators are composed of $f^\dagger f$ pairs
which do not change the total $f$-electron number].  

The strategy for evaluating the trace over itinerant-electron states is the
same as in Section~\ref{sec: dyn_charge}.  The trace involves operator averages 
that are not time-translation invariant because the $\chi_\tau$ field is not a
function of $\tau^\prime-\tau^{\prime\prime}$ only.  We start with the
auxiliary Green's function in Eq.~(\ref{eq: gaux_mats_path}) and note that
the double Fourier transform of the $\chi_\tau$ field [analogous to 
Eq.~(\ref{eq: gaux_mats})] becomes
\begin{eqnarray}
\chi_\tau(i\nu_l)&=&T\int_0^\beta d\tau^\prime e^{i\nu_l\tau^\prime}\chi_\tau(
\tau^\prime)\cr
&=&
\Biggr\{
\begin{array}[c]{ll}
\frac{U}{i\nu_l\beta}(1-e^{i\nu_l\tau}),&l\ne 0;\\
-\frac{U\tau}{\beta},&l=0,
\end{array}
\label{eq: chi_tau_ft}
\end{eqnarray}
where our choice for normalizing the Fourier transform has an extra factor of
$T$ for convenience here [in other words $\chi_\tau(i\nu_l)$ has dimensions
of energy] and the bosonic Matusbara frequency arises from the difference of
two fermionic Matsubara frequencies.
Since $\chi_\tau(i\nu_0)\ne 0$, we now have $\mathcal{Z}^{aux}(-\chi_\tau)
=[1+\exp(\beta\mu-U\tau) ]^{2s+1}$.
The partition function can also be expressed as the determinant of the
matrix that appears in the exponent in Eq.~(\ref{eq: gaux_mats_path}).  In
the general case, we can write the partition function as an infinite product
\begin{equation}
\mathcal{Z}^{aux}(-\chi_\tau)=\left [ 2e^{\beta(\mu+\chi_0)/2}\prod_{n=-\infty}^{\infty}
\frac{i\omega_n+\mu+\chi_0}{i\omega_n}\right ]^{2s+1},
\label{eq: zaux_general_prod}
\end{equation}
which is equal to the product of the diagonal elements of the corresponding
matrix
\begin{equation}
M_{mn}=(i\omega_m+\mu)\delta_{mn}+\chi_{m-n},
\label{eq: mdef}
\end{equation}
up to an overall normalization factor.  \textit{Hence, the determinant of
the matrix in Eq.~(\ref{eq: mdef}) is simply a product of the diagonal
elements!}  This is surprising, since one would naively expect that the
determinant depended on the higher Fourier modes of $\chi$.  We can show this
result explicitly, by checking that
\begin{equation}
\phi^{(m^\prime)}_m=T\int_0^\beta d\tau e^{(i\omega_m-i\omega_{m^\prime})\tau}
\exp\{\int_0^\tau d\tau^\prime [\chi(\tau^\prime)-\chi_0]\}
\label{eq: m_eigenvector}
\end{equation}
is the $m^\prime$-th eigenvector, with eigenvalue $i\omega_{m^\prime}+\mu+\chi_0$.
The verification, is straightforward, provided one first writes $\chi_{m-n}$ 
in terms of the integral over $\tau$, and notes that $\chi(\tau)-\chi_0$ times the
last exponential factor in Eq.~(\ref{eq: m_eigenvector}) is equal to the
negative $\tau$ derivative of the exponential factor.  The eigenvalue 
can then be determined by evaluating the integral by parts; we need to subtract
the $\chi_0$ factor from the exponent in order for the boundary terms to vanish
when one performs the integration by parts.

Knowing an explicit form for the eigenvectors then allows us to determine the
auxiliary Green's function in the general case.  As before, we imagine adding
an infinitesimal field $\bar\chi_{mn}=\bar\chi \delta_{mm_0}\delta_{nn_0}$
to the matrix $M_{mn}$.  The auxiliary Green's function is determined by
taking the logarithmic derivative of the partition function with respect
to $\bar\chi$: $g^{aux}_\sigma(i\omega_{n_0},i\omega_{m_0})=
\partial \ln \mathcal{Z}^{aux}(-\chi_\tau)
/\partial\bar\chi$. This derivative is easy to calculate,
because it says we need only determine the shift in each of the eigenvalues
for $\mathcal{Z}^{aux}(-\chi_\tau)$ 
to first order in $\bar\chi$.  In the presence of $\bar\chi$, it is easy to
show that
\begin{equation}
E^{(m^\prime)}=i\omega_{m^\prime}+\mu+\chi_0+\bar\chi
\phi^{(m^\prime)*}_{m_0}\phi^{(m^\prime)}_{n_0}+O(\bar\chi^2),
\label{eq: eigenvalue_shift}
\end{equation}
which leads then to
\begin{eqnarray}
g^{aux}_\sigma(i\omega_{n},i\omega_m)&=&
\sum_{m^\prime=-\infty}^{\infty}\frac{\phi^{(m^\prime)*}_m\phi^{(m^\prime)}_n}
{i\omega_{m^\prime}+\mu+\chi_0}\cr
&=&T\int_0^\beta d\tau\int_0^\beta d\tau^\prime
e^{i\omega_n\tau-i\omega_m\tau^\prime}\cr
&\times&e^{\int_0^\tau d\bar\tau [\chi(\bar\tau)-
\chi_0]}e^{\int_0^{\tau^\prime} d\tilde\tau [\chi^*(\tilde\tau)-\chi^*_0]}\cr
&\times&T \sum_{m^\prime=-\infty}^{\infty}\frac{e^{-i\omega_{m^\prime}(\tau-
\tau^\prime)}}{i\omega_{m^\prime}+\mu+\chi_0}.
\label{eq: gaux_general}
\end{eqnarray}
Note that if we choose $\chi(\tau)=\chi(i\nu_l)\exp(-i\nu_l
\tau)$, with $\chi(i\nu_l)$
real, then evaluating Eq.~(\ref{eq: gaux_general}) to first order in
$\chi(i\nu_l)$ reproduces Eq.~(\ref{eq: gaux_final}) as it must.  For our case,
we are interested in the real $\chi$ field of Eq.~(\ref{eq: chi_tau_def}).
Substituting this field into Eq.~(\ref{eq: gaux_general}) and performing some
tedious algebra, then yields the results for the diagonal and off-diagonal
auxiliary Green's function shown in Table~\ref{table: gaux}.  These results
generalize those of~\textcite{brandt_urbanek_1992} off of half filling and 
correct some typos in~\textcite{zlatic_review_2001}.

\begin{table*}
\caption{Auxiliary Green's function $g^{aux}_\sigma(i\omega_n,i\omega_m)=
T\xi_0 A_{nm}+T(\xi_0-1) B_{nm}$ for $n\ne m$ and
$g^{aux}_\sigma(i\omega_n,i\omega_n)=
T\xi_0 C_{nn}+T(\xi_0-1) D_{nn}$ for $n=m$. Here 
$\xi_0=1/[1+\exp(U\tau-\beta\mu)]$.
\label{table: gaux}
}
\begin{ruledtabular}
\begin{tabular}{ll}
$A_{nm}$&$\frac{1}{ i(\omega_m-\omega _n)(i\omega_m+\mu -U) }
    -
     \frac{ e^{i(\omega_n+\mu)\tau -\beta \mu } }
          { (i\omega _m+\mu )(i\omega_n+\mu ) }
+  \frac{  (e^{-(i\omega_m+\mu )\tau }+e^{-\beta \mu})
            (e^{(i\omega_n+\mu )\tau }-e^{U\tau}) }
          { (i\omega_m+\mu )(i\omega _n+\mu -U) }
-  \frac{1}{ i(\omega_m-\omega_n)(i\omega_n+\mu ) }
    +
     \frac{ e^{-i(\omega_m-\omega _n)\tau } }
          { i(\omega_m-\omega_n)(i\omega_m+\mu ) }$\\
&$ +  \frac{ e^{-(i\omega_m+\mu)\tau}[i(\omega_m-\omega_n)
            e^{U\tau }-(i\omega_m+\mu-U)
            e^{(i\omega_n+\mu)\tau }] }
          { i(\omega_m-\omega_n)(i\omega_m+\mu-U)(i\omega_n+\mu-U) }
$      \\
$B_{nm}$&$\frac{-1}{ i(\omega_m-\omega_n)(i\omega_n+\mu-U) }
    -
     \frac{  e^{-i(\omega_m-\omega_n)\tau } }
         { i(\omega_m-\omega_n)(i\omega_n+\mu) }
- \frac{ -e^{i\omega_n\tau}(e^{-i\omega _m\tau }-e^{(\mu-U)\tau })
            -e^{\beta \mu }(e^{-(i\omega _m+\mu )\tau }
            -e^{-U\tau }) }
        { (i\omega _m+\mu-U)(i\omega _n+\mu) }
+
      \frac{  i\omega_n+\mu-i(\omega_m-\omega_n)
             e^{\beta\mu-(i\omega_m+\mu )\tau } }
           {  i(\omega_m-\omega_n)(i\omega_m+\mu )(i\omega_n+\mu )}$\\
&$ +
      \frac{  e^{-i(\omega_m-\omega_n)\tau }[i\omega_n+\mu-U
             +i(\omega_m-\omega _n)
              e^{(i\omega _m+\mu -U)\tau }]}
           { i(\omega_m-\omega_n)(i\omega_m+\mu-U)(i\omega_n+\mu -U)}$\\
\colrule
$C_{nn}$&$\frac{  (i\omega _n+\mu -U)\tau
               - 1
               + e^{-(i\omega _n+\mu -U)\tau } }
             { (i\omega _n+\mu -U)^2 }
+      \frac{ 1-e^{-(i\omega _n+\mu -U)\tau}
                + e^{-\beta\mu }(e^{(i\omega_n+\mu)\tau}
                - e^{U\tau})}
             { (i\omega _n+\mu )(i\omega _n+\mu -U)}
+      \frac{  (i\omega _n+\mu )(\beta -\tau )
                - 1
                - e^{-\beta \mu +(i\omega_n+\mu)\tau }}
              {  (i\omega _n+\mu )^2}$            \\
$D_{nn}$&$\frac{ -(i\omega _n+\mu -U)\tau
               - 1
               + e^{(i\omega_n+\mu-U)\tau} }
             {(i\omega _n+\mu -U)^2}
+     \frac{ 1-e^{(i\omega_n+\mu-U)\tau}
               + e^{\beta\mu}
                (e^{-(i\omega_n+\mu )\tau }- e^{-U\tau })}
             { (i\omega _n+\mu )(i\omega _n+\mu -U) }
+     \frac{ (i\omega _n+\mu )(\tau -\beta )-1
                -e^{\beta\mu-(i\omega_n+\mu)\tau }}
             { (i\omega _n+\mu )^2 }$\\
\end{tabular}
\end{ruledtabular}
\end{table*}

Once $g^{aux}_\sigma$ is known, then we follow the same procedure as before
to calculate the relevant averages.  The object of interest is
$\textrm{Tr}_c\mathcal{T}_ \tau
\langle \exp(-\beta\mathcal{H}_{0})S(\lambda-\chi_\tau)\rangle$, which
can be viewed as the partition function corresponding to a particle moving
in both the $\lambda$ and $\chi_\tau$ fields.  Since the partition function
is the determinant of the inverse of the Green's function, and since 
the Green's function $g_\sigma(i\omega_n,i\omega_m)$ associated with the
$\lambda-\chi_\tau $ field is related to the auxiliary Green's function
via
\begin{equation}
[g_\sigma]^{-1}=[g_\sigma^{aux}]^{-1}-\lambda_{n\sigma}\delta_{nm},
\label{eq: g_gaux_lambda}
\end{equation}
we immediately conclude that
\begin{eqnarray}
&~&\textrm{Tr}_c\mathcal{T}_ \tau
\langle \exp(-\beta\mathcal{H}_{0})S(\lambda-\chi_\tau)\rangle\cr
&~&=
\mathcal{Z}^{aux}(-\chi_\tau)\prod_{\sigma=1}^{2s+1} \textrm{Det}[
\delta_{nm}-g^{aux}_\sigma(i\omega_n,i\omega_m)\lambda_{m\sigma}] .
\label{eq: final_partfcn}
\end{eqnarray}
Substituting into Eq.~(\ref{eq: f_green_trace}) yields our final result
for the $f$-electron Green's function
\begin{eqnarray}
F_\eta(\tau)&=&-\frac{1}{\mathcal{Z}_{imp}}e^{-(E_{f\eta}-\mu_f)\tau}\mathcal{Z}^{aux}(-\chi_\tau)
\cr
&\prod_{\sigma=1}^{2s+1}&
\textrm{Det} [ \delta_{nm}-g^{aux}_\sigma(i\omega_n,i\omega_m)\lambda_{m\sigma} ].
\label{eq: f_green_final}
\end{eqnarray}
One needs to recall that $\mathcal{Z}^{aux}(-\chi_\tau)$ changes with $\tau$ because the
$\chi_\tau$ field depends on $\tau$.
One can easily verify in the case where $E_{f\eta}$ has no $\eta$ dependence, 
and in the limit where $\tau\rightarrow 0^+$, then
$g^{aux}_\sigma\rightarrow \delta_{nm}/(i\omega_n+\mu)$ and the 
$f$-electron Green's function can be evaluated explicitly, producing
$F_\eta(0^+)=-w_0$ as it must.  Similarly, for $\tau\rightarrow\beta^-$
one finds $g^{aux}_\sigma\rightarrow \delta_{nm}/(i\omega_n+\mu-U)$ and
$F_\eta(\beta^-)=-w_1$.

\begin{figure}[htbf]
\epsfxsize=1.2in
\centerline{\epsffile{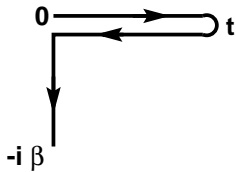}}
\caption{\label{fig: keldysh_contour} Contour $c$ used in integrating
the action in the Keldysh formalism.}
\end{figure}

One is also interested in the $f$-electron Green's function on the real
axis. Unfortunately, there is no obvious way to analytically continue
Eq.~(\ref{eq: f_green_final}) from the imaginary to the real axis.  Instead,
one is forced to re-evaluate the $f$-electron Green's function directly on
the real axis.  Since there is no time-translation invariance to the 
effective fields, it is most convenient to employ a Keldysh formalism, even
though the $f$-electron is in equilibrium.  In the Keldysh formalism, we need
to perform the path integral in the time domain over the contour illustrated
in Fig.~\ref{fig: keldysh_contour} to determine the contour-ordered
Green's function.
The Keldysh (greater) Green's function for real time $t$ becomes
\begin{equation}
F^>_\eta(t)=-\textrm{Tr}_{cf}\frac{\langle 
e^{-\beta {\mathcal H}_{imp}}S_c(\lambda)
f_\eta(t)f^\dagger_\eta(0)\rangle}{\mathcal{Z}_{imp}}
\label{eq: f_green_t_def}
\end{equation}
with $f_\eta(t)=\exp(it{\mathcal H}_{imp})f_\eta(0)\exp(-it{\mathcal H}_{imp})$
and
\begin{equation}
S_c(\lambda)=\mathcal{T}_{c}
\exp\left [ \int_c d\bar t \int_c d\bar t^\prime
\sum_\sigma c_\sigma^\dagger(\bar t )\lambda(\bar t, \bar t^\prime)c_\sigma
(\bar t^\prime)\right ].
\label{eq: s_c}
\end{equation}
One can directly determine $\lambda(\omega)$ on the real axis.  Using this
function one then finds
\begin{eqnarray}
\lambda(\bar t, \bar t^\prime)&=&-\frac{1}{\pi}\int_{-\infty}^{\infty}d\omega
\textrm{Im} \lambda(\omega)\exp [-i\omega(\bar t-\bar t^\prime)]\cr
&\times&[f(\omega)- \theta_c(\bar t-\bar t^\prime)]
\label{eq: lambda_t}
\end{eqnarray}
where $\theta_c(\bar t-\bar t^\prime)=0$ if $\bar t^\prime$ is in front of
$\bar t$ on the contour $c$ and 1 if it is behind.  

Using an equation of motion, similar to Eq.~(\ref{eq: f_eom}), leads us to 
introduce a time-dependent field
\begin{equation}
\chi_t(\bar t,\bar t^\prime)=-iU\theta_c(t -\bar t)\delta_c(\bar t
-\bar t^\prime).
\label{eq: chi_t}
\end{equation}
The derivation of the Green's function follows as before resulting in the
analogue of Eq.~(\ref{eq: f_green_final})
\begin{eqnarray}
F^>_\eta(t)&=&-\frac{1}{\mathcal{Z}_{imp}}e^{-i(E_{f\eta}-\mu)t}
\mathcal{Z}^{aux}(-\chi_t)\cr
&\times&\prod_{\sigma=1}^{2s+1}\textrm{Det}[\delta(\bar t-\bar t^\prime) \cr
&+&
\int_c d\bar t^{\prime\prime}g_\sigma^{aux}(\bar t,
\bar t^{\prime\prime})\lambda_\sigma(\bar t^{\prime\prime},\bar t^\prime)],
\label{eq: f_greater_t}
\end{eqnarray}
except now, the determinant is of a \textit{continuous} matrix operator.

\begin{table}[htb]
\caption{$g_\sigma^{aux}(\bar t,\bar t^\prime)$ for different orderings of
$t$, $\bar t$, and $\bar t^\prime$ along the contour $c$. The symbol
$\bar\xi_0$ satisfies $\bar\xi_0=1/[1+\exp(iUt-\beta\mu)]$.
\label{table: g_aux_t}}
\begin{ruledtabular}
\begin{tabular}{ll}
$\bar\xi_0\exp[i\mu(\bar t-\bar t^\prime )]$& $t<\bar t<\bar t^\prime$\\
$\bar\xi_0\exp[iUt+i(\mu-U)\bar t -i\mu\bar t^\prime]$
& $\bar t< t<\bar t^\prime$\\  
$\bar\xi_0\exp[i(\mu-U)(\bar t-\bar t^\prime )]$& $\bar t<\bar t^\prime<t$\\  
$(\bar\xi_0-1)\exp[i\mu(\bar t-\bar t^\prime)]$& $t<\bar t^\prime<\bar t$\\  
$(\bar\xi_0-1)\exp[-iUt+i\mu\bar t-i(\mu-U)\bar t^\prime]$
& $\bar t^\prime<t<\bar t$\\  
$(\bar\xi_0-1)\exp[i(\mu-U)(\bar t-\bar t^\prime)]$
& $\bar t^\prime<\bar t<t$\\  
\end{tabular}
\end{ruledtabular}
\end{table}    

The function $g_\sigma^{aux}(\bar t,\bar t^\prime)$ can be found directly
from its operator definition
\begin{equation}
g_\sigma^{aux}(\bar t,\bar t^\prime)=-\frac{1}{\mathcal{Z}^{aux}(-\chi_t)}
\textrm{Tr}_{c}\langle \mathcal{T}_c
e^{-\beta \mathcal{H}_{0}}S_c(-\chi_t)c_{\sigma}(\bar t)
c_\sigma^\dagger(\bar t^\prime)\rangle
\label{eq: g_aux_real_t}
\end{equation}
noting that we order the times along the contour $c$ and 
that $\mathcal{Z}^{aux}(-\chi_t)=[1+\exp(\beta\mu-iUt)]^{2s+1}$.
The result is shown in Table~\ref{table: g_aux_t}
for the six different possible orderings of $t$, $\bar t$, and $\bar t^\prime$
along the contour $c$.
\textcite{brandt_urbanek_1992} show how to calculate the
discretized determinant in an efficient manner.  If we use a quadrature
rule
\begin{equation}
\int_c dt I(t)=\sum_i W_i I(t_i)
\label{eq: quadrature}
\end{equation}
with weights $W_i$ for the discrete set of times $\{t_i\}$ on the contour $c$,
then the continuous determinant can be approximated by the discrete
determinant
\begin{equation}
\textrm{Det} [W_i\{\frac{\delta_{ij}}{\Delta t_c}+\sum_kg_\sigma^{aux}
(t_i,t_k)W_k\lambda_\sigma(t_k,t_j)\}]
\label{eq: det_f_cont_approx}
\end{equation}
for each $\sigma$ ($1/\Delta t_c$ is the approximation to the delta function
on contour $c$
with $\Delta t_c$ the width of the interval that includes the delta function;
for a (midpoint) rectangular quadrature rule, one takes $\Delta t_c=1/W_i$).

Now the Keldysh Green's function satisfies the following spectral formula
\begin{equation}
F^>_\eta(t)=\int_{-\infty}^{\infty}d\omega e^{-i\omega t}[f(\omega)-1]
A_\eta^f(\omega)
\label{eq: f_spectral}
\end{equation}
because it involves the product of the spectral function $A^f_\eta(
\omega)$ with the relevant distribution function $[f(\omega)-1]$.  The
(greater) Green's function obviously satisfies
\begin{equation}
F^>_\eta(t)=F^{>*}_\eta(-t).
\label{eq: f_cc}
\end{equation}
We break the spectral density into its even and odd pieces:
$A^{fe}_\eta(\omega)=[A^f_\eta(\omega)+A^f_\eta(-\omega)]/2$ and
$A^{fo}_\eta(\omega)=[A^f_\eta(\omega)-A^f_\eta(-\omega)]/2$.  Then
we use Eq.~(\ref{eq: f_cc}) to show that 
\begin{eqnarray}
A^{fe}_\eta(\omega)&=&-\frac{1}{\pi}\int_{0}^{\infty}dt [(\textrm{cosh}
\beta\omega+1)\textrm{Re} \{ F^>_\eta(t)\}\cos (\omega t)
\cr
&-&\textrm{sinh}(\beta\omega)\textrm{Im}\{F^>(t)\}\sin(\omega t)]
\label{eq: f_spectral_even}
\end{eqnarray}
and
\begin{eqnarray}
A^{fo}_\eta(\omega)&=&-\frac{1}{\pi}\int_{0}^{\infty}dt [-\textrm{sinh}         
(\beta\omega)\textrm{Re} \{ F^>_\eta(t)\}\cos (\omega t)
\cr
&+&(\textrm{cosh}\beta\omega+1)\textrm{Im}\{F^>(t)\}\sin(\omega t)].
\label{eq: f_spectral_odd}
\end{eqnarray}    
At half filling, we have that $A^{fo}_\eta(\omega)=0$, so that 
Eqs.~(\ref{eq: f_spectral_even}) and (\ref{eq: f_spectral_odd}) reduce to
\begin{equation}
A^{f}_\eta(\omega)=-\frac{2}{\pi}\int_{0}^{\infty}dt \textrm{Re} \{ F^>_\eta
(t)\}\cos (\omega t).
\label{eq: f_spectral_final}
\end{equation}
This then determines the localized electron spectral function on the 
real axis.

\subsection{Spontaneous Hybridization}

Sham and coworkers proposed that correlations can lead to a spontaneous
ferroelectricity via the creation of a dynamically correlated
hybridization in the FK model~\cite{portengen_oestreich_1996a,%
portengen_oestreich_1996b}.  Such an effect is necessarily a subtle one, 
because it is well known that Elitzur's theorem~\cite{elitzur_1975}
applies to the FK model~\cite{subrahmanyam_barma_1988} and no such
spontaneous hybridization can occur at any finite temperature.  Numerical
and analytical calculations in finite-dimensions~\cite{farkasovsky_1997,%
farkasovsky_1999,farkasovsky_2002} and in infinite-dimensions~\cite{czycholl_1999}
indicated that such spontaneous hybridization did not occur in the FK model,
but they did not cover all of the available parameter space.  An alternate way to
test these ideas is to directly calculate the susceptibility toward spontaneous
hybridization formation~\cite{zlatic_review_2001}.

Because the $f$-electron dynamics are local, only the local susceptibility
toward spontaneous hybridization is relevant.  We will restrict ourselves to
the spinless case for simplicity. The local spontaneous hybridization
susceptibility is defined by
\begin{eqnarray}
\chi_{hyb}&=&-\int_0^\beta d\tau\frac{1}{\mathcal{Z}_{imp}}\cr
&\times&\textrm{Tr}_{cf}\langle\mathcal{T}_\tau
\exp(-\beta\mathcal{H}_{imp})
S(\lambda)f(\tau)d^\dagger(\tau)d(0)f^\dagger(0)\rangle.\cr
&&
\label{eq: chi_hyb}
\end{eqnarray}
If we introduce the time-dependent field $\lambda(\tau,\tau^\prime)$ that
couples the itinerant electron at $\tau^\prime$ to its Hermitian conjugate
at time $\tau$ [in equilibrium, we have $\lambda(\tau,\tau^\prime)=\lambda(\tau-
\tau^\prime)$], then we can express the susceptibility in terms of a functional
derivative
\begin{eqnarray}
\chi_{hyb}&=&-\int_0^\beta d\tau\frac{\delta F(\tau)}{\delta \lambda(\tau,0)}\cr
&=&\int_0^\beta d\tau\frac{e^{-(E_f-\mu_f-U)\tau}
\mathcal{Z}^{aux}(-\chi_\tau)}{\mathcal{Z}_{imp}}
\frac{\delta \textrm{Det} [1-g^{aux}\lambda]}{\delta \lambda(\tau,0)}.\cr
&&
\label{eq: chi_hyb_fcnl}
\end{eqnarray}
Noting that $\textrm{Det} A=\exp[\textrm{Tr}(\ln A)]$, allows us to immediately
compute
\begin{equation}
\frac{\delta \textrm{Det} [1-g^{aux}\lambda]}{\delta \lambda(\tau,0)}=
-\textrm{Det} [1-g^{aux}\lambda][(1-g^{aux}\lambda)^{-1}g^{aux}]_{0\tau},
\label{eq: determ_ident}
\end{equation}
where matrix multiplication is understood.  Substituting 
Eq.~(\ref{eq: determ_ident}) into Eq.~(\ref{eq: chi_hyb_fcnl}) then immediately
yields
\begin{equation}
\chi_{hyb}=\int_0^\beta d\tau\int_0^\beta d\tau^\prime F(\tau)
(1-g^{aux}\lambda)^{-1}_{0\tau^\prime}g^{aux}_{\tau^\prime\tau},
\label{eq: chi_hyb_final_tau}
\end{equation}
where one must recall that $g^{aux}$ must be recomputed for each $\tau$ value,
since it depends explicitly on $\chi_\tau$.  Reexpressing 
Eq.~(\ref{eq: chi_hyb_final_tau}) in terms of Fourier transformed quantities
produces
\begin{equation}
\chi_{hyb}=\int_0^\beta d\tau F(\tau) T\sum_{mn}(1-g^{aux}\lambda)^{-1}_{mn}
g^{aux}(i\omega_n,\tau),
\label{eq: chi_hyb_final}
\end{equation}
where the partial Fourier transform of the auxiliary Green's function
satisfies
\begin{equation}
g^{aux}(i\omega_n,\tau)=T\sum_mg^{aux}(i\omega_n,i\omega_m)e^{i\omega_m\tau}.
\label{eq: partial_ft}
\end{equation}
Note that this final result for the spontaneous hybridization susceptibility
in Eq.~(\ref{eq: chi_hyb_final}) requires the matrix inverse of 
$[1-g^{aux}\lambda]$ which is straightforward to compute numerically for a
finite truncation of the matrix.

\section{Analysis of Solutions}

\subsection{Charge-Density-Wave Order and Phase Separation}

The first problem examined in the FK model with DMFT was the problem of
ordering into a two-sublattice charge density wave (CDW) at half filling on a
hypercubic lattice~\cite{brandt_mielsch_1989}.  Because the hypercubic lattice
is bipartite, we expect a transition at finite temperature for all 
$U$~\cite{kennedy_lieb_1986} and indeed this is true.   Since the original work,
a number of other studies of CDW order and phase separation
followed~\cite{brandt_mielsch_1990,%
brandt_mielsch_1991,vandongen_vollhardt_1990,vandongen_1992,freericks_1993a,%
freericks_1993b,freericks_gruber_macris_1999,letfulov_1999,%
freericks_lemanski_2000,gruber_macris_royer_2001} on both the hypercubic and
the Bethe lattices.  We concentrate on the spinless FK model here.

Since the DMFT is in the thermodynamic limit, we can determine the transition
temperature for a continuous transition by simply finding the temperature
at which the susceptibility diverges.  Using the results of 
Table~\ref{table: susceptibilities}, we find that the susceptibility diverges
whenever the denominator of $\gamma(\textbf{q})$ vanishes.  In 
Fig.~\ref{fig: cdw_tc}, we plot the $T_c$ for checkerboard CDW order ($X=-1$)
and the spinodal temperature for phase separation $(X=1)$ near half filling (a)
and near the band edge (b).  Note that the transition temperature has a maximum
that is about 1/40th of the effective bandwidth ($\approx 4t^*$) and at half 
filling for large $U$
it behaves like $T_c\approx t^{*2}/4U$ [as expected for the equivalent (Ising) 
spin model] and for small $U$ it appears to grow like $T_c\approx U^2|\ln U|$
[which is different from the expected exponentially growing behavior
that occurs in most nested systems at weak coupling]
\cite{vandongen_vollhardt_1990,vandongen_1992}.  As we move away from
half filling, the checkerboard phase is suppressed (disappearing at small and 
large $U$) and the phase separation is enhanced (especially at large $U$). Near
the band edge [panel (b)], there is no checkerboard instability, but the 
phase separation becomes even stronger. We see that the
spinodal temperature for phase separation does not depend too strongly on
the electron filling, and surprisingly, the curves cross as a function
of $U$.

\begin{figure}[htb]
\epsfxsize=3.0in
\epsffile{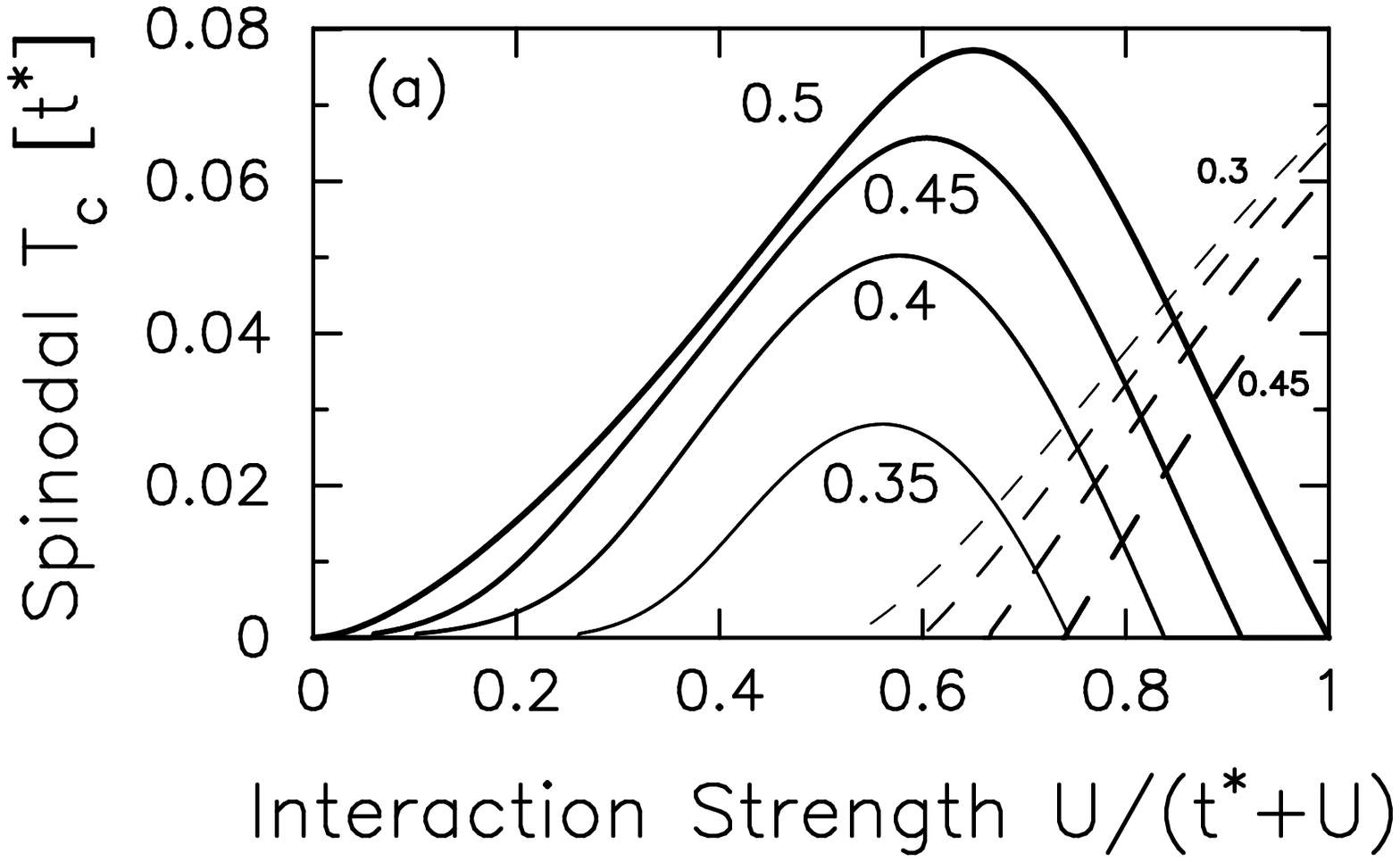}
\epsfxsize=3.0in
\epsffile{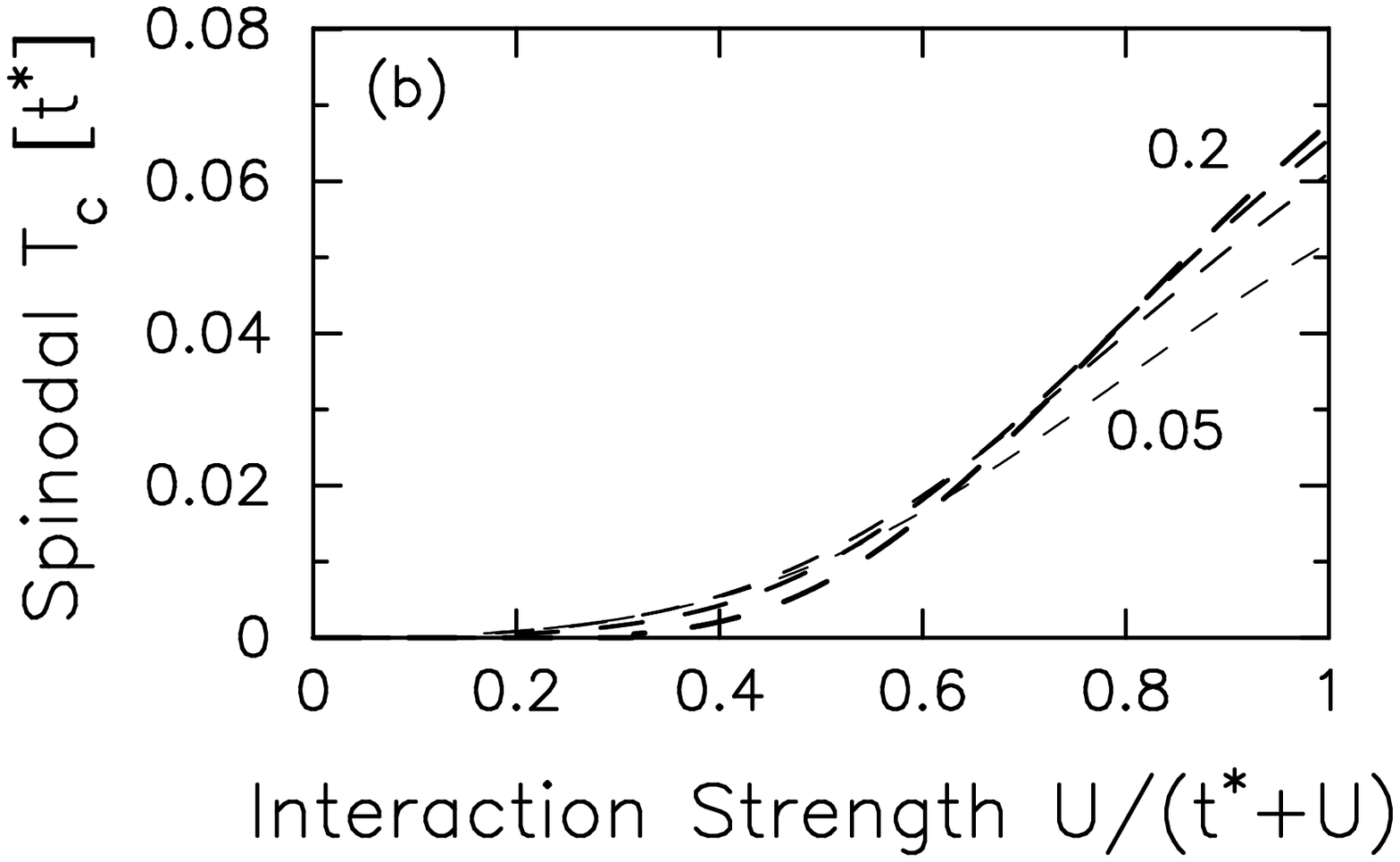}  
\caption{$T_c$ for CDW order in the spinless FK model (a) near half filling
and (b) near the band edge.  The solid lines denote checkerboard order
and the dashed lines denote the spinodal decomposition temperature for
phase separation (the true first-order transition temperature is always higher
than the spinodal temperature). The labels denote the electron filling $\rho_e$
[which runs from top to bottom for the dashed lines in panel (a)
as 0.3, 0.35, 0.4, and 0.45,
the spinodal temperature for 0.5 vanishes; in panel (b) the lines are for
0.05, 0.1, 0.15, and 0.2---note how the curves cross as a function of $U$]; 
the ion filling is fixed at $\rho_f=0.5$.
\label{fig: cdw_tc}}
\end{figure}

We can also examine the transition temperature for fixed $U$ as a function
of electron concentration (with $\rho_f=0.5$ again) as shown in 
Fig.~\ref{fig: cdw_tc2}. For small $U$ [panel (a)], we see that the
region of the checkerboard phase increases as $U$ increases, as does the region
of phase separation.  For large $U$ [panel (b)], we see that the $T_c$ curve
for the checkerboard phase develops a cusp at half 
filling~\cite{freericks_lemanski_2000}, and that
the region of stability shrinks as $U$ increases further.  Phase
separation dominates for large $U$, as expected.

\begin{figure}[htb]
\epsfxsize=3.0in
\epsffile{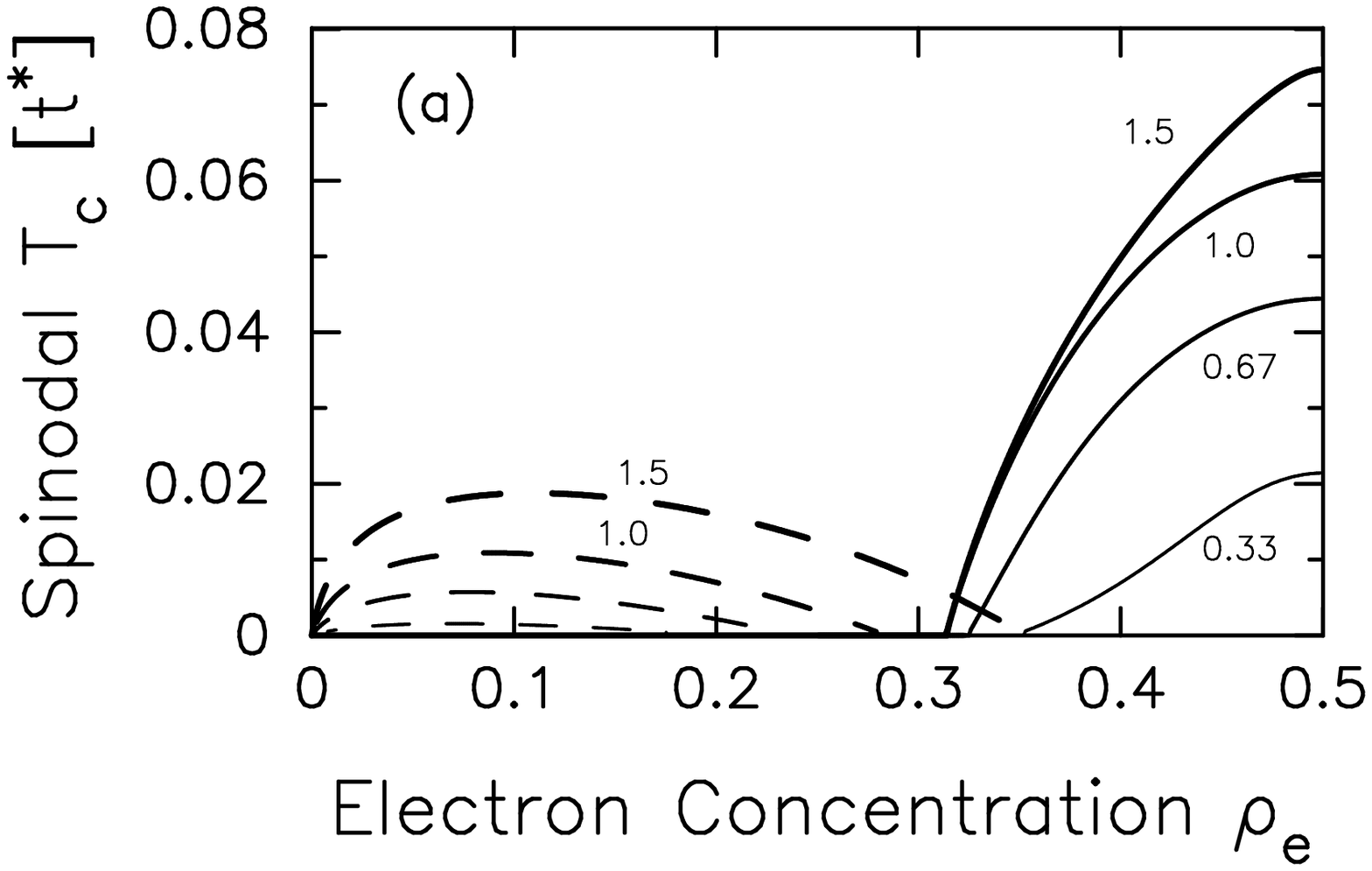}
\epsfxsize=3.0in
\epsffile{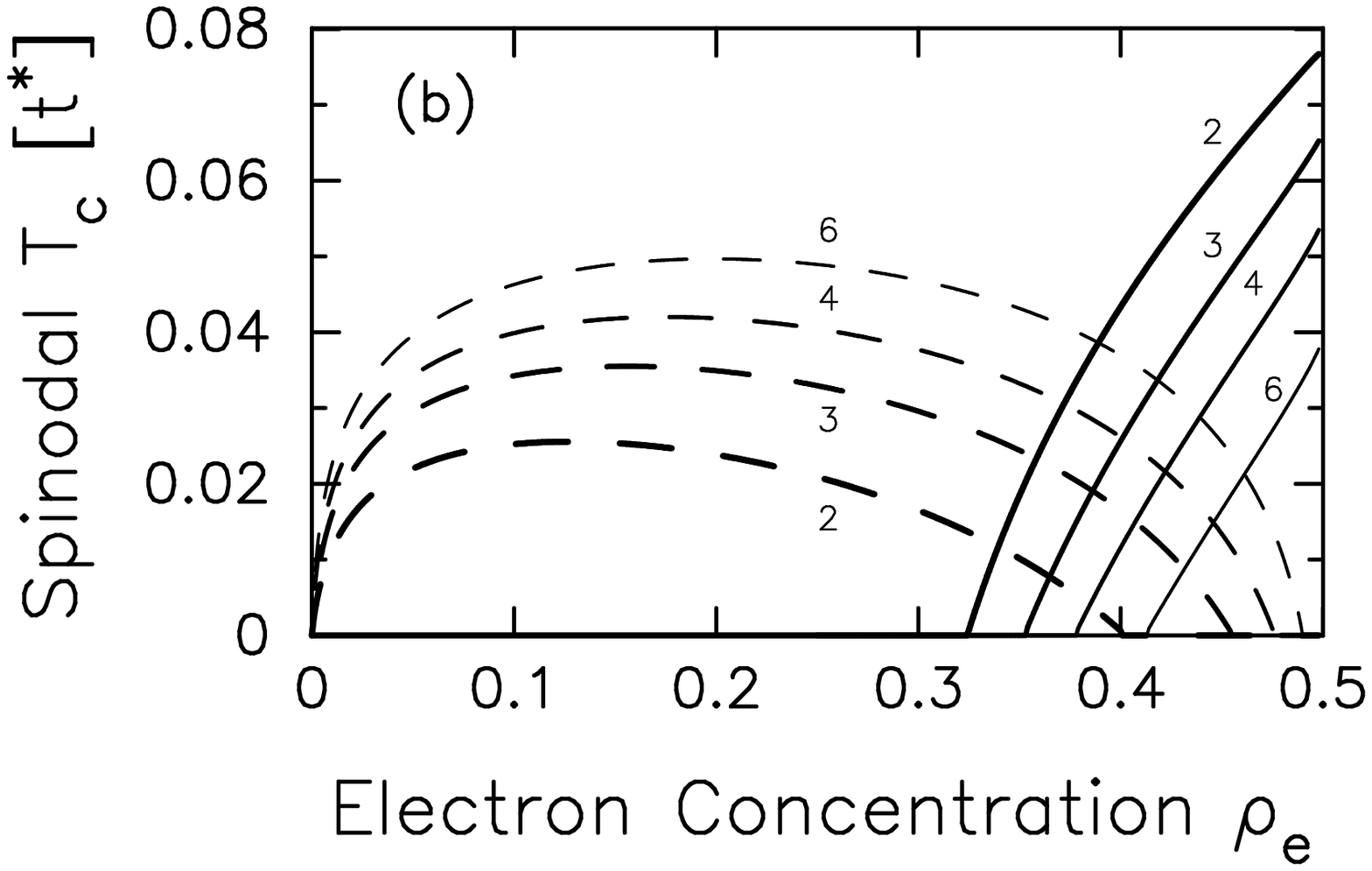}
\caption{$T_c$ for CDW order in the spinless FK model (a) for small $U$
and (b) for large $U$.  The solid lines denote checkerboard order
and the dashed lines denote the spinodal decomposition temperature for
phase separation (the true first-order transition temperature is always higher
than the spinodal temperature). The labels denote the value of $U$;
the ion filling is fixed at $\rho_f=0.5$.
\label{fig: cdw_tc2}}
\end{figure}

It turns out, that for small $U$, the system also has instabilities to CDW
phases with incommensurate values of $X$ (i.e., $X$ changes continuously with
the electron concentration).  These results are summarized with a 
projected phase diagram~\cite{freericks_1993a} in Fig.~\ref{fig: cdw_phase}, 
which plots the regions of
stability for different CDW phases as determined by the ordering wavevector
of the \textit{initial ordered phase} as $T$ is lowered to the first 
(continuous) instability
at $T_c$ (as determined by the divergence of the relevant 
susceptibility).  This is an approximation of the zero-temperature phase 
diagram---the phase boundaries may change as one reduces the temperature 
from $T_c$ to
zero, but we are not able to study the ordered phase of incommensurate states.
They may also change if there are first-order phase transitions.
Note how the system segregates at large $U$ for all fillings except $\rho_e=0.5$
where it is degenerate with the checkerboard phase.

\begin{figure}[htb]
\epsfxsize=3.0in
\epsffile{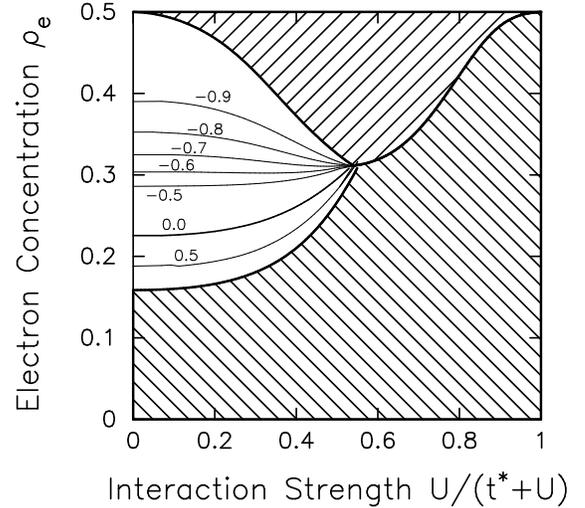} 
\caption{Projected phase diagram for the spinless FK model with $\rho_f=0.5$.
The shaded region near half filling denotes the region of stability of the
checkerboard phase $(X=-1)$, the shaded region near the band edge denotes
the region of stability for the segregated phase (as determined by the
spinodal decomposition temperature), and the white region shows the
incommensurate phases at small $U$ (the solid lines are lines of constant
$X$).
\label{fig: cdw_phase}}
\end{figure}

In order to understand the incommensurate order better, we plot $T_c(X)$
for different electron fillings at $U=0.5$ in Fig.~\ref{fig: tc_X}.  Near
half filling, these curves are peaked at $X=-1$, but as the system is doped 
sufficiently far from half filling, the curves develop a peak at an intermediate
value of $X$ which evolves towards $X=0$ as $T_c\rightarrow 0$ ($\rho_e=0.35$
and 0.3).  Then, if
doped further from half filling, $T_c$ rises again and the ordering wavevector
evolves smoothly toward $X=1$ ($\rho_e=0.2$ and 0.15).  Note that the region of 
stability for positive
(incommensurate) $X$ is much smaller than that of negative (incommensurate) $X$.
In the large $U$ case, the maximum always appears at $X=-1$ or $X=1$ and
the incommensurate order disappears (see Fig.~\ref{fig: cdw_phase}).

\begin{figure}[htb]
\epsfxsize=3.0in
\epsffile{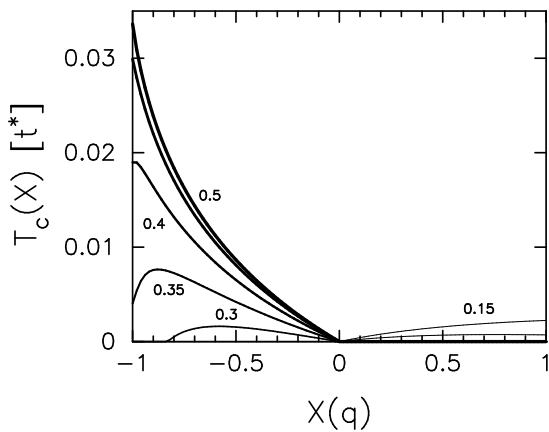}
\caption{CDW transition temperature as a function of the ordering wavevector
described by $X$.  The numbers label the electron filling  with the filling 
changing in steps of 0.05 ($U=0.5$ and $\rho_f=0.5$). The results for $\rho_e=0.25$
are too low to be seen on this figure.
\label{fig: tc_X}}
\end{figure}

It is interesting to also examine the dynamical charge susceptibility~\cite{%
shvaika_2000,freericks_miller_2000,shvaika_2002}.  We don't expect
the dynamical susceptibility to show any signs of the CDW instability because
it is an isolated susceptibility, and the divergence of the static
susceptibility arises solely from the coupling to the static electrons which
produced an additional contribution to the (isothermal)
static susceptibility.  Indeed this
is the case, as can be seen for the checkerboard ($X=-1$) dynamical charge
susceptibility at $U=4$ shown in Fig.~\ref{fig: dyn_susc}.  We plot the
imaginary part of the susceptibility only.  Note how at high temperature
there is a low-energy peak and a charge-transfer peak (centered at 
$\nu\approx U=4$), but as the temperature
is lowered, the low-energy peak rapidly disappears and the susceptibility
has little further temperature dependence in the homogeneous
phase (even though we have passed through
$T_c$ which occurs at 0.0547).  We also find that as we move from the zone corner
toward the zone center, the higher-energy (charge-transfer peak) loses spectral
weight but does not have much dispersion or peak narrowing, while the low-energy
peak shows an appreciable dispersion toward lower energy and narrows as we
approach the zone center (not shown).  At the zone center, the dynamical charge 
susceptibility vanishes.

\begin{figure}[htb]
\epsfxsize=3.0in
\epsffile{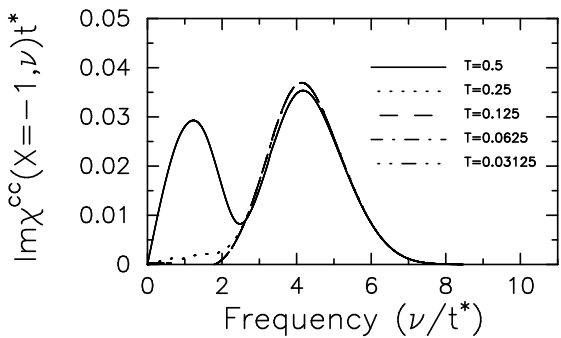}
\caption{Imaginary part of the dynamical charge susceptibility at $X=-1$ and
$U=4$. Note the low-energy spectral weight which rapidly disappears as $T$ is
lowered, and note further that there is no signal of the CDW order (which
sets in at $T_c=0.0547$) in the isolated susceptibility.
\label{fig: dyn_susc}}
\end{figure}

The results on the Bethe lattice are similar to those on the hypercubic lattice
except for the incommensurate order.  We will concentrate on discussing
the Bethe lattice case for the remainder of this section.  We begin by
examining the ordered checkerboard phase at half filling.  In 
Fig.~\ref{fig: cdw_order}, we plot the electronic density of states for
$U=1$ above $T_c$, and below $T_c$ in steps down to $T=0$.  Inset
in the figure is a plot of the ``normalized''
order parameter $w_1^A-w_1^B$ as a function
of temperature.  The order parameter takes a BCS-like form, being flat for
low temperatures and having a square-root dependence as $T_c$ is approached
[this behavior changes as $U\rightarrow 0$, see \textcite{vandongen_1992}].
The density of states evolves with temperature in the ordered phase because
of this temperature dependence in the order parameter.  Each sublattice has
a sharp peak in the density of states (which becomes singular at $T=0$)
as expected.  Note the interesting behavior of the subgap states, which split
into two subbands then shrink and disappear as $T$ is reduced. In the limit as 
$U\rightarrow\infty$, the bandwidth of the subbands goes to zero, and the
density of states becomes a delta function on each 
sublattice~\cite{vandongen_1992}.

\begin{figure}[htb]
\epsfxsize=3.0in
\epsffile{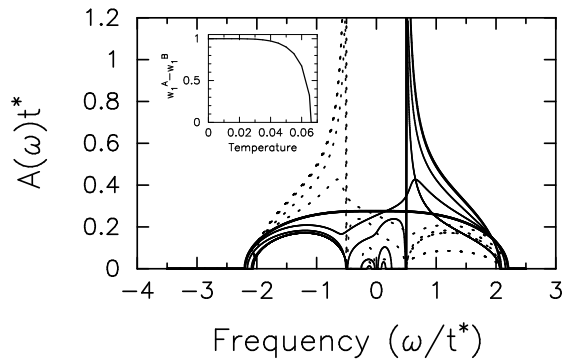}
\caption{Density of states above and below $T_c=0.06614$ in the spinless FK model
at half filling
on a Bethe lattice with $U=1$.  The DOS is plotted for $T=0.07$, $T=0.065$,
$T=0.06$, $T=0.045$, $T=0.03$,
and $T=0$.  Note that in the ordered phase, there are two DOS plots, one for each
sublattice (solid and dotted lines); these DOS are mirror images of each
other.  The inset plots the ``normalized'' order parameter as 
a function of temperature, which has a BCS-like shape.  The exact form is known
for $U\rightarrow\infty$ and $U\rightarrow 0$~\cite{vandongen_1992}, and one 
expects, for smaller $U$, that the shape will change dramatically since the
order parameter becomes very small in the range $0.5T_c<T<T_c$.
\label{fig: cdw_order}}
\end{figure}

In addition to the checkerboard phase, we can also examine the phase-separated
(segregated) phase, where the itinerant electrons avoid the localized ions.
If we take the limit $U\rightarrow\infty$, where the phase separation is the
strongest, we find the plot projected onto the $w_1=(1-\rho_e)/2$
plane (which corresponds to relative half filling of the lower Hubbard
band) shown in Fig.~\ref{fig: phase}.  The solid line denotes the first-order
phase transition line calculated by performing a Maxwell construction on
the free energy~\cite{freericks_gruber_macris_1999}.  The dotted line is
the spinodal decomposition temperature calculated from the divergence of
the susceptibility.  Note how the two temperatures track well with one another
and how they meet at the maximal $T_c$ as they must because that is the
temperature where the first-order phase transition becomes second order.

\begin{figure}[htb]
\epsfxsize=3.0in
\epsffile{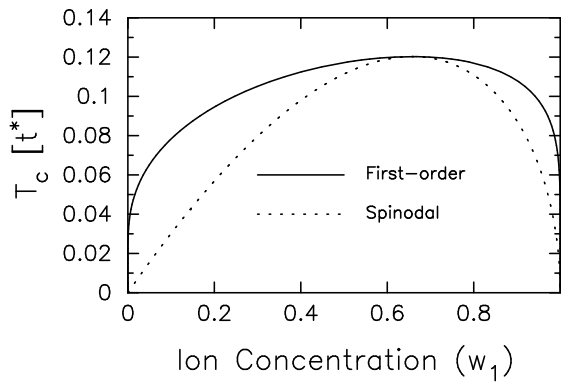}
\caption{Phase diagram for the first-order phase separation transition on
a Bethe lattice with $w_1=(1-\rho_e)/2$ and $U\rightarrow\infty$.  The solid
line is the first-order transition temperature and the dotted line is the
spinodal temperature.
\label{fig: phase}}
\end{figure}

Finally, we examine higher-period ordered phases on the Bethe 
lattice~\cite{gruber_macris_royer_2001}.  Since the Bethe lattice does
not have the same periodicity properties of regular lattices, one must
construct higher-period ordered phases with care.  This is done by working 
directly in real space and repeating quasi-one-dimensional patterns of
charge density waves according to the different levels of the tree that
forms the Bethe lattice.  
A period-two phase is the conventional checkerboard phase with alternating 
charge densities on each level.  There 
is a region of parameter space where the period-three phase has been shown to
be stable.  This is illustrated in Fig.~\ref{fig: three_phase}~(a), which is
a restricted phase diagram at $T=0$ that compares the ground-state energy
of the segregated phase, the period-two phase, the period-three phase
and the homogeneous phase.  Note the wide region of stability for the
period-three phase (this region can shrink as additional phases are added to
the phase diagram; in particular, we conjecture that the homogeneous phase stability
will disappear as higher-period phases are added to the diagram).  To
examine in more detail, we plot the free energy of the homogeneous phase, the
period-two phase, and the period three phase for $U=3$, $\rho_f=2/3$, and
$\mu=0.84861$ ($\rho_e\approx 0.332$).  Note how at low temperature the
period-three phase is lowest in energy, but how there must be a higher-period 
phase that intervenes between the period-two and period three phases
for the free energy
to be continuous.  We expect the phase transition from the period-three
phase to be first order.  Note further how the period-two phase free energy
smoothly joins the homogeneous-phase free energy, as expected for a
continuous (second-order) phase transition.

\begin{figure}[htb]
\epsfxsize=3.0in
\epsffile{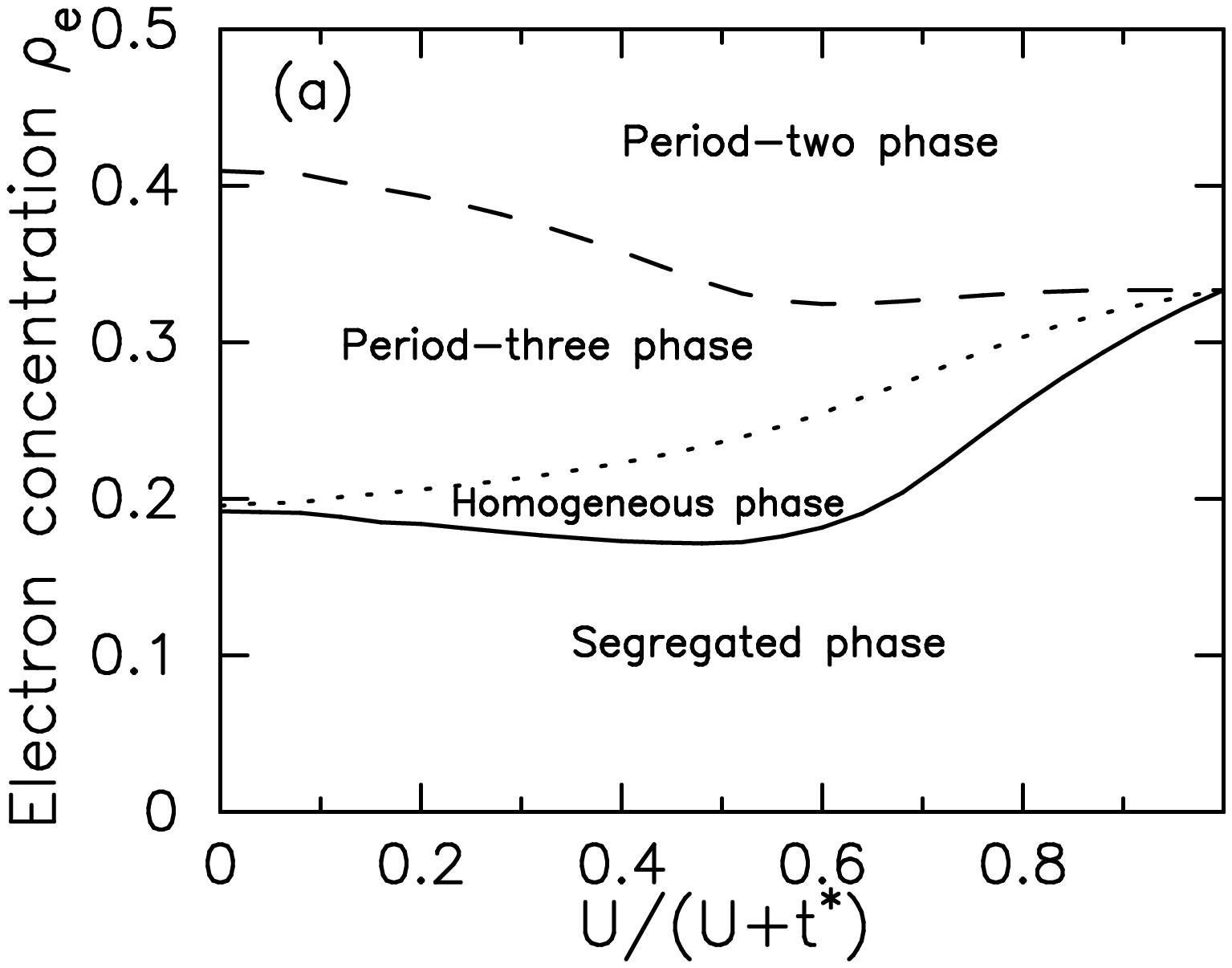}
\epsfxsize=3.25in
\epsffile{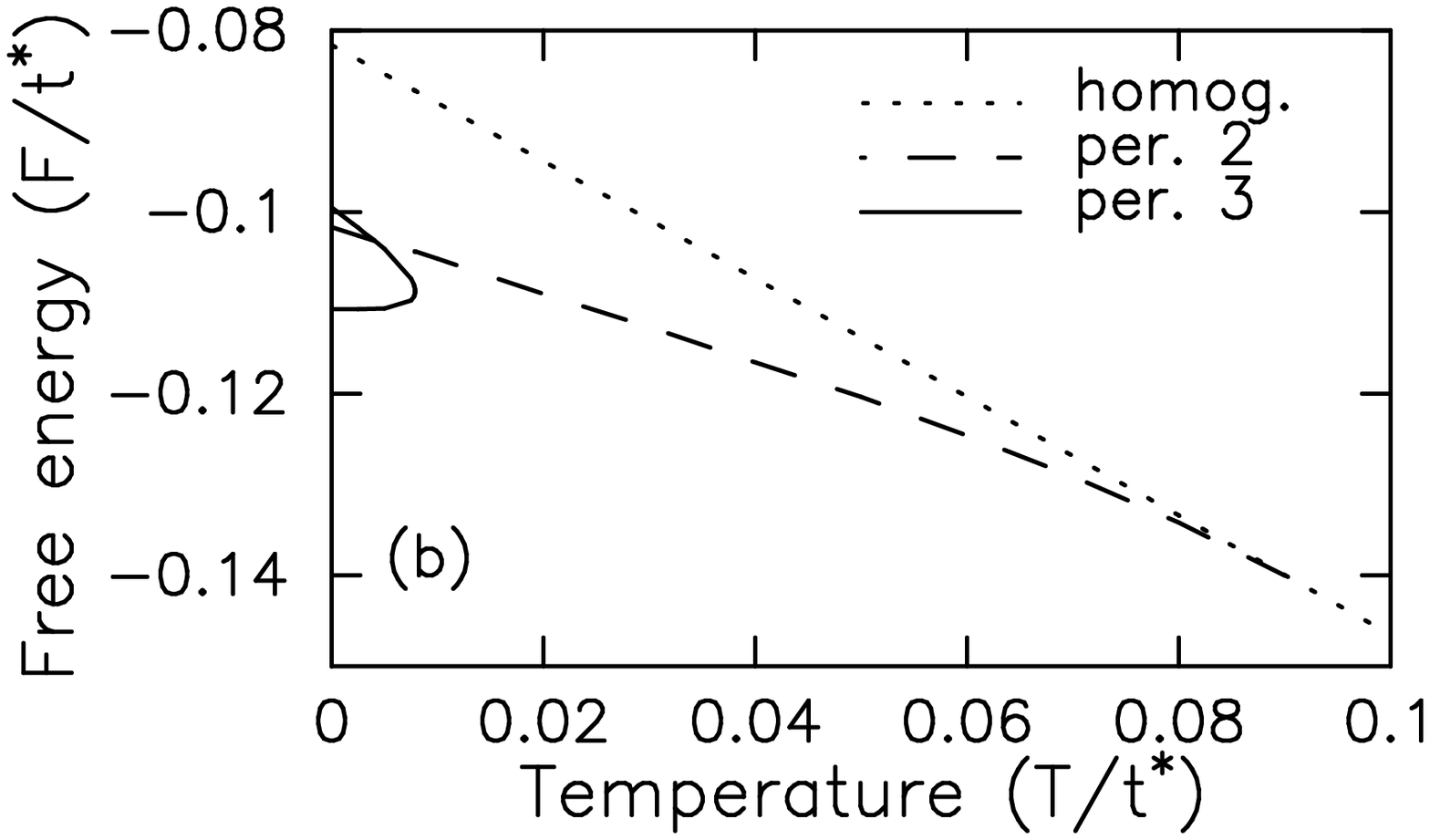}
\caption{(a) Restricted phase diagram on the Bethe lattice for $\rho_f=2/3$ and
(b) free energy versus temperature for $\mu=0.84861$.  Note the large
region of stability of the period-three phase in (a) and how the shape of the
free-energy curves in (b) suggest that phase transitions to higher-period
phases will be \textit{first order}.
\label{fig: three_phase}}
\end{figure}

\subsection{Mott-like Metal-Insulator Transitions}

\begin{figure}[htb]
\epsfxsize=3.0in
\epsffile{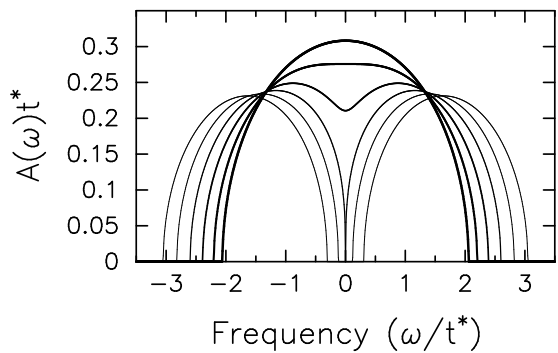}
\caption{Itinerant electron DOS for the spinless FK model
on the Bethe lattice for different
values of $U$ ranging from 0.5 to 3.0 in steps of 0.5 ($\rho_e=\rho_f=0.5$).  
The metal-insulator
transition occurs at U=2.  It is preceded by a pseudogap phase for
$(1<U<2)$.
\label{fig: dos_mott}}
\end{figure}

The FK model is not a fermi liquid~\cite{si_kotliar_1992}
whenever $\rho_f$ is not equal to 0 or 1 (when $\rho_f=0$ or 1, the FK model is a
noninteracting fermi gas).  This is because the $f$-electrons
appear like disorder scatterers (with an annealed averaging rather than the
conventional quenched averaging).  These scatterers always produce a finite
lifetime at the fermi surface, so rigorously quasiparticles do not exist and the
system is a non-fermi-liquid.  If $U$ is small, one can still view the
system as a ``dirty fermi liquid'', but the system rapidly changes character
as $U$ increases further.  The non-fermi-liquid character can also be
seen in the self energy on the real axis.  The imaginary part does have
a quadratic behavior, but the curvature is the \textit{wrong sign} and the intercept
at $\omega=0$ \textit{does not go to zero}
as $T\rightarrow 0$ as it must in a fermi
liquid.  Similarly, the real part of the self energy is linear, but the
slope \textit{has the wrong sign} near $\omega=0$.

It is known in a fermi liquid with a local self energy that the fermi surface
and the DOS at the fermi level (at $T=0$) are 
unchanged as the correlations increase~\cite{mueller-hartmann_1989c}. But 
because the FK model is not a fermi liquid there is no restriction on the DOS.
\textcite{vandongen_leinung_1997} studied the Mott-like metal-insulator
transition on the Bethe lattice in detail; in this study, one continues the
homogeneous phase down to $T=0$, ignoring any possible CDW
phases.  They found a simple cubic
equation for the interacting DOS on the Bethe lattice which produces
the results shown in Fig.~\ref{fig: dos_mott}.  One can see that for 
weak coupling, the DOS is essentially unchanged from the noninteracting case,
but as $U$ is increased further, the DOS first develops a pseudogap, where
spectral weight is depleted near the chemical potential, and then it gets 
fully reduced to zero.  As $U$ is increased further, the system splits into 
lower and upper Hubbard bands, and the bandwidth of each subband decreases
as $U$ increases, while the separation increases between the two bands.  This is
precisely the kind of metal-insulator transition that Mott envisioned,
where the DOS within a single band is suppressed to zero, but unlike Mott's
prediction that the transition would generically be a discontinuous first-order
transition, here it is a continuous second-order transition.

The metal-insulator transition occurs precisely at the point where the
self energy develops a pole and diverges at $\omega=0$.  This forces the
DOS at the chemical potential to equal 0 on any lattice.  On lattices where
the noninteracting DOS has band edges, the interacting DOS can remain zero
within a correlation-induced gap.  But on the hypercubic lattice, which has
an infinite bandwidth, with an exponentially small DOS for large frequency, 
there is no precise gap, rather the DOS is exponentially small in the
``gap region'' and then becomes of order one within the Hubbard subbands.
Indeed, on the hypercubic lattice, the DOS looks much like that in 
Fig.~\ref{fig: dos_mott}, but the metal-insulator transition occurs at 
$U\approx 1.5$ for $\rho_e=\rho_f=0.5$.

The DOS of the FK model does have a curious property---it is independent of
temperature in the homogeneous phase~\cite{vandongen_1992} (which was proved
by mapping the FK model onto a coordination-three noninteracting Bethe
lattice problem, which then has a temperature-independent DOS). The
temperature-independence holds only in canonical formulations, where both
$\rho_e$ and $\rho_f$ are separately fixed as functions of $T$ [ignoring the
trivial shift of the DOS with $\mu(T)$].  In cases where
the total electron concentration is fixed, but can vary between itinerant
and localized electrons, the DOS will vary with $T$ since $\rho_f$ generally
varies with $T$.

One can also examine the spectral function of the localized 
electrons~\cite{brandt_urbanek_1992}, which is temperature dependent!
The DOS shares similar behavior with the conduction-electron DOS, having a gap
when the correlation energy is large.  We plot results 
for $U=2.5$ where the localized
electron DOS develops a pseudogap and then a gap
region as $T$ is lowered. Note how
the gap develops as $T$ is reduced below $0.4$.

\begin{figure}[htbf]
\epsfxsize=3.0in
\centerline{\epsffile{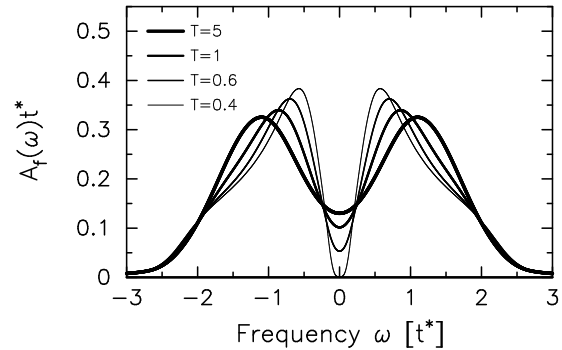}}
\caption{\label{fig: f_spectral} Single particle DOS of the $f$-electrons
for the spinless
FK model at half filling on a Bethe lattice with 
$U=2.5$.  Note how the
gap opens up as $T$ is reduced even though the calculations are above
the CDW transition temperature (which occurs at $T=0.113746$).  The DOS is 
symmetric due to particle-hole symmetry.}
\end{figure}

\subsection{Falicov-Kimball-like Metal-Insulator Transitions}

The original work of Falicov, Kimball, and Ramirez~\cite{falicov_kimball_1969,%
ramirez_falicov_kimball_1970} studied a different type of metal-insulator
transition---one where the character of the electronic states was unchanged,
but their statistical occupancy fluctuated with temperature or pressure.
As occupancy shifted from the itinerant to localized bands, the system
underwent a charge-transfer
metal-insulator transition.  One of the main points of interest
of this model was that the presence of a Coulomb interaction $U$ between
the two types of electrons could make the transition become discontinuous
(first-order).  The original approximations used a mean-field theory
which showed these first-order transitions.  Later, \textcite{plischke_1972}
proposed that when the coherent-potential approximation was applied to
the FK model, the first-order transitions disappeared.  This was refuted
by \textcite{dasilva_falicov_1972}.

In the Falicov and Kimball approach to the metal-insulator transition, one works
with spin-one-half electrons and fixes the total electron concentration
(usually at $\rho_{total}=\rho_e+\rho_f=1$).  Alternatively, one can perform 
a partial
particle-hole transformation on the $f$-electrons, and have a system where
$\rho_e=\rho_f$ and $U$ is negative (attractive).  There is a gap $\Delta$ to 
creating an electron-hole pair. We will concentrate on the
electronic picture here (where one has $E_f=U-\Delta-2t^*$) and the choice
$\Delta=t^*$ which most closely represents the original FK problem. We
work on an infinite-coordination Bethe lattice.

\begin{figure}[htb]
\epsfxsize=3.2in
\epsffile{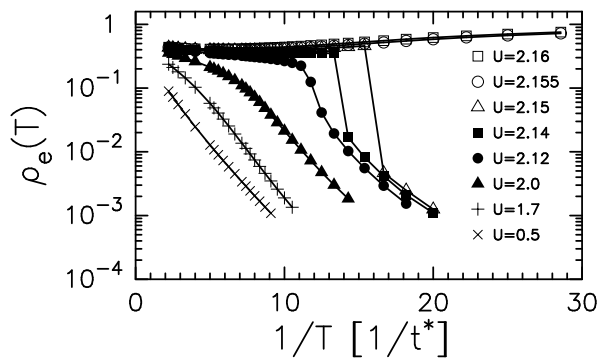}
\caption{Conduction electron density as a function of $1/T$ for the spin-one-half
FK model on a Bethe lattice with $\Delta=1$.  The different curves correspond to
different values of $U$.
\label{fig: fkmit_n}}
\end{figure}    

This model for metal-insulator transitions was solved in infinite dimensions
by~\textcite{chung_freericks_1998}.  It requires a fine tuning of the
Coulomb interaction in order to have a first-order phase transition.
This is illustrated in Fig.~\ref{fig: fkmit_n}, where we plot the
conduction electron density $\rho_e$ versus inverse temperature $1/T$
for different values of $U$ (recall $E_f$ also varies with $U$, since $\Delta$
is fixed at 1).  The states that have $\rho_e\rightarrow 0$ as $T\rightarrow 0$
are insulators, while those with finite $\rho_e$ are metallic.  We say a
metal-insulator transition occurs when there is a discontinuous jump 
in the electron density, which goes from the metallic phase at high temperature
to the insulating phase at low $T$.  This occurs for $2.12<U<2.155$ when 
$\Delta=1$.  The shape of these curves is remarkably similar to those found
in the mean-field-theory solution of~\textcite{ramirez_falicov_kimball_1970}
and verifies that first-order metal-insulator transition do occur within the
CPA.

\begin{figure}[htb]
\epsfxsize=3.1in
\epsffile{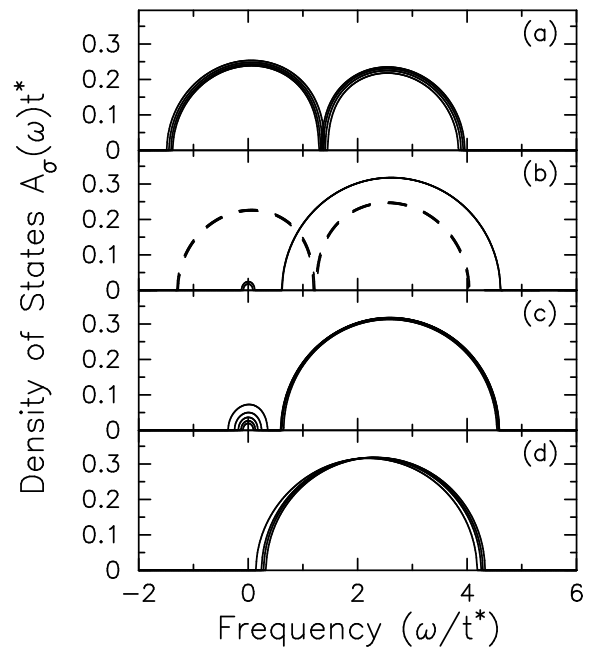}
\caption{Conduction-electron DOS for the spin-one-half FK model on a Bethe lattice
with $\Delta=1$.  The different panels are for different values of $U$, and the
results are plotted for a number of different temperatures [the dashed lines
are used in panel (b) for clarity].  The DOS has $T$
dependence because $\rho_f$ varies with $T$.
\label{fig: fkmit_dos}}
\end{figure}

The interacting electronic DOS is plotted in Fig.~\ref{fig: fkmit_dos} for
four different values of $U$~\cite{chung_freericks_1998}.  
In panel (a) we show the metallic case
$U=2.16$, where the DOS is large at the chemical potential ($\omega=0$)
and does not depend strongly on $T$.  As $U$ is reduced to 2.15, where
there is a discontinuous metal-insulator phase transition, we see a 
huge reconstruction of the DOS at the transition, where the DOS at the
chemical potential becomes small and decreases to zero as $T\rightarrow 0$.
In panel (c), we see the evolution of the insulating phase for the strongly
correlated insulator $U=2.12$.  Here the conduction electron occupation
does not evolve in a simple exponential fashion, but has a much sharper increase
as $T$ increases.  The DOS has upper and lower Hubbard bands, with the
lower band losing spectral weight as $T\rightarrow 0$.  Finally, in panel (d),
we show the DOS for a weakly correlated insulator.  Here the conduction
electron density evolves in an exponential fashion and there is no DOS
at the chemical potential, since all of the electronic occupation is
thermally activated.

\subsection{Intermediate-Valence}

The FK model does not have any hybridization between the itinerant and 
localized electrons, so it cannot have any quantum-mechanical mixed valence
at finite temperature~\cite{subrahmanyam_barma_1988} which is quantified
by having a nonzero average $\langle c^\dagger f\rangle$.  Nevertheless,
one can have a \textit{classical} intermediate-valence state, where the
average $f$-electron occupancy lies between 0 and 1 for $T\rightarrow 0$.
In many such cases, one would expect CDW order (or phase separation)
to take over in the ground state,
but there are regions of parameter space where the system appears to remain
in a homogeneous classical intermediate-valence state all the way to $T=0$.
We study such systems here.

The intermediate-valence problem, and the possibility of the FK model having
an instability to a spontaneously generated hybridization, was first
proposed by~\textcite{portengen_oestreich_1996a,portengen_oestreich_1996b}.
Calculations in infinite-dimensions~\cite{czycholl_1999} and one
dimension~\cite{farkasovsky_1997,farkasovsky_1999} showed that a spontaneously 
generated
hybridization was unlikely over a wide range of parameter space, but they did
not rule out the possibility everywhere.   

\begin{figure}[htb]
\epsfxsize=3.0in
\epsffile{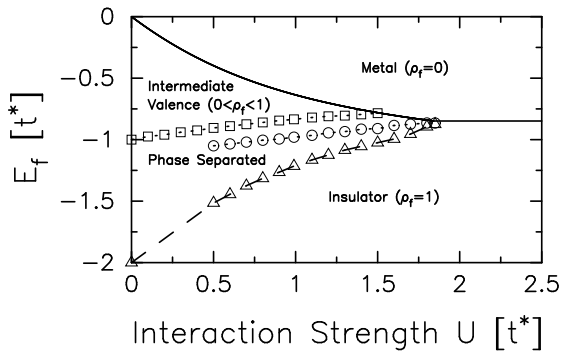}
\caption{Intermediate-valence phase diagram for the spin-one-half FK model on
a Bethe lattice with $\rho_e+\rho_f=1$. Note how the classical intermediate
valence state occupies only a small part of the phase diagram, being
taken over by phase-separated, metallic, and insulating phases over most
of the phase diagram.
\label{fig: iv_phase}}
\end{figure}   

The first step to examining the intermediate-valence problem is to study the
phase diagram of the FK model to see what different types of phases occur.
We pick the spin-one-half model on the Bethe lattice again.  We also
pick $\rho_e+\rho_f=1$ for the total electron concentration.  Since the conduction
electron fermi level is at 0 when there are no $f$-electrons, if we pick
$-2<E_f<0$, then the fermi level is pinned to $E_f$ at $U=0$, and one of
two things can occur as $U$ is increased: (i) either the classical
intermediate-valence
state survives, or (ii) it phase separates into a mixture of a state with
$\rho_f=0$ and $\rho_f=1$.  The former occurs for $E_f>-1$ and the latter for
$E_f<-1$.  As the interaction strength is increased further, then we
find only two different stable solutions, a metallic phase, where $\rho_f=0$ and
an insulating phase where $\rho_f=1$.  The zero-temperature phase diagram
is summarized in Fig.~\ref{fig: iv_phase}~\cite{chung_freericks_2000}. In the
phase-separated region, there are more possibilities: either the system remains
phase separated in mixtures of two integer-valent states, or the phase
separation has at least one intermediate-valence state in its mixture.  The 
line of circles represents an approximate crossover line between these two
possibilities (the intermediate-valence mixtures lie above the circles). Note
how the metallic and insulator phases take over the phase diagram as $U$ 
increases.  In fact, the classical intermediate-valence state only occupies
a small region of phase space, because it is unstable to phase separation over
a wide region.  This could explain why there are not too many classical
intermediate-valence states seen in real materials.

\begin{figure}[htb]
\epsfxsize=3.0in
\epsffile{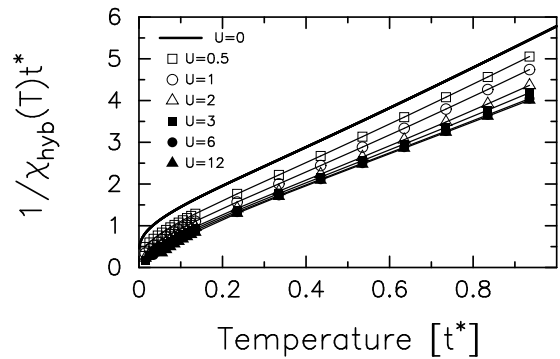}
\caption{Inverse of the hybridization susceptibility for the spinless FK model 
on a Bethe lattice with $\rho_e+\rho_f=0.5$ and $E_f=-0.75$. Note how the 
susceptibility
monotonically increases with increasing $U$ which suggests that it continues
to diverge at $T=0$.
\label{fig: hyb_susc}}
\end{figure} 

We conclude this section with a discussion about the possibility for 
spontaneous hybridization~\cite{zlatic_review_2001}.  We confine our 
discussion to the spinless model with $\rho_e+\rho_f=0.5$
for simplicity.  If we repeat the above analysis, then the region of
stability for intermediate valence at small $U$ is the same as above
$-1<E_f<0$.  But as $U$ is increased, it appears that the intermediate-valence
state should be stable for all $U$ when $E_f=-0.75$ (but we have not performed
a free-energy analysis to rule out the possibility of phase separation).
We plot the inverse of the hybridization susceptibility from 
Eq.~(\ref{eq: chi_hyb_final}) in 
Fig.~\ref{fig: hyb_susc}.  Note that it remains finite for all $T$.  But we 
can get an analytic form when $U=0$~\cite{zlatic_review_2001}, which says that
the susceptibility has a term proportional to $|\ln T|$, which does diverge at
$T=0$.  But this divergence does not guarantee that the ground state has 
spontaneous hybridization, in fact, it is known in this case that the ground
state has $\langle c^\dagger f\rangle=0$~\cite{farkasovsky_2002}, even though
the susceptibility diverges.  Here we find, down to the lowest temperature
that we can calculate, that the hybridization susceptibility increases with
$U$, which would imply that it continues to diverge at $T=0$ for all $U$.
In this case, we do not know whether the ground state would possess a 
spontaneous hybridization, or not, but the results of
\textcite{si_kotliar_1992} suggest that spontaneous hybridization is
indeed possible.  In any case, since a real material will
always have a nonzero hybridization (although it may be quite small), one
could expect a quantum-mechanical intermediate valence transition to occur
at low temperature in regions of parameter space where the hybridization
susceptibility is large for the FK model. Alternatively, spontaneous 
hybridization may also occur if the $f$-electrons are allowed to 
hop~\cite{batista_2002}.

\subsection{Transport Properties}

\begin{figure}[htb]
\epsfxsize=3.0in
\epsffile{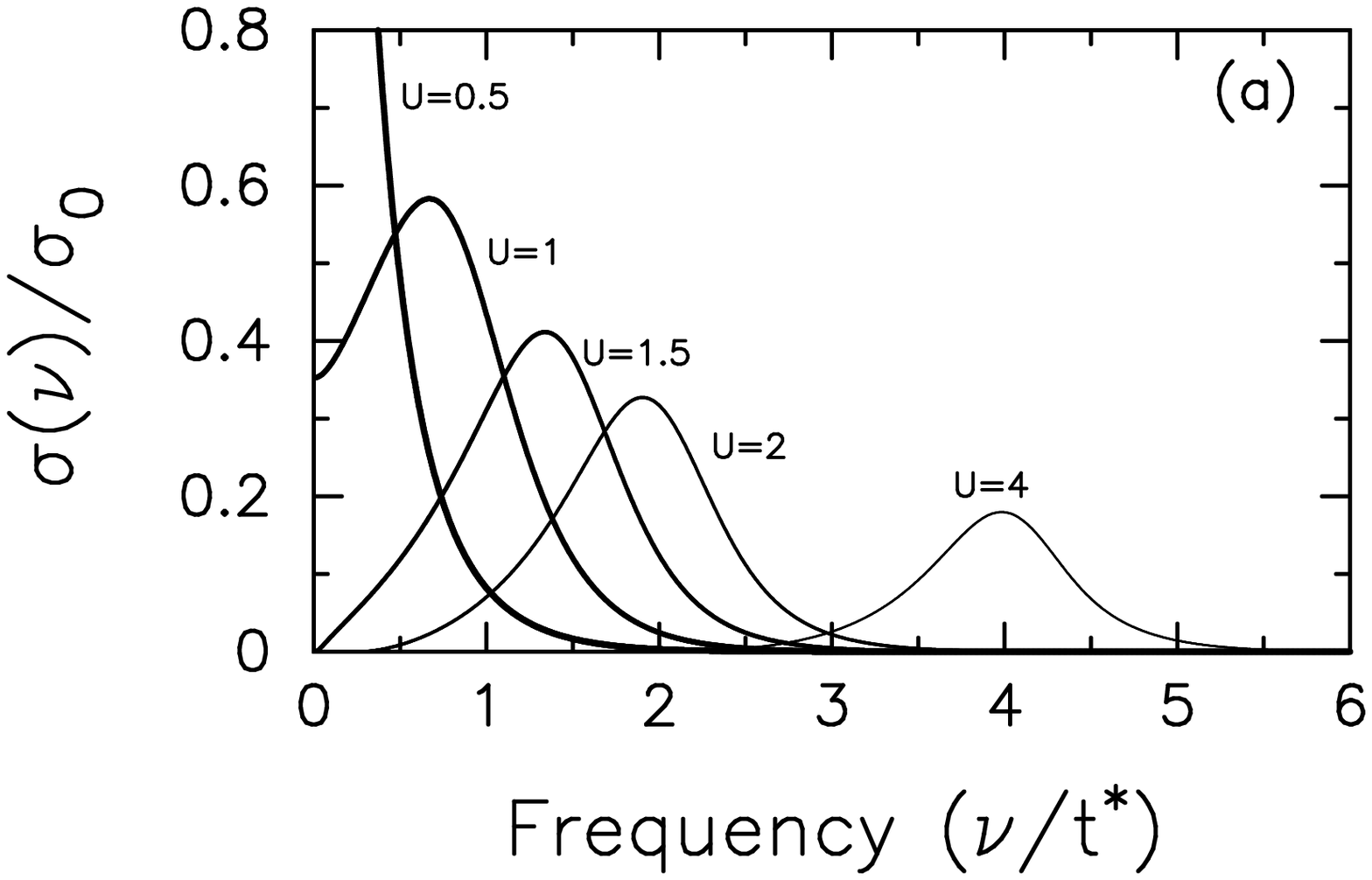}
\epsfxsize=3.0in
\epsffile{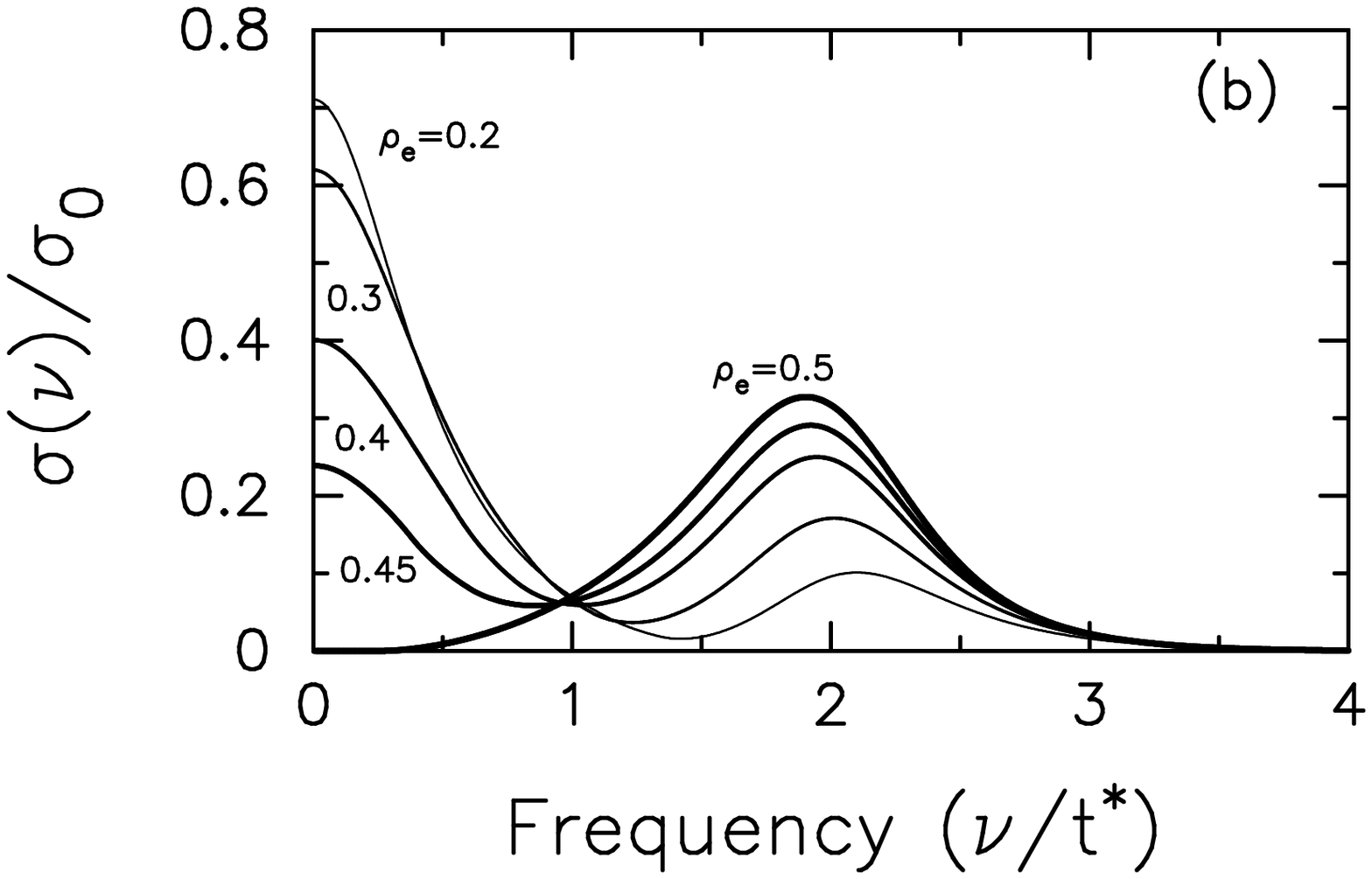}
\caption{Optical conductivity of the spinless FK model with $w_1=0.5$ on
a hypercubic lattice at $T=0.005$: panel (a) is the half filling case 
$(\rho_e=0.5$) with different values of $U$ and panel (b) is the
$U=2$ case doped away from half filling.  The numbers in panel (a) label the 
value of $U$, while the numbers in panel (b) show $\rho_e$.
\label{fig: oc_hyp}}
\end{figure}  

The optical conductivity, calculated from Eq.~(\ref{eq: oc_integrated}), 
is plotted in Fig.~\ref{fig: oc_hyp}, for the spinless FK model on the
hypercubic lattice with $w_1=0.5$.  Panel (a) is the half-filled case
($\rho_e=0.5$) and panel (b) shows the behavior for doping away from
half filling.  Both calculations are at low temperature $T=0.005$ in
the homogeneous phase (ignoring any possible CDW phases).

At half filling, the data behaves as expected.  For small $U$, the system 
has a Drude like peak, whose width is determined by the scattering rate
at low temperature (note the scattering rate does not vanish even at
$T=0$ because the FK model is not a fermi liquid).  
As $U$ increases, we see a charge-transfer peak
develop, centered at $\nu\approx U$, and the low-energy spectral weight
is suppressed because the system becomes a correlated (Mott-like) insulator
(the metal-insulator transition occurs at $U\approx
1.5$ here).  One can see that the
transition is continuous, with $\sigma_{dc}$ smoothly approaching zero as
the correlations increase.

We concentrate on the $U=2$ case as we dope away from half filling.  One can 
see that as the system is doped, it becomes metallic, because the chemical
potential now lies within the lower Hubbard band.  As a result there is a
transfer of spectral weight from the charge-transfer peak down to a Drude-like
feature as the system is doped.  What is remarkable, is that there is an
isosbestic point present (which is defined to be a point where the 
optical conductivity is independent of doping and all the curves cross
and typically occurs at $\nu\approx U/2$).
The occurrence of such isosbestic points is present in a wide
variety of models, but its origin is not well understood.

\begin{figure}[htb]
\epsfxsize=3.0in
\epsffile{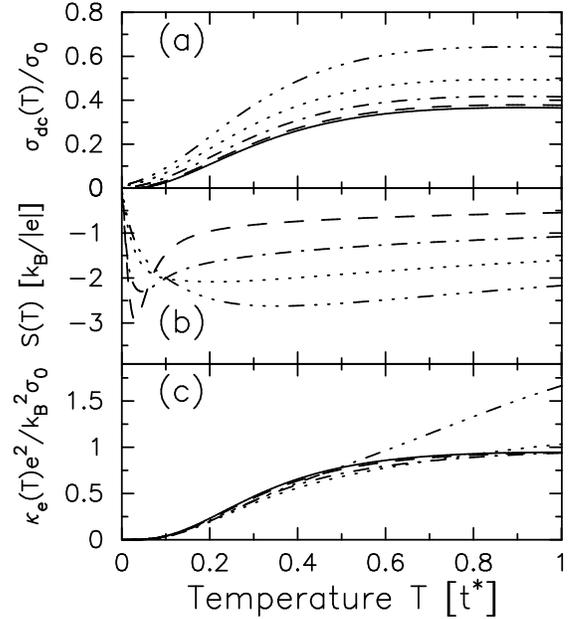}
\caption{(a) DC conductivity, (b) thermopower, and (c) electronic contribution
to the thermal conductivity, for the spinless FK model with $\rho_e=1-w_1$ and
$U=2$. Five fillings are shown: (i) $w_1=0.5$ (solid line); (ii) $w_1=0.4$
(dashed line); (iii) $w_1=0.3$ (chain-dotted); (iv) $w_1=0.2$ (dotted);
and (v) $w_1=0.1$ (chain-triple-dotted).
\label{fig: transport}}
\end{figure}  

We also plot the dc-conductivity, thermopower, and electronic contribution 
to the thermal conductivity, as derived in Eqs.~(\ref{eq: sigmadc_l}--%
\ref{eq: l22}) for a correlated system ($U=2$, solid line) with $\rho_e=1-w_1$
and five different $w_1$ values~\cite{freericks_zlatic_2002}.
For these parameters, there is always a region of exponentially small
DOS near the chemical potential at low temperature [but in this region 
$\tau(\omega)$
decreases only as a power law].  As the localized electron concentration $w_1$
moves away from 0.5, the high-temperature
thermopower increases in magnitude due to the asymmetry in the
DOS (it must vanish at 0.5 due to particle-hole symmetry) and the
low-temperature thermopower shows a sharp peak for fillings close to half
filling (the sign is hole-like, because the DOS from the lower Hubbard band
dominates the transport coefficients at low temperature); the dc conductivity 
and thermal conductivity both vanish at low $T$
due to the ``gap'' as well. The thermoelectric figure-of-merit
$ZT=T\sigma_{dc}S^2/\kappa_{el}$ is plotted in Fig.~\ref{fig: zt_l}---we find 
it is larger than one at high $T$ for $w_1<0.22$, and for fillings
close to half filling, there is a low-temperature spike in $ZT$ that can 
become larger than one over a narrow temperature range.
The spike at low $T$ is due to the large peak in $S$ and the small
thermal conductivity; but the phonon contribution to the thermal conductivity
can sharply reduce $ZT$ if the phonon thermal conductivity is much larger
than the electronic thermal
conductivity (this all electronic calculation provides only an upper
bound to $ZT$).  The Lorenz number is also plotted in Fig.~\ref{fig: zt_l}.
It gets huge at half filling, but becomes more metallic ($\approx \pi^2/3$)
as the filling
moves further away from half filling.  It is not a constant even at low
temperature because the system is not a fermi liquid.

\begin{figure}[htb]
\epsfxsize=3.0in
\epsffile{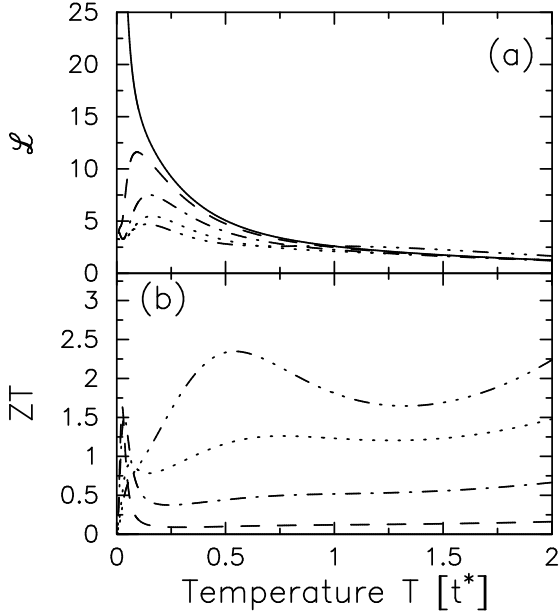}
\caption{(a) Lorenz number $\mathcal{L}k_B^2/e^2=\kappa_{el}/\sigma_{dc}T$ and 
(b) electronic thermoelectric figure of merit $ZT=T\sigma_{dc}S^2/\kappa_{el}$
for the spinless FK model with $\rho_e=1-w_1$ and
$U=2$. Five fillings are shown: (i) $w_1=0.5$ (solid line); (ii) $w_1=0.4$
(dashed line); (iii) $w_1=0.3$ (chain-dotted); (iv) $w_1=0.2$ (dotted);
and (v) $w_1=0.1$ (chain-triple-dotted).
\label{fig: zt_l}}
\end{figure}  

\subsection{Magnetic-Field Effects}
 
The magnetic field brings new features to the solution,
which we illustrate for the model with a fixed total number of 
particles and relatively large Falicov-Kimball interaction 
on a hypercubic lattice.
We restrict the number of $f$-particles per site to less than one, 
taking the limit $U^{ff}\longrightarrow \infty$, and   
choose the position of the $f$-level such that there is 
a rather sharp crossover from the high-temperature state with
a large concentration of $f$-electrons and a gap in the
single particle DOS
to the low-temperature state with a metallic conduction
band (i.e., a fermi gas)
and no $f$-electrons. That is, model parameters are such that
the renormalized $f$-electron level is slightly above the chemical potential
at $T=0$.  This is the regime that yields an anomalous magnetic 
response~\cite{freericks_zlatic_1998} and is closely related to 
experimental materials like YbInCu$_4$ 
which exhibit a valence-change transition.

The average of the z-component of the $f$-electron magnetization is 
\begin{equation}
\langle  m_f^z \rangle   
=
\textrm{Tr}_{cf}
\frac{
     ( 
        e^{-\beta [\mathcal{H}-\mu N -\mu_f N_f ]} m_f^z
     )  
     }
{\mathcal{Z}_L}
\end{equation}
where only the states with one $f$-electron in the presence of 
the magnetic field are considered. This lattice trace can be evaluated by
using the cavity method again, where the lattice trace becomes equal to
an impurity trace in the presence of an additional time-dependent
dynamical mean field
\begin{equation}
\langle m_f^z \rangle 
=\frac{\mathcal{Z}_{\textrm{cavity}}}{\mathcal{Z}_L}\textrm{Tr}_{cf}
( e^{-\beta[\mathcal{H}_{imp}-\mu N -\mu_f N_f ]}S(\lambda)m_f^z)
\label{eq: mf_trace_cavity}
\end{equation}
where $S(\lambda)$ is defined in Eq.~(\ref{eq: evolution}).
The trace is performed  by using 
the basis set $|\eta\rangle$ which diagonalizes simultaneously 
$\mathcal{H}_{imp}$ and $m_f^z$. 
Since $m_f^z$ has no matrix elements in the subspace without 
the $f$-particles and $\mathcal{Z}_{imp}=\mathcal{Z}_L/
\mathcal{Z}_{\textrm{cavity}}$, we obtain 
\begin{eqnarray}
\langle m_f^z \rangle
&=&
\frac{ \prod_{\sigma=1}^2\mathcal{Z}_{0\sigma}(\mu+g\mu_BH-U) }
     {\mathcal{Z}_{imp}}
\cr
&\times&
\sum_{\eta}^{}  \langle \eta| m_f^z |\eta \rangle 
e^{ -\beta (E_{f\eta}(H) - \mu_f)}
\cr&=&
\frac{\rho_f}{\mathcal{Z}_f}
\sum_{\eta}^{}  \langle \eta| m_f^z |\eta \rangle 
e^{ -\beta (E_{f\eta}(H) - \mu_f)}, 
\label{eq: mag}
\end{eqnarray}
where $\mathcal{Z}_f=\sum_\eta \exp[-\beta \{ E_{f\eta}(H)-\mu_f\}]$ 
is the partition function of an isolated $f$-ion, 
$E_{f\eta}(H)$ are the field-dependent eigenstates, and $\mu_f=\mu$ is the 
common chemical potential for the itinerant and localized electrons.
The average magnetization is the product of a single-ion 
response and the average $f$-filling $0\le \rho_f\le 1$, i.e., 
the magnetic response of independent 
$f$-ions~\cite{dzero_2002a,dzero_2000,dzero_2002b} is modified by 
the interaction between the $f$-electrons and the conduction sea.  
This reduces the high-temperature Curie constant and introduces 
additional temperature and field dependences. 

We illustrate these features for the spin-1/2 model $(s=S=1/2)$ 
and plot in Fig.~\ref{fig: nf_h} the temperature dependence of $\rho_f$ 
as a function of various parameters (including the magnetic field).  
\begin{figure}[htb]
\epsfxsize=3.0in
\epsffile{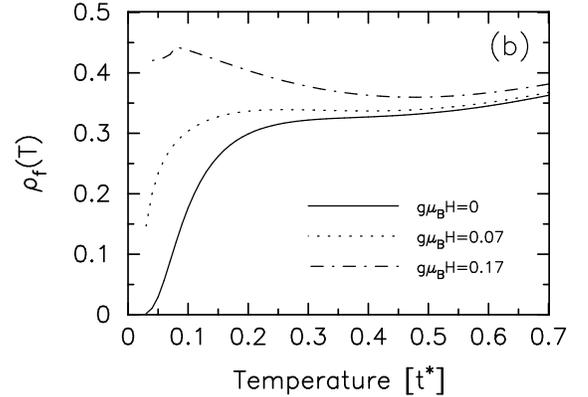}
\caption{The effect of the magnetic field on the $f$-electron 
concentration is shown for the spin 1/2 FK model with $\rho_{\textrm{total}}
=1.5$, $U=3$, and $E_f=-0.5$, with $g\mu_BH=0.07$ (dotted line) and 
$g\mu_BH=0.17$ (chain-dotted line). 
\label{fig: nf_h} }
\end{figure}
The results can be explained by noting that the renormalized $f$-level 
is just above the chemical potential and that a finite temperature 
induces an entropy driven ``transition'' (or crossover)
into the magnetically degenerate state (the crossover temperature is
denoted $T_v$).
The field pushes the renormalized $f$-level closer to the chemical 
potential, which enhances the $f$-occupation, reduces the cross-over 
temperature and makes the transition sharper. For large enough fields, 
the concentration of $f$-electrons remains finite down to T=0, 
i.e., the system goes through a field-induced metamagnetic transition.
\begin{figure}[htb]
\epsfxsize=3.0in
\centerline{\epsffile{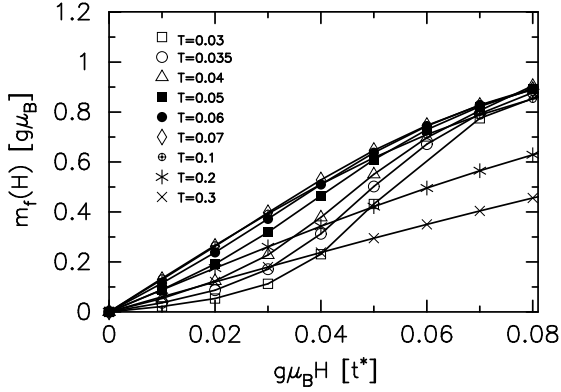}}
\caption{
Localized electron magnetization for the spin 1/2 FK model with
$\rho_{total}=1.5$, $U=4$ and $E_f=-0.5$ plotted as a function
of magnetic field for various values of temperature, as indicated
in the figure.
\label{fig: mf_h}
}
\end{figure}
This is shown in Fig.~\ref{fig: mf_h} where the magnetization 
obtained from Eq.~(\ref{eq: mag}) is plotted versus magnetic field 
for various temperatures. Below the crossover temperature, 
the low-field response is negligibly small (the Pauli susceptibility 
of the conduction electrons is neglected) but at large enough fields 
the magnetization curves go through an inflection point, 
which indicates a crossover to a magnetic state 
(and is a metamagnetic transition).
Above the crossover, the curvature of the magnetization is 
positive and typical of a well-defined local moment.
The effect of the magnetic field on transport properties is equally 
drastic. The magnetoresistance is plotted as a function of field 
in Fig.~\ref{fig: resistance_H} and it shows that the low-temperature 
metallic state is destroyed above some critical field $H_c$. 
Thus, the metamagnetic transition in the $f$-subsystem is accompanied 
by metal-insulator transition in the conduction band.
\begin{figure}[htb]
\epsfxsize=3.0in
\centerline{\epsffile{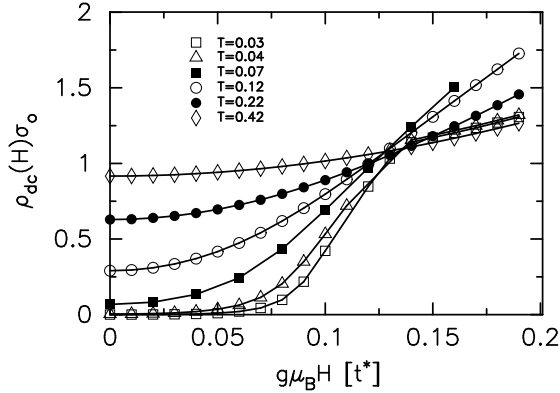}}
\caption{Magnetoresistance as a function of magnetic field
plotted for several values of temperature.
\label{fig: resistance_H}}
\end{figure}       
Taking the inflection point of the magnetization or the magnetoresistance, 
calculated for several values of $U$ and $E_f$ as an estimate of $H_c(T)$, 
we obtain the phase boundary which is shown in Fig.~\ref{fig: Hc_vs_T}, 
together with the analytic form $H_c(T)/H_c(0)=\sqrt{1-(T/T_v^*)^2}$.  
The crossover temperature is renormalized by $T_v^*=T_v/2$.
Note that the $T_v^*$ values in Fig.~\ref{fig: Hc_vs_T} differ by more than
an order of magnitude, while the ratio $k_B T_v^*/\mu_B H_c(0)$
is only weakly parameter dependent.
\begin{figure}[htb]
\epsfxsize=3.0in
\centerline{\epsffile{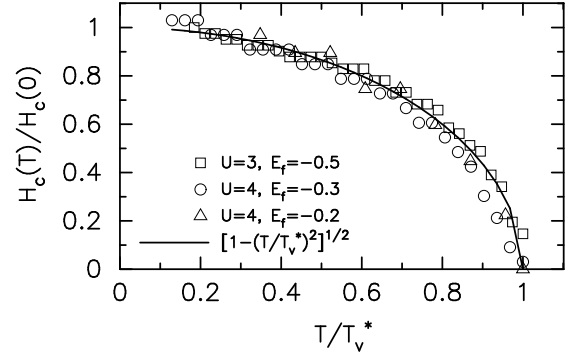}}
\caption{Normalized critical field plotted
as a function of reduced temperature $T/T_v^*$ 
for several values of $E_f$ and $U$.
The solid line represents the analytic form
$\sqrt{1-(T/T_v^*)^2}$ with $T_v^*=T_v/2$.
\label{fig: Hc_vs_T}}
\end{figure}

\subsection{Static Holstein Model}

The static Holstein model~\cite{holstein_1959,millis_littlewood_shraiman_1995,%
millis_mueller_shraiman_1996} can be viewed as a generalization of the
FK model to the continuous-spin case.  Like the FK model, it also displays
metal-insulator transitions and CDW order phase transitions.  We discuss both
possibilities here.

The phonon distribution function $w(x)$ in Eq.~(\ref{eq: w_x}) becomes
sharply peaked as $T\rightarrow 0$.  If $g_{ep}$ is small enough, it approaches
$\delta(x)$, and the ground state is a noninteracting fermi gas (at half filling,
the critical $g_{ep}$ is 0.8432 on the hypercubic lattice and 1.0854 on the
Bethe lattice).  Beyond
this critical value of $g_{ep}$, the phonon distribution function shows a 
double-peaked structure (becoming two delta functions at $T=0$), and 
the ``quasiparticles'' scatter off of the 
local phonon even at $T=0$.  This creates a non-fermi-liquid state, and the 
DOS develops a pseudogap.  As $g_{ep}$ is increased further, the pseudogap
phase becomes fully gapped (when the self energy develops a pole), and a 
metal-insulator transition takes place.
These results are illustrated for the spin-one-half model on the hypercubic
lattice in Fig.~\ref{fig: holst_mit} [similar to what was done 
in~\textcite{millis_mueller_shraiman_1996}].

\begin{figure}[htb]
\epsfxsize=3.0in
\epsffile{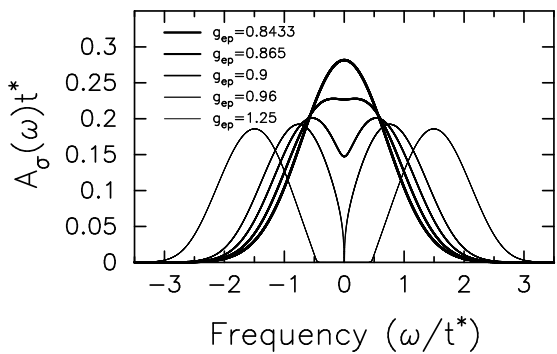}
\caption{Plot of the electronic DOS (per spin) for the harmonic static 
Holstein model at half filling on the hypercubic lattice at $T=0$. Different
values of $g_{ep}$ have different thicknesses of the lines. Note how the 
metal-insulator transition is continuous here too.
\label{fig: holst_mit}}
\end{figure}  

The other area of interest is the CDW ordered phase at half filling.  Since
a static model does not superconduct, we need not worry about that order at
all.  One interesting puzzle in real materials is that the ratio of 
twice the CDW gap to the transition temperature is surprisingly large,
usually much larger than the BCS prediction of 3.5~\cite{blawid_millis_2001}.
The CDW transition temperature can be calculated by performing ordered-phase
calculations and extrapolating them to the point where the order disappears to
produce $T_c$~\cite{ciuchi_depasquale_1999}.  These calculations show that
even in the weak coupling limit, one finds a large ratio of $2\Delta/k_BT_c$.
Since CDW systems involve a distortion of the lattice to produce the ordered
phase, one expects that anharmonic terms may play an important role in
describing the physics behind them.  So it is important to also examine 
what happens in the presence of anharmonic interactions.  

A surprising result was found when this system was analyzed on the hypercubic
lattice.  There was 
a universal scaling law for the transition temperature, when $T_c$ was
plotted against an extrapolated approximation to the wavefunction renormalization
parameter (also called the quasiparticle $Z$ factor, when one is in a fermi 
liquid), which is defined on the imaginary axis by
\begin{equation}
Z(0)=1-\frac{3}{2}\frac{\textrm{Im}\Sigma(i\pi T)}{\pi T}+
\frac{1}{2}\frac{\textrm{Im}\Sigma(3i\pi T)}{3\pi T}.
\label{eq: zfactor}
\end{equation}
The scaling law is plotted in 
Fig.~\ref{fig: holst_scale}~(a)~\cite{freericks_zlatic_jarrell_2000}.
The results for $T_c$ for a variety of coupling strengths and even for the
case of quantum-mechanical phonons, all collapse onto the same curve.  The
results for an attractive Hubbard model (X symbols) do not, indicating that this
scaling curve breaks down when the phonon frequency is made large enough.

\begin{figure}[htb]
\epsfxsize=3.0in
\epsffile{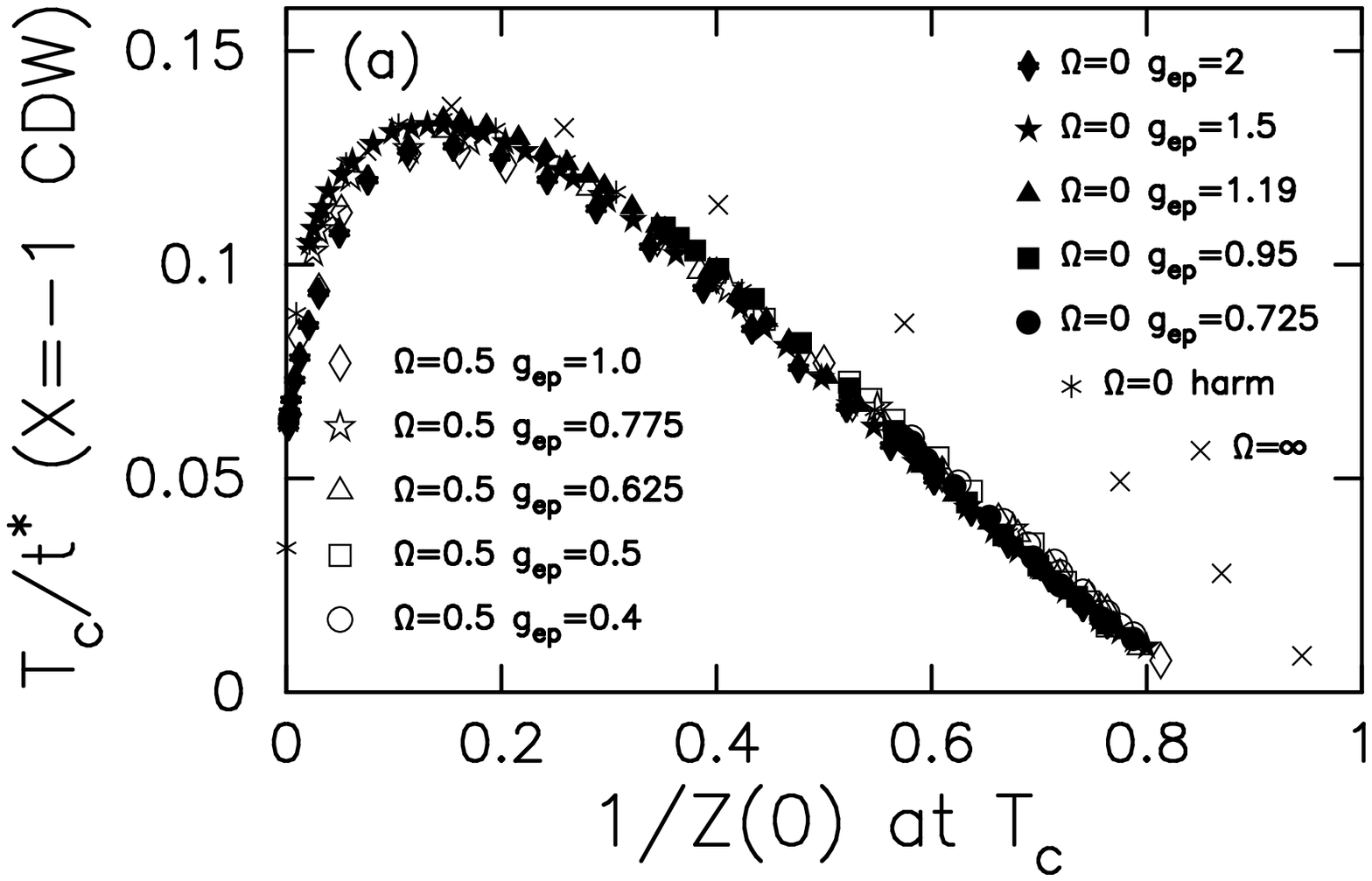}
\epsfxsize=3.0in
\epsffile{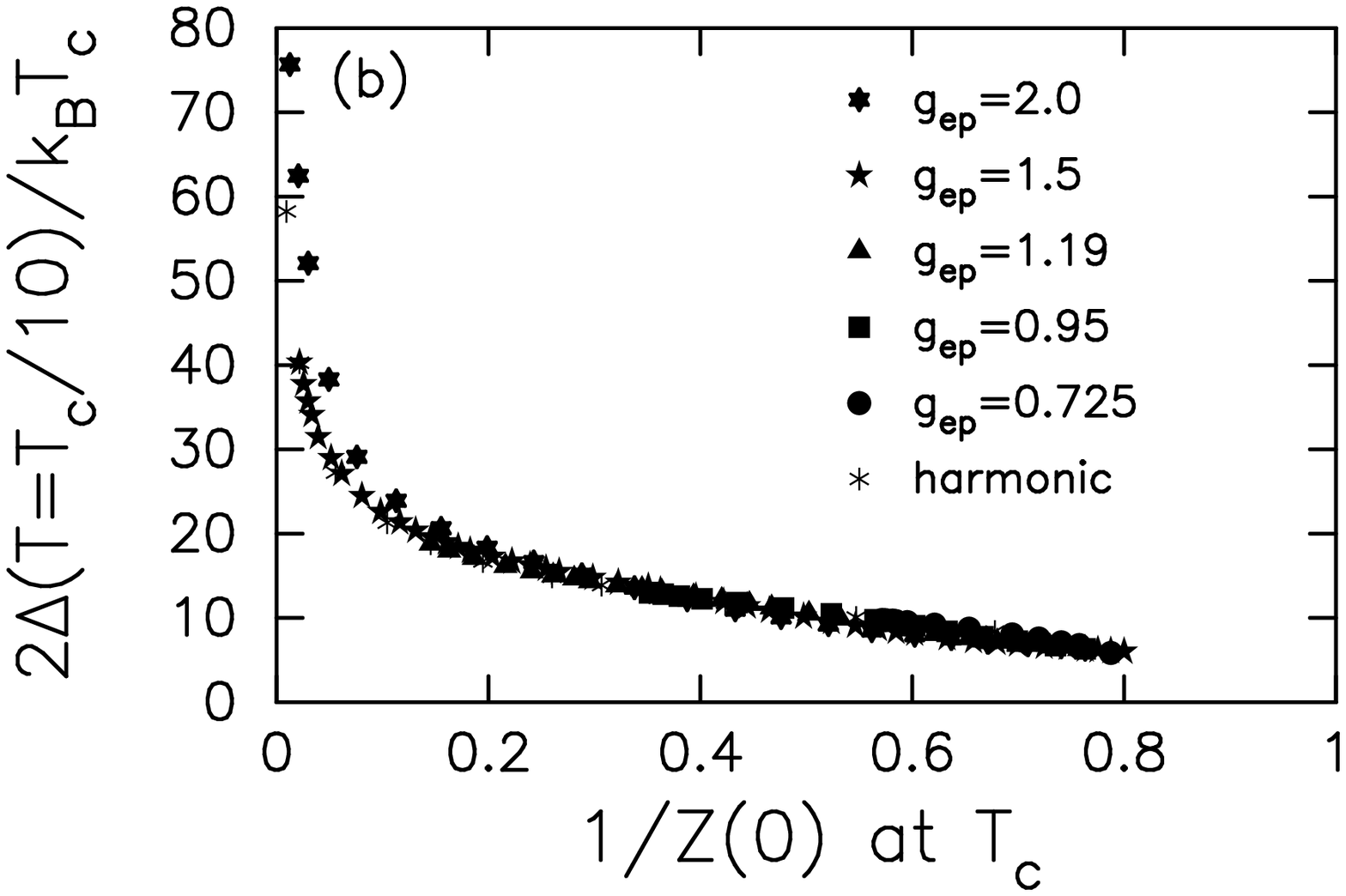}
\caption{Scaling plot of (a) the CDW transition temperature and (b) the 
CDW gap ratio versus the wavefunction renormalization extrapolated at $T_c$ for 
the static Holstein model at half filling on the hypercubic lattice. In panel 
(a), we have also included some results for a quantum Holstein model, with
a phonon frequency equal to 0.5, and for the attractive Hubbard model (which
is the infinite-phonon-frequency limit).
\label{fig: holst_scale}}
\end{figure}

The other quantity of interest to examine is the ratio of the CDW order parameter
at $T=0$ to the transition temperature.  Since one might have expected the
scaling theory for $T_c$ would have produced a universal plot there, that
result might not have been so surprising.  But, when anharmonicity is included,
its effects should be stronger at low temperature, since the phonon distortion
won't be allowed to be as large in the anharmonic case, and hence one would
naively
expect a smaller order parameter.  But, as shown in panel (b), we see that
the ratio of $2\Delta/k_BT_c$ still satisfies an approximate
scaling law when anharmonicity
is included~\cite{freericks_zlatic_2001a}.  
Hence, one should expect, generically, that this ratio will
be much larger than 3.5 except in extremely weak coupling cases (since it
does approach 3.5 as $g_{ep}\rightarrow 0$).

\section{Comparison with Experiment}

\subsection{Valence-Change Materials \label{sec: valence}}

The FK model can be used to describe the anomalous features of
rare earth intermetallic compounds that have an isostructural 
valence-change transition, as observed in the YbInCu$_4$ and
EuNi$_2$(Si$_{1-x}$Ge$_x$)$_2$ family of compounds.
These intermetallics have been attracting a lot of attention recently
~\cite{felner_novik_1986,levin_1990,figueroa_1998,%
sarrao_1999,wada_1997,garner_2000,zhang_2002}
and we describe the most typical features briefly.

The temperature-dependent properties of YbInCu$_4$, 
which have been studied most thoroughly and which we take 
as our example, are dominated at ambient pressure by 
a first-order valence-change transition at about 40 K. 
The valence of Yb ions changes abruptly from Yb$^{3+}$ above the 
transition temperature, $T_v$, to Yb$^{2.85+}$ below $T_v$
~\cite{felner_novik_1986,dallera_2002}.
The specific heat data shows at $T_v$ a first-order transition 
with an entropy change of about $\Delta S \simeq$~R~ln~8 corresponding 
to a complete loss of magnetic degeneracy in the ground 
state~\cite{sarrao_1999}. Neutron scattering does not provide any evidence
for long-range order below $T_v$~\cite{lawrence_1997}.
At the transition, the $f$-occupation becomes non-integral~\cite{dallera_2002}
and the lattice expands by about 5\%.
The crystal structure remains in the C15(b) class and
the volume expansion estimated from the known atomic radii of Yb$^{3+}$
and Yb$^{2+}$ ions is compatible with the valence change estimated from
the $L_{III}$-edge data~\cite{felner_novik_1986,cornelius_1997}.

The low-temperature phase shows anomalies typical of a 
valence-fluctuating intermetallic compound. 
The electronic specific heat and the susceptibility are 
enhanced~\cite{sarrao_1999}, the electrical resistance and the Hall 
constant are small and metallic, and the low-temperature slope of the 
thermoelectric power is large~\cite{ocko_2002}. 
The optical conductivity is Drude like, with an additional structure 
in the mid-infrared range which appears quite suddenly 
at $T_v~$\cite{garner_2000}.
The ESR data~\cite{rettori_1997} indicate a large density of states
at the fermi level $E_F$.
Neither the susceptibility, nor the resistivity, nor the Hall constant show
any temperature dependence below $T_v$, i.e., the system behaves 
as a fermi liquid with a characteristic energy scale $T_{FL}\gg T_v$. 
The magnetic moment of the rare earth ions is quenched 
in the ground state by the $f-d$ hybridization but the onset of the 
high-entropy phase cannot be explained by the usual Anderson model 
in which the low- and high-temperature scales are the same and the 
spin degeneracy is not expected to be recovered below $T_{FL}$. 
The valence-change systems, however, recover the  $f$-moment at 
$T_{v}$ which is a much lower temperature than $T_{FL}$. 

The high-$T$ phase of YbInCu$_4$ sets in at $T_v$ and is also anomalous. 
The Yb ions are in the stable ${3+}$ configuration with one $f$-hole 
and with the magnetic moment close to the free ion value 
$g_L\sqrt{J(J+1)}\mu_B=4.53 \mu_B$
($g_L=8/7$ is the Land\'e factor and $J=7/2$ is the total angular
momentum of the $4f^{13}$ hole).
The magnetic response is Curie-Weiss like with a small Curie--Weiss
temperature $\Theta$, which does not seem to be connected with $T_v$
in any simple way~\cite{felner_novik_1986,sarrao_1999} (of course one does
not expect a first-order transition temperature to be related to the
Curie-Weiss temperature, but this observation shows that the two phenomena
are not governed by the same microscopic physics).
The dynamical susceptibility obtained from neutron scattering
data~\cite{goremychkin_1993} is typical of isolated local moments, 
with well resolved crystal-field excitations~\cite{murani_2002}.
However, neither the line shape of the dynamical response nor the 
temperature dependence of the static susceptibility can be explained 
by the Kondo model assuming $T_K\simeq T_v$. 
The Hall constant is large and negative, typical of a 
semi-metal~\cite{figueroa_1998}, the electrical resistance is also very 
large and not changed much by magnetic field up to 30~T \cite{immer_1997}.
In typical Kondo systems, on the other hand,  one expects a logarithmic 
behavior on the scale $T/T_K$ and large negative magnetoresistance. 
The high temperature optical conductivity~\cite{garner_2000} shows 
a pronounced maximum of the optical spectral weight at a charge-transfer 
peak near 1~eV and a strongly suppressed Drude peak. 
The high-temperature ESR data for Gd$^{3+}$ embedded in YbInCu$_4$ resemble
those found in integer-valence semi-metallic or insulating
hosts~\cite{altschuler_1995}. 

The anomalous properties of the high-temperature phase become most 
transparent if the valence-fluctuating phase is suppressed 
completely by pressure or doping as in Yb$_{1-x}$Y$_x$InCu$_4$, where 
a substitution of 15\% Y ions stabilizes the high-temperature phase
down to T=0~K~\cite{ocko_2002,mitsuda_2002,zhang_2002}. 
The experimental results for the resistivity, susceptibility
and the thermopower~\cite{ocko_2002} are shown in
Fig.~\ref{fig: Ocko_YbY_resistivity}.
The susceptibility data show that the Curie-Weiss temperature for all
the samples is about the same and much less than $T_v$, i.e.,
the magnetic response of the high-temperature phase can be represented
by a single universal curve, provided one scales the data by an effective
concentration of magnetic $f$-ions, $\rho_f$, which is always smaller than
the nominal concentration of $f$-ions. The functional form of the 
susceptibility above 10~K agrees well with the ``single-ion'' crystal 
field (CF) theory.
At lower temperatures the susceptibility deviates appreciably
from the CF theory and shows a significant reduction of the Curie constant
$\rho_f$ which is not of the Kondo type.

\begin{figure}[htb]
\epsfxsize=3.0in
\epsffile{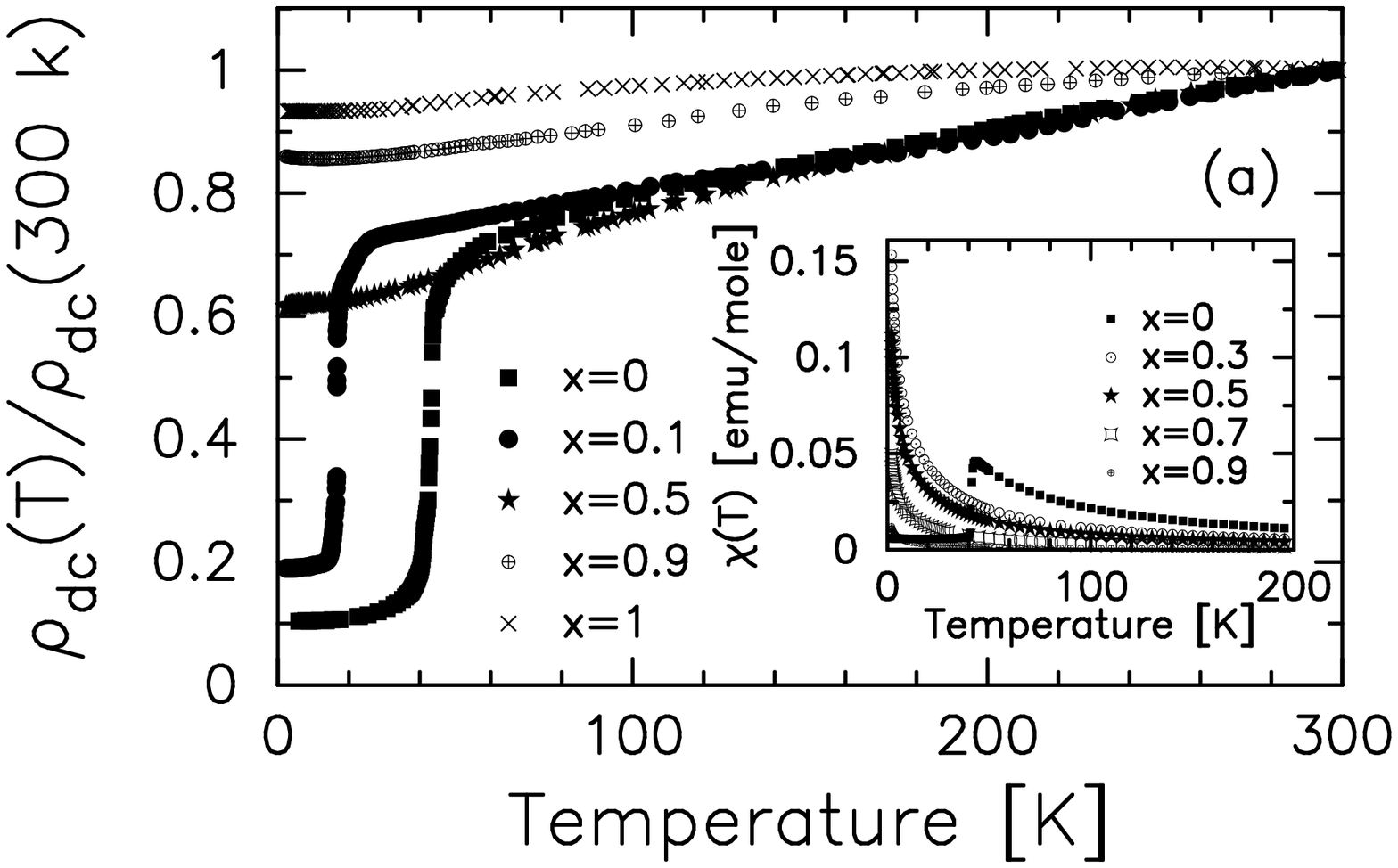}
\epsfxsize=3.0in
\epsffile{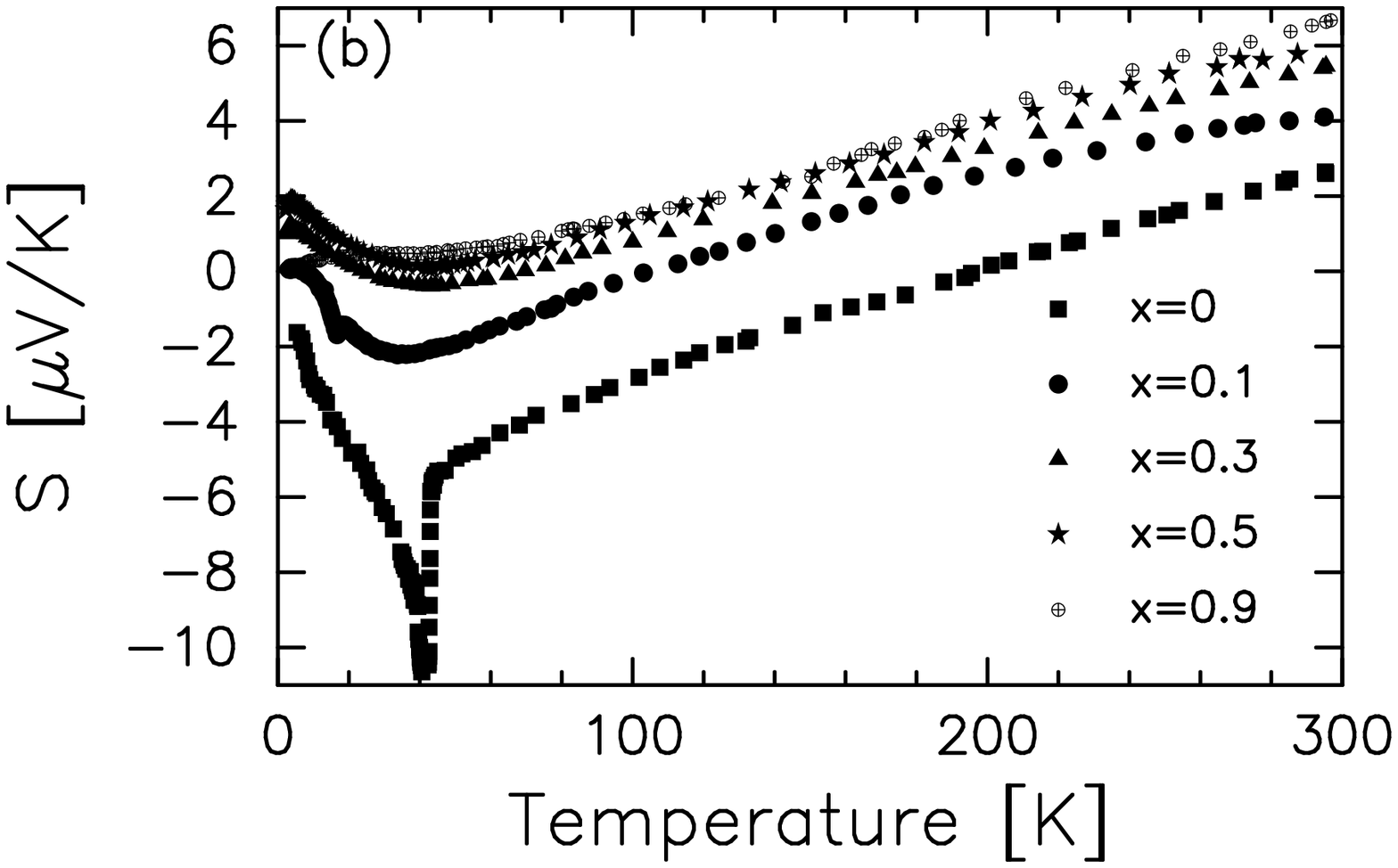}
\caption{
Panel (a) shows the resistivity and the magnetic
susceptibility of Yb$_{1-x}$Y$_x$InCu$_4$ as functions of
temperature for various concentrations of Y ions~\cite{ocko_2002}.
Note, all the ``high-temperature'' data can be collapsed onto a
single universal curve, by normalizing the susceptibility
with respect to an effective Yb-concentration (not shown).
Panel (b) shows the thermopower of Yb$_{1-x}$Y$_x$InCu$_4$
as a function of temperature for various concentrations of
Y ions~\cite{ocko_2002}. 
\label{fig: Ocko_YbY_resistivity}
}
\end{figure}

The resistivity in the high-temperature phase of Yb$_{1-x}$Y$_x$InCu$_4$
exhibits a weak maximum and the thermopower has a minimum above 100 K
but neither quantity shows much structure at low temperatures, where the
susceptibility drops below the single-ion CF values.
That is, despite the presence of the well-defined local moments,
there are no Kondo-like anomalies in transport or thermodynamic properties
of the high-temperature phase.
On general grounds, one can argue that
the discontinuity of the thermoelectric power
at the valence transition is a trivial consequence of the different
thermoelectric properties of the two phases: the thermopower of the
valence-fluctuating phase has an enhanced slope and grows rapidly up
to $T_v$, where it suddenly drops to the values characteristic of the
high-temperature phase.

The hydrostatic pressure, doping and the magnetic field also have, 
like the temperature, a strong effect on the properties 
of the valence-change materials. 
The critical temperature of  YbInCu$_4$ decreases with
pressure~\cite{immer_1997} but the data are difficult
to explain with the Kondo volume collapse model~\cite{sarrao_1999}.
Doping the Yb sites with Y$^{3+}$ or Lu$^{3+}$ ions reduces $T_v$ 
despite the fact that Y has a bigger and Lu a smaller ionic 
radius~\cite{zhang_2002}; doping the In sites by smaller Ag ions 
enhances $T_v$ as in YbIn$_{1-x}$Ag$_x$Cu$_4$
for $x\leq 0.3$~\cite{cornelius_1997,lawrence_1999}.
Thus, doping cannot be explained in terms of a chemical pressure. 
The low-temperature phase is easily destabilized by an external magnetic 
field: a critical field $H_c(T)$ induces a metamagnetic transition
which is clearly seen in the magnetoresistance and the
magnetization data. The experimental values of $H_c(T)$ define
the $H-T$ phase boundary and the analysis of many systems with different
$T_v^0$ (zero-field transition temperature) and
$H_c(0)$ (zero-temperature critical field) reveal the following universal
behavior: $H_c(T)=H_c(0)\sqrt{1-(2T/T_v^0)^2}$.
Despite the differences in $T_v$ and $H_c$, the data give a
constant ratio $k_B T_v^0/\mu_B H_c(0)=1.8$.

In summary, the Yb systems switch at the valence-change transition 
from a low-entropy valence-fluctuating phase to a high-entropy magnetic 
semi-metallic phase. The transition is accompanied by the transfer
of holes from the conduction band to the 4f shell and a metal-insulator 
transition (or crossover).
The high-temperature phase has degenerate local moments
which interact with the conduction band (and, hence, get reduced)
but the interaction is not of the usual Kondo type and no logarithmic
anomalies are seen.
Such a behavior can be qualitatively understood by assuming that 
(i) the chemical potential of the metallic phase is close 
to the location of the dip or gap in the DOS, which arises from 
many-body  interactions, and that 
(ii) the energy of the localized magnetic configuration is above 
the non-magnetic one as $T\rightarrow 0$.
The metallic phase is destabilized by magnetic fluctuations 
when the entropy gain due to additional (localized) magnetic states 
overcomes the energy loss due to the change of the ionic 
configuration and the entropy loss due to the reduction of the 
number of holes in the conduction band.
Once the chemical potential is brought close enough to the band edge,
by doping, pressure, temperature or magnetic field, the entropy of 
the band states is suppressed and the magnetic entropy of the localized 
states becomes sufficient to compensate the energy loss
and destabilize the metallic phase.
This behavior is observed in many systems and it is not likely to 
be due to band-structure effects alone. Rather, the gap or pseudogap 
which is a necessary ingredient of the above scenario seems to be 
due to many-body interactions.

The valence-change transition is also found in many Eu-based systems,  
and EuNi$_2$(Si$_{1-x}$Ge$_x$)$_2$~\cite{wada_1997} provides a typical 
and well-studied example. 
The thermodynamic anomalies are similar as in Yb systems but the transition 
temperature is higher and Eu ions undergo an almost complete valence change 
between the high-temperature, $f^7$ and the low-temperature $f^6$ 
configuration. That is, the transition is from a free-spin system at 
high temperatures to a simple metal at low temperatures. 
The electron transport is  similar as in YbInCu$_4$ but the intrinsic 
data are more difficult to measure because the volume change at the transition 
is large and samples typically crack when thermally cycled.  
In EuNi$_2$(Si$_{1-x}$Ge$_x$)$_2$~\cite{sakurai_2000} 
the sign of the thermopower is reversed with respect to Yb systems, 
which indicates an electron-like rather than hole-like transport. 

The high-temperature behavior of YbInCu$_4$ and
EuNi$_2$(Si$_{1-x}$Ge$_x$)$_2$-like compounds can be well 
described by the FK model~\cite{freericks_zlatic_1998,zlatic_freericks_2001a,%
zlatic_review_2001,zlatic_freericks_2003}. We describe the $f$-ions by 
two energetically 
close configurations which differ in their $f$-count by one, 
we take a common chemical potential for the conduction and localized 
electrons (which is 
adjusted at each temperature to conserve the total number of electrons), 
we take the FK Coulomb repulsion between the conduction electrons (or holes) 
and the additional $f$-electron (or hole) large enough 
to open a gap in the conduction band. For realistic modeling  
we should also take into account the actual crystal-field structure including
the splittings and the degeneracy of the ionic energy levels. 
For Yb ions, the state with no $f$-holes has unit degeneracy, while the
single-hole case has a degeneracy of 8 (corresponding to $J=7/2$).
The two-hole case is forbidden due to the mutual repulsion of two
holes being too large.
The 8-fold degeneracy of a single $f$-hole is further reduced by
CF splittings and in a cubic environment we expect
4 doublets (unless there is some accidental degeneracy).
The external field further splits these CF states.
For Eu ions, we take into consideration the 4$f^6$ (3$^+$) configuration
with a non-magnetic ground state and two excited magnetic states,
and the magnetic 4$f^7$ (2$^+$) configurations. Since these states
are pure spins states, there are Zeeman splittings but no CF splittings.
All other states of Eu ions are higher in energy and neglected.

To find the thermodynamic and transport properties we need the
weights of the ionic 4$f$ configurations, i.e. we need the partition
function of a more general FK model. Since the trace over $f$ states
still separates into a sum over the states which differ by one $f$-electron,
the end result can be written as,
\begin{equation}
\mathcal{Z}_{imp}
=
\mathcal{Z}_{n_f^0}
\prod_{\sigma}\mathcal{Z}_{0\sigma}(\mu)
+\mathcal{Z}_{n_f^0\pm 1}
\prod_{\sigma}\mathcal{Z}_{0\sigma}(\mu-U)
\label{eq: z_imp3}
\end{equation}
where $\mathcal{Z}_{0\sigma}(\mu)$ is defined by Eq.~(~\ref{eq: z0})
and  $\mathcal{Z}_{n_f^0}$ and  $\mathcal{Z}_{n_f^0\pm 1}$ are
the partition functions of an isolated $f$-ion with  ${n_f^0}$ 
and  ${n_f^0 \pm 1}$ $f$-electrons, respectively.
In the case of Yb ions, the $n_f^0={14}$ configuration is  non-magnetic, 
such that $\mathcal{Z}_{n_f^0=14} = 1$, and the $n_f^0-1$ configuration 
has one magnetic hole in the $J=7/2$ spin-orbit state, 
such that the partition function becomes,
\begin{equation}
\mathcal{Z}_{n_f^0=13}
=
\sum_{\eta=1}^{8} e^{-\beta [E_{f\eta}(H)-\mu_f]}.
\label{eq: Zf13}
\end{equation}
The excitation energies $E_{f\eta}(H)$ of a single magnetic hole
are split into multiplets belonging to different irreducible 
representations of the crystal.

In the case of Eu ions, the $f^6$ configuration has a non-magnetic
ground state and two magnetic $S=1$ and $S=2$ excitations. Assuming 
the usual Zeeman coupling with the magnetic field gives the ionic 
partition function, 
\begin{eqnarray}
\mathcal{Z}_{n_f^0=6}
=
1
&+&
e^{-\beta E_{S=1}}
\sum_{m=-1}^{1} e^{-\beta g\mu_BHm}
\cr
&+&
e^{-\beta E_{S=2}}
\sum_{m=-2}^{2} e^{-\beta g\mu_BHm}.
\label{eq: zfn6}
\end{eqnarray}
The $f^7$ configuration is the magnetic Hund's rule state
($S=7/2$) and its partition function is simply
\begin{equation}
\mathcal{Z}_{n_f^0=7}
=
e^{-\beta E_{S=7/2}}
\sum_{m_s=-7/2}^{7/2} e^{-\beta [g\mu_BHm_s-\mu_f]}.
\label{eq: zfn7}
\end{equation}
The weights of the two ionic configurations are given by 
\begin{equation}
w_{0}
=
\frac{\mathcal{Z}_{n_f^0}}{\mathcal{Z}_{imp}}
\prod_{\sigma}\mathcal{Z}_{0\sigma}(\mu)
\end{equation}
and
\begin{equation}
w_{1}
=
\frac{\mathcal{Z}_{n_f^0\pm 1}}{\mathcal{Z}_{imp}}
\prod_{\sigma}\mathcal{Z}_{0\sigma}(\mu-U)
\end{equation}
which enter the Green's function formalism.

\begin{figure}[htb]
\epsfxsize=3.5in
\centerline{\epsffile{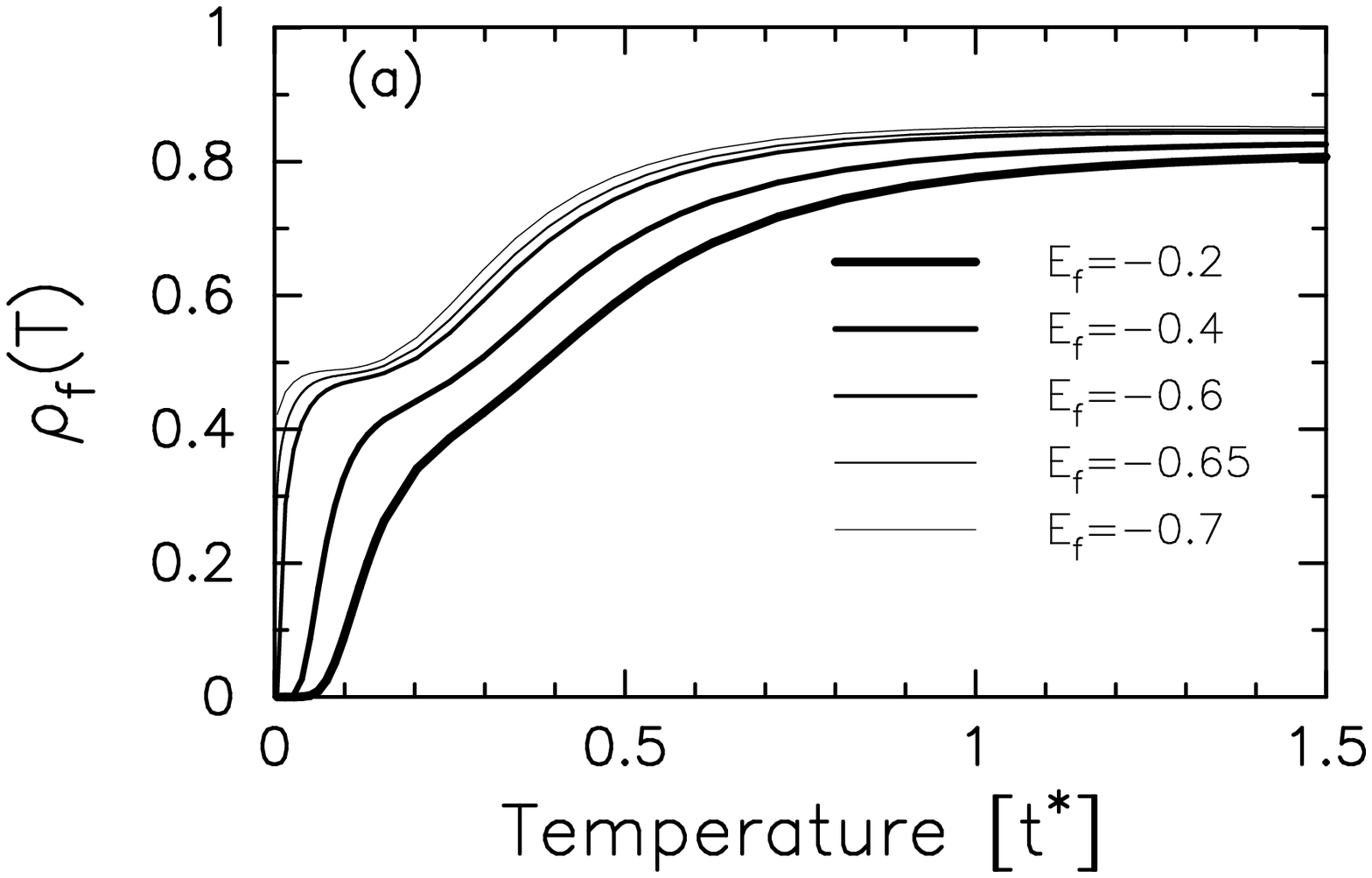}}
\centerline{\epsffile{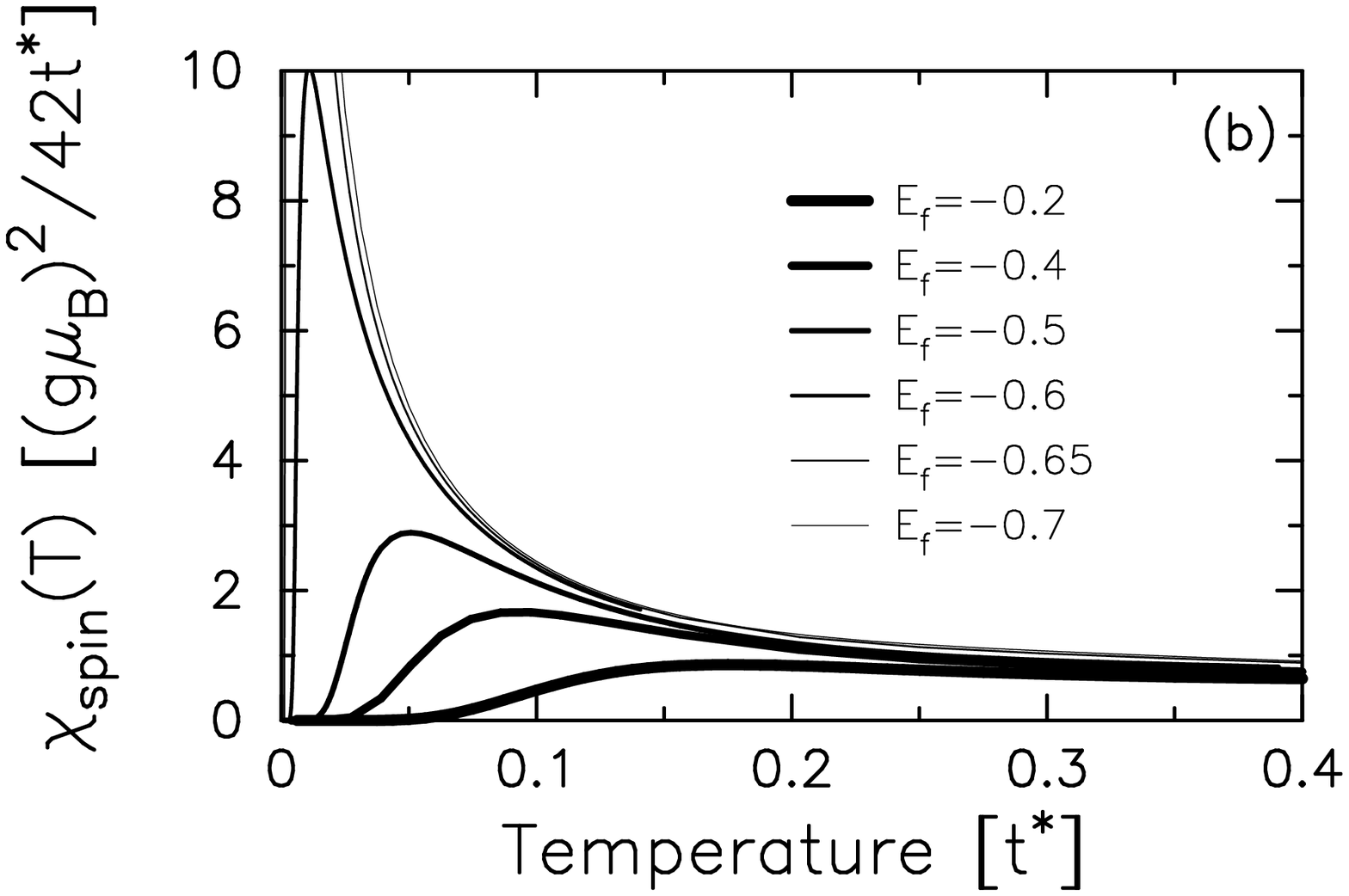}}
\caption{(a) Localized electron filling and  (b) normalized spin susceptibility
for eightfold degenerate $(S=7/2)$ Falicov-Kimball model with 
$\rho_{\textrm{total}}=1.5$, $U=2$, and various $E_f$.  
\label{fig: nf_spin}
}
\end{figure}
In what follows, we discuss a case that is similar to that of Yb: the case of 
an $S=7/2$ magnetic particle in the high-temperature phase and of a
magnetically inert unoccupied state 
in the low-temperature phase; we neglect CF splitting 
but do consider Zeeman splitting. 
The generalization to Yb ions with the magnetic $f^{13}$ states split 
by the cubic CF, or to Eu ions with excited magnetic states 
in the low-temperature $f^6$ configuration, is straightforward. 
We choose the Coulomb repulsion to be $U=2$, which is large enough to open 
a small gap in the interaction DOS (see below) and we tune the bare $f$-level 
to bring the renormalized $f$-level slightly above the chemical potential.
Thus, at $T=0$, there are no $f$-electrons in the ground state but  
as $T$ increases, the $f$-occupancy increases, producing a local-moment 
response. Since the filling of the $f$ electrons is entropically driven, 
the filling increases rapidly with $T$ as the degeneracy increases. 
In Fig.~\ref{fig: nf_spin}, we plot the average $f$-occupation 
and the spin susceptibility  (normalized by the total $m_z^2$ 
which equals 42 for spin-seven-halves). 
\begin{figure}[htb]
\epsfxsize=3.0in
\centerline{\epsffile{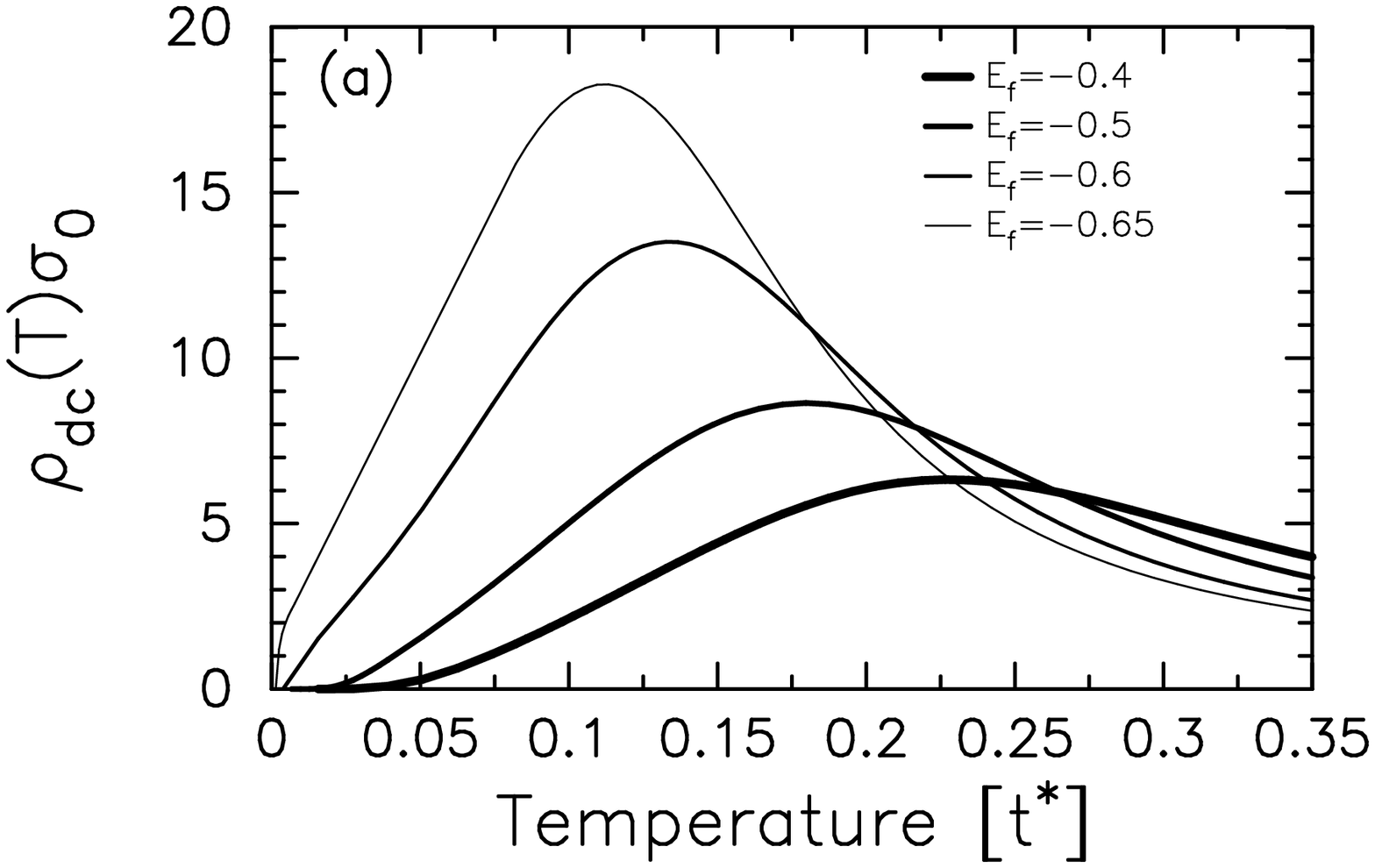}}
\centerline{\epsffile{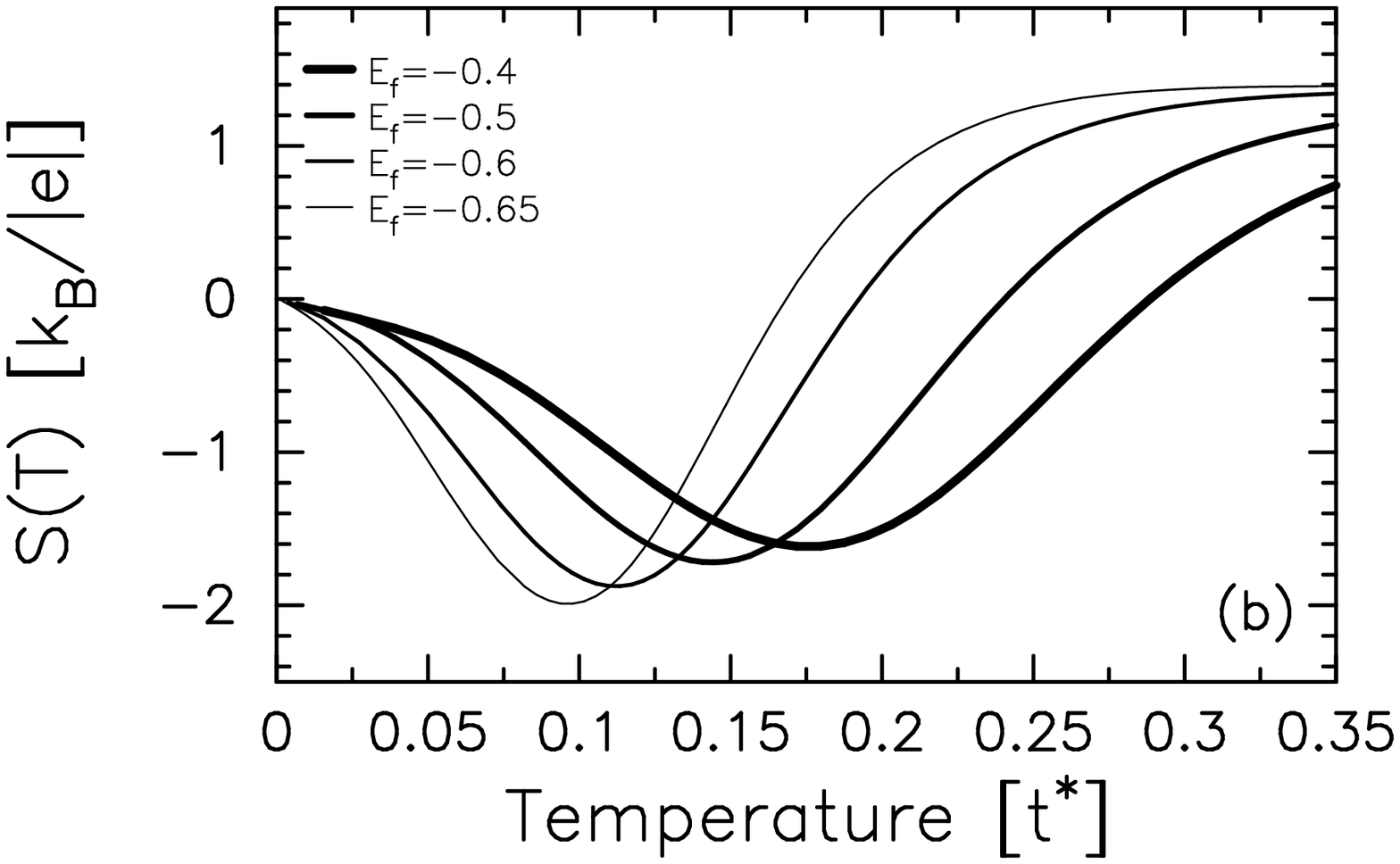}}
\caption{(a) DC resistivity and  (b) thermopower
for the ($S=7/2$) Falicov-Kimball model with $\rho_{\textrm{total}}=1.5$ and 
$U=2$.  The different curves correspond to different values of $E_f$. 
\label{fig: transport_iv}
}
\end{figure} 
Panel (a) of Fig.~\ref{fig: nf_spin} shows that as $E_f$ 
is reduced, the high temperature value of $\rho_f$ increases, 
the ``transition temperature'' decreases, and the transition becomes sharper. 
Panel (b) shows the spin susceptibility, which is Curie-like 
but with a temperature-dependent concentration of local $f$-moments.  
Hence it has a peaked form, with a sharp reduction of the magnetic 
response as $T\rightarrow 0$.  
Defining $T_v$ as the temperature at which the spin susceptibility drops 
to half of the maximum value, we find $T_v=0.05, 0.03, 0.007$ and $0.001$ 
for $E_f=-0.4, -0.5, -0.6$, and $-0.65$, respectively. In all these cases 
we are dealing with a crossover rather than a sharp phase transition. 
In the case $E_f=-0.2$ the peak is too broad for the transition to be defined, 
and for $E_f=-0.7$ the occupation of $f$-states remains finite down to 
lowest temperatures. A first-order phase transition is possible for $E_f$ 
between $-0.65$ and $-0.7$ but details of this transition cannot be 
obtained by the iterative numerical procedure, which becomes unstable. 
\begin{figure}[htb]
\epsfxsize=3.5in
\centerline{\epsffile{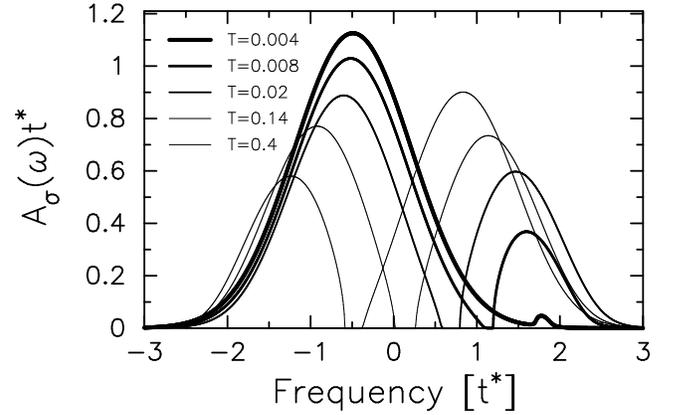}}
\caption{Interacting DOS
for the $(S=7/2$) Falicov-Kimball model with $\rho_{\textrm{total}}=1.5$, 
$U=2$, $E_f=-0.6$ and for various temperatures. 
\label{fig:dos-T}
}
\end{figure}
The effect of the $f$-level position on transport properties is shown 
in Fig.~\ref{fig: transport_iv}. 
Panel (a) shows that the resistivity has a maximum at an effective 
temperature $T^*$ which is much larger than $T_v$ and that the peak 
sharpens and moves to somewhat lower temperatures as $E_f$ is reduced. 
Effects at $T_v$ are not visible, 
except for $E_f=-0.65$, which lies close to the first-order transition. 
The thermopower results shown in panel (b) are somewhat similar, 
except that the transition at $T_v$ cannot be detected.  In addition,
there is a sign change in $S$ that occurs at higher temperature.
The transport anomalies are due to the development of a gap and 
the renormalization of the single-particle DOS at high temperature, 
driven by the increased $f$-electron occupation and the Falicov-Kimball 
interaction. 

The theoretical results shown in Figs.~\ref{fig: nf_spin} and 
\ref{fig: transport_iv}, obtained for $\rho_{\textrm{total}}=1.5$, $U=2$ and 
$E_f$ close to $-0.6$, exhibit most of the qualitative features seen 
in the experimental data. 
However, the model parameters and the absolute value 
of the temperature scale used for the static response functions 
cannot be determined unless one compares the dynamic properties 
of the model to experimental data. 

\begin{figure}[htb]
\epsfxsize=3.5in
\centerline{\epsffile{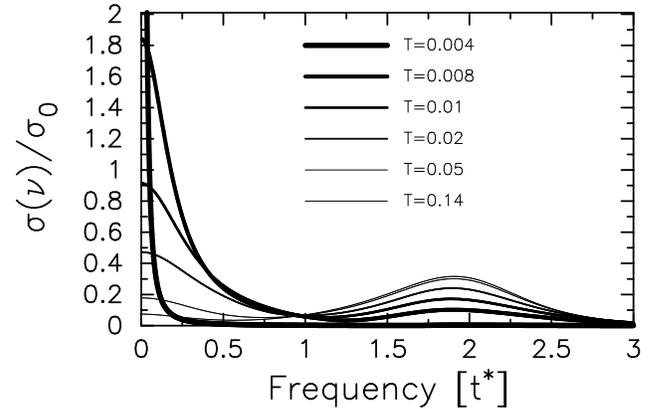}}
\caption{Optical conductivity for the ($S=7/2$) Falicov-Kimball 
model with $\rho_{\textrm{total}}=1.5$, $U=2$, $E_f=-0.6$, and  for various 
temperatures. 
\label{fig:optical-T}
}
\end{figure}
The interacting conduction DOS is plotted in 
Fig.~\ref{fig:dos-T} as a function of frequency, for the case 
$\rho_{\textrm{total}}=1.5$, $U=2$, $E_f=-0.6$,  and for various temperatures. 
Note, the temperature-induced shift in the chemical potential from the 
metallic region, with a high density of states, into the gap-region. 
Unfortunately, direct comparison with the photoemission spectra is 
not possible because the bulk and surface effects in valence-fluctuating 
systems are difficult to separate.

Fig.~\ref{fig:optical-T} shows the optical conductivity. We find a 
characteristic high-frequency hump at $\nu=U$, which is almost 
$E_f$-independent,  
and a temperature-induced transfer of spectral weight between the high- 
and low-frequency regions. A Drude peak grows as 
$\nu\rightarrow 0$ for $T\leq T_v$. Thus, while the static 
transport is dominated by a pseudo-gap on the order of $T^*$, 
the optical conductivity is sensitive to the valence-change transition. 
The high-frequency peak in $\sigma(\nu)$ allows an estimate of 
the value of the FK Coulomb interaction, yielding 
$U\approx$~1~eV~\cite{garner_2000} for YbInCu$_4$. Using $T_v=42$~K we find 
$U/T_v\approx 250$ which can be used to narrow the choice of the 
model parameters. The experimental features can be reproduced by
taking the $f$-level 
position at $E_f=-0.6$ but the procedure is not completely unique, 
because $U/T_v$ is not independent of the total number of particles. 
As an additional constraint on the parameters one should demand that 
the theory reproduces the peaks in the high-temperature transport 
properties in the right temperature range. 
Clearly, a proper quantitative description 
should also incorporate the CF splittings of J=7/2 Yb states, which would  
provide the field-induced anisotropies of the magnetization. This has not
yet been worked out in detail for either the Yb or Eu compounds.

\subsection{Electronic Raman Scattering}

\begin{figure}[htb]
\epsfxsize=3.0in
\epsffile{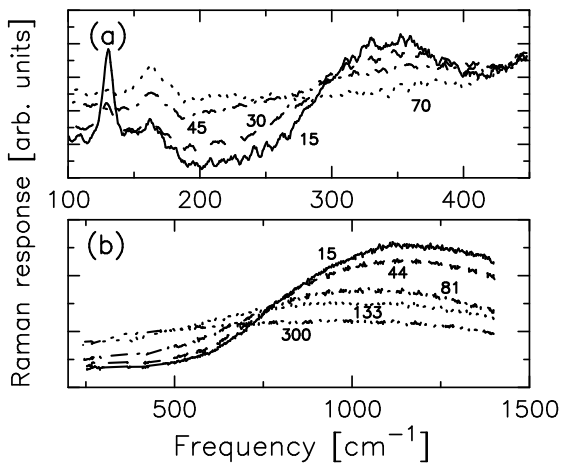}
\caption{Experimental Raman scattering results for (a) SmB$_6$ 
\cite{nyhus_cooper_fisk_1995a,nyhus_cooper_fisk_1997} and (b)
FeSi \cite{nyhus_cooper_fisk_1995b}.  The labels mark the different temperatures.
\label{fig: raman_exp}}
\end{figure}   

Experimental Raman scattering data on a wide variety of different correlated
insulators shows three distinctive features: (i) as the temperature increases,
there is a sudden transfer of spectral weight from a high-energy charge-transfer
peak to lower energies; (ii) there is an isosbestic point, where the Raman
response is independent of temperature at a special frequency, and all
Raman scattering curves cross; and (iii) if one takes the ratio of twice the
spectral range where spectral weight is depleted at low temperature 
(representative of the insulating gap) to the temperature at which the
low-energy spectral weight is restored (representative of $T_c$), then
$2\Delta/k_BT_c\gg 3.5$.  These features are shown in Fig.~\ref{fig: raman_exp}
where we plot the Raman response for (a) SmB$_6$ 
\cite{nyhus_cooper_fisk_1995a,nyhus_cooper_fisk_1997}} and (b)
FeSi \cite{nyhus_cooper_fisk_1995b}.

Using the results of Tables~\ref{table: xray_response} and 
\ref{table: chi0_raman} at $X=1$ allows us to calculate the Raman
response in the $A_{\textrm{1g}}$ and $B_{\textrm{1g}}$ channels.  Since the
Raman response ultimately is a complicated functional of the interacting
density of states, and since all insulators have the same qualitative
feature of a gap near the chemical potential, we expect the Raman response
to depend only weakly on the microscopic features of the insulating
phase.  Hence, we perform our calculations for the simplest possible 
system that goes through a Mott-like metal-insulator transition: the
spinless FK model at half filling ($\rho_e=w_1=0.5$) on a hypercubic 
lattice~\cite{freericks_devereaux_2001a,freericks_devereaux_2001b,%
freericks_devereaux_2001c}.  

\begin{figure}[htb]
\epsfxsize=3.0in
\epsffile{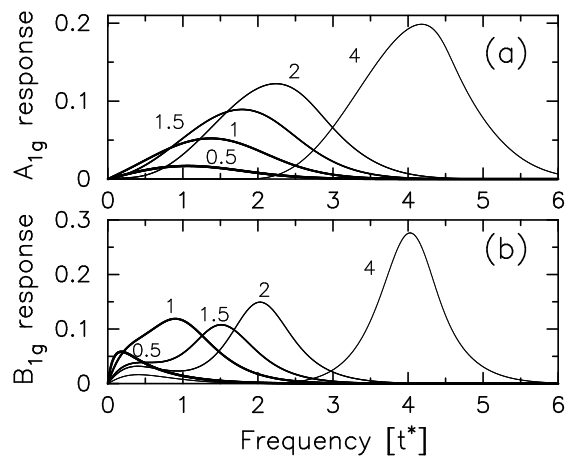}
\caption{Theoretical calculation of the nonresonant Raman response in the (a) 
$A_{\textrm{1g}}$ channel and the (b) $B_{\textrm{1g}}$ channel at $T=0.5$
and for various values of $U$.
\label{fig: raman_t}}
\end{figure} 

We begin with a plot of the Raman response at fixed temperature $T=0.5$ and
for a variety of values of $U$ in the two symmetry channels 
(Fig.~\ref{fig: raman_t}).  Note how for
weak correlations we have a linear onset, and a higher-energy cutoff (from
the band) as expected.  As the correlations increase, we see the response
separates into a charge-transfer peak centered at $U$ and a low-energy peak
(in the $B_{\textrm{1g}}$ sector only).  The vertex corrections in the
$A_{\textrm{1g}}$ sector suppress the low-energy features and we simply see
the evolution of the charge-transfer peak, which is more asymmetric in shape.  

\begin{figure*}[htb]
\epsfxsize=5.5in
\epsffile{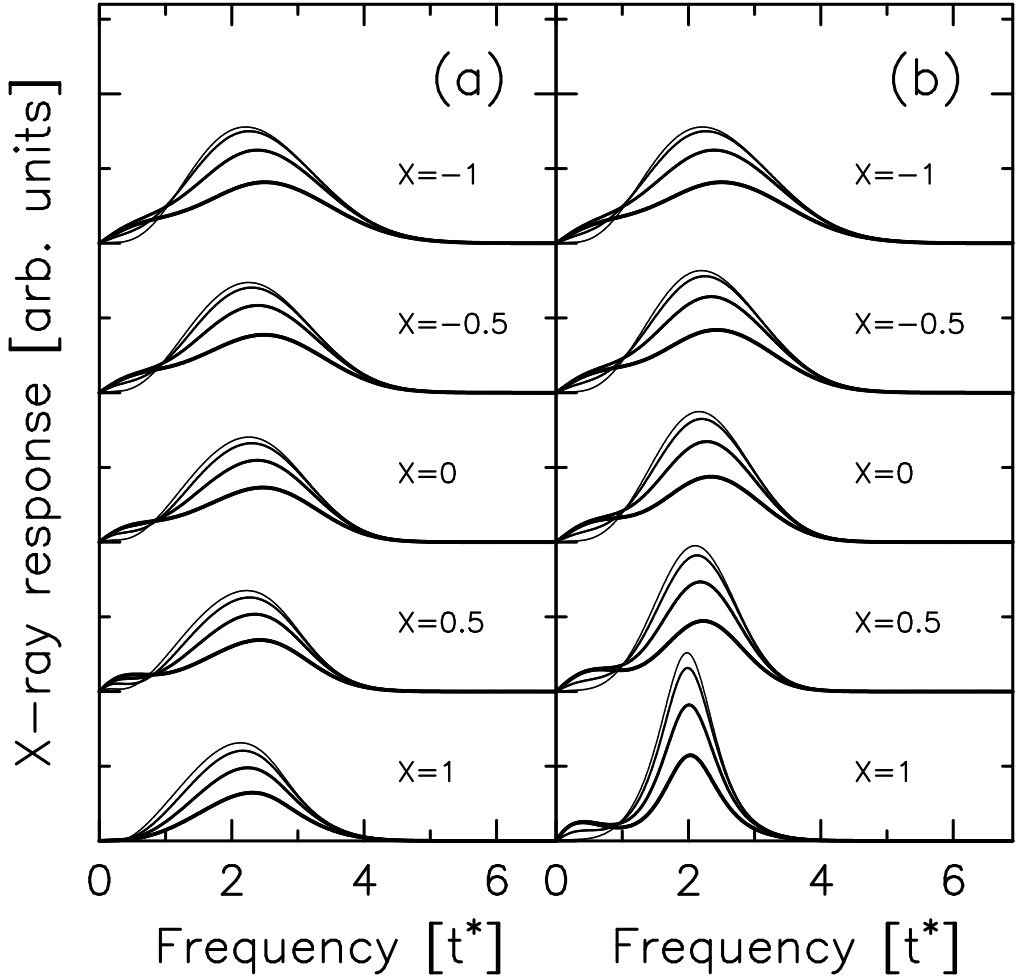}
\caption{Theoretical calculation of nonresonant 
inelastic X-ray scattering  in the (a)
$A_{\textrm{1g}}$ channel and the (b) $B_{\textrm{1g}}$ channel for $U=2$
along the zone diagonal for a range of different temperatures.  The thickness
of the lines determines the temperature which ranges from 1.0 to 0.5 to 0.25 to 
0.1.  The individual curves for different $X$ values have been shifted for
clarity. The $X=1$ curves correspond to conventional electronic Raman
scattering.
\label{fig: raman_u=2}}
\end{figure*}

We choose to examine the inelastic light
scattering for $U=2$ in more detail, because it 
is a correlated insulator that lies close to the metal-insulator
transition.  We plot in Fig.~\ref{fig: raman_u=2} the inelastic X-ray
scattering response function along the zone diagonal for a variety of different
temperatures~\cite{devereaux_mccormack_freericks_2003}.  The 
response at the zone center is the Raman response.
We find that it displays all of the three features seen in experiment:
(i) there is a low-temperature depletion of low-energy spectral weight,
(ii) there is an isosbestic point, and (iii) the ratio of $2\Delta/k_BT_c$
(as extracted from the Raman response curves) is on the order of 10--20.
These three features are generic to the plots at all $X$, except for $X=1$
in the $A_{\textrm{1g}}$ sector, where the low-energy spectral weight disappears.
This occurs because the finite-\textbf{q} vector mixes the different symmetries
and the low-energy spectral weight is seen in the other symmetries.  Note
that the results are independent of symmetry channel for $X=-1$, which 
makes for the possibility of
an interesting experimental probe of nonlocal correlations.
In general, we see a small amount of dispersion of the peaks and a generic
broadening of the peaks as we move from zone center to zone boundary, but
the isosbestic behavior remains for all \textbf{q}.  So far resonant
inelastic X-ray scattering experiments have been performed mainly at
room temperature and with polarizers only on the incident beam of light,
so direct comparison with theory is not yet feasible.

\subsection{Josephson Junctions}

Another area where the FK model has been applied to real materials is in
the field of Josephson junctions [see \textcite{freericks_nikolic_miller_2002}
for a review]. A Josephson junction is a sandwich of two superconductors 
surrounding a barrier material that can be a normal metal, an insulator,
or something in between~\cite{josephson_1962}.
In rapid single flux quantum (RSFQ) logic~\cite{likharev_2000}, one tries to
maximize the switching speed of the Josephson junction, while maintaining
a nonhysteretic (single-valued) $I-V$ characteristic.  Since the integral
of a voltage pulse over time is equal to a flux quantum for a Josephson
junction, the height of the voltage pulse is inversely proportional to the
width of the pulse; hence one wants to maximize the characteristic voltage to
achieve the fastest switching speeds.  The characteristic voltage is a product
of the critical current at zero voltage $I_c$ with the normal-state resistance
$R_n$ (slope of the $I-V$ characteristic at high voltage).  

In conventional tunnel junctions, the barrier material is an insulator, so
$I_c$ is low and $R_n$ is high; in proximity-effect junctions, the barrier
material is a normal metal, so $I_c$ is high and $R_n$ is low.  Is it possible
that one can maximize the product of $I_cR_n$ by choosing a material to lie
close to the metal-insulator transition, where both $I_c$ and $R_n$ can be
large?  We can examine this question by describing the barrier material
with the spin-one-half FK model on a cubic lattice in the local approximation
with $w_1=0.5$ and $\rho_e=1$.  This system has a metal-insulator transition
in the bulk at $U\approx 4.9t$.  To simulate the properties of a Josephson
junction, we must solve an inhomogeneous DMFT problem as first done by
\textcite{potthoff_nolting_1999} and generalized to the superconducting state
by \textcite{miller_freericks_2001}.  Using the FK model to describe the
barrier material was performed in \textcite{freericks_nikolic_miller_2001a,%
freericks_nikolic_miller_2003}
and has been reviewed elsewhere~\cite{freericks_nikolic_miller_2002}.

\begin{figure}[htb]
\epsfxsize=3.0in
\epsffile{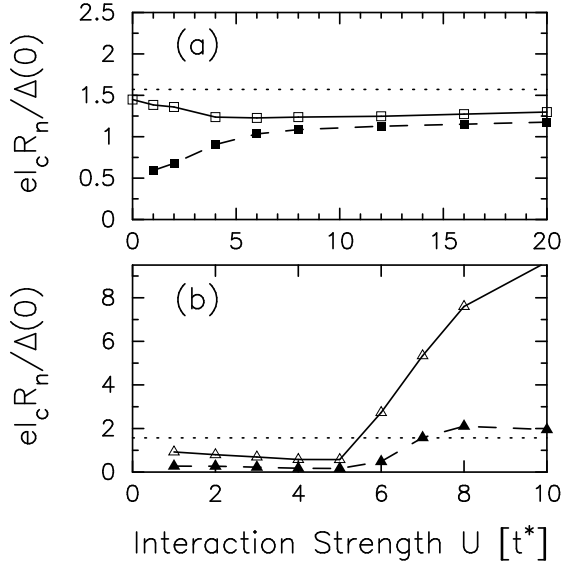}
\caption{Figure of merit as a function of the FK correlation strength in
a Josephson junction with a barrier of (a) one and (b) five planes.  The
dotted line is the Ambegaokar-Baratoff prediction for an ideal tunnel junction.
The plots are at two temperatures $T=T_c/11$ (open symbols) and $T=T_c/2$
(filled symbols). $\Delta(0)$ is the superconducting gap at $T=0$.
\label{fig: jj_icrn}}
\end{figure}

In Fig.~\ref{fig: jj_icrn}, we plot the figure of merit $I_cR_n$ versus $U$
for two cases: (a) a single-plane barrier and (b) a five-plane barrier.
Notice how the $I_cR_n$ is maximized in a ballistic metal for $N=1$ at
low $T$, but then is maximized in the tunnel junctions as $T$ increases.
The flatness of the curves in panel (a) for $U>5$ is a verification of
the \textcite{ambegaokar_baratoff_1963} analysis 
which says $I_cR_n$ is independent
of the properties of the barrier for thin tunnel junctions.  Our result
is somewhat lower than the Ambegaokar-Baratoff prediction due to 
fermi-surface effects, proximity and inverse proximity effects, etc.
What is interesting is that in panel (b) there is a marked increase in the
figure of merit at low temperature
near the metal-insulator transition.  Indeed, we find 
$I_cR_n$ is maximized on the insulating side of the metal-insulator
transition, and the optimization remains for a wide range of temperature
(the reduction in $I_cR_n$ arises mainly from the temperature dependence
of $R_n$ for a correlated insulator).
These results are consistent with experiments performed on
junctions made out of NbTiN for the
superconductor and Ta$_x$N for the barrier~\cite{newman_2001}.  
As tantalum is removed from TaN,
it creates tantalum vacancies, which are strongly interacting with the
conduction electrons and can trap them at the vacancy site.  Such physics
is described well by the FK model, with the tantalum vacancy sites serving as
the ``$f$-electrons'' that scatter the conduction electrons.

\begin{figure}[htb]
\epsfxsize=3.0in
\epsffile{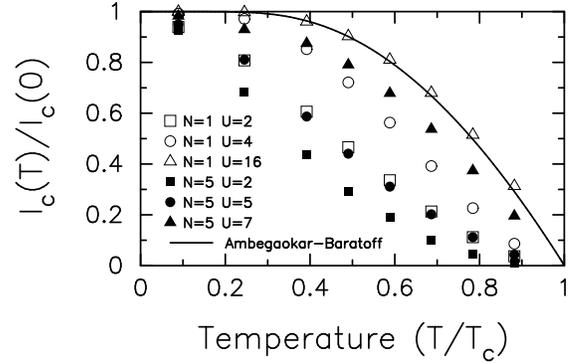}
\caption{Critical current as a function of temperature for a variety of different
Josephson junctions. Note how the correlated insulator (solid triangles) is
parallel to the thin insulator (solid line) for a wide range of $T$.
\label{fig: jj_ic}}
\end{figure} 

In Fig.~\ref{fig: jj_ic}, we plot the critical current as a function of
temperature.  Once again, our calculations produce the 
\textcite{ambegaokar_baratoff_1963} results for thin tunnel junctions.
As junctions are made more metallic, or thicker, the critical current
drops more rapidly as a function of temperature.  But the correlated
barrier, just on the insulating side of the transition, has a thermal
slope $dI_c(T)/dT$ that is essentially equal to that of the thin tunnel
junction in the range $0.4T_c<T<0.7T_c$ which is a typical operating
range for a junction.  Hence these correlated metal barriers (or SCmS junctions)
may be the optimal choice for the barrier in a Josephson junction, enabling
ultrafast superconductor-based digital electronics.

\subsection{Resistivity Saturation}

The semiclassical Boltzmann equation approach to transport indicates that
the resistivity of a material should continue to rise as the temperature
rises.  However, the theory breaks down once the mean free path of the
electrons becomes shorter than the interatomic spacing.  Hence, there is
an expectation for the resistivity to slow its increase once the mean free path
becomes too small.  An interesting set of materials that have been investigated
a quarter century ago are the so-called A15 compounds such as V$_3$Si and
Nb$_3$Ge~\cite{fisk_webb_1976}.  These materials sparked much interest as
being the highest temperature superconductors of their day, and they were
widely studied, but a number of features about these materials remain
unsolved.

In conventional electron-phonon scattering metals, the resistivity behaves
like
\begin{equation}
\rho(T)=A_\rho T+B_\rho
\label{eq: rho_high_t}
\end{equation}
at high temperature.  Here $A_\rho$ is proportional to the electron-phonon
coupling strength and $B_\rho$ is proportional to the impurity concentration.
What was found in the A15 materials was that $A_\rho$ was much smaller than
expected, given the known strength of the electron-phonon coupling and
$B_\rho$ was sizable, even in very pure samples.  This behavior was
called resistivity saturation, since it implies a slow turnover of the
high-temperature resistivity.  

In 1999, \textcite{millis_hu_dassarma_1999} solved an important puzzle
in the resistivity saturation problem.  They examined the static harmonic
Holstein model at high temperature and saw the characteristic shape
seen in resistivity saturation for strongly coupled systems.  This analysis
neglects the quantum-mechanical nature of the phonons, since their kinetic
energy is neglected, but this should be a good approximation if the
phonons are at a temperature much higher than the Einstein frequency of the
Holstein model (estimated to be about the Debye frequency of a real material).
Note that the resistivity formula used in \textcite{millis_hu_dassarma_1999}
does not properly have the velocity factors for the Bethe lattice.
This does not change the results much near half filling, but would have
a larger effect for fillings closer to the band edge.

\begin{figure}[htb]
\epsfxsize=3.0in
\epsffile{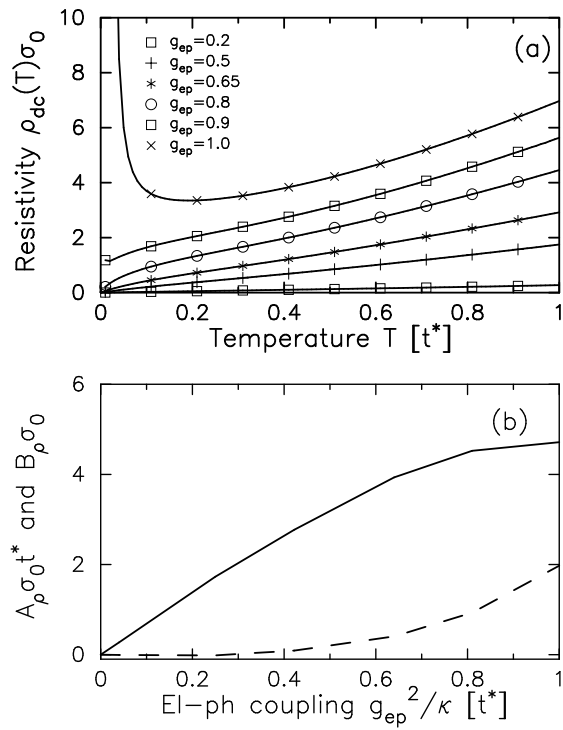}
\caption{(a) Resistivity as a function of temperature for the half filled
static harmonic Holstein model on the hypercubic lattice and (b) parameters
$A_\rho$ (solid line) and B$_\rho$ (dashed line)
fit for the range of temperature $0.2<T<1$.  Note how
there is a wide range of temperature where
the resistivity appears linear, but with a large intercept.
Note further how $A_\rho$ grows slower than $g_{ep}^2\propto \lambda$.
\label{fig: saturate}}
\end{figure}

We illustrate this phenomenon in Fig.~\ref{fig: saturate}, where we plot
the resistivity at half filling $\rho_e=1$ for the spin-one-half static
harmonic Holstein model on the hypercubic lattice.  We examine coupling strengths
ranging from a weakly coupled system (that is a fermi gas at $T=0$) to
a strongly coupled insulator, that develops a gap in the conduction
electron DOS (see Fig.~\ref{fig: holst_mit}).

The theory behind the resistivity form in Eq.~(\ref{eq: rho_high_t}) depends
on three assumptions: (i) the temperature is much higher than the phonon
energy scale, so the ions can be approximated by classical oscillators
and the equipartition theorem says $\langle x_i^2\rangle\approx T/\kappa$;
(ii) the electron-phonon interaction can be treated in second-order perturbation
theory so the scattering rate is $g_{ep}^2\langle x_i^2\rangle$; and (iii)
Boltzmann transport theory can relate the scattering rate to the resistivity.
Assumptions (i) and (ii) imply that the scattering rate is linear in 
temperature.  \textcite{millis_hu_dassarma_1999} find that the scattering
rate actually increases like $\sqrt{T}$ due to the breakdown of the second-order
perturbation theory for strong coupling.  This is the key to resistivity
saturation, as can be seen in Fig.~\ref{fig: saturate}. Indeed, forcing the
square root behavior to fit to a linear form necessarily produces all of the
observed resistivity saturation phenomena.  Note further that even for
an insulating phase, we see the same kind of ``resistivity saturation'' at
high enough temperatures.

It is interesting as well that there are a variety of materials (mainly the
doped fullerenes and the high-$T_c$ superconductors) where the resistivity
obeys the linear form of Eq.~(\ref{eq: rho_high_t}) with small or
vanishing $B_\rho$ up to very high temperatures,
where a naive estimate of the mean free path is much less than a lattice
spacing.  The DMFT analysis given here does not shed light onto that
puzzle, but recent work using other techniques has made much 
progress~\cite{calandra_gunnarsson_2001}.

\subsection{Pressure-Induced Metal-Insulator Transitions}

Nickel iodide (NiI$_2$) is a transition-metal halide that undergoes
an isostructural metal-insulator transition at room temperature as
a function of pressure~\cite{pasternak_1990}.  Nickel iodide crystallizes in the
CdCl$_2$ structure, which consists of alternating hexagonal planes of
nickel and iodine.  The nickel ions have a $2+$ valence and the iodine ions
have a $1-$ valence; the neutral sandwich I$^-$-Ni$^{++}$-I$^-$ is stacked
vertically to form the crystal. In the insulating state, 
the $S=1$ nickel spins order in a helical spin density wave, that is closely
approximated by ferromagnetic nickel planes, stacked in an antiferromagnetic
fashion.  At ambient pressure $T_N=75$~K, and $T_N$ increases by a factor of
four to 310~K at 19~GPa, where the system undergoes a metal-insulator
transition and becomes metallic.  The physical picture is as follows:
the $d$-electrons in the nickel band are strongly correlated and form a
Mott-like insulating state (which can be approximated by dispersionless
localized electrons), and the iodine $p$-bands are completely filled by
the transfer of an electron to each iodine.  As pressure increases, the
relative position of the iodine $p$-bands and the nickel $d$-bands changes,
with the iodine $p$-bands moving closer to the fermi level.  Once the
$p$-bands reach the fermi level, electrons spill from the $p$-bands to the
nickel $d$-bands, quenching the magnetic moment (changing Ni$^{++}$ to
Ni$^+$), stopping the antiferromagnetic transition,
 and allowing hole conduction within the $p$-bands.  Hence there
is a transfer of charge from the iodine to the nickel as the pressure
increases causing a metal-insulator transition.  This is precisely the kind
of transition envisioned in the original Falicov-Kimball model and optical
experiments confirm this picture~\cite{chen_yu_taylor_1993}.

\begin{figure}[htb]
\epsfxsize=3.0in
\epsffile{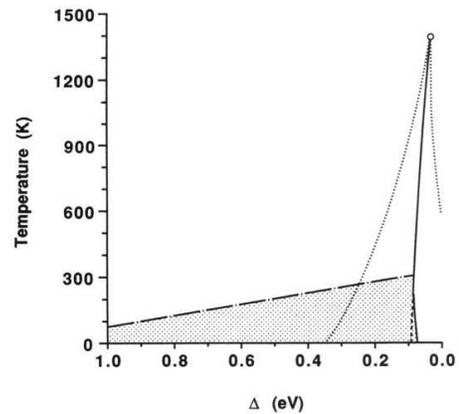}
\caption{Mean-field theory phase diagram for NiI$_2$ calculated within the
Falicov-Kimball model picture~\cite{freericks_falicov_1992}.  The shaded 
region is the antiferromagnetic 
insulator, the region above the chain-dotted line is the paramagnetic insulator.
The paramagnetic metallic phase lies to the right of the nearly vertical
solid line.  The first order transition line from an insulator to a metal
ends in a classical critical point (above that temperature, one can continuously
crossover
from the metal to the insulator).  The symbol $\Delta$ denotes the difference
in energy from the localized Ni electrons to the bottom of the I bands.  It
varies with an external parameter like pressure.  The dotted lines denote
the regions of phase space where three solutions of the mean-field theory 
equations are stabilized.
\label{fig: nii2}}
\end{figure}

Indeed, one can add spin interaction terms to the FK model, to more accurately
approximate this system, and perform an analysis of the thermodynamics and
map out a phase diagram.  This has already been done with a mean-field-theory
calculation~\cite{freericks_falicov_1992}, which shows antiferromagnetic
insulating phases, paramagnetic insulating phases, and metallic phases
and is depicted in Fig.~\ref{fig: nii2}.
The phase diagram also shows a classical critical point, where the
first-order metal-insulator transition disappears.  Unfortunately, the
critical point is estimated to lie around 1400~K, which is beyond the
disintegration temperature for NiI$_2$.  It would be interesting to 
repeat the analysis of NiI$_2$ with DMFT, to see if the estimate of the
critical point is reduced in temperature and to examine properties like
optical conductivity which have been measured.

\section{New Directions}

\subsection{$1/d$ Corrections}

The dynamical mean field theory is exact in the infinite-dimensional
limit.  But the real world is finite dimensional, and so it is important
to understand how the infinite-dimensional limit relates to finite
dimensions.  The simplest approximation for finite dimensions that one
can take is the so-called local approximation, where one performs the
DMFT, but uses the correct finite-dimensional noninteracting DOS in the
relevant Hilbert transform.  Such an approach was adopted for the
FK model by \textcite{freericks_1993b}.  Since we expect Hartree-Fock theory
to be asymptotically exact in the limit as $U\rightarrow 0$ (for $d>1$), and 
since the
DMFT uses the correct noninteracting susceptibility in determining CDW order,
the local approximation is asymptotically exact at weak coupling.  Deviations
are expected to be much stronger in the large coupling limit (where the
model can be mapped onto an effective Ising model at half filling),
due to the renormalization of $T_c$ from spatial spin-wave fluctuations
that are neglected in the local approximation.  In fact, the situation
at weak coupling is somewhat more complicated, because it is well known
that the Hartree-Fock approximation to the Hubbard model is renormalized
due to local quantum fluctuations by factors on the order of three to 
five~\cite{vandongen_1991b,vandongen_1994}.  The $1/d$ corrections to
this renormalization are typically small for the Hubbard model (on the order
of a few percent in two and three dimensions).  This weak-coupling 
perturbative analysis has not been carried out for the FK model, and there
may be surprises there, because the DMFT $T_c$ at weak coupling does not
take the conventional exponential form.  But the general thrust of this
analysis is that one expects DMFT and the local approximation to be
reasonably good at weak coupling, with larger deviations as the coupling
is made stronger.

The first attempt at a systematic expansion in $1/d$ that can include
nonperturbative effects was made by \textcite{schiller_ingersent_1995}.
They constructed a generic self-consistent two-site impurity problem,
which had feedback into an auxiliary single-site impurity problem
to calculate both the local and nearest-neighbor contributions to the
self energy (i.e. the self energy is allowed to have an additional
momentum dependence proportional to $\epsilon_{\textbf{k}}$).  This
approach suffered from two difficulties.  First, their self-consistent equations
encountered convergence problems when $U$ was made large, and they were only
able to achieve converged results for small $U$ (where one does not expect
there to be large corrections).  Second, one can immediately see that in
regions where the local contribution to the self energy is small, one
might lose causality (the imaginary part of the self energy can have the
wrong sign) if the coefficient of the $\epsilon_{\textbf{k}}$ term is
too large.  Indeed, the numerical calculations suffered from negative
DOS at the band edges as well.

A number of different approximate methods were also attempted to improve 
upon the situation.  The first method is called the dynamical cluster
approximation (DCA)~\cite{hettler_tahvildarzadeh_jarrell_1998,%
hettler_mukherjee_jarrell_2000} and we will describe it in detail
below.  The second one is based on a truncation of a memory function
expansion (which uses the Liouville operator space and continued-fraction
techniques to determine Green's functions)~\cite{tran_1998,tran_1999}.  
This technique is approximate and it includes static nonlocal correlations
and dynamic local correlations, which results in some pathologic behavior,
such as the interacting DOS remains temperature independent in the canonical
ensemble, just like DMFT.  The third method is based on a moment analysis of
the Green's function and uses a self-consistent feedback of the $f$-electron
charge susceptibility onto the conduction electron Green's 
function~\cite{laad_bossche_2000}.  This
technique also requires a number of uncontrolled approximations in order
to have numerical tractability.  One of the controversial predictions of
the moment approach is that the FK model in two dimensions has the Mott-like
metal-insulator transition occur at $U=0$. The first two techniques predict
that it occurs at a finite value of $U$ on the order of half of the
bandwidth.

We will concentrate our discussion here on the dynamical cluster approximation,
which is a systematic technique in the thermodynamic limit for
incorporating nonlocal correlations into the many-body problem.  It is
not a formal expansion in $1/d$, but rather is an expansion in $1/N_c$,
where $N_c$ is the size of the self-consistent cluster employed in
the computational algorithm.  From a physical point of view, one should
view the DCA as an expansion in the spatial size over which spatial 
fluctuations are included, so it allows large momentum (short-range)
spatial fluctuations, but does not properly describe long wavelength 
fluctuations.  The basic idea is to allow the self energy to have momentum
dependence in a coarse-grained fashion.  The Brillouin zone is divided into
$N_c$ equally sized regions, and the self energy assumes constant values
within each of these regions (but can vary from region to region).  It
turns out that one can guarantee causality is preserved with this technique.

We now describe the DCA algorithm in detail.  We begin by dividing the 
Brillouin zone into $N_c$ coarse-grained cells, labeled by the central 
point of the cell $\textbf{K}$.  In this fashion, every wavevector in
the Brillouin zone can be written as $\textbf{k}=\textbf{K}+\tilde{\textbf{k}}$
with $\tilde{\textbf{k}}$ ranging over the coarse-grained cell.  The $\textbf{K}$
points are chosen to correspond to the wavevectors of the $N_c$-site
cluster with periodic boundary conditions.  The algorithm proceeds then as
follows: (i) we choose our initial guess for the coarse-grained self
energy $\Sigma(\textbf{K},\omega)$; (ii) we construct the coarse-grained
Green's function via
\begin{equation}
\bar G(\textbf{K},\omega)=\frac{N_c}{N}
\sum_{\tilde{\textbf{k}}}\frac{1}{\omega+\mu-
\Sigma(\textbf{K},\omega)-\epsilon_{\textbf{K}+\tilde{\textbf{k}}}};
\label{eq: g_coarse}
\end{equation}
(iii) we extract the effective medium for the cluster $G_0^{-1}(\textbf{K},
\omega)=\bar G^{-1}(\textbf{K},\omega)+\Sigma(\textbf{K},\omega)$; (iv) we
solve the cluster problem for the cluster Green's functions $G_c$ given the
effective medium $G_0$ and determine the cluster self energy via
$\Sigma_c(\textbf{K},\omega)=G_0^{-1}(\textbf{K},\omega)-
G_c^{-1}(\textbf{K},\omega)$; (v) we equate the cluster self energy with the
coarse-grained self energy $\Sigma(\textbf{K},\omega)=
\Sigma_c(\textbf{K},\omega)$ and substitute into step (ii) to repeat the
process.  The algorithm is iterated until it converges.  Once converged,
one can then evaluate the irreducible vertex functions on the cluster
using analogous Dyson equations; equating the cluster irreducible vertex
functions to the coarse-grained irreducible vertex functions on the lattice
then allows one to compute susceptibilities.

Under the assumption that the technique used to solve the cluster Green's
functions given the effective medium $G_0$ is a causal
procedure, then the cluster Green's functions and self energies are
manifestly causal.  Forming the coarse-grained Green's function on the
lattice from the causal cluster self energy maintains the causality of
the Green's function.  The only place where noncausality could enter
the algorithm is in the step where we extract the effective medium, if
the imaginary part of $\bar G^{-1}$ is smaller in magnitude than the
imaginary part of the coarse-grained self energy.  But one can prove that
this never occurs, hence the algorithm is manifestly causal.

The DCA was applied to the FK model using two different techniques.  When
the cluster size was small enough ($N_c\le 16$), one could use exact
enumeration to determine the weights for every possible configuration
of localized electrons on the lattice, and since the action is
quadratic in the fermions, one can determine the partition function
exactly.  For larger clusters, one needs to determine the weights via a 
statistical sampling procedure, including the weights that are largest
in the partition sums and neglecting those that are too small.  We
will describe the exact enumeration method only here.

We let $f$ denote a configuration $\{n_1^f,n_2^f,...,n_{N_c}^f\}$ of the
localized electrons (we consider only the spinless FK model for simplicity).
Then, the partition function for the cluster becomes
\begin{eqnarray}
\mathcal{Z}_c&=&\sum_{\{f\}} 2^{N_c}e^{\beta\mu N_c/2}
e^{-\beta(E_f-\mu_f)N_c^f}\cr
&\times&\prod_{n=-\infty}^{\infty}
\frac{\textrm{Det} [ G_0^{-1}(i,j,i\omega_n)-Un_i^f\delta_{ij}]}
{[i\omega_n]^{N_c}}
\label{eq: zee_cluster}
\end{eqnarray}
where the symbol $N_c^f$ denotes the total number of localized electrons
in the configuration $f$ and $i$ and $j$ are spatial indices on the
cluster.  The determinant is over the spatial indices $i$ and $j$.
The weight for the configuration $f$ is the
corresponding term in Eq.~(\ref{eq: zee_cluster}) divided by $\mathcal{Z}_c$;
we denote that weight by $w_f$.  The cluster Green's function on the imaginary
or real axis is then simply
\begin{equation}
G_c(i,j,z)=\sum_{\{f\}}w_f [ G_0^{-1}(i,j,z)-Un_i^f\delta_{ij}]^{-1}.
\label{eq: green_cluster}
\end{equation}
The weights are determined solely from the Green's functions evaluated on
the imaginary axis, while the Green's functions on the real axis are
trivial to determine once the weights are known. The convergence of the
iterative algorithm is much slower on the real axis though.

\begin{figure}[htb]
\epsfxsize=3.0in
\epsffile{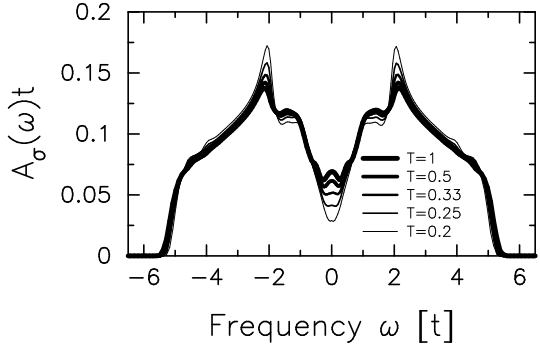}
\caption{Many-body density of states for the half-filled FK model on
a square lattice using the DCA on a $4\times 4$ cluster with 
$U=4$~\cite{hettler_mukherjee_jarrell_2000}.
Note how the DOS now has temperature dependence, and how a pseudogap
develops and deepens as $T$ is lowered.  Once a CDW opens, the system 
will open a true gap.
\label{fig: dca_dos}}
\end{figure}

One of the interesting features of the DCA is that it restores temperature
dependence to the conduction-electron DOS.  This is shown in 
Fig.~\ref{fig: dca_dos} where we plot the DOS at five different temperatures
(all above the CDW $T_c$) for an intermediate value of $U=4t$ on a
$4\times 4$ square-lattice cluster at half 
filling~\cite{hettler_mukherjee_jarrell_2000}.  Note how the pseudogap 
deepens as the temperature
is lowered.  Of course a true gap will develop in the CDW phase at the
lowest temperatures.  There is no obvious way of ``turning off'' the CDW
phase below $T_c$ anymore, because the $f$ configurations with the highest
weight will be those that correspond to the CDW ordered state below $T_c$.

\begin{figure}[htb]
\epsfxsize=3.0in
\epsffile{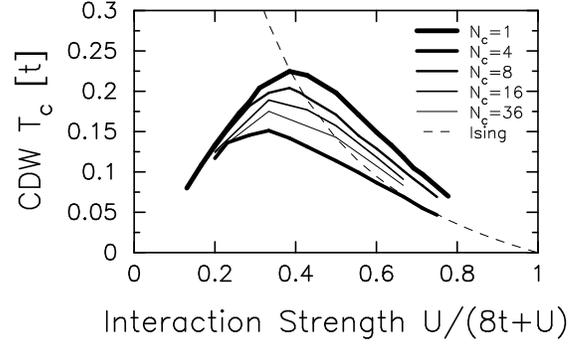}
\caption{Phase diagram for the half-filled FK model on a square lattice
with the DCA~\cite{hettler_mukherjee_jarrell_2000}.  The 
different lines correspond to different size clusters.
The $2\times 2$ cluster determines $T_c$ from the ordered phase, while the
other clusters employ a susceptibility analysis.  Also included in the figure
is the Ising model $T_c$ for the effective spin model with $J=t^2/2U$.
\label{fig: dca_phase}}
\end{figure}

We also plot the phase diagram in Fig.~\ref{fig: dca_phase} at half 
filling~\cite{hettler_mukherjee_jarrell_2000}.
Included in the plot are the local approximation, cluster calculations from
susceptibilities on clusters up to $N_c=36$, and an Ising model $T_c$ for
$J=2t^2/U$ which the FK model maps onto at strong coupling; note that the
$2\times 2$ lattice result comes from an ordered-phase calculation with
exact enumeration, rather than a susceptibility calculation.  There are
a number of interesting comments to make about this phase diagram.  First,
as expected we find the $T_c$ is hardly changed on the weak-coupling
side of the diagram, indicating that the local approximation determines the
majority of the many-body correlations in this regime.  In the large-$U$ 
regime, one can see that the DCA does converge towards the correct Ising 
model result, albeit slowly.  The $2\times 2$ data is surprising, as it
is much lower than the results for other clusters, but one can perhaps 
understand why this would be by realizing that a four-site square lattice
cluster with periodic boundary conditions is identical to a four-site
one-dimensional chain cluster, except for a change in the overall normalization
of the hopping.  Since the $T_c$ for a one-dimensional chain is equal to
zero, the suppression of $T_c$ for the four-site cluster could be arising
from the two possible interpretations of the cluster.  
\textcite{hettler_mukherjee_jarrell_2000} argue in a similar vein, where they
observe that the fermi-surface points on a four-site cluster occur at
the van-Hove singularities, which have uncommonly large scattering rates
and hence are unfavorable for CDW order.

It is possible to explore dynamical susceptibilities within the DCA as well.
One simply needs to extract the frequency and coarse-grained momentum
irreducible susceptibilities and use the relevant Bethe-Salpeter equation
to calculate the full susceptibility.  Unlike in the DMFT, where the 
irreducible vertex only renormalizes symmetry channels that have the full
symmetry of the lattice, the vertex now renormalizes all channels, because
it contains momentum dependence.  This remains one of the open problems
of interest for the DCA.

\subsection{Hybridization and $f$-Electron Hopping}

The FK model in infinite dimensions admits a simple solution,
because the Hilbert space of the impurity problem  generated by
the DMFT mapping is a direct product of invariant subspaces with
a fixed $f$-occupation.
In each subspace we can easily find the itinerant electron
Green's function and the generalized partition function
[the generating functional of Kadanoff and Baym, $\mathcal{Z}=\textrm{Det}
(G_0^{-1})$].
In the absence of quantum fluctuations in the $f$-particle number
each $f$-state has an infinite lifetime but the $f$-spectral function
acquires a finite width due to the $f-c$ coupling which acts
as an effective time-dependent potential for the $f$-electrons
and the broadening is due to the statistical averaging over all possible
states.
In this respect the FK problem is a lattice generalization of the X-ray
edge problem. Actually, as long as the coherent scattering of conduction
electrons on $f$-electrons is neglected, we can drop the self-consistency
condition Eq.(31) and use the X-ray edge model solution for the lattice 
problem~\cite{si_kotliar_1992}.
At lower temperatures, the system becomes coherent and the single-site
X-ray solution has to be replaced by the DMFT self-consistency (effective
medium), so as to keep track of all the other $f$-sites on the lattice.
Despite its simplicity, the DMFT solution of the FK model has some
interesting features and can be used to describe
physical systems in which quantum fluctuations can be neglected,
like the high-temperature phase of Yb- and Eu-compounds with
a valence-change transition or the charge-transfer metal-insulator
transition in NiI$_2$.

However, if the $f$-particle dynamics are important, which is often the
case at low enough temperature, one should consider a more general model.
For example, in the Yb-systems discussed in Section~\ref{sec: valence}, the 
valence of Yb ions below $T_v$ is much larger than predicted by the
FK model and the reduction of the local moment does not follow
from $\rho_f(T)$  but is due to quantum fluctuations and lifetime effects.
If the ground state is a mixture of 2+ and 3+ states one should
not neglect the $f-c$ hybridization and a better description would be
provided by a periodic Anderson model with an additional FK term.
The actual situation pertaining to Yb ions is quite complicated,
as one should consider an extremely asymmetric limit of the Anderson model,
in which there is no Kondo resonance (the ground state is not Kondo-like)
and there is no single universal energy scale which is relevant at all
temperatures~\cite{Krishnamurti_1980}.
We believe such a generalized model would behave as the FK model
at high temperatures and as a periodic Anderson model at low temperatures.
Indeed, incoherent scattering of conduction electrons on the FK ions
favors the gap opening even in the presence of hybridization; and
if the width of the $f$-level is large due to the FK interaction,
the additional effects due to quantum mixing should be irrelevant.
This is supported by the scaling arguments~\cite{withoff_1990}
and the numerical renormalization group analysis
of various impurity models~\cite{vojta_2002}
which show that, as long as the chemical potential is close to the gap and
the hybridization is below some critical value, the $f$-level decouples
from the conduction sea.
On the other hand, at low temperature, the chemical potential shifts
away from the gap in the DOS.
For a flat conduction band and the $f$-level close to the chemical potential,
the hybridization effects are important and they can drive the system
towards a valence-fluctuating fixed point. The most likely effect of the FK
correlation at low temperatures is to renormalize the parameters of
the Anderson model.

A generalized FK model might also be needed to describe the electronic
ferroelectrics proposed by 
\textcite{portengen_oestreich_1996a,portengen_oestreich_1996b}.
The Bose-Einstein condensation (BEC) of $f-c$ excitons and a spontaneous
polarization can be obtained by a mean-field treatment of
the hybridized model.
This picture is further supported by the examination of the spontaneous
hybridization for the spinless FK model on the Bethe lattice, where it is
found  that as the temperature is lowered the 
system appears to have a logarithmic divergence
in the spontaneous hybridization susceptibility at $T=0$. Normally we
cannot reach such a state because the system will have a phase transition
to either a phase separated state or a charge-density wave, but we can tune
the system so that it remains in a classical intermediate-valence state down
to $T=0$.  When this occurs, effects of even a small hybridization
will take the system away from the FK fixed point at low enough
temperature.

Another generalization is obtained by allowing direct $f-f$ hopping,
$t_{ij}^f$, which leads, for $t_{ij}^f=t_{ij}^c$, to the rich physics
of the Hubbard model~\cite{hubbard_I_1963}.
The case $t_{ij}^f \ll t_{ij}^c$ is also interesting, because
even a small $f-f$ hopping induces an $f-c$ coherence and gives a ground
state (for large $U$)
which is either a BEC of electron-hole pairs ($f-c$ excitons) or
an orbitally ordered state~\cite{batista_2002}.
Unfortunately, as soon as the $f$-electron dynamics is restored,
the evolution operator connects all sectors of the Hilbert space
and the resulting DMFT equations describe a two-level system with
each level coupled to an arbitrary external field.
This problem (or the equivalent Anderson impurity problem) is much
more difficult to solve than the FK impurity problem with a single external
field coupled to an electron.

\subsection{Nonequilibrium Effects}

Equilibrium properties in the DMFT have been studied extensively.  Much less
effort has been placed on nonequilibrium calculations, where the correlated
many-body system is driven away from equilibrium by large amplitude external
fields or driving voltages.  The formalism for DMFT in a nonequilibrium
situation has been addressed by \textcite{schmidt_monien_2002}.  They considered
the case of a Hubbard model that interacts with a large amplitude 
electromagnetic field with a constant driving frequency $\omega$.  The
formalism employs Keldysh techniques to derive the three Green's functions
of relevance: the advanced, retarded, and so-called lesser Green's function.
They find that in the steady state, one can perform calculations in frequency 
space, if the driving frequency is uniform.  They applied their calculations
to the Hubbard model, and used the iterated perturbation theory to evaluate
the impurity-problem dynamics.  Nevertheless, their formalism is general,
and can be used for FK model problems as well.  In addition, one could choose
to perform calculations purely in the time domain.  This would allow
one to calculate the response of the system to large amplitude, and fast
(i.e. femtosecond) electromagnetic pulses.  In this case, one would be
interested in the transient response that brings the system back to the 
equilibrium state after being disturbed.

There is another class of problems that are ideal for being considered with
Keldysh techniques.  Namely, the problem of the steady-state current
response to a voltage applied over an inhomogeneous device.  One can imagine
stacking correlated planes (described by the FK model) in between a 
semi-infinite
number of noninteracting metallic planes (above and below) that form ballistic
current
leads for the device.  One enforces current conservation from plane to plane,
which then determines the change in the voltage from plane to plane in the
device.  In this fashion, one can calculate the self-consistent current-voltage
characteristic of a nonlinear device that contains correlated materials
(and also the voltage profile throughout the device).  
The leads can be described by a wide class of Hamiltonians, including the 
Hartree-Fock description of a ferromagnet, or a Bardeen-Cooper-Schrieffer
superconductor, or a diffusive FK metal, and the barrier material (of 
arbitrary width) can be described by the FK model tuned through the 
metal-insulator transition.  Then, since the total action is quadratic
in the fermionic variables (although it is not time-translation-invariant)
the path integral over the Keldysh contour can be evaluated by taking the
determinant of the corresponding continuous 
(matrix) operator.  Discretizing the operator
as was done when we described the localized electron spectral function,
will then allow for the response to be solved exactly.  We believe this will
be the first example of a correlated electronic system that can undergo a
metal-insulator transition, and have it's nonequilibrium
response determined exactly.  The calculation is nontrivial, but is likely to
produce a series of interesting results.  One can apply such a formalism to
a wide variety of devices such as Josephson junctions, spin-valve transistors,
ballistic spin filters, or thermoelectric coolers.  The choices are endless,
as long as one always restricts the effective actions to remain quadratic
in the fermionic variables.  Generalization to more complicated systems is
also possible, but usually requires one to resort to approximate methods
to solve the corresponding impurity problems.

\section{Conclusions}

It has been over thirty years since Falicov and Kimball introduced their
model to describe the physics behind rare-earth and transition-metal
metal-insulator transitions.  The model has a rich history and much effort
has been devoted to solving the model and illustrating its properties.  
The field was dramatically advanced fourteen years ago when Brandt and 
Mielsch showed that the model could be solved exactly in the limit of
large dimensions. Since then much work has been done on investigating the
solutions of the model.

In this review we have covered the exact solution of the Falicov-Kimball model
(and the related static Holstein model) with dynamical mean-field theory.  Our 
focus was on developing the formalism with a path-integral approach that
concentrated on the Matsubara frequency representation of the fermions. 
Falicov-Kimball model physics 
is a mature field where nearly all thermodynamic properties of the solutions
have been worked out in the limit of large spatial dimensions.  This includes
charge-density-wave transitions, metal-insulator transitions, phase separation,
electrical and thermal transport, inelastic light scattering, and so on.
The model has also been employed in more applied problems such as
investigating thermodynamic properties of Josephson junctions.

These exact
solutions are useful for two reasons.  First, there is a growing list of 
materials that appear to be able to be described by FK model physics,
including YbInCu$_4$, EuNi$_2$(Si$_{1-x}$Ge$_x$)$_2$, Ta$_x$N, and
NiI$_2$.  Similar models have been also applied to the doped manganites
and to diluted magnetic semiconductors. Second, exact solutions are
useful benchmarks for illustrating the phenomena associated with strong
correlations and for testing different approximation methods for
their accuracy.

The newer frontiers that will be examined in the future include a thorough
investigation of equilibrium properties in finite-dimensional systems,
the addition of hybridization or itineracy of the $f$-electrons
and the examination of nonequilibrium effects in all dimensions.  The
success to date with this model, indicates that the time is ripe for more
applications in the near future.  It seems likely the FK model can be
employed to determine a number of interesting applications in correlated
devices, which can lay the groundwork for later efforts in
more complicated correlated systems.  We look forward to seeing how the
field will develop in the ensuing years and hope that this review will
encourage others to enter the field and rapidly contribute to its future 
development.

\acknowledgments

Our work on the Falicov-Kimball model and related models has been supported
over the past few years by the National Science Foundation under grants
DMR-9973225 and DMR-0210717 and by the Office of Naval Research under grants
N00014-96-1-0828 and N00014-99-1-0328.  
V.Z. acknowledges support from the
Swiss National Science Foundation grant no. 7KRPJ65554, the Alexander von 
Humboldt foundation and the Ministry of Science of Croatia.
In addition, we are grateful to the many 
useful discussions and
collaborations with colleagues on the properties of the Falicov-Kimball
model.  While we cannot list everyone who has influenced us, we especially would
like to thank P. Anderson,
J. Annett, I. Aviani, D. Belitz, U. Brandt, A. Chattopadhyay, A. Chen,
W. Chung, S.L. Cooper, G. Czycholl, T. Devereaux, L. Falicov, 
P. Farsakov\v sk\'y,
Z. Fisk, C. Geibel, A. Georges, Ch. Gruber, R. Hackl, Z. Hasan, M. Hettler,
J. Hirsch, V. Janis, M. Jarrell, J. Jedrezejewski, T. Kirkpatrick,
T. Klapwijk, M. Klein, G. Kotliar, R. Lema\'nski, E. Lieb, N. Macris,
G. Mahan, J. Mannhart, M. Mihjak, P. Miller, A. Millis, E. M\" uller-Hartmann,
N. Newman, B. Nikoli\'c, M. O\v cko, P. Nozi\'eres, 
Th. Pruschke, J. Rowell, J. Sarrao,
D. Scalapino, J. Serene, L. Sham, S. Shastry, Z.-X. Shen, A. Schiller,
A. Shvaika, F. Steglich, N. Tahvildar-Zadeh, D. Ueltschi, P. van Dongen, 
T. van Duzer, and D. Vollhardt.
We also thank P. van Dongen for a critical reading of this manuscript.

\section*{List of Symbols}

\begin{table}[h]
\begin{ruledtabular}
\begin{tabular}{ll} 
$A_{\textrm{1g}},B_{\textrm{1g}},B_{\textrm{2g}}$&symmetries for
inelastic light scattering\\
$A_\rho$ ($B_\rho$)&coefficients for resistivity saturation\\
$A_\sigma(\omega)$&interacting conduction electron DOS\\
$A_\sigma(\epsilon,\omega)$&spectral function\\
Det&determinant\\
$E_{f\eta}$&localized electron site energy\\
$E_f^*$&renormalized $f$-level\\
$F_{Helm}$&Helmholz free energy\\
$F_\eta(z)$&localized electron Green's function\\
$F_\infty(z)$&Hilbert transform\\
$G(z)$&local conduction electron Green's function\\
$G_0(z)$&effective medium\\
$H$&magnetic field\\
$\bar K$& cluster momentum\\
$L_{ij}$&transport coefficients\\
$S$&thermopower\\
$2S+1$&number of $\eta$ states\\
$S_L$ ($S_{imp}$)&lattice (impurity) action\\
$T$&temperature\\
$T_v$&valence-change transition temperature\\
$T_\tau$&time-ordering operator\\
Tr&trace\\
$U$&FK interaction\\
$U^{ff}_{\eta\eta^\prime}$&$f-f$ Coulomb interaction\\
$V$&number of lattice sites\\
$W_i$&integral weight factors\\
$X(\textbf{q})$&momentum parameter\\
$Z_{n\sigma}$& argument $(i\omega_n+\mu-\Sigma_{n\sigma})$\\
$ZT$&thermal transport figure-of-merit\\
$Z(0)$&wavefunction renormalization factor\\
$c^\dagger_{i\sigma}~(c_{i\sigma})$&creation (annihilation) operator of
itinerant\\
& electron at site $i$ with spin $\sigma$\\
$e$&electric charge\\
$\exp$&exponential\\
$f^\dagger_{i\eta}~(f_{i\eta})$&creation (annihilation) operator of
localized\\
& electron at site $i$ with spin $\eta$\\
$f(\omega)$&Fermi-Dirac distribution\\
$g$&gyromagnetic ratio (conduction electrons)\\
$g^{aux}$&auxiliary Green's function\\
$g_{ep}$&electron-phonon coupling (deformation potential)\\
$g_f$&gyromagnetic ratio (itinerant electrons)\\
$j_n$ ($j_Q$)&number (heat) current operator\\
$k_B$&Boltzmann constant\\
$\ln$&natural logarithm\\
\end{tabular}
\end{ruledtabular}
\end{table}

\begin{table}[h]
\begin{ruledtabular}
\begin{tabular}{ll}
$m_\sigma$ $(m_\eta)$&$z$-component of spin of electrons\\
$2s+1$&number of $\sigma$ states\\
$t$&hopping\\
$t^*$&rescaled hopping\\
$v(\epsilon)$&velocity operator\\
$w(x)$&phonon distribution function\\
$w_0=1-w_1$&empty site density\\
$w_1$&localized electron (ion) density\\
$x_i$&phonon coordinate at site $i$\\
$\Gamma^{cc}$&irreducible charge vertex\\   
$\Delta$&superconducting gap at $T=0$\\
$\Sigma$&self energy\\
$\Omega_{uc}$&unit cell volume\\
$\alpha_{an}$&anharmonic phonon potential (quartic)\\
$\beta$&1/T\\
$\beta_{an}$&anharmonic phonon potential (cubic)\\
$\gamma(\textbf{q})$&parameter in charge susceptibility\\
$\eta(\textbf{q})$&deviation from local susceptibility\\
$\xi_0$ ($\bar\xi_0$)&parameters for $F_\eta$ calculation\\
$\kappa$&spring constant\\
$\lambda(z)$&dynamical mean field\\
$\mu$&chemical potential, conduction electrons\\
$\mu_B$&Bohr magneton\\
$\mu_f$&chemical potential, localized electrons\\
$i\nu_l=i \pi T 2l$&bosonic Matsubara frequency\\
$\psi$ ($\bar\psi$)&fermionic Grassman variable\\
$\rho(\epsilon)$&bare conduction electron DOS\\
$\rho_{dc}$&dc resistivity\\
$\rho_e$&itinerant electron density\\
$\rho_f$&localized electron density\\
$\sigma(\nu)$&optical conductivity\\
$\sigma_0$&conductivity unit\\
$\sigma_{dc}$&dc conductivity\\
$\tau$&imaginary time\\
$\tau_\sigma(\omega)$&relaxation time\\
$\chi^\prime$&spin susceptibility\\
$\chi^{cc}$&conduction electron charge susceptibility\\
$\bar{\chi}^{cc}$&pair-field susceptibility\\
$\chi^{cf}$&mixed charge susceptibility\\
$\chi^{ff}$&localized electron charge susceptibility\\
$\chi_{hyb}$&hybridization susceptibility\\
$\chi_0$&bare susceptibility\\
$\chi_\tau$&time-dependent conduction electron field\\
$i\omega_n=i\pi T(2n+1)$&fermionic Matsubara frequency\\
$\mathcal{H}$&Hamiltonian\\
$\mathcal{L}$&Lorenz number\\
$\mathcal{Z}$&partition function\\
\end{tabular}
\end{ruledtabular}
\end{table}

\bibliography{fk_dmft.bib}

\begin{thebibliography}{175}
\expandafter\ifx\csname natexlab\endcsname\relax\def\natexlab#1{#1}\fi
\expandafter\ifx\csname bibnamefont\endcsname\relax
  \def\bibnamefont#1{#1}\fi
\expandafter\ifx\csname bibfnamefont\endcsname\relax
  \def\bibfnamefont#1{#1}\fi
\expandafter\ifx\csname citenamefont\endcsname\relax
  \def\citenamefont#1{#1}\fi
\expandafter\ifx\csname url\endcsname\relax
  \def\url#1{\texttt{#1}}\fi
\expandafter\ifx\csname urlprefix\endcsname\relax\def\urlprefix{URL }\fi
\providecommand{\bibinfo}[2]{#2}
\providecommand{\eprint}[2][]{\url{#2}}

\bibitem[{\citenamefont{Allen and Martin}(1982)}]{allen_martin_1982}
\bibinfo{author}{\bibnamefont{Allen}, \bibfnamefont{J.~W.}}, and
  \bibinfo{author}{\bibfnamefont{R.~M.} \bibnamefont{Martin}},
  \bibinfo{year}{1982}, \bibinfo{journal}{Phys. Rev. Lett.}
  \textbf{\bibinfo{volume}{49}}, \bibinfo{pages}{1106}.

\bibitem[{\citenamefont{Allub and Alascio}(1996)}]{allub_alascio_1996}
\bibinfo{author}{\bibnamefont{Allub}, \bibfnamefont{R.}}, and
  \bibinfo{author}{\bibfnamefont{B.}~\bibnamefont{Alascio}},
  \bibinfo{year}{1996}, \bibinfo{journal}{Solid St. Commun.}
  \textbf{\bibinfo{volume}{99}}, \bibinfo{pages}{66}.

\bibitem[{\citenamefont{Allub and Alascio}(1997)}]{allub_alascio_1997}
\bibinfo{author}{\bibnamefont{Allub}, \bibfnamefont{R.}}, and
  \bibinfo{author}{\bibfnamefont{B.}~\bibnamefont{Alascio}},
  \bibinfo{year}{1997}, \bibinfo{journal}{Phys. Rev. B}
  \textbf{\bibinfo{volume}{55}}, \bibinfo{pages}{14113}.

\bibitem[{\citenamefont{Altshuler} \emph{et~al.}(1995)\citenamefont{Altshuler,
  Bresler, Schlott, Elschner, and Graty}}]{altschuler_1995}
\bibinfo{author}{\bibnamefont{Altshuler}, \bibfnamefont{T.}},
  \bibinfo{author}{\bibfnamefont{M.}~\bibnamefont{Bresler}},
  \bibinfo{author}{\bibfnamefont{M.}~\bibnamefont{Schlott}},
  \bibinfo{author}{\bibfnamefont{B.}~\bibnamefont{Elschner}}, and
  \bibinfo{author}{\bibfnamefont{E.}~\bibnamefont{Graty}},
  \bibinfo{year}{1995}, \bibinfo{journal}{Z. Phys. B}
  \textbf{\bibinfo{volume}{99}}, \bibinfo{pages}{57}.

\bibitem[{\citenamefont{Ambegaokar and
  Baratoff}(1963)}]{ambegaokar_baratoff_1963}
\bibinfo{author}{\bibnamefont{Ambegaokar}, \bibfnamefont{V.}}, and
  \bibinfo{author}{\bibfnamefont{A.}~\bibnamefont{Baratoff}},
  \bibinfo{year}{1963}, \bibinfo{journal}{Phys. Rev. Lett.}
  \textbf{\bibinfo{volume}{10}}, \bibinfo{pages}{486}.

\bibitem[{\citenamefont{Anderson}(1959)}]{anderson_1959}
\bibinfo{author}{\bibnamefont{Anderson}, \bibfnamefont{P.~W.}},
  \bibinfo{year}{1959}, \bibinfo{journal}{J. Phys. Chem. Solids}
  \textbf{\bibinfo{volume}{11}}, \bibinfo{pages}{26}.

\bibitem[{\citenamefont{Ashcroft and Mermin}(1976)}]{ashcroft_mermin_1976}
\bibinfo{author}{\bibnamefont{Ashcroft}, \bibfnamefont{N.~W.}}, and
  \bibinfo{author}{\bibfnamefont{N.~D.} \bibnamefont{Mermin}},
  \bibinfo{year}{1976}, \emph{\bibinfo{title}{Solid State Physics}}
  (\bibinfo{publisher}{Holt, Rinehart and Winston},
  \bibinfo{address}{Philadelphia}).

\bibitem[{\citenamefont{Batista}(2002)}]{batista_2002}
\bibinfo{author}{\bibnamefont{Batista}, \bibfnamefont{C.~D.}},
  \bibinfo{year}{2002}, \bibinfo{journal}{Phys. Rev. Lett.}
  \textbf{\bibinfo{volume}{89}}, \bibinfo{pages}{166403}.

\bibitem[{\citenamefont{Baym}(1962)}]{baym_1962}
\bibinfo{author}{\bibnamefont{Baym}, \bibfnamefont{G.}}, \bibinfo{year}{1962},
  \bibinfo{journal}{Phys. Rev.} \textbf{\bibinfo{volume}{127}},
  \bibinfo{pages}{1391}.

\bibitem[{\citenamefont{Bergmann and Rainer}(1974)}]{bergmann_rainer_1974}
\bibinfo{author}{\bibnamefont{Bergmann}, \bibfnamefont{G.}}, and
  \bibinfo{author}{\bibfnamefont{D.}~\bibnamefont{Rainer}},
  \bibinfo{year}{1974}, \bibinfo{journal}{Z. Phys.}
  \textbf{\bibinfo{volume}{174}}, \bibinfo{pages}{445}.

\bibitem[{\citenamefont{Blawid and Millis}(2000)}]{blawid_millis_2000}
\bibinfo{author}{\bibnamefont{Blawid}, \bibfnamefont{S.}}, and
  \bibinfo{author}{\bibfnamefont{A.~J.} \bibnamefont{Millis}},
  \bibinfo{year}{2000}, \bibinfo{journal}{Phys. Rev. B}
  \textbf{\bibinfo{volume}{62}}, \bibinfo{pages}{2424}.

\bibitem[{\citenamefont{Blawid and Millis}(2001)}]{blawid_millis_2001}
\bibinfo{author}{\bibnamefont{Blawid}, \bibfnamefont{S.}}, and
  \bibinfo{author}{\bibfnamefont{A.~J.} \bibnamefont{Millis}},
  \bibinfo{year}{2001}, \bibinfo{journal}{Phys. Rev. B}
  \textbf{\bibinfo{volume}{63}}, \bibinfo{pages}{115114}.

\bibitem[{\citenamefont{Brandt and
  Fledderjohann}(1992)}]{brandt_fledderjohann_1992}
\bibinfo{author}{\bibnamefont{Brandt}, \bibfnamefont{U.}}, and
  \bibinfo{author}{\bibfnamefont{A.}~\bibnamefont{Fledderjohann}},
  \bibinfo{year}{1992}, \bibinfo{journal}{Z. Phys. B}
  \textbf{\bibinfo{volume}{87}}, \bibinfo{pages}{111}.

\bibitem[{\citenamefont{Brandt} \emph{et~al.}(1990)\citenamefont{Brandt,
  Fledderjohann, and Hulsenbeck}}]{brandt_fledderjohann_1990}
\bibinfo{author}{\bibnamefont{Brandt}, \bibfnamefont{U.}},
  \bibinfo{author}{\bibfnamefont{A.}~\bibnamefont{Fledderjohann}}, and
  \bibinfo{author}{\bibfnamefont{G.}~\bibnamefont{Hulsenbeck}},
  \bibinfo{year}{1990}, \bibinfo{journal}{Z. Phys. B}
  \textbf{\bibinfo{volume}{81}}, \bibinfo{pages}{409}.

\bibitem[{\citenamefont{Brandt and Mielsch}(1989)}]{brandt_mielsch_1989}
\bibinfo{author}{\bibnamefont{Brandt}, \bibfnamefont{U.}}, and
  \bibinfo{author}{\bibfnamefont{C.}~\bibnamefont{Mielsch}},
  \bibinfo{year}{1989}, \bibinfo{journal}{Z. Phys. B}
  \textbf{\bibinfo{volume}{75}}, \bibinfo{pages}{365}.

\bibitem[{\citenamefont{Brandt and Mielsch}(1990)}]{brandt_mielsch_1990}
\bibinfo{author}{\bibnamefont{Brandt}, \bibfnamefont{U.}}, and
  \bibinfo{author}{\bibfnamefont{C.}~\bibnamefont{Mielsch}},
  \bibinfo{year}{1990}, \bibinfo{journal}{Z. Phys. B}
  \textbf{\bibinfo{volume}{79}}, \bibinfo{pages}{295}.

\bibitem[{\citenamefont{Brandt and Mielsch}(1991)}]{brandt_mielsch_1991}
\bibinfo{author}{\bibnamefont{Brandt}, \bibfnamefont{U.}}, and
  \bibinfo{author}{\bibfnamefont{C.}~\bibnamefont{Mielsch}},
  \bibinfo{year}{1991}, \bibinfo{journal}{Z. Phys. B}
  \textbf{\bibinfo{volume}{82}}, \bibinfo{pages}{37}.

\bibitem[{\citenamefont{Brandt and Schmidt}(1986)}]{brandt_schmidt_1986}
\bibinfo{author}{\bibnamefont{Brandt}, \bibfnamefont{U.}}, and
  \bibinfo{author}{\bibfnamefont{R.}~\bibnamefont{Schmidt}},
  \bibinfo{year}{1986}, \bibinfo{journal}{Z. Phys. B}
  \textbf{\bibinfo{volume}{63}}, \bibinfo{pages}{45}.

\bibitem[{\citenamefont{Brandt and Schmidt}(1987)}]{brandt_schmidt_1987}
\bibinfo{author}{\bibnamefont{Brandt}, \bibfnamefont{U.}}, and
  \bibinfo{author}{\bibfnamefont{R.}~\bibnamefont{Schmidt}},
  \bibinfo{year}{1987}, \bibinfo{journal}{Z. Phys. B}
  \textbf{\bibinfo{volume}{67}}, \bibinfo{pages}{43}.

\bibitem[{\citenamefont{Brandt and Urbanek}(1992)}]{brandt_urbanek_1992}
\bibinfo{author}{\bibnamefont{Brandt}, \bibfnamefont{U.}}, and
  \bibinfo{author}{\bibfnamefont{M.~P.} \bibnamefont{Urbanek}},
  \bibinfo{year}{1992}, \bibinfo{journal}{Z. Phys. B}
  \textbf{\bibinfo{volume}{89}}, \bibinfo{pages}{297}.

\bibitem[{\citenamefont{Calandra and
  Gunnarsson}(2001)}]{calandra_gunnarsson_2001}
\bibinfo{author}{\bibnamefont{Calandra}, \bibfnamefont{M.}}, and
  \bibinfo{author}{\bibfnamefont{O.}~\bibnamefont{Gunnarsson}},
  \bibinfo{year}{2001}, \bibinfo{journal}{Phys. Rev. Lett.}
  \textbf{\bibinfo{volume}{87}}, \bibinfo{pages}{266601}.

\bibitem[{\citenamefont{Chattopadhyay}
  \emph{et~al.}(2001)\citenamefont{Chattopadhyay, {Das Sarma}, and
  Millis}}]{chattopadhyay_dassarma_millis_2001}
\bibinfo{author}{\bibnamefont{Chattopadhyay}, \bibfnamefont{A.}},
  \bibinfo{author}{\bibfnamefont{S.}~\bibnamefont{{Das Sarma}}}, and
  \bibinfo{author}{\bibfnamefont{A.~J.} \bibnamefont{Millis}},
  \bibinfo{year}{2001}, \bibinfo{journal}{Phys. Rev. Lett.}
  \textbf{\bibinfo{volume}{87}}, \bibinfo{pages}{227202}.

\bibitem[{\citenamefont{Chattopadhyay}
  \emph{et~al.}(2000)\citenamefont{Chattopadhyay, Millis, and {Das
  Sarma}}}]{chattopadhyay_millis_dassarma_2000}
\bibinfo{author}{\bibnamefont{Chattopadhyay}, \bibfnamefont{A.}},
  \bibinfo{author}{\bibfnamefont{A.~J.} \bibnamefont{Millis}}, and
  \bibinfo{author}{\bibfnamefont{S.}~\bibnamefont{{Das Sarma}}},
  \bibinfo{year}{2000}, \bibinfo{journal}{Phys. Rev. B}
  \textbf{\bibinfo{volume}{61}}, \bibinfo{pages}{10738}.

\bibitem[{\citenamefont{Chen} \emph{et~al.}(1993)\citenamefont{Chen, Yu, and
  Talyor}}]{chen_yu_taylor_1993}
\bibinfo{author}{\bibnamefont{Chen}, \bibfnamefont{A.~L.}},
  \bibinfo{author}{\bibfnamefont{P.~Y.} \bibnamefont{Yu}}, and
  \bibinfo{author}{\bibfnamefont{R.~D.} \bibnamefont{Talyor}},
  \bibinfo{year}{1993}, \bibinfo{journal}{Phys. Rev. Lett.}
  \textbf{\bibinfo{volume}{71}}, \bibinfo{pages}{4011}.

\bibitem[{\citenamefont{Chester and Thellung}(1961)}]{chester_thellung_1961}
\bibinfo{author}{\bibnamefont{Chester}, \bibfnamefont{G.~V.}}, and
  \bibinfo{author}{\bibfnamefont{A.}~\bibnamefont{Thellung}},
  \bibinfo{year}{1961}, \bibinfo{journal}{Proc. Phys. Soc. London}
  \textbf{\bibinfo{volume}{77}}, \bibinfo{pages}{1005}.

\bibitem[{\citenamefont{Chung and Freericks}(1998)}]{chung_freericks_1998}
\bibinfo{author}{\bibnamefont{Chung}, \bibfnamefont{W.}}, and
  \bibinfo{author}{\bibfnamefont{J.~K.} \bibnamefont{Freericks}},
  \bibinfo{year}{1998}, \bibinfo{journal}{Phys. Rev. B}
  \textbf{\bibinfo{volume}{57}}, \bibinfo{pages}{11955}.

\bibitem[{\citenamefont{Chung and Freericks}(2000)}]{chung_freericks_2000}
\bibinfo{author}{\bibnamefont{Chung}, \bibfnamefont{W.}}, and
  \bibinfo{author}{\bibfnamefont{J.~K.} \bibnamefont{Freericks}},
  \bibinfo{year}{2000}, \bibinfo{journal}{Phys. Rev. Lett.}
  \textbf{\bibinfo{volume}{84}}, \bibinfo{pages}{2461}.

\bibitem[{\citenamefont{Ciuchi and {de
  Pasquale}}(1999)}]{ciuchi_depasquale_1999}
\bibinfo{author}{\bibnamefont{Ciuchi}, \bibfnamefont{S.}}, and
  \bibinfo{author}{\bibfnamefont{F.}~\bibnamefont{{de Pasquale}}},
  \bibinfo{year}{1999}, \bibinfo{journal}{Phys. Rev. B}
  \textbf{\bibinfo{volume}{59}}, \bibinfo{pages}{5431}.

\bibitem[{\citenamefont{Cornelius} \emph{et~al.}(1997)\citenamefont{Cornelius,
  Lawrence, , Sarrao, Fisk, Hundley, Kwei, Thompson, Booth, and
  Bridges}}]{cornelius_1997}
\bibinfo{author}{\bibnamefont{Cornelius}, \bibfnamefont{A.~L.}},
  \bibinfo{author}{\bibfnamefont{J.~M.} \bibnamefont{Lawrence}}, ,
  \bibinfo{author}{\bibfnamefont{J.}~\bibnamefont{Sarrao}},
  \bibinfo{author}{\bibfnamefont{Z.}~\bibnamefont{Fisk}},
  \bibinfo{author}{\bibfnamefont{M.~F.} \bibnamefont{Hundley}},
  \bibinfo{author}{\bibfnamefont{G.~H.} \bibnamefont{Kwei}},
  \bibinfo{author}{\bibfnamefont{J.~D.} \bibnamefont{Thompson}},
  \bibinfo{author}{\bibfnamefont{C.~H.} \bibnamefont{Booth}}, and
  \bibinfo{author}{\bibfnamefont{F.}~\bibnamefont{Bridges}},
  \bibinfo{year}{1997}, \bibinfo{journal}{Phys. Rev. B}
  \textbf{\bibinfo{volume}{56}}, \bibinfo{pages}{7993}.

\bibitem[{\citenamefont{Czycholl}(1999)}]{czycholl_1999}
\bibinfo{author}{\bibnamefont{Czycholl}, \bibfnamefont{G.}},
  \bibinfo{year}{1999}, \bibinfo{journal}{Phys. Rev. B}
  \textbf{\bibinfo{volume}{59}}, \bibinfo{pages}{2642}.

\bibitem[{\citenamefont{Dallera} \emph{et~al.}(2002)\citenamefont{Dallera,
  Grioni, Shukla, Vank\'o, Sarrao, Rueff, and Cox}}]{dallera_2002}
\bibinfo{author}{\bibnamefont{Dallera}, \bibfnamefont{C.}},
  \bibinfo{author}{\bibfnamefont{M.}~\bibnamefont{Grioni}},
  \bibinfo{author}{\bibfnamefont{A.}~\bibnamefont{Shukla}},
  \bibinfo{author}{\bibfnamefont{G.}~\bibnamefont{Vank\'o}},
  \bibinfo{author}{\bibfnamefont{J.~L.} \bibnamefont{Sarrao}},
  \bibinfo{author}{\bibfnamefont{J.~P.} \bibnamefont{Rueff}}, and
  \bibinfo{author}{\bibfnamefont{D.~L.} \bibnamefont{Cox}},
  \bibinfo{year}{2002}, \bibinfo{journal}{Phys. Rev. Lett.}
  \textbf{\bibinfo{volume}{88}}, \bibinfo{pages}{196403}.

\bibitem[{\citenamefont{{de Vries}} \emph{et~al.}(1993)\citenamefont{{de
  Vries}, Michielsen, and {De Raedt}}}]{devries_michielsen_deraedt_1993}
\bibinfo{author}{\bibnamefont{{de Vries}}, \bibfnamefont{P.}},
  \bibinfo{author}{\bibfnamefont{K.}~\bibnamefont{Michielsen}}, and
  \bibinfo{author}{\bibfnamefont{H.}~\bibnamefont{{De Raedt}}},
  \bibinfo{year}{1993}, \bibinfo{journal}{Phys. Rev. Lett.}
  \textbf{\bibinfo{volume}{70}}, \bibinfo{pages}{2463}.

\bibitem[{\citenamefont{{de Vries}} \emph{et~al.}(1994)\citenamefont{{de
  Vries}, Michielsen, and {De Raedt}}}]{devries_michielsen_deraedt_1994}
\bibinfo{author}{\bibnamefont{{de Vries}}, \bibfnamefont{P.}},
  \bibinfo{author}{\bibfnamefont{K.}~\bibnamefont{Michielsen}}, and
  \bibinfo{author}{\bibfnamefont{H.}~\bibnamefont{{De Raedt}}},
  \bibinfo{year}{1994}, \bibinfo{journal}{Z. Phys. B}
  \textbf{\bibinfo{volume}{95}}, \bibinfo{pages}{475}.

\bibitem[{\citenamefont{Devereaux} \emph{et~al.}(2003)\citenamefont{Devereaux,
  McCormack, and Freericks}}]{devereaux_mccormack_freericks_2003}
\bibinfo{author}{\bibnamefont{Devereaux}, \bibfnamefont{T.~P.}},
  \bibinfo{author}{\bibfnamefont{G.~E.~D.} \bibnamefont{McCormack}}, and
  \bibinfo{author}{\bibfnamefont{J.~K.} \bibnamefont{Freericks}},
  \bibinfo{year}{2003}, \bibinfo{journal}{Phys. Rev. Lett.}
  \textbf{\bibinfo{volume}{90}}, \bibinfo{pages}{XXXXXX}.

\bibitem[{\citenamefont{van Dongen}(1991{\natexlab{a}})}]{vandongen_1991}
\bibinfo{author}{\bibnamefont{van Dongen}, \bibfnamefont{P.~G.~J.}},
  \bibinfo{year}{1991}{\natexlab{a}}, \bibinfo{journal}{Mod. Phys. Lett. B}
  \textbf{\bibinfo{volume}{5}}, \bibinfo{pages}{861}.

\bibitem[{\citenamefont{van Dongen}(1991{\natexlab{b}})}]{vandongen_1991b}
\bibinfo{author}{\bibnamefont{van Dongen}, \bibfnamefont{P.~G.~J.}},
  \bibinfo{year}{1991}{\natexlab{b}}, \bibinfo{journal}{Phys. Rev. Lett.}
  \textbf{\bibinfo{volume}{67}}, \bibinfo{pages}{757}.

\bibitem[{\citenamefont{van Dongen}(1992)}]{vandongen_1992}
\bibinfo{author}{\bibnamefont{van Dongen}, \bibfnamefont{P.~G.~J.}},
  \bibinfo{year}{1992}, \bibinfo{journal}{Phys. Rev. B}
  \textbf{\bibinfo{volume}{45}}, \bibinfo{pages}{2267}.

\bibitem[{\citenamefont{van Dongen}(1994)}]{vandongen_1994}
\bibinfo{author}{\bibnamefont{van Dongen}, \bibfnamefont{P.~G.~J.}},
  \bibinfo{year}{1994}, \bibinfo{journal}{Phys. Rev. B}
  \textbf{\bibinfo{volume}{50}}, \bibinfo{pages}{14016}.

\bibitem[{\citenamefont{van Dongen and
  {Bl\"umer}}(2002)}]{vandongen_blumer_2002}
\bibinfo{author}{\bibnamefont{van Dongen}, \bibfnamefont{P.~G.~J.}}, and
  \bibinfo{author}{\bibfnamefont{N.}~\bibnamefont{{Bl\"umer}}},
  \bibinfo{year}{2002}, \bibinfo{note}{unpublished}.

\bibitem[{\citenamefont{van Dongen and Leinung}(1997)}]{vandongen_leinung_1997}
\bibinfo{author}{\bibnamefont{van Dongen}, \bibfnamefont{P.~G.~J.}}, and
  \bibinfo{author}{\bibfnamefont{C.}~\bibnamefont{Leinung}},
  \bibinfo{year}{1997}, \bibinfo{journal}{Ann. Phys. (Leipzig)}
  \textbf{\bibinfo{volume}{6}}, \bibinfo{pages}{45}.

\bibitem[{\citenamefont{van Dongen and
  Vollhardt}(1990)}]{vandongen_vollhardt_1990}
\bibinfo{author}{\bibnamefont{van Dongen}, \bibfnamefont{P.~G.~J.}}, and
  \bibinfo{author}{\bibfnamefont{D.}~\bibnamefont{Vollhardt}},
  \bibinfo{year}{1990}, \bibinfo{journal}{Phys. Rev. Lett.}
  \textbf{\bibinfo{volume}{65}}, \bibinfo{pages}{1663}.

\bibitem[{\citenamefont{Dzero}(2002)}]{dzero_2002b}
\bibinfo{author}{\bibnamefont{Dzero}, \bibfnamefont{M.~O.}},
  \bibinfo{year}{2002}, \bibinfo{journal}{J. Phys.: Condens. Matter}
  \textbf{\bibinfo{volume}{14}}, \bibinfo{pages}{631}.

\bibitem[{\citenamefont{Dzero} \emph{et~al.}(2000)\citenamefont{Dzero, Gor'kov,
  and Zvezdin}}]{dzero_2000}
\bibinfo{author}{\bibnamefont{Dzero}, \bibfnamefont{M.~O.}},
  \bibinfo{author}{\bibfnamefont{L.}~\bibnamefont{Gor'kov}}, and
  \bibinfo{author}{\bibfnamefont{A.~K.} \bibnamefont{Zvezdin}},
  \bibinfo{year}{2000}, \bibinfo{journal}{J. Phys.: Condens. Matter}
  \textbf{\bibinfo{volume}{12}}, \bibinfo{pages}{L711}.

\bibitem[{\citenamefont{Dzero} \emph{et~al.}(2002)\citenamefont{Dzero, Gor'kov,
  and Zvezdin}}]{dzero_2002a}
\bibinfo{author}{\bibnamefont{Dzero}, \bibfnamefont{M.~O.}},
  \bibinfo{author}{\bibfnamefont{L.}~\bibnamefont{Gor'kov}}, and
  \bibinfo{author}{\bibfnamefont{A.~K.} \bibnamefont{Zvezdin}},
  \bibinfo{year}{2002}, \bibinfo{journal}{Physica B}
  \textbf{\bibinfo{volume}{312 \& 313}}, \bibinfo{pages}{321}.

\bibitem[{\citenamefont{Economou}(1983)}]{economou_1983}
\bibinfo{author}{\bibnamefont{Economou}, \bibfnamefont{E.~N.}},
  \bibinfo{year}{1983}, \emph{\bibinfo{title}{Green's Functions in Quantum
  Physics}} (\bibinfo{publisher}{Springer-Verlag}, \bibinfo{address}{Berlin}).

\bibitem[{\citenamefont{Elitzur}(1975)}]{elitzur_1975}
\bibinfo{author}{\bibnamefont{Elitzur}, \bibfnamefont{S.}},
  \bibinfo{year}{1975}, \bibinfo{journal}{Phys. Rev. D}
  \textbf{\bibinfo{volume}{12}}, \bibinfo{pages}{3978}.

\bibitem[{\citenamefont{Falicov and Kimball}(1969)}]{falicov_kimball_1969}
\bibinfo{author}{\bibnamefont{Falicov}, \bibfnamefont{L.~M.}}, and
  \bibinfo{author}{\bibfnamefont{J.~C.} \bibnamefont{Kimball}},
  \bibinfo{year}{1969}, \bibinfo{journal}{Phys. Rev. Lett.}
  \textbf{\bibinfo{volume}{22}}, \bibinfo{pages}{997}.

\bibitem[{\citenamefont{{Farkasov\v sk\' y}}(1997)}]{farkasovsky_1997}
\bibinfo{author}{\bibnamefont{{Farkasov\v sk\' y}}, \bibfnamefont{P.}},
  \bibinfo{year}{1997}, \bibinfo{journal}{Z. Phys. B}
  \textbf{\bibinfo{volume}{104}}, \bibinfo{pages}{553}.

\bibitem[{\citenamefont{{Farkasov\v sk\' y}}(1999)}]{farkasovsky_1999}
\bibinfo{author}{\bibnamefont{{Farkasov\v sk\' y}}, \bibfnamefont{P.}},
  \bibinfo{year}{1999}, \bibinfo{journal}{Phys. Rev. B}
  \textbf{\bibinfo{volume}{59}}, \bibinfo{pages}{9707}.

\bibitem[{\citenamefont{{Farkasov\v sk\' y}}(2002)}]{farkasovsky_2002}
\bibinfo{author}{\bibnamefont{{Farkasov\v sk\' y}}, \bibfnamefont{P.}},
  \bibinfo{year}{2002}, \bibinfo{journal}{Phys. Rev. B}
  \textbf{\bibinfo{volume}{65}}, \bibinfo{pages}{081102}.

\bibitem[{\citenamefont{Felner and Novik}(1986)}]{felner_novik_1986}
\bibinfo{author}{\bibnamefont{Felner}, \bibfnamefont{I.}}, and
  \bibinfo{author}{\bibfnamefont{I.}~\bibnamefont{Novik}},
  \bibinfo{year}{1986}, \bibinfo{journal}{Phys. Rev. B}
  \textbf{\bibinfo{volume}{33}}, \bibinfo{pages}{617}.

\bibitem[{\citenamefont{Fetter and Walecka}(1971)}]{fetter_walecka_1971}
\bibinfo{author}{\bibnamefont{Fetter}, \bibfnamefont{A.~L.}}, and
  \bibinfo{author}{\bibfnamefont{J.~D.} \bibnamefont{Walecka}},
  \bibinfo{year}{1971}, \emph{\bibinfo{title}{Quantum Theory of Many-Particle
  Systems}} (\bibinfo{publisher}{McGraw-Hill}, \bibinfo{address}{New York}).

\bibitem[{\citenamefont{Figueroa} \emph{et~al.}(1998)\citenamefont{Figueroa,
  Lawrence, Sarrao, Fisk, Hundley, and Thompson}}]{figueroa_1998}
\bibinfo{author}{\bibnamefont{Figueroa}, \bibfnamefont{E.}},
  \bibinfo{author}{\bibfnamefont{J.~M.} \bibnamefont{Lawrence}},
  \bibinfo{author}{\bibfnamefont{J.}~\bibnamefont{Sarrao}},
  \bibinfo{author}{\bibfnamefont{Z.}~\bibnamefont{Fisk}},
  \bibinfo{author}{\bibfnamefont{M.~F.} \bibnamefont{Hundley}}, and
  \bibinfo{author}{\bibfnamefont{J.~D.} \bibnamefont{Thompson}},
  \bibinfo{year}{1998}, \bibinfo{journal}{Solid State Commun.}
  \textbf{\bibinfo{volume}{106}}, \bibinfo{pages}{347}.

\bibitem[{\citenamefont{Fisk and Webb}(1976)}]{fisk_webb_1976}
\bibinfo{author}{\bibnamefont{Fisk}, \bibfnamefont{Z.}}, and
  \bibinfo{author}{\bibfnamefont{G.~W.} \bibnamefont{Webb}},
  \bibinfo{year}{1976}, \bibinfo{journal}{Phys. Rev. Lett.}
  \textbf{\bibinfo{volume}{36}}, \bibinfo{pages}{1084}.

\bibitem[{\citenamefont{Freericks}(1993{\natexlab{a}})}]{freericks_1993a}
\bibinfo{author}{\bibnamefont{Freericks}, \bibfnamefont{J.~K.}},
  \bibinfo{year}{1993}{\natexlab{a}}, \bibinfo{journal}{Phys. Rev. B}
  \textbf{\bibinfo{volume}{47}}, \bibinfo{pages}{9263}.

\bibitem[{\citenamefont{Freericks}(1993{\natexlab{b}})}]{freericks_1993b}
\bibinfo{author}{\bibnamefont{Freericks}, \bibfnamefont{J.~K.}},
  \bibinfo{year}{1993}{\natexlab{b}}, \bibinfo{journal}{Phys. Rev. B}
  \textbf{\bibinfo{volume}{48}}, \bibinfo{pages}{14797}.

\bibitem[{\citenamefont{Freericks and
  Devereaux}(2001{\natexlab{a}})}]{freericks_devereaux_2001a}
\bibinfo{author}{\bibnamefont{Freericks}, \bibfnamefont{J.~K.}}, and
  \bibinfo{author}{\bibfnamefont{T.~P.} \bibnamefont{Devereaux}},
  \bibinfo{year}{2001}{\natexlab{a}}, \bibinfo{journal}{Condens. Mat. Phys.}
  \textbf{\bibinfo{volume}{4}}, \bibinfo{pages}{149}.

\bibitem[{\citenamefont{Freericks and
  Devereaux}(2001{\natexlab{b}})}]{freericks_devereaux_2001b}
\bibinfo{author}{\bibnamefont{Freericks}, \bibfnamefont{J.~K.}}, and
  \bibinfo{author}{\bibfnamefont{T.~P.} \bibnamefont{Devereaux}},
  \bibinfo{year}{2001}{\natexlab{b}}, \bibinfo{journal}{Phys. Rev. B}
  \textbf{\bibinfo{volume}{64}}, \bibinfo{pages}{125110}.

\bibitem[{\citenamefont{Freericks}
  \emph{et~al.}(2001{\natexlab{a}})\citenamefont{Freericks, Devereaux, and
  Bulla}}]{freericks_devereaux_2001c}
\bibinfo{author}{\bibnamefont{Freericks}, \bibfnamefont{J.~K.}},
  \bibinfo{author}{\bibfnamefont{T.~P.} \bibnamefont{Devereaux}}, and
  \bibinfo{author}{\bibfnamefont{R.}~\bibnamefont{Bulla}},
  \bibinfo{year}{2001}{\natexlab{a}}, \bibinfo{journal}{Acta Phys. Pol. B}
  \textbf{\bibinfo{volume}{32}}, \bibinfo{pages}{3219}.

\bibitem[{\citenamefont{Freericks and Falicov}(1990)}]{freericks_falicov_1990}
\bibinfo{author}{\bibnamefont{Freericks}, \bibfnamefont{J.~K.}}, and
  \bibinfo{author}{\bibfnamefont{L.~M.} \bibnamefont{Falicov}},
  \bibinfo{year}{1990}, \bibinfo{journal}{Phys. Rev. B}
  \textbf{\bibinfo{volume}{41}}, \bibinfo{pages}{2163}.

\bibitem[{\citenamefont{Freericks and Falicov}(1992)}]{freericks_falicov_1992}
\bibinfo{author}{\bibnamefont{Freericks}, \bibfnamefont{J.~K.}}, and
  \bibinfo{author}{\bibfnamefont{L.~M.} \bibnamefont{Falicov}},
  \bibinfo{year}{1992}, \bibinfo{journal}{Phys. Rev. B}
  \textbf{\bibinfo{volume}{45}}, \bibinfo{pages}{1896}.

\bibitem[{\citenamefont{Freericks} \emph{et~al.}(1996)\citenamefont{Freericks,
  Gruber, and Macris}}]{freericks_gruber_macris_1996}
\bibinfo{author}{\bibnamefont{Freericks}, \bibfnamefont{J.~K.}},
  \bibinfo{author}{\bibfnamefont{C.}~\bibnamefont{Gruber}}, and
  \bibinfo{author}{\bibfnamefont{N.}~\bibnamefont{Macris}},
  \bibinfo{year}{1996}, \bibinfo{journal}{Phys. Rev. B}
  \textbf{\bibinfo{volume}{53}}, \bibinfo{pages}{16189}.

\bibitem[{\citenamefont{Freericks} \emph{et~al.}(1999)\citenamefont{Freericks,
  Gruber, and Macris}}]{freericks_gruber_macris_1999}
\bibinfo{author}{\bibnamefont{Freericks}, \bibfnamefont{J.~K.}},
  \bibinfo{author}{\bibfnamefont{C.}~\bibnamefont{Gruber}}, and
  \bibinfo{author}{\bibfnamefont{N.}~\bibnamefont{Macris}},
  \bibinfo{year}{1999}, \bibinfo{journal}{Phys. Rev. B}
  \textbf{\bibinfo{volume}{60}}, \bibinfo{pages}{1617}.

\bibitem[{\citenamefont{Freericks and
  Lema\'nski}(2000)}]{freericks_lemanski_2000}
\bibinfo{author}{\bibnamefont{Freericks}, \bibfnamefont{J.~K.}}, and
  \bibinfo{author}{\bibfnamefont{R.}~\bibnamefont{Lema\'nski}},
  \bibinfo{year}{2000}, \bibinfo{journal}{Phys. Rev. B}
  \textbf{\bibinfo{volume}{61}}, \bibinfo{pages}{13438}.

\bibitem[{\citenamefont{Freericks}
  \emph{et~al.}(2002{\natexlab{a}})\citenamefont{Freericks, Lieb, and
  Ueltschi}}]{freericks_lieb_ueltschi_2002a}
\bibinfo{author}{\bibnamefont{Freericks}, \bibfnamefont{J.~K.}},
  \bibinfo{author}{\bibfnamefont{E.~H.} \bibnamefont{Lieb}}, and
  \bibinfo{author}{\bibfnamefont{D.}~\bibnamefont{Ueltschi}},
  \bibinfo{year}{2002}{\natexlab{a}}, \bibinfo{journal}{Phys. Rev. Lett.}
  \textbf{\bibinfo{volume}{88}}, \bibinfo{pages}{106401}.

\bibitem[{\citenamefont{Freericks}
  \emph{et~al.}(2002{\natexlab{b}})\citenamefont{Freericks, Lieb, and
  Ueltschi}}]{freericks_lieb_ueltschi_2002b}
\bibinfo{author}{\bibnamefont{Freericks}, \bibfnamefont{J.~K.}},
  \bibinfo{author}{\bibfnamefont{E.~H.} \bibnamefont{Lieb}}, and
  \bibinfo{author}{\bibfnamefont{D.}~\bibnamefont{Ueltschi}},
  \bibinfo{year}{2002}{\natexlab{b}}, \bibinfo{journal}{Commun. Math. Phys.}
  \textbf{\bibinfo{volume}{227}}, \bibinfo{pages}{243}.

\bibitem[{\citenamefont{Freericks and Miller}(2000)}]{freericks_miller_2000}
\bibinfo{author}{\bibnamefont{Freericks}, \bibfnamefont{J.~K.}}, and
  \bibinfo{author}{\bibfnamefont{P.}~\bibnamefont{Miller}},
  \bibinfo{year}{2000}, \bibinfo{journal}{Phys. Rev. B}
  \textbf{\bibinfo{volume}{62}}, \bibinfo{pages}{10022}.

\bibitem[{\citenamefont{Freericks}
  \emph{et~al.}(2001{\natexlab{b}})\citenamefont{Freericks, Nikoli\'c, and
  Miller}}]{freericks_nikolic_miller_2001a}
\bibinfo{author}{\bibnamefont{Freericks}, \bibfnamefont{J.~K.}},
  \bibinfo{author}{\bibfnamefont{B.~N.} \bibnamefont{Nikoli\'c}}, and
  \bibinfo{author}{\bibfnamefont{P.}~\bibnamefont{Miller}},
  \bibinfo{year}{2001}{\natexlab{b}}, \bibinfo{journal}{Phys. Rev. B}
  \textbf{\bibinfo{volume}{64}}, \bibinfo{pages}{054511}.

\bibitem[{\citenamefont{Freericks}
  \emph{et~al.}(2002{\natexlab{c}})\citenamefont{Freericks, Nikoli\'c, and
  Miller}}]{freericks_nikolic_miller_2002}
\bibinfo{author}{\bibnamefont{Freericks}, \bibfnamefont{J.~K.}},
  \bibinfo{author}{\bibfnamefont{B.~N.} \bibnamefont{Nikoli\'c}}, and
  \bibinfo{author}{\bibfnamefont{P.}~\bibnamefont{Miller}},
  \bibinfo{year}{2002}{\natexlab{c}}, \bibinfo{journal}{Int. J. Modern Phys. B}
  \textbf{\bibinfo{volume}{16}}, \bibinfo{pages}{531}.

\bibitem[{\citenamefont{Freericks} \emph{et~al.}(2003)\citenamefont{Freericks,
  Nikoli\'c, and Miller}}]{freericks_nikolic_miller_2003}
\bibinfo{author}{\bibnamefont{Freericks}, \bibfnamefont{J.~K.}},
  \bibinfo{author}{\bibfnamefont{B.~N.} \bibnamefont{Nikoli\'c}}, and
  \bibinfo{author}{\bibfnamefont{P.}~\bibnamefont{Miller}},
  \bibinfo{year}{2003}, \bibinfo{journal}{Appl. Phys. Lett.}
  \textbf{\bibinfo{volume}{XX}}, \bibinfo{pages}{XXXXX}.

\bibitem[{\citenamefont{Freericks and Zlati\'c}(1998)}]{freericks_zlatic_1998}
\bibinfo{author}{\bibnamefont{Freericks}, \bibfnamefont{J.~K.}}, and
  \bibinfo{author}{\bibfnamefont{V.}~\bibnamefont{Zlati\'c}},
  \bibinfo{year}{1998}, \bibinfo{journal}{Phys. Rev. B}
  \textbf{\bibinfo{volume}{58}}, \bibinfo{pages}{322}.

\bibitem[{\citenamefont{Freericks and
  Zlati\'c}(2001{\natexlab{a}})}]{freericks_zlatic_2001a}
\bibinfo{author}{\bibnamefont{Freericks}, \bibfnamefont{J.~K.}}, and
  \bibinfo{author}{\bibfnamefont{V.}~\bibnamefont{Zlati\'c}},
  \bibinfo{year}{2001}{\natexlab{a}}, \bibinfo{journal}{Phys. Rev. B}
  \textbf{\bibinfo{volume}{64}}, \bibinfo{pages}{073109}.

\bibitem[{\citenamefont{Freericks and
  Zlati\'c}(2001{\natexlab{b}})}]{freericks_zlatic_2001b}
\bibinfo{author}{\bibnamefont{Freericks}, \bibfnamefont{J.~K.}}, and
  \bibinfo{author}{\bibfnamefont{V.}~\bibnamefont{Zlati\'c}},
  \bibinfo{year}{2001}{\natexlab{b}}, \bibinfo{journal}{Phys. Rev. B}
  \textbf{\bibinfo{volume}{64}}, \bibinfo{pages}{245118}.

\bibitem[{\citenamefont{Freericks and
  Zlati\'c}(2002{\natexlab{a}})}]{freericks_zlatic_2001_erratum}
\bibinfo{author}{\bibnamefont{Freericks}, \bibfnamefont{J.~K.}}, and
  \bibinfo{author}{\bibfnamefont{V.}~\bibnamefont{Zlati\'c}},
  \bibinfo{year}{2002}{\natexlab{a}}, \bibinfo{journal}{Phys. Rev. B}
  \textbf{\bibinfo{volume}{66}}, \bibinfo{pages}{249901},
  \bibinfo{note}{erratum}.

\bibitem[{\citenamefont{Freericks and
  Zlati\'c}(2002{\natexlab{b}})}]{freericks_zlatic_2002}
\bibinfo{author}{\bibnamefont{Freericks}, \bibfnamefont{J.~K.}}, and
  \bibinfo{author}{\bibfnamefont{V.}~\bibnamefont{Zlati\'c}},
  \bibinfo{year}{2002}{\natexlab{b}}, \eprint{cond-mat/0211385}.

\bibitem[{\citenamefont{Freericks} \emph{et~al.}(2000)\citenamefont{Freericks,
  Zlati\'c, and Jarrell}}]{freericks_zlatic_jarrell_2000}
\bibinfo{author}{\bibnamefont{Freericks}, \bibfnamefont{J.~K.}},
  \bibinfo{author}{\bibfnamefont{V.}~\bibnamefont{Zlati\'c}}, and
  \bibinfo{author}{\bibfnamefont{M.}~\bibnamefont{Jarrell}},
  \bibinfo{year}{2000}, \bibinfo{journal}{Phys. Rev. B}
  \textbf{\bibinfo{volume}{61}}, \bibinfo{pages}{838}.

\bibitem[{\citenamefont{Garner} \emph{et~al.}(2000)\citenamefont{Garner,
  Hancock, Rodriguez, Schlesinger, Bucher, Fisk, and Sarrao}}]{garner_2000}
\bibinfo{author}{\bibnamefont{Garner}, \bibfnamefont{S.~R.}},
  \bibinfo{author}{\bibfnamefont{J.}~\bibnamefont{Hancock}},
  \bibinfo{author}{\bibfnamefont{Y.}~\bibnamefont{Rodriguez}},
  \bibinfo{author}{\bibfnamefont{Z.}~\bibnamefont{Schlesinger}},
  \bibinfo{author}{\bibfnamefont{B.}~\bibnamefont{Bucher}},
  \bibinfo{author}{\bibfnamefont{Z.}~\bibnamefont{Fisk}}, and
  \bibinfo{author}{\bibfnamefont{J.~L.} \bibnamefont{Sarrao}},
  \bibinfo{year}{2000}, \bibinfo{journal}{Phys. Rev. B}
  \textbf{\bibinfo{volume}{63}}, \bibinfo{pages}{R4778}.

\bibitem[{\citenamefont{Georges} \emph{et~al.}(1996)\citenamefont{Georges,
  Kotliar, Krauth, and Rozenberg}}]{georges_kotliar_krauth_rozenberg_1996}
\bibinfo{author}{\bibnamefont{Georges}, \bibfnamefont{A.}},
  \bibinfo{author}{\bibfnamefont{G.}~\bibnamefont{Kotliar}},
  \bibinfo{author}{\bibfnamefont{W.}~\bibnamefont{Krauth}}, and
  \bibinfo{author}{\bibfnamefont{M.~J.} \bibnamefont{Rozenberg}},
  \bibinfo{year}{1996}, \bibinfo{journal}{Rev. Mod. Phys.}
  \textbf{\bibinfo{volume}{68}}, \bibinfo{pages}{13}.

\bibitem[{\citenamefont{{Gon\c calves~da~Silva} and
  Falicov}(1972)}]{dasilva_falicov_1972}
\bibinfo{author}{\bibnamefont{{Gon\c calves~da~Silva}},
  \bibfnamefont{C.~E.~T.}}, and \bibinfo{author}{\bibfnamefont{L.~M.}
  \bibnamefont{Falicov}}, \bibinfo{year}{1972}, \bibinfo{journal}{J. Phys. C:
  Solid State Phys.} \textbf{\bibinfo{volume}{5}}, \bibinfo{pages}{906}.

\bibitem[{\citenamefont{Goremychkin and Osborn}(1993)}]{goremychkin_1993}
\bibinfo{author}{\bibnamefont{Goremychkin}, \bibfnamefont{E.~A.}}, and
  \bibinfo{author}{\bibfnamefont{R.}~\bibnamefont{Osborn}},
  \bibinfo{year}{1993}, \bibinfo{journal}{Phys. Rev. B}
  \textbf{\bibinfo{volume}{47}}, \bibinfo{pages}{14280}.

\bibitem[{\citenamefont{Greenwood}(1958)}]{greenwood_1958}
\bibinfo{author}{\bibnamefont{Greenwood}, \bibfnamefont{D.~A.}},
  \bibinfo{year}{1958}, \bibinfo{journal}{Proc. Phys. Soc. (London)}
  \textbf{\bibinfo{volume}{71}}, \bibinfo{pages}{585}.

\bibitem[{\citenamefont{Gruber}(1999)}]{gruber_1999}
\bibinfo{author}{\bibnamefont{Gruber}, \bibfnamefont{C.}},
  \bibinfo{year}{1999}, \emph{\bibinfo{title}{Mathematical Results in
  Statistical Mechanics}} (\bibinfo{publisher}{World Scientific},
  \bibinfo{address}{Singapore}), p. \bibinfo{pages}{199}.

\bibitem[{\citenamefont{Gruber and Macris}(1996)}]{gruber_macris_1996}
\bibinfo{author}{\bibnamefont{Gruber}, \bibfnamefont{C.}}, and
  \bibinfo{author}{\bibfnamefont{N.}~\bibnamefont{Macris}},
  \bibinfo{year}{1996}, \bibinfo{journal}{Helv. Phys. Acta}
  \textbf{\bibinfo{volume}{69}}, \bibinfo{pages}{850}.

\bibitem[{\citenamefont{Gruber} \emph{et~al.}(2001)\citenamefont{Gruber,
  Macris, Royer, and Freericks}}]{gruber_macris_royer_2001}
\bibinfo{author}{\bibnamefont{Gruber}, \bibfnamefont{C.}},
  \bibinfo{author}{\bibfnamefont{N.}~\bibnamefont{Macris}},
  \bibinfo{author}{\bibfnamefont{P.}~\bibnamefont{Royer}}, and
  \bibinfo{author}{\bibfnamefont{J.~K.} \bibnamefont{Freericks}},
  \bibinfo{year}{2001}, \bibinfo{journal}{Phys. Rev. B}
  \textbf{\bibinfo{volume}{63}}, \bibinfo{pages}{165111}.

\bibitem[{\citenamefont{Gruber} \emph{et~al.}(1994)\citenamefont{Gruber,
  Ueltschi, and Jedrzejewski}}]{gruber_ueltschi_jedrezejewski_1994}
\bibinfo{author}{\bibnamefont{Gruber}, \bibfnamefont{C.}},
  \bibinfo{author}{\bibfnamefont{D.}~\bibnamefont{Ueltschi}}, and
  \bibinfo{author}{\bibfnamefont{J.}~\bibnamefont{Jedrzejewski}},
  \bibinfo{year}{1994}, \bibinfo{journal}{J. Stat. Phys.}
  \textbf{\bibinfo{volume}{76}}, \bibinfo{pages}{125}.

\bibitem[{\citenamefont{Haller}(2000)}]{haller_2000}
\bibinfo{author}{\bibnamefont{Haller}, \bibfnamefont{K.}},
  \bibinfo{year}{2000}, \bibinfo{journal}{Commun. Math. Phys.}
  \textbf{\bibinfo{volume}{210}}, \bibinfo{pages}{703}.

\bibitem[{\citenamefont{Haller and Kennedy}(2001)}]{haller_kennedy_2001}
\bibinfo{author}{\bibnamefont{Haller}, \bibfnamefont{K.}}, and
  \bibinfo{author}{\bibfnamefont{T.}~\bibnamefont{Kennedy}},
  \bibinfo{year}{2001}, \bibinfo{journal}{J. Stat. Phys.}
  \textbf{\bibinfo{volume}{102}}, \bibinfo{pages}{15}.

\bibitem[{\citenamefont{Hettler} \emph{et~al.}(2000)\citenamefont{Hettler,
  Mukherjee, Jarrell, and Krishnamurthy}}]{hettler_mukherjee_jarrell_2000}
\bibinfo{author}{\bibnamefont{Hettler}, \bibfnamefont{M.~H.}},
  \bibinfo{author}{\bibfnamefont{M.}~\bibnamefont{Mukherjee}},
  \bibinfo{author}{\bibfnamefont{M.}~\bibnamefont{Jarrell}}, and
  \bibinfo{author}{\bibfnamefont{H.~R.} \bibnamefont{Krishnamurthy}},
  \bibinfo{year}{2000}, \bibinfo{journal}{Phys. Rev. B}
  \textbf{\bibinfo{volume}{61}}, \bibinfo{pages}{12739}.

\bibitem[{\citenamefont{Hettler} \emph{et~al.}(1998)\citenamefont{Hettler,
  {Tahvildar-Zadeh}, Jarrell, Pruschke, and
  Krishnamurthy}}]{hettler_tahvildarzadeh_jarrell_1998}
\bibinfo{author}{\bibnamefont{Hettler}, \bibfnamefont{M.~H.}},
  \bibinfo{author}{\bibfnamefont{N.}~\bibnamefont{{Tahvildar-Zadeh}}},
  \bibinfo{author}{\bibfnamefont{M.}~\bibnamefont{Jarrell}},
  \bibinfo{author}{\bibfnamefont{T.}~\bibnamefont{Pruschke}}, and
  \bibinfo{author}{\bibfnamefont{H.~R.} \bibnamefont{Krishnamurthy}},
  \bibinfo{year}{1998}, \bibinfo{journal}{Phys. Rev. B}
  \textbf{\bibinfo{volume}{58}}, \bibinfo{pages}{R7475}.

\bibitem[{\citenamefont{Hirsch}(1993)}]{hirsch_1993}
\bibinfo{author}{\bibnamefont{Hirsch}, \bibfnamefont{J.~E.}},
  \bibinfo{year}{1993}, \bibinfo{journal}{Phys. Rev. B}
  \textbf{\bibinfo{volume}{47}}, \bibinfo{pages}{5351}.

\bibitem[{\citenamefont{Holstein}(1959)}]{holstein_1959}
\bibinfo{author}{\bibnamefont{Holstein}, \bibfnamefont{T.}},
  \bibinfo{year}{1959}, \bibinfo{journal}{Ann. Phys. (New York)}
  \textbf{\bibinfo{volume}{8}}, \bibinfo{pages}{325}.

\bibitem[{\citenamefont{Hubbard}(1963)}]{hubbard_I_1963}
\bibinfo{author}{\bibnamefont{Hubbard}, \bibfnamefont{J.}},
  \bibinfo{year}{1963}, \bibinfo{journal}{Proc. Roy. Soc. (London)}
  \textbf{\bibinfo{volume}{A276}}, \bibinfo{pages}{238}.

\bibitem[{\citenamefont{Hwang} \emph{et~al.}(2002)\citenamefont{Hwang, Millis,
  and {Das Sarma}}}]{hwang_millis_dassarma_2002}
\bibinfo{author}{\bibnamefont{Hwang}, \bibfnamefont{E.~H.}},
  \bibinfo{author}{\bibfnamefont{A.~J.} \bibnamefont{Millis}}, and
  \bibinfo{author}{\bibfnamefont{S.}~\bibnamefont{{Das Sarma}}},
  \bibinfo{year}{2002}, \textbf{\bibinfo{volume}{65}}, \bibinfo{pages}{233206}.

\bibitem[{\citenamefont{Immer} \emph{et~al.}(1997)\citenamefont{Immer, Sarrao,
  Fisk, Lacerda, Mielke, and Thompson}}]{immer_1997}
\bibinfo{author}{\bibnamefont{Immer}, \bibfnamefont{C.~D.}},
  \bibinfo{author}{\bibfnamefont{J.}~\bibnamefont{Sarrao}},
  \bibinfo{author}{\bibfnamefont{Z.}~\bibnamefont{Fisk}},
  \bibinfo{author}{\bibfnamefont{A.}~\bibnamefont{Lacerda}},
  \bibinfo{author}{\bibfnamefont{C.}~\bibnamefont{Mielke}}, and
  \bibinfo{author}{\bibfnamefont{J.~D.} \bibnamefont{Thompson}},
  \bibinfo{year}{1997}, \bibinfo{journal}{Phys. Rev. B}
  \textbf{\bibinfo{volume}{56}}, \bibinfo{pages}{71}.

\bibitem[{\citenamefont{Janis}(1994)}]{janis_1994}
\bibinfo{author}{\bibnamefont{Janis}, \bibfnamefont{V.}}, \bibinfo{year}{1994},
  \bibinfo{journal}{Phys. Rev. B} \textbf{\bibinfo{volume}{49}},
  \bibinfo{pages}{1612}.

\bibitem[{\citenamefont{Jarrell}(1992)}]{jarrell_1992}
\bibinfo{author}{\bibnamefont{Jarrell}, \bibfnamefont{M.}},
  \bibinfo{year}{1992}, \bibinfo{journal}{Phys. Rev. Lett.}
  \textbf{\bibinfo{volume}{69}}, \bibinfo{pages}{168}.

\bibitem[{\citenamefont{Jonson and Mahan}(1980)}]{jonson_mahan_1980}
\bibinfo{author}{\bibnamefont{Jonson}, \bibfnamefont{M.}}, and
  \bibinfo{author}{\bibfnamefont{G.~D.} \bibnamefont{Mahan}},
  \bibinfo{year}{1980}, \bibinfo{journal}{Phys. Rev. B}
  \textbf{\bibinfo{volume}{21}}, \bibinfo{pages}{4223}.

\bibitem[{\citenamefont{Jonson and Mahan}(1990)}]{jonson_mahan_1990}
\bibinfo{author}{\bibnamefont{Jonson}, \bibfnamefont{M.}}, and
  \bibinfo{author}{\bibfnamefont{G.~D.} \bibnamefont{Mahan}},
  \bibinfo{year}{1990}, \bibinfo{journal}{Phys. Rev. B}
  \textbf{\bibinfo{volume}{42}}, \bibinfo{pages}{9350}.

\bibitem[{\citenamefont{Josephson}(1962)}]{josephson_1962}
\bibinfo{author}{\bibnamefont{Josephson}, \bibfnamefont{B.~D.}},
  \bibinfo{year}{1962}, \bibinfo{journal}{Phys. Lett.}
  \textbf{\bibinfo{volume}{1}}, \bibinfo{pages}{251}.

\bibitem[{\citenamefont{Kaul} \emph{et~al.}(2001)\citenamefont{Kaul, Whitely,
  {van Duzer}, Yu, Newman, and Rowell}}]{newman_2001}
\bibinfo{author}{\bibnamefont{Kaul}, \bibfnamefont{A.}},
  \bibinfo{author}{\bibfnamefont{S.}~\bibnamefont{Whitely}},
  \bibinfo{author}{\bibfnamefont{T.}~\bibnamefont{{van Duzer}}},
  \bibinfo{author}{\bibfnamefont{L.}~\bibnamefont{Yu}},
  \bibinfo{author}{\bibfnamefont{N.}~\bibnamefont{Newman}}, and
  \bibinfo{author}{\bibfnamefont{J.}~\bibnamefont{Rowell}},
  \bibinfo{year}{2001}, \bibinfo{journal}{Appl. Phys. Lett.}
  \textbf{\bibinfo{volume}{78}}, \bibinfo{pages}{99}.

\bibitem[{\citenamefont{Kennedy}(1994)}]{kennedy_1994}
\bibinfo{author}{\bibnamefont{Kennedy}, \bibfnamefont{T.}},
  \bibinfo{year}{1994}, \bibinfo{journal}{Rev. Math. Phys.}
  \textbf{\bibinfo{volume}{6}}, \bibinfo{pages}{901}.

\bibitem[{\citenamefont{Kennedy}(1998)}]{kennedy_1998}
\bibinfo{author}{\bibnamefont{Kennedy}, \bibfnamefont{T.}},
  \bibinfo{year}{1998}, \bibinfo{journal}{J. Stat. Phys.}
  \textbf{\bibinfo{volume}{91}}, \bibinfo{pages}{829}.

\bibitem[{\citenamefont{Kennedy and Lieb}(1986)}]{kennedy_lieb_1986}
\bibinfo{author}{\bibnamefont{Kennedy}, \bibfnamefont{T.}}, and
  \bibinfo{author}{\bibfnamefont{E.}~\bibnamefont{Lieb}}, \bibinfo{year}{1986},
  \bibinfo{journal}{Physica} \textbf{\bibinfo{volume}{138A}},
  \bibinfo{pages}{320}.

\bibitem[{\citenamefont{Khurana}(1990)}]{khurana_1990}
\bibinfo{author}{\bibnamefont{Khurana}, \bibfnamefont{A.}},
  \bibinfo{year}{1990}, \bibinfo{journal}{Phys. Rev. Lett.}
  \textbf{\bibinfo{volume}{64}}, \bibinfo{pages}{1990}.

\bibitem[{\citenamefont{Krishnamurty}
  \emph{et~al.}(1980)\citenamefont{Krishnamurty, Wilkins, and
  Wilson}}]{Krishnamurti_1980}
\bibinfo{author}{\bibnamefont{Krishnamurty}, \bibfnamefont{H.~B.}},
  \bibinfo{author}{\bibfnamefont{J.~W.} \bibnamefont{Wilkins}}, and
  \bibinfo{author}{\bibfnamefont{K.~G.} \bibnamefont{Wilson}},
  \bibinfo{year}{1980}, \bibinfo{journal}{Phys. Rev. B}
  \textbf{\bibinfo{volume}{21}}, \bibinfo{pages}{1044}.

\bibitem[{\citenamefont{Kubo}(1957)}]{kubo_1957}
\bibinfo{author}{\bibnamefont{Kubo}, \bibfnamefont{R.}}, \bibinfo{year}{1957},
  \bibinfo{journal}{J. Phys. Soc. Japan} \textbf{\bibinfo{volume}{12}},
  \bibinfo{pages}{570}.

\bibitem[{\citenamefont{Laad and {van den Bossche}}(2000)}]{laad_bossche_2000}
\bibinfo{author}{\bibnamefont{Laad}, \bibfnamefont{M.~S.}}, and
  \bibinfo{author}{\bibfnamefont{M.}~\bibnamefont{{van den Bossche}}},
  \bibinfo{year}{2000}, \bibinfo{journal}{J. Phys.: Conden. Matter}
  \textbf{\bibinfo{volume}{12}}, \bibinfo{pages}{2209}.

\bibitem[{\citenamefont{Lawrence} \emph{et~al.}(1999)\citenamefont{Lawrence,
  Osborn, Sarrao, and Fisk}}]{lawrence_1999}
\bibinfo{author}{\bibnamefont{Lawrence}, \bibfnamefont{J.~M.}},
  \bibinfo{author}{\bibfnamefont{R.}~\bibnamefont{Osborn}},
  \bibinfo{author}{\bibfnamefont{J.~L.} \bibnamefont{Sarrao}}, and
  \bibinfo{author}{\bibfnamefont{Z.}~\bibnamefont{Fisk}}, \bibinfo{year}{1999},
  \bibinfo{journal}{Phys. Rev.} \textbf{\bibinfo{volume}{59}},
  \bibinfo{pages}{1134}.

\bibitem[{\citenamefont{Lawrence} \emph{et~al.}(1997)\citenamefont{Lawrence,
  Shapiro, Sarrao, and Fisk}}]{lawrence_1997}
\bibinfo{author}{\bibnamefont{Lawrence}, \bibfnamefont{J.~M.}},
  \bibinfo{author}{\bibfnamefont{S.~M.} \bibnamefont{Shapiro}},
  \bibinfo{author}{\bibfnamefont{J.~L.} \bibnamefont{Sarrao}}, and
  \bibinfo{author}{\bibfnamefont{Z.}~\bibnamefont{Fisk}}, \bibinfo{year}{1997},
  \bibinfo{journal}{Phys. Rev.} \textbf{\bibinfo{volume}{55}},
  \bibinfo{pages}{14467}.

\bibitem[{\citenamefont{Lemberger}(1992)}]{lemberger_1992}
\bibinfo{author}{\bibnamefont{Lemberger}, \bibfnamefont{P.}},
  \bibinfo{year}{1992}, \bibinfo{journal}{J. Phys. A}
  \textbf{\bibinfo{volume}{25}}, \bibinfo{pages}{715}.

\bibitem[{\citenamefont{Letfulov}(1999)}]{letfulov_1999}
\bibinfo{author}{\bibnamefont{Letfulov}, \bibfnamefont{B.~M.}},
  \bibinfo{year}{1999}, \bibinfo{journal}{Europhys. J.}
  \textbf{\bibinfo{volume}{B11}}, \bibinfo{pages}{423}.

\bibitem[{\citenamefont{Letfulov and
  Freericks}(2001)}]{letfulov_freericks_2001}
\bibinfo{author}{\bibnamefont{Letfulov}, \bibfnamefont{B.~M.}}, and
  \bibinfo{author}{\bibfnamefont{J.~K.} \bibnamefont{Freericks}},
  \bibinfo{year}{2001}, \bibinfo{journal}{Phys. Rev. B}
  \textbf{\bibinfo{volume}{64}}, \bibinfo{pages}{174409}.

\bibitem[{\citenamefont{Levin} \emph{et~al.}(1990)\citenamefont{Levin, Kuzhel,
  Bodak, Belan, and Stets}}]{levin_1990}
\bibinfo{author}{\bibnamefont{Levin}, \bibfnamefont{E.}},
  \bibinfo{author}{\bibfnamefont{B.~S.} \bibnamefont{Kuzhel}},
  \bibinfo{author}{\bibfnamefont{O.}~\bibnamefont{Bodak}},
  \bibinfo{author}{\bibfnamefont{B.~D.} \bibnamefont{Belan}}, and
  \bibinfo{author}{\bibfnamefont{I.}~\bibnamefont{Stets}},
  \bibinfo{year}{1990}, \bibinfo{journal}{Phys. Stat. Sol. (b)}
  \textbf{\bibinfo{volume}{161}}, \bibinfo{pages}{783}.

\bibitem[{\citenamefont{Lieb}(1986)}]{lieb_1986}
\bibinfo{author}{\bibnamefont{Lieb}, \bibfnamefont{E.}}, \bibinfo{year}{1986},
  \bibinfo{journal}{Physica} \textbf{\bibinfo{volume}{140A}},
  \bibinfo{pages}{240}.

\bibitem[{\citenamefont{Likharev}(2000)}]{likharev_2000}
\bibinfo{author}{\bibnamefont{Likharev}, \bibfnamefont{K.}},
  \bibinfo{year}{2000}, \emph{\bibinfo{title}{Applications of
  Superconductivity}} (\bibinfo{publisher}{Kluwer},
  \bibinfo{address}{Dordrecht}), chapter~\bibinfo{chapter}{5}.

\bibitem[{\citenamefont{Luttinger and Ward}(1960)}]{luttinger_ward_1960}
\bibinfo{author}{\bibnamefont{Luttinger}, \bibfnamefont{J.~M.}}, and
  \bibinfo{author}{\bibfnamefont{J.~C.} \bibnamefont{Ward}},
  \bibinfo{year}{1960}, \bibinfo{journal}{Phys. Rev.}
  \textbf{\bibinfo{volume}{118}}, \bibinfo{pages}{1417}.

\bibitem[{\citenamefont{Maldague}(1977)}]{maldague_1977}
\bibinfo{author}{\bibnamefont{Maldague}, \bibfnamefont{P.~F.}},
  \bibinfo{year}{1977}, \bibinfo{journal}{Phys. Rev. B}
  \textbf{\bibinfo{volume}{16}}, \bibinfo{pages}{2437}.

\bibitem[{\citenamefont{Metzner}(1991)}]{metzner_1991}
\bibinfo{author}{\bibnamefont{Metzner}, \bibfnamefont{W.}},
  \bibinfo{year}{1991}, \bibinfo{journal}{Phys. Rev. B}
  \textbf{\bibinfo{volume}{43}}, \bibinfo{pages}{8549}.

\bibitem[{\citenamefont{Metzner and Vollhardt}(1989)}]{metzner_vollhardt_1989}
\bibinfo{author}{\bibnamefont{Metzner}, \bibfnamefont{W.}}, and
  \bibinfo{author}{\bibfnamefont{D.}~\bibnamefont{Vollhardt}},
  \bibinfo{year}{1989}, \bibinfo{journal}{Phys. Rev. Lett.}
  \textbf{\bibinfo{volume}{62}}, \bibinfo{pages}{324}.

\bibitem[{\citenamefont{Michielsen}(1993)}]{michielsen_1993}
\bibinfo{author}{\bibnamefont{Michielsen}, \bibfnamefont{K.}},
  \bibinfo{year}{1993}, \bibinfo{journal}{Int. J. Mod. Phys. B}
  \textbf{\bibinfo{volume}{7}}, \bibinfo{pages}{2571}.

\bibitem[{\citenamefont{Miller and Freericks}(2001)}]{miller_freericks_2001}
\bibinfo{author}{\bibnamefont{Miller}, \bibfnamefont{P.}}, and
  \bibinfo{author}{\bibfnamefont{J.~K.} \bibnamefont{Freericks}},
  \bibinfo{year}{2001}, \bibinfo{journal}{J. Phys.: Conden. Matter}
  \textbf{\bibinfo{volume}{13}}, \bibinfo{pages}{3187}.

\bibitem[{\citenamefont{Millis} \emph{et~al.}(1999)\citenamefont{Millis, Hu,
  and {Das Sarma}}}]{millis_hu_dassarma_1999}
\bibinfo{author}{\bibnamefont{Millis}, \bibfnamefont{A.~J.}},
  \bibinfo{author}{\bibfnamefont{J.}~\bibnamefont{Hu}}, and
  \bibinfo{author}{\bibfnamefont{S.}~\bibnamefont{{Das Sarma}}},
  \bibinfo{year}{1999}, \bibinfo{journal}{Phys. Rev. Lett.}
  \textbf{\bibinfo{volume}{82}}, \bibinfo{pages}{2354}.

\bibitem[{\citenamefont{Millis} \emph{et~al.}(1995)\citenamefont{Millis,
  Littlewood, and Shraiman}}]{millis_littlewood_shraiman_1995}
\bibinfo{author}{\bibnamefont{Millis}, \bibfnamefont{A.~J.}},
  \bibinfo{author}{\bibfnamefont{P.~B.} \bibnamefont{Littlewood}}, and
  \bibinfo{author}{\bibfnamefont{B.~I.} \bibnamefont{Shraiman}},
  \bibinfo{year}{1995}, \bibinfo{journal}{Phys. Rev. Lett.}
  \textbf{\bibinfo{volume}{74}}, \bibinfo{pages}{5144}.

\bibitem[{\citenamefont{Millis} \emph{et~al.}(1996)\citenamefont{Millis,
  Mueller, and Shraiman}}]{millis_mueller_shraiman_1996}
\bibinfo{author}{\bibnamefont{Millis}, \bibfnamefont{A.~J.}},
  \bibinfo{author}{\bibfnamefont{R.}~\bibnamefont{Mueller}}, and
  \bibinfo{author}{\bibfnamefont{B.~I.} \bibnamefont{Shraiman}},
  \bibinfo{year}{1996}, \bibinfo{journal}{Phys. Rev. B}
  \textbf{\bibinfo{volume}{54}}, \bibinfo{pages}{5389}.

\bibitem[{\citenamefont{{Minh-Tien}}(1998)}]{tran_1998}
\bibinfo{author}{\bibnamefont{{Minh-Tien}}, \bibfnamefont{T.}},
  \bibinfo{year}{1998}, \bibinfo{journal}{Phys. Rev. B}
  \textbf{\bibinfo{volume}{58}}, \bibinfo{pages}{R15965}.

\bibitem[{\citenamefont{Mitsuda} \emph{et~al.}(2002)\citenamefont{Mitsuda,
  Goto, Yoshimura, Zhang, Sato, Kosige, and Wada}}]{mitsuda_2002}
\bibinfo{author}{\bibnamefont{Mitsuda}, \bibfnamefont{A.}},
  \bibinfo{author}{\bibfnamefont{T.}~\bibnamefont{Goto}},
  \bibinfo{author}{\bibfnamefont{K.}~\bibnamefont{Yoshimura}},
  \bibinfo{author}{\bibfnamefont{W.}~\bibnamefont{Zhang}},
  \bibinfo{author}{\bibfnamefont{N.}~\bibnamefont{Sato}},
  \bibinfo{author}{\bibfnamefont{K.}~\bibnamefont{Kosige}}, and
  \bibinfo{author}{\bibfnamefont{H.}~\bibnamefont{Wada}}, \bibinfo{year}{2002},
  \bibinfo{journal}{Phys. Rev. Lett.} \textbf{\bibinfo{volume}{88}},
  \bibinfo{pages}{137204}.

\bibitem[{\citenamefont{Moeller} \emph{et~al.}(1992)\citenamefont{Moeller,
  Ruckenstein, and Schmitt-Rink}}]{moeller_ruckenstein_schmittrink_1992}
\bibinfo{author}{\bibnamefont{Moeller}, \bibfnamefont{G.}},
  \bibinfo{author}{\bibfnamefont{A.}~\bibnamefont{Ruckenstein}}, and
  \bibinfo{author}{\bibfnamefont{S.}~\bibnamefont{Schmitt-Rink}},
  \bibinfo{year}{1992}, \bibinfo{journal}{Phys. Rev. B}
  \textbf{\bibinfo{volume}{46}}, \bibinfo{pages}{7427}.

\bibitem[{\citenamefont{{M\"uller-Hartmann}}(1989{\natexlab{a}})}]{mueller-har%
tmann_1989a}
\bibinfo{author}{\bibnamefont{{M\"uller-Hartmann}}, \bibfnamefont{E.}},
  \bibinfo{year}{1989}{\natexlab{a}}, \bibinfo{journal}{Z. Phys. B}
  \textbf{\bibinfo{volume}{74}}, \bibinfo{pages}{507}.

\bibitem[{\citenamefont{{M\"uller-Hartmann}}(1989{\natexlab{b}})}]{mueller-har%
tmann_1989c}
\bibinfo{author}{\bibnamefont{{M\"uller-Hartmann}}, \bibfnamefont{E.}},
  \bibinfo{year}{1989}{\natexlab{b}}, \bibinfo{journal}{Int J. Mod. Phys. B}
  \textbf{\bibinfo{volume}{3}}, \bibinfo{pages}{2169}.

\bibitem[{\citenamefont{Murani} \emph{et~al.}(2002)\citenamefont{Murani,
  Richard, and Bewley}}]{murani_2002}
\bibinfo{author}{\bibnamefont{Murani}, \bibfnamefont{A.}},
  \bibinfo{author}{\bibfnamefont{D.}~\bibnamefont{Richard}}, and
  \bibinfo{author}{\bibfnamefont{R.}~\bibnamefont{Bewley}},
  \bibinfo{year}{2002}, \bibinfo{journal}{Physica B}
  \textbf{\bibinfo{volume}{312 \& 313}}, \bibinfo{pages}{346}.

\bibitem[{\citenamefont{Nyhus}
  \emph{et~al.}(1995{\natexlab{a}})\citenamefont{Nyhus, Cooper, and
  Fisk}}]{nyhus_cooper_fisk_1995b}
\bibinfo{author}{\bibnamefont{Nyhus}, \bibfnamefont{P.}},
  \bibinfo{author}{\bibfnamefont{S.}~\bibnamefont{Cooper}}, and
  \bibinfo{author}{\bibfnamefont{Z.}~\bibnamefont{Fisk}},
  \bibinfo{year}{1995}{\natexlab{a}}, \bibinfo{journal}{Phys. Rev. B}
  \textbf{\bibinfo{volume}{51}}, \bibinfo{pages}{R15626}.

\bibitem[{\citenamefont{Nyhus}
  \emph{et~al.}(1995{\natexlab{b}})\citenamefont{Nyhus, Cooper, Fisk, and
  Sarrao}}]{nyhus_cooper_fisk_1995a}
\bibinfo{author}{\bibnamefont{Nyhus}, \bibfnamefont{P.}},
  \bibinfo{author}{\bibfnamefont{S.}~\bibnamefont{Cooper}},
  \bibinfo{author}{\bibfnamefont{Z.}~\bibnamefont{Fisk}}, and
  \bibinfo{author}{\bibfnamefont{J.}~\bibnamefont{Sarrao}},
  \bibinfo{year}{1995}{\natexlab{b}}, \bibinfo{journal}{Phys. Rev. B}
  \textbf{\bibinfo{volume}{52}}, \bibinfo{pages}{R14308}.

\bibitem[{\citenamefont{Nyhus} \emph{et~al.}(1997)\citenamefont{Nyhus, Cooper,
  Fisk, and Sarrao}}]{nyhus_cooper_fisk_1997}
\bibinfo{author}{\bibnamefont{Nyhus}, \bibfnamefont{P.}},
  \bibinfo{author}{\bibfnamefont{S.}~\bibnamefont{Cooper}},
  \bibinfo{author}{\bibfnamefont{Z.}~\bibnamefont{Fisk}}, and
  \bibinfo{author}{\bibfnamefont{J.}~\bibnamefont{Sarrao}},
  \bibinfo{year}{1997}, \bibinfo{journal}{Phys. Rev. B}
  \textbf{\bibinfo{volume}{55}}, \bibinfo{pages}{12488}.

\bibitem[{\citenamefont{{O\v cko} and Sarrao}(2002)}]{ocko_2002}
\bibinfo{author}{\bibnamefont{{O\v cko}}, \bibfnamefont{M.}}, and
  \bibinfo{author}{\bibfnamefont{J.~L.} \bibnamefont{Sarrao}},
  \bibinfo{year}{2002}, \bibinfo{journal}{Physica B}
  \textbf{\bibinfo{volume}{312 \& 313}}, \bibinfo{pages}{341}.

\bibitem[{\citenamefont{Pasternak} \emph{et~al.}(1990)\citenamefont{Pasternak,
  Taylor, Chen, Meade, Falicov, Giesekus, Jeanloz, and Yu}}]{pasternak_1990}
\bibinfo{author}{\bibnamefont{Pasternak}, \bibfnamefont{M.~P.}},
  \bibinfo{author}{\bibfnamefont{R.~D.} \bibnamefont{Taylor}},
  \bibinfo{author}{\bibfnamefont{A.}~\bibnamefont{Chen}},
  \bibinfo{author}{\bibfnamefont{C.}~\bibnamefont{Meade}},
  \bibinfo{author}{\bibfnamefont{L.~M.} \bibnamefont{Falicov}},
  \bibinfo{author}{\bibfnamefont{A.}~\bibnamefont{Giesekus}},
  \bibinfo{author}{\bibfnamefont{R.}~\bibnamefont{Jeanloz}}, and
  \bibinfo{author}{\bibfnamefont{P.~Y.} \bibnamefont{Yu}},
  \bibinfo{year}{1990}, \bibinfo{journal}{Phys. Rev. Lett.}
  \textbf{\bibinfo{volume}{65}}, \bibinfo{pages}{790}.

\bibitem[{\citenamefont{Plischke}(1972)}]{plischke_1972}
\bibinfo{author}{\bibnamefont{Plischke}, \bibfnamefont{M.}},
  \bibinfo{year}{1972}, \bibinfo{journal}{Phys. Rev. Lett.}
  \textbf{\bibinfo{volume}{28}}, \bibinfo{pages}{361}.

\bibitem[{\citenamefont{Portengen}
  \emph{et~al.}(1996{\natexlab{a}})\citenamefont{Portengen, Oestreich, and
  Sham}}]{portengen_oestreich_1996a}
\bibinfo{author}{\bibnamefont{Portengen}, \bibfnamefont{T.}},
  \bibinfo{author}{\bibfnamefont{T.}~\bibnamefont{Oestreich}}, and
  \bibinfo{author}{\bibfnamefont{L.~J.} \bibnamefont{Sham}},
  \bibinfo{year}{1996}{\natexlab{a}}, \bibinfo{journal}{Phys. Rev. Lett.}
  \textbf{\bibinfo{volume}{76}}, \bibinfo{pages}{2284}.

\bibitem[{\citenamefont{Portengen}
  \emph{et~al.}(1996{\natexlab{b}})\citenamefont{Portengen, Oestreich, and
  Sham}}]{portengen_oestreich_1996b}
\bibinfo{author}{\bibnamefont{Portengen}, \bibfnamefont{T.}},
  \bibinfo{author}{\bibfnamefont{T.}~\bibnamefont{Oestreich}}, and
  \bibinfo{author}{\bibfnamefont{L.~J.} \bibnamefont{Sham}},
  \bibinfo{year}{1996}{\natexlab{b}}, \bibinfo{journal}{Phys. Rev. B}
  \textbf{\bibinfo{volume}{54}}, \bibinfo{pages}{17452}.

\bibitem[{\citenamefont{Potthoff and Nolting}(1999)}]{potthoff_nolting_1999}
\bibinfo{author}{\bibnamefont{Potthoff}, \bibfnamefont{M.}}, and
  \bibinfo{author}{\bibfnamefont{W.}~\bibnamefont{Nolting}},
  \bibinfo{year}{1999}, \bibinfo{journal}{Phys. Rev. B}
  \textbf{\bibinfo{volume}{59}}, \bibinfo{pages}{2549}.

\bibitem[{\citenamefont{Pruschke}
  \emph{et~al.}(1993{\natexlab{a}})\citenamefont{Pruschke, Cox, and
  Jarrell}}]{pruschke_cox_jarrell_1993a}
\bibinfo{author}{\bibnamefont{Pruschke}, \bibfnamefont{T.}},
  \bibinfo{author}{\bibfnamefont{D.~L.} \bibnamefont{Cox}}, and
  \bibinfo{author}{\bibfnamefont{M.}~\bibnamefont{Jarrell}},
  \bibinfo{year}{1993}{\natexlab{a}}, \bibinfo{journal}{Europhys. Lett.}
  \textbf{\bibinfo{volume}{21}}, \bibinfo{pages}{593}.

\bibitem[{\citenamefont{Pruschke}
  \emph{et~al.}(1993{\natexlab{b}})\citenamefont{Pruschke, Cox, and
  Jarrell}}]{pruschke_cox_jarrell_1993b}
\bibinfo{author}{\bibnamefont{Pruschke}, \bibfnamefont{T.}},
  \bibinfo{author}{\bibfnamefont{D.~L.} \bibnamefont{Cox}}, and
  \bibinfo{author}{\bibfnamefont{M.}~\bibnamefont{Jarrell}},
  \bibinfo{year}{1993}{\natexlab{b}}, \bibinfo{journal}{Phys. Rev. B}
  \textbf{\bibinfo{volume}{47}}, \bibinfo{pages}{3553}.

\bibitem[{\citenamefont{Pruschke} \emph{et~al.}(1995)\citenamefont{Pruschke,
  Jarrell, and Freericks}}]{pruschke_jarrell_freericks_1995}
\bibinfo{author}{\bibnamefont{Pruschke}, \bibfnamefont{T.}},
  \bibinfo{author}{\bibfnamefont{M.}~\bibnamefont{Jarrell}}, and
  \bibinfo{author}{\bibfnamefont{J.~K.} \bibnamefont{Freericks}},
  \bibinfo{year}{1995}, \bibinfo{journal}{Adv. Phys.}
  \textbf{\bibinfo{volume}{44}}, \bibinfo{pages}{187}.

\bibitem[{\citenamefont{Ramakrishnan}(2003)}]{ramakrishnan_2002}
\bibinfo{author}{\bibnamefont{Ramakrishnan}, \bibfnamefont{T.~V.}},
  \bibinfo{year}{2003}, \emph{\bibinfo{title}{Concepts in Electron
  Correlation}} (\bibinfo{publisher}{Kluwer}, \bibinfo{address}{Dordrecht}),
  volume~\bibinfo{volume}{XX} of \emph{\bibinfo{series}{Nato Science Series II.
  Mathematics, Physics, and Chemistry}}, p. \bibinfo{pages}{XXX}.

\bibitem[{\citenamefont{Ramirez and Falicov}(1971)}]{ramirez_falicov_1971}
\bibinfo{author}{\bibnamefont{Ramirez}, \bibfnamefont{R.}}, and
  \bibinfo{author}{\bibfnamefont{L.~M.} \bibnamefont{Falicov}},
  \bibinfo{year}{1971}, \bibinfo{journal}{Phys. Rev. B}
  \textbf{\bibinfo{volume}{3}}, \bibinfo{pages}{2425}.

\bibitem[{\citenamefont{Ramirez} \emph{et~al.}(1970)\citenamefont{Ramirez,
  Falicov, and Kimball}}]{ramirez_falicov_kimball_1970}
\bibinfo{author}{\bibnamefont{Ramirez}, \bibfnamefont{R.}},
  \bibinfo{author}{\bibfnamefont{L.~M.} \bibnamefont{Falicov}}, and
  \bibinfo{author}{\bibfnamefont{J.~C.} \bibnamefont{Kimball}},
  \bibinfo{year}{1970}, \bibinfo{journal}{Phys. Rev. B}
  \textbf{\bibinfo{volume}{2}}, \bibinfo{pages}{3383}.

\bibitem[{\citenamefont{Rettori} \emph{et~al.}(1997)\citenamefont{Rettori,
  Oseroff, Rao, Pagliuso, Barberis, Sarrao, Fisk, and Hundley}}]{rettori_1997}
\bibinfo{author}{\bibnamefont{Rettori}, \bibfnamefont{C.}},
  \bibinfo{author}{\bibfnamefont{S.~B.} \bibnamefont{Oseroff}},
  \bibinfo{author}{\bibfnamefont{D.}~\bibnamefont{Rao}},
  \bibinfo{author}{\bibfnamefont{P.~G.} \bibnamefont{Pagliuso}},
  \bibinfo{author}{\bibfnamefont{G.~E.} \bibnamefont{Barberis}},
  \bibinfo{author}{\bibfnamefont{J.}~\bibnamefont{Sarrao}},
  \bibinfo{author}{\bibfnamefont{Z.}~\bibnamefont{Fisk}}, and
  \bibinfo{author}{\bibfnamefont{M.~F.} \bibnamefont{Hundley}},
  \bibinfo{year}{1997}, \bibinfo{journal}{Phys. Rev. B}
  \textbf{\bibinfo{volume}{56}}, \bibinfo{pages}{7993}.

\bibitem[{\citenamefont{Sakurai} \emph{et~al.}(2000)\citenamefont{Sakurai,
  Fukuda, Mitsuda, Wada, and Shiga}}]{sakurai_2000}
\bibinfo{author}{\bibnamefont{Sakurai}, \bibfnamefont{J.}},
  \bibinfo{author}{\bibfnamefont{S.}~\bibnamefont{Fukuda}},
  \bibinfo{author}{\bibfnamefont{A.}~\bibnamefont{Mitsuda}},
  \bibinfo{author}{\bibfnamefont{H.}~\bibnamefont{Wada}}, and
  \bibinfo{author}{\bibfnamefont{M.}~\bibnamefont{Shiga}},
  \bibinfo{year}{2000}, \bibinfo{journal}{Physica B}
  \textbf{\bibinfo{volume}{281 \& 282}}, \bibinfo{pages}{134}.

\bibitem[{\citenamefont{Sarrao}(1999)}]{sarrao_1999}
\bibinfo{author}{\bibnamefont{Sarrao}, \bibfnamefont{J.~L.}},
  \bibinfo{year}{1999}, \bibinfo{journal}{Physica B}
  \textbf{\bibinfo{volume}{259 \& 261}}, \bibinfo{pages}{129}.

\bibitem[{\citenamefont{Schiller}(1999)}]{schiller_1999}
\bibinfo{author}{\bibnamefont{Schiller}, \bibfnamefont{A.}},
  \bibinfo{year}{1999}, \bibinfo{journal}{Phys. Rev. B}
  \textbf{\bibinfo{volume}{60}}, \bibinfo{pages}{15660}.

\bibitem[{\citenamefont{Schiller and
  Ingersent}(1995)}]{schiller_ingersent_1995}
\bibinfo{author}{\bibnamefont{Schiller}, \bibfnamefont{A.}}, and
  \bibinfo{author}{\bibfnamefont{K.}~\bibnamefont{Ingersent}},
  \bibinfo{year}{1995}, \bibinfo{journal}{Phys. Rev. Lett.}
  \textbf{\bibinfo{volume}{75}}, \bibinfo{pages}{113}.

\bibitem[{\citenamefont{Schmidt and Monien}(2002)}]{schmidt_monien_2002}
\bibinfo{author}{\bibnamefont{Schmidt}, \bibfnamefont{P.}}, and
  \bibinfo{author}{\bibfnamefont{H.}~\bibnamefont{Monien}},
  \bibinfo{year}{2002}, \eprint{cond-mat/0202046}.

\bibitem[{\citenamefont{Schwartz and Siggia}(1972)}]{schwartz_siggia_1972}
\bibinfo{author}{\bibnamefont{Schwartz}, \bibfnamefont{L.}}, and
  \bibinfo{author}{\bibfnamefont{E.}~\bibnamefont{Siggia}},
  \bibinfo{year}{1972}, \bibinfo{journal}{Phys. Rev. B}
  \textbf{\bibinfo{volume}{5}}, \bibinfo{pages}{383}.

\bibitem[{\citenamefont{Shastry and Shraiman}(1990)}]{shastry_shraiman_1990}
\bibinfo{author}{\bibnamefont{Shastry}, \bibfnamefont{B.~S.}}, and
  \bibinfo{author}{\bibfnamefont{B.~I.} \bibnamefont{Shraiman}},
  \bibinfo{year}{1990}, \bibinfo{journal}{Phys. Rev. Lett.}
  \textbf{\bibinfo{volume}{65}}, \bibinfo{pages}{1068}.

\bibitem[{\citenamefont{Shastry and Shraiman}(1991)}]{shastry_shraiman_1991}
\bibinfo{author}{\bibnamefont{Shastry}, \bibfnamefont{B.~S.}}, and
  \bibinfo{author}{\bibfnamefont{B.~I.} \bibnamefont{Shraiman}},
  \bibinfo{year}{1991}, \bibinfo{journal}{Int. J. Mod. Phys. B}
  \textbf{\bibinfo{volume}{5}}, \bibinfo{pages}{365}.

\bibitem[{\citenamefont{Shvaika}(2000)}]{shvaika_2000}
\bibinfo{author}{\bibnamefont{Shvaika}, \bibfnamefont{A.~M.}},
  \bibinfo{year}{2000}, \bibinfo{journal}{Physica C}
  \textbf{\bibinfo{volume}{341-348}}, \bibinfo{pages}{177}.

\bibitem[{\citenamefont{Shvaika}(2001)}]{shvaika_2002}
\bibinfo{author}{\bibnamefont{Shvaika}, \bibfnamefont{A.~M.}},
  \bibinfo{year}{2001}, \bibinfo{journal}{J. Phys. Stud.}
  \textbf{\bibinfo{volume}{5}}, \bibinfo{pages}{349}.

\bibitem[{\citenamefont{Shvaika}(2002)}]{shvaika_2002b}
\bibinfo{author}{\bibnamefont{Shvaika}, \bibfnamefont{A.~M.}},
  \bibinfo{year}{2002}, \eprint{cond-mat/0205322}.

\bibitem[{\citenamefont{Shvaika and Freericks}(2002)}]{shvaika_freericks_2002}
\bibinfo{author}{\bibnamefont{Shvaika}, \bibfnamefont{A.~M.}}, and
  \bibinfo{author}{\bibfnamefont{J.~K.} \bibnamefont{Freericks}},
  \bibinfo{year}{2002}, \eprint{cond-mat/0211451}.

\bibitem[{\citenamefont{Si} \emph{et~al.}(1992)\citenamefont{Si, Kotliar, and
  Georges}}]{si_kotliar_1992}
\bibinfo{author}{\bibnamefont{Si}, \bibfnamefont{Q.}},
  \bibinfo{author}{\bibfnamefont{G.}~\bibnamefont{Kotliar}}, and
  \bibinfo{author}{\bibfnamefont{A.}~\bibnamefont{Georges}},
  \bibinfo{year}{1992}, \bibinfo{journal}{Phys. Rev. B}
  \textbf{\bibinfo{volume}{46}}, \bibinfo{pages}{1261}.

\bibitem[{\citenamefont{Soven}(1967)}]{soven_1967}
\bibinfo{author}{\bibnamefont{Soven}, \bibfnamefont{P.}}, \bibinfo{year}{1967},
  \bibinfo{journal}{Phys. Rev.} \textbf{\bibinfo{volume}{156}},
  \bibinfo{pages}{809}.

\bibitem[{\citenamefont{Subrahmanyam and
  Barma}(1988)}]{subrahmanyam_barma_1988}
\bibinfo{author}{\bibnamefont{Subrahmanyam}, \bibfnamefont{V.}}, and
  \bibinfo{author}{\bibfnamefont{M.}~\bibnamefont{Barma}},
  \bibinfo{year}{1988}, \bibinfo{journal}{J. Phys. C}
  \textbf{\bibinfo{volume}{21}}, \bibinfo{pages}{L19}.

\bibitem[{\citenamefont{Tran}(1999)}]{tran_1999}
\bibinfo{author}{\bibnamefont{Tran}, \bibfnamefont{M.-T.}},
  \bibinfo{year}{1999}, \bibinfo{journal}{Phys. Rev. B}
  \textbf{\bibinfo{volume}{60}}, \bibinfo{pages}{16371}.

\bibitem[{\citenamefont{Velick\'y}(1969)}]{velicky_1969}
\bibinfo{author}{\bibnamefont{Velick\'y}, \bibfnamefont{B.}},
  \bibinfo{year}{1969}, \bibinfo{journal}{Phys. Rev.}
  \textbf{\bibinfo{volume}{184}}, \bibinfo{pages}{614}.

\bibitem[{\citenamefont{Velick\'y} \emph{et~al.}(1968)\citenamefont{Velick\'y,
  Kirkpatrick, and Ehrenreich}}]{velicky_kirkpatrick_ehrenreich_1968}
\bibinfo{author}{\bibnamefont{Velick\'y}, \bibfnamefont{B.}},
  \bibinfo{author}{\bibfnamefont{S.}~\bibnamefont{Kirkpatrick}}, and
  \bibinfo{author}{\bibfnamefont{H.}~\bibnamefont{Ehrenreich}},
  \bibinfo{year}{1968}, \bibinfo{journal}{Phys. Rev.}
  \textbf{\bibinfo{volume}{175}}, \bibinfo{pages}{747}.

\bibitem[{\citenamefont{Vojta and Bulla}(2002)}]{vojta_2002}
\bibinfo{author}{\bibnamefont{Vojta}, \bibfnamefont{M.}}, and
  \bibinfo{author}{\bibfnamefont{R.}~\bibnamefont{Bulla}},
  \bibinfo{year}{2002}, \bibinfo{journal}{Phys. Rev. B}
  \textbf{\bibinfo{volume}{65}}, \bibinfo{pages}{014511}.

\bibitem[{\citenamefont{Wada} \emph{et~al.}(1997)\citenamefont{Wada, Nakamura,
  Mitsuda, Shiga, Tanaka, Mitamura, and Goto}}]{wada_1997}
\bibinfo{author}{\bibnamefont{Wada}, \bibfnamefont{H.}},
  \bibinfo{author}{\bibfnamefont{A.}~\bibnamefont{Nakamura}},
  \bibinfo{author}{\bibfnamefont{A.}~\bibnamefont{Mitsuda}},
  \bibinfo{author}{\bibfnamefont{M.}~\bibnamefont{Shiga}},
  \bibinfo{author}{\bibfnamefont{T.}~\bibnamefont{Tanaka}},
  \bibinfo{author}{\bibfnamefont{H.}~\bibnamefont{Mitamura}}, and
  \bibinfo{author}{\bibfnamefont{T.}~\bibnamefont{Goto}}, \bibinfo{year}{1997},
  \bibinfo{journal}{J. Phys.: Condens. Matter} \textbf{\bibinfo{volume}{9}},
  \bibinfo{pages}{7913}.

\bibitem[{\citenamefont{Watson and Lema\'nski}(1995)}]{watson_lemanski_1995}
\bibinfo{author}{\bibnamefont{Watson}, \bibfnamefont{G.}}, and
  \bibinfo{author}{\bibfnamefont{R.}~\bibnamefont{Lema\'nski}},
  \bibinfo{year}{1995}, \bibinfo{journal}{J. Phys. Conden. Matter}
  \textbf{\bibinfo{volume}{7}}, \bibinfo{pages}{9521}.

\bibitem[{\citenamefont{Wilcox}(1968)}]{wilcox_1968}
\bibinfo{author}{\bibnamefont{Wilcox}, \bibfnamefont{R.~M.}},
  \bibinfo{year}{1968}, \bibinfo{journal}{Phys. Rev.}
  \textbf{\bibinfo{volume}{174}}, \bibinfo{pages}{624}.

\bibitem[{\citenamefont{Withoff and Fradkin}(1990)}]{withoff_1990}
\bibinfo{author}{\bibnamefont{Withoff}, \bibfnamefont{D.}}, and
  \bibinfo{author}{\bibfnamefont{E.}~\bibnamefont{Fradkin}},
  \bibinfo{year}{1990}, \bibinfo{journal}{Phys. Rev. Lett.}
  \textbf{\bibinfo{volume}{64}}, \bibinfo{pages}{1835}.

\bibitem[{\citenamefont{Zhang} \emph{et~al.}(2002)\citenamefont{Zhang, Sato,
  Yoshimura, Mitsuda, Goto, and Kosuge}}]{zhang_2002}
\bibinfo{author}{\bibnamefont{Zhang}, \bibfnamefont{W.}},
  \bibinfo{author}{\bibfnamefont{N.}~\bibnamefont{Sato}},
  \bibinfo{author}{\bibfnamefont{K.}~\bibnamefont{Yoshimura}},
  \bibinfo{author}{\bibfnamefont{A.}~\bibnamefont{Mitsuda}},
  \bibinfo{author}{\bibfnamefont{T.}~\bibnamefont{Goto}}, and
  \bibinfo{author}{\bibfnamefont{K.}~\bibnamefont{Kosuge}},
  \bibinfo{year}{2002}, \bibinfo{journal}{Phys. Rev. B}
  \textbf{\bibinfo{volume}{66}}, \bibinfo{pages}{024112}.

\bibitem[{\citenamefont{Zlati\'c and
  Freericks}(2001{\natexlab{a}})}]{zlatic_freericks_2001b}
\bibinfo{author}{\bibnamefont{Zlati\'c}, \bibfnamefont{V.}}, and
  \bibinfo{author}{\bibfnamefont{J.~K.} \bibnamefont{Freericks}},
  \bibinfo{year}{2001}{\natexlab{a}}, \bibinfo{journal}{Acta Phys. Pol. B}
  \textbf{\bibinfo{volume}{32}}, \bibinfo{pages}{3253}.

\bibitem[{\citenamefont{Zlati\'c and
  Freericks}(2001{\natexlab{b}})}]{zlatic_freericks_2001a}
\bibinfo{author}{\bibnamefont{Zlati\'c}, \bibfnamefont{V.}}, and
  \bibinfo{author}{\bibfnamefont{J.~K.} \bibnamefont{Freericks}},
  \bibinfo{year}{2001}{\natexlab{b}}, \emph{\bibinfo{title}{Open Problems in
  Strongly Correlated Electron Systems}} (\bibinfo{publisher}{Kluwer},
  \bibinfo{address}{Dordrecht}), volume~\bibinfo{volume}{15} of
  \emph{\bibinfo{series}{Nato Science Series II. Mathematics, Physics, and
  Chemistry}}, p. \bibinfo{pages}{371}.

\bibitem[{\citenamefont{Zlati\'c and Freericks}(2003)}]{zlatic_freericks_2003}
\bibinfo{author}{\bibnamefont{Zlati\'c}, \bibfnamefont{V.}}, and
  \bibinfo{author}{\bibfnamefont{J.~K.} \bibnamefont{Freericks}},
  \bibinfo{year}{2003}, \emph{\bibinfo{title}{Concepts in Electron
  Correlation}} (\bibinfo{publisher}{Kluwer}, \bibinfo{address}{Dordrecht}),
  volume~\bibinfo{volume}{XX} of \emph{\bibinfo{series}{Nato Science Series II.
  Mathematics, Physics, and Chemistry}}, p. \bibinfo{pages}{XXX}.

\bibitem[{\citenamefont{Zlati\'c} \emph{et~al.}(2001)\citenamefont{Zlati\'c,
  Freericks, Lema\'nski, and Czycholl}}]{zlatic_review_2001}
\bibinfo{author}{\bibnamefont{Zlati\'c}, \bibfnamefont{V.}},
  \bibinfo{author}{\bibfnamefont{J.~K.} \bibnamefont{Freericks}},
  \bibinfo{author}{\bibfnamefont{R.}~\bibnamefont{Lema\'nski}}, and
  \bibinfo{author}{\bibfnamefont{G.}~\bibnamefont{Czycholl}},
  \bibinfo{year}{2001}, \bibinfo{journal}{Phil. Mag. B}
  \textbf{\bibinfo{volume}{81}}, \bibinfo{pages}{1443}.

\bibitem[{\citenamefont{Zlati\'c and Horvati\'c}(1990)}]{zlatic_horvatic_1990}
\bibinfo{author}{\bibnamefont{Zlati\'c}, \bibfnamefont{V.}}, and
  \bibinfo{author}{\bibfnamefont{B.}~\bibnamefont{Horvati\'c}},
  \bibinfo{year}{1990}, \bibinfo{journal}{Solid State Commun.}
  \textbf{\bibinfo{volume}{75}}, \bibinfo{pages}{263}.

\end{thebibliography}

\end{document}